\documentclass[final,reqno]{amsart}

\usepackage{amsaddr}
\usepackage{amssymb,amsfonts,amsmath,amsthm}
\usepackage{bm}
\usepackage{enumerate}
\usepackage{physics}
\usepackage{bbold}
\usepackage{tikz}
\usetikzlibrary{cd}
\usepackage{accents}
\usepackage{mathrsfs}
\usepackage{xcolor}

\usepackage{hyperref}
\hypersetup{
  colorlinks=true,
  allcolors=blue
}
\usepackage[capitalise]{cleveref}

\DeclareMathOperator{\ran}{ran}
\let\Span\relax 
\DeclareMathOperator{\Span}{Span}
\let\div\relax
\DeclareMathOperator{\div}{div}
\DeclareMathOperator{\supp}{supp}
\DeclareRobustCommand\bar[1]{\accentset{\rule{.4em}{.3pt}}{#1}}

\newtheorem{theorem}{Theorem}[section]
\newtheorem{lemma}[theorem]{Lemma}
\crefname{lemma}{Lemma}{Lemmas}
\Crefname{lemma}{Lemma}{Lemmas}
\newtheorem{corollary}[theorem]{Corollary}
\crefname{corollary}{Corollary}{Corollaries}
\Crefname{corollary}{Corollary}{Corollaries}
\newtheorem{proposition}[theorem]{Proposition}
\crefname{proposition}{Proposition}{Propositions}
\Crefname{proposition}{Proposition}{Propositions}
\crefname{equation}{}{}

\theoremstyle{remark}
\newtheorem{remark}{Remark}
\crefname{remark}{Remark}{Remarks}
\Crefname{remark}{Remark}{Remarks}

\newcommand{\RR}{\mathbb{R}}
\newcommand{\CC}{\mathbb{C}}
\newcommand{\ZZ}{\mathbb{Z}}
\newcommand{\NN}{\mathbb{N}}

\newcommand{\wconv}{\rightharpoonup}
\newcommand{\iden}{\mathbb{1}}
\renewcommand{\i}{\mathrm{i}}

\newcommand{\dua}[2]{\langle #1, #2 \rangle}

\newcommand{\ol}[1]{\overline{#1}}
\newcommand{\wt}[1]{\widetilde{#1}}

\newcommand{\HC}{\mathcal{H}}
\newcommand{\NC}{\mathcal{N}}
\newcommand{\SC}{\mathcal{S}}

\newcommand{\SymG}{S}

\newcommand{\AF}{\bm{\mathfrak{A}}}
\newcommand{\BF}{\bm{\mathfrak{B}}}
\newcommand{\EF}{\bm{\mathfrak{E}}}
\newcommand{\Af}{\mathfrak{A}}
\newcommand{\Ef}{\mathfrak{E}}
\newcommand{\vjp}{\mathbf{j}^{\mathrm{p}}}
\newcommand{\vjph}{\mathbf{j}^{\mathrm{ph}}}
\newcommand{\vjc}{\mathbf{j}^{\mathrm{c}}}
\newcommand{\vjcmodel}[1]{\mathbf{j}^{\mathrm{c},#1}}
\newcommand{\vjcdip}{\mathbf{j}^{\mathrm{c},\dip}}

\newcommand{\LL}{\mathrm{PFLL}}
\newcommand{\LI}{\mathrm{PFL}}
\newcommand{\VR}{\mathrm{VR}}
\newcommand{\PF}{\mathrm{PF}}

\newcommand{\ph}{\mathrm{ph}}
\newcommand{\el}{\mathrm{el}}
\newcommand{\rad}{\mathrm{rad}}
\newcommand{\dip}{\mathrm{d}}
\newcommand{\fin}{\mathrm{fin}}
\newcommand{\sol}{\mathrm{sol}}

\newcommand{\rdens}{\mathcal{I}}
\newcommand{\rcdens}{\mathcal{J}}

\newcommand{\Fock}{\mathcal{F}}
\newcommand{\Hil}{\mathfrak{H}}

\newcommand{\dm}{\mathcal{D}}

\let\phi=\varphi 

\newcommand{\vsigma}{{\bm{\sigma}}}
\newcommand{\vzeta}{{\bm{\zeta}}}
\newcommand{\vxi}{{\bm{\xi}}}
\newcommand{\vgamma}{{\bm{\gamma}}}
\newcommand{\vepsilon}{{\bm{\epsilon}}}
\newcommand{\vomega}{{\bm{\omega}}}
\newcommand{\vnu}{{\bm{\nu}}}
\newcommand{\vtheta}{{\bm{\theta}}}

\newcommand{\vGamma}{{\bm{\Gamma}}}

\renewcommand{\va}{\mathbf{a}}
\newcommand{\ve}{\mathbf{e}}

\renewcommand{\vu}{\mathbf{u}}
\newcommand{\vz}{\mathbf{z}}

\newcommand{\vq}{\mathbf{q}}
\newcommand{\vw}{\mathbf{w}}
\newcommand{\vy}{\mathbf{y}}
\newcommand{\vc}{\mathbf{c}}
\newcommand{\vk}{\mathbf{k}}
\newcommand{\vx}{\mathbf{x}}
\newcommand{\lapl}{\bm{\Delta}}
\newcommand{\vt}{\mathbf{t}}
\newcommand{\vG}{\mathbf{G}}
\newcommand{\vv}{\mathbf{v}}
\newcommand{\vj}{\mathbf{j}}
\newcommand{\vf}{\mathbf{f}}
\newcommand{\vg}{\mathbf{g}}
\newcommand{\vT}{\mathbf{T}}
\newcommand{\vK}{\mathbf{K}}
\newcommand{\vM}{{\mathbf{M}}}
\newcommand{\vA}{{\mathbf{A}}}
\newcommand{\vP}{{\mathbf{P}}}
\newcommand{\vX}{{\mathbf{X}}}
\newcommand{\vZ}{{\mathbf{Z}}}
\newcommand{\vD}{{\mathbf{D}}}
\newcommand{\vR}{{\mathbf{R}}}
\newcommand{\vY}{{\mathbf{Y}}}
\let\vb\relax
\newcommand{\vb}{{\mathbf{b}}}

\newcommand{\wh}[1]{\widehat{#1}}

\renewcommand{\epsilon}{\varepsilon}

\title[Density functionals for the Pauli--Fierz Hamiltonian]{Density functionals for the\\ Pauli--Fierz Hamiltonian}

\author[M.\,A. Csirik, A. Laestadius, M. Penz, and E.\,I. Tellgren]{Mih\'aly~A.~Csirik$^{1,*}$, Andre~Laestadius$^{1,2,3}$, Markus~Penz$^{3,4,5}$, \and Erik~I.~Tellgren$^{2}$}

\address{$^1$Department of Mathematical Sciences, Chalmers University of
Technology and the University of Gothenburg, G\"oteborg, Sweden}
\address{$^2$Hylleraas Centre for Quantum Molecular Sciences, Department of
Chemistry, University of Oslo, Norway}
\address{$^3$Department of Computer Science, Oslo Metropolitan University,
Norway}
\address{$^4$Arnold Sommerfeld Center for Theoretical Physics,
Ludwig-Maximilians-Universit\"at M\"unchen, Germany}
\address{$^5$Max Planck Institute for the Structure and Dynamics of Matter
and Center for Free-Electron Laser Science, Hamburg, Germany}

\thanks{$^*$Corresponding author, email: \href{mailto:csirik@gmail.com}{\texttt{csirik@gmail.com}}}


\begin{document}
\begin{abstract}
We develop a density-functional theory for the Pauli--Fierz Hamiltonian of
nonrelativistic quantum electrodynamics in the spirit of
the convex-analytic formulation of Lieb. We consider finitely many electrons,
subject to external scalar and vector potentials, coupled to the full
continuum of massless transverse photons with an ultraviolet cutoff.
The natural basic variables are the standard electron density and a combined current
density, the sum of the paramagnetic current density and a
field-induced current density generated by the quantized vector
potential.  We prove representability results for these density--current
pairs and show
that the mixed-state constrained-search functional attains its infimum
and coincides with the Lieb functional obtained by convex conjugation.
Since arbitrarily soft photons purify mixed states at a negligible energy
cost without changing their density and combined current density, the
pure-state Levy--Lieb functional equals the Lieb functional and is
therefore convex, even though its infimum need not be attained.
In addition, we establish the zero-coupling limit, which recovers the
mixed-state Vignale--Rasolt functional of current-density-functional theory.
Finally, we consider the long-wavelength (dipole) limit and show that
the limiting functional can be written as a constrained search for the
well-known length-gauge dipole Hamiltonian.
\end{abstract}

\maketitle

\clearpage
\enlargethispage{\baselineskip}
\tableofcontents

\clearpage
\section{Introduction}\label{sec-intro}

Strong coupling of molecules and solids to the quantized electromagnetic
field of an optical cavity has been reported to modify chemical and
material properties~\cite{ebbesen2016hybrid,Basov2020,garciavidal2021manipulating,Bloch2022}.
At the energy scales of chemistry and condensed-matter physics, such
effects are described by
nonrelativistic quantum electrodynamics (QED) in the form of the
Pauli--Fierz Hamiltonian~\cite{pauli1938theorie,spohn2004dynamics}, in
which electrons are minimally coupled to the quantized transverse vector
potential.  To treat realistic systems, density-functional theory (DFT)
has been extended to this setting as quantum-electrodynamical DFT
(QEDFT)~\cite{tokatly2013time,ruggenthaler2014quantum,Ruggenthaler2017}.
See \cite{flick2015kohn,schafer2021making} for approximate functionals
and \cite{ruggenthaler2018quantum,ruggenthaler2023understanding} for
reviews.  While the Pauli--Fierz Hamiltonian itself is well understood
mathematically~\cite{hiroshima2002selfadjoint,hasler2008selfadjoint,bach1999spectral,griesemer2001ground,lieb2003existence},
the density-functional side has been analyzed in mathematical detail only
for quantum-optical model systems such as the Rabi and Dicke
models~\cite{Novokreschenov2023,bakkestuen2025dicke,bakkestuen2025rabi},
in which the system is reduced to two-level systems coupled to finitely many
field modes.

The mathematical structure of DFT was clarified by Lieb~\cite{Lieb1983},
who placed Levy's
constrained search~\cite{levy1979universal} into the framework of convex
analysis (see~\cite{lewin2023universal,penz2023structureI} for recent
accounts).  In the presence of magnetic fields, the density is supplemented by the
paramagnetic current density $\vjp$, which couples linearly to the
external vector potential $\vA$ in the current-density-functional theory
(CDFT) of Vignale and Rasolt~\cite{vignale1987density,vignale1988current}.
Absorbing the diamagnetic term $\frac12|\vA|^2$ into the scalar potential
makes the ground-state energy jointly concave in the
potentials~\cite{tellgren2012choice}, which also leads to a convex formulation
of CDFT~\cite{penz2023structureII}.  Several cornerstones of DFT are more
delicate here: the density--current pair does not in general determine the potentials
uniquely~\cite{capelle2002nonuniqueness,laestadius2014hohenberg}.
Furthermore, only sufficient conditions for $N$-representability are known, and it is
open whether pure- and mixed-state $N$-representability
coincide~\cite{lieb2013current,tellgren2014fermion,csirik2024thermodynamic}.
Lower semicontinuity of the (mixed-state) Vignale--Rasolt functional was proved only
recently~\cite{kvaal2021lower}.

In this article, we develop density-functional theory for the
Pauli--Fierz Hamiltonian itself, in the spirit of Lieb's analysis.  We
consider $N$ spinless electrons in $\RR^3$ coupled to the full continuum
of massless transverse photons, with a fixed ultraviolet cutoff but no
infrared cutoff, in an external scalar potential $v$ and an external
classical vector potential $\vA$.  The photon mode $\vk$ couples proportionally
to $e\wh\chi(\vk)$, where $e\geq0$ is the coupling
constant and the form factor $\wh\chi$ is any function satisfying
\cref{cutoff-assumption}, such as the hard cutoff
$\iden_{|\vk|\leq\Lambda}/\sqrt{|\vk|}$.  We indicate these choices by the superscripts $e,\chi$.  A new feature
compared with CDFT is that a finite-energy state may contain arbitrarily many photons of arbitrarily small energy.  This
infrared freedom causes the main technical difficulties, but it is also
behind some of the most striking results.

\emph{The combined current density.}  For a finite-energy state $\Psi$,
\begin{equation*}
 \dua{\Psi}{H^{v,\vA}\Psi}=\dua{\Psi}{H^{0,0}\Psi}
 +\dua{v+\tfrac12|\vA|^2}{\rho_\Psi}+\dua{\vA}{\vjc_\Psi},
 \qquad
 \vjc_\Psi=\vjp_\Psi+\vjph_\Psi,
\end{equation*}
where $H^{0,0}$ is the Pauli--Fierz Hamiltonian without external
potentials \eqref{q00-def}, $\rho_\Psi$ is the electronic density, and
$\vjph_\Psi(\vx)=-eN\Re\dua{\Psi}{\AF(\vx)\Psi}_{\vx}$ is the
\emph{field-induced current density}, that is, $-e$ times the
density-weighted mean of the quantized vector potential $\AF$ at the
position $\vx$ of an electron, the partial inner product being taken in
all other variables.  Thus the
external potentials see a state only through $\rho_\Psi$ and the
\emph{combined current density} $\vjc_\Psi$, and these are our basic
variables.  The
combined current density obeys a kinetic-energy bound similar to that for $\vjp$
(\cref{prop-current-bounds}), but although $\div\AF=0$, its
field-induced part need not be divergence-free for entangled
electron--photon states (\cref{prop-div-jph}).  Ground states carry the
magnetostatic fields of their own physical current density
$\vjc+\rho\vA$ (\cref{prop-ground-state-fields}).

\emph{Representability.}  The photon field purifies the electrons: the
pairs $(\rho_\Psi,\vjp_\Psi)$ of pure electron--photon states with
finite energy are exactly
those of electronic \emph{mixed} states with finite kinetic energy
(\cref{thm-photon-purification}).  For $e>0$,
coherent photon displacements depending on the electron positions produce
field-induced currents $-e\rho\va$, where the vector field $\va$ need not
be transverse.  Hence, every $(\rho,\rho\vv)$ with
$\int_{\RR^3}\rho=N$, $\sqrt\rho\in H^1(\RR^3)$ and
$\int_{\RR^3}\rho(|\vv|^2+|\grad\vv|^2)<\infty$ is a pure-state
$N$-representable density--combined-current pair
(\cref{thm-photonic-representability,cor-photonic-combined-class}).

\emph{Universal functionals.}  The Pauli--Fierz Levy--Lieb (PFLL)
functional is the pure-state constrained search
\begin{equation*}
 F_\LL^{e,\chi}(\rho,\vjc)
 =\inf_{\substack{\|\Psi\|=1\\ \rho_\Psi=\rho\\ \vjc_\Psi=\vjc}}
 \dua{\Psi}{H^{0,0}\Psi},
\end{equation*}
and the Pauli--Fierz Lieb (PFL) functional $F_\LI^{e,\chi}(\rho,\vjc)$ is the
convex conjugate of the ground-state energy, reparametrized as
in~\cite{tellgren2012choice}.  As in Lieb's theory, it is the lower
semicontinuous convex envelope of $F_\LL^{e,\chi}(\rho,\vjc)$.  Our first main
result (\cref{thm-unrestricted-attainment}) is the counterpart
of~\cite{Lieb1983,kvaal2021lower}:
the mixed-state constrained-search functional is
lower semicontinuous, equals $F_\LI^{e,\chi}(\rho,\vjc)$, and attains its infimum
at a density matrix of finite mean photon number.  Since an energy bound
does not control the photon number, the proof removes soft photons and
restores the combined current density exactly, at a controlled energy
cost.  Our second main result (\cref{thm-many-electron-soft-purification})
points in the opposite direction and sets QEDFT apart from DFT or CDFT:
soft photons purify any finite-energy
mixed state at arbitrarily small energy cost without changing
$(\rho,\vjc)$, so that
\[
 F_\LL^{e,\chi}(\rho,\vjc)=F_\LI^{e,\chi}(\rho,\vjc).
\]
Hence, unlike the Levy--Lieb functional of electronic DFT, which is not
convex~\cite{Lieb1983}, the PFLL functional is convex and lower
semicontinuous, and pure- and mixed-state $N$-representability coincide
(\cref{cor-many-electron-pure-functional}).  The price to pay is that the
pure-state infimum is in general not attained. This is true already for one electron
and at every coupling (\cref{prop-pure-nonattainment}).  Optimizers need
not be ground states, and in \cref{sec-optimizer-properties} we discuss
which ground-state properties carry over to optimizers.  
Mixed-state optimizers contain few soft photons: for the hard cutoff, the expected
number and energy of photons with frequency below $\kappa$ are $O(\kappa)$
and $O(\kappa^2)$, respectively, so the probability of finding any such
photon vanishes as $\kappa\downarrow0$
(\cref{prop-optimizer-soft-spectrum}).

\emph{Zero-coupling and dipole limits.}  At $e=0$, the photons still
purify electronic mixed states, so that
$F_\LL^{0,\chi}(\rho,\vjp)=F_\LI^{0,\chi}(\rho,\vjp)=F_\VR(\rho,\vjp)$, the mixed-state Vignale--Rasolt
functional (\cref{thm-soft-photon-relaxation}), and $F_\LL^{e,\chi}$
converges to $F_\VR$ pointwise and in the sense of Mosco as $e\downarrow0$
(\cref{thm-zero-coupling-convergence}).  The long-wavelength (dipole)
approximation, ubiquitous in practical
QEDFT~\cite{flick2015kohn,rokaj2018lightmatter,schafer2020relevance},
applies when the system size is small compared to the wavelength of the
relevant photon modes, so that the vector potential is effectively uniform across the system.
We realize this limit by instead rescaling the system:
let $F_{\LI,\varepsilon}$ be the PFL functional of the model in which
$\AF(\vx)$ is replaced by $\AF(\varepsilon\vx)$, so that $\varepsilon=1$
is the original model and $\varepsilon$ plays the role of the ratio of system size to wavelength.
As $\varepsilon\downarrow0$, these functionals converge
in the sense of Mosco to the dipole functional $F_\dip$
(\cref{thm-dipole-limit}), whose current argument $\vjcdip$ is the
combined current density formed with the spatially uniform field
$\AF(0)$.  In the length gauge, $F_\dip$ is the constrained search for
the usual dipole Hamiltonian (\cref{cor-dipole-length-gauge}), and
\cref{prop-dipole-electronic-comparison} gives $F_\dip\to F_\VR$ as
$e\downarrow0$.  For fixed $N$ and form factor, these relations are
summarized in the diagram
\begin{equation*}
\begin{tikzcd}[column sep=large,row sep=large]
 F_\LL^{e,\chi}(\rho,\vjc)=F_\LI^{e,\chi}(\rho,\vjc)
   \arrow[r,"\vjc=0"]
   \arrow[d,"\text{dipole, }\varepsilon\downarrow0"]
   \arrow[dd,bend right=60,"e\downarrow0"']
 & F_\LL^{e,\chi}(\rho,0)=F_\LI^{e,\chi}(\rho,0)
   \arrow[dd,"e\downarrow0"]
 \\
 F_\dip(\rho,\vjcdip)
   \arrow[d,"e\downarrow0"]
 & \\
 F_\LI^{0,\chi}(\rho,\vjp)=F_\VR(\rho,\vjp)
   \arrow[r,"\vjp=0"]
 & F_\LI^{0,\chi}(\rho,0)=F_{\mathrm L}(\rho)
\end{tikzcd}
\end{equation*}
where $F_{\mathrm L}$ is the Lieb functional~\cite{Lieb1983}.  The arrow $\varepsilon\downarrow0$
denotes Mosco convergence, and the arrows $e\downarrow0$ denote pointwise
convergence, which for the curved arrow also holds in the Mosco sense.

\emph{Organization of the paper.}  \Cref{sec-prelim,sec-nrqed}
introduce the Fock-space setting and the Pauli--Fierz Hamiltonian.  The
main results are stated in \cref{sec-main-results} and proved in
\cref{sec-proofs}.  The appendices collect transformation properties of
the model (dilations, gauge covariance, Weyl displacements, and the
length-gauge transformation, among others) and further
technical tools.

\section*{Acknowledgements}

M.A.Cs., A.L., and M.P. acknowledge funding from the European Research Council
through ERC-2021-STG Grant Agreement No.~101041487 (REGAL).
A.L. acknowledges support through QSTAR (Centre for Quantum Computing and Applications), 
funded by the Research Council of Norway (RCN) under project No. 361617. 
A.L. and E.I.T. also acknowledge support through the Hylleraas 
Centre for Quantum Molecular Sciences, funded by the RCN under its 
Centres of Excellence scheme, project No. 262695.

\section*{Disclosure of AI use and Lean verification}
The authors made substantial use of the large language models Claude
Opus 5.5 (Anthropic) and GPT 6 Astra (OpenAI) to edit the text, to
generate and refine statements and proofs, to check arguments and find
errors, and to write the Lean formalization described below.  All output
was reviewed and edited by the authors, who take full responsibility for
the content.

A Lean~4 formalization in the directory
\texttt{gpt\_lean}\footnote{\url{https://github.com/xyzw/qedft_pf_pub}}
verifies all new results except those on the long-wavelength limit
(\cref{sec-dipole-limit}).  Standard results not involving the
Pauli--Fierz model enter as explicit hypotheses: the canonical
commutation relations, Weyl operators, Fock-space factorization and the
Wiener--It\^o--Segal isomorphism; normal and partial traces and closed
quadratic forms; completed tensor powers; Sobolev representatives, a
local Sobolev embedding, the Hardy and fractional Hardy inequalities,
Rellich compactness for $H^{1/2}$ and density in $\dot H^{-1/2}$; and
$L^p$ duality and the Fenchel--Moreau theorem.

\clearpage

\section{Preliminary notions}\label{sec-prelim}

\subsection{\texorpdfstring{$N$-particle spaces and the bosonic Fock space}{N-particle spaces and the bosonic Fock space}}\label{sec-fock-space}

Let $d\in\NN$ be the number of spatial dimensions, $N\in\NN$ the number of
particles, and $q\in\NN$, $q\geq1$, the number of internal states (for
example, $q=2$ for spin-$1/2$ particles or for the two transverse polarizations of a photon).
We denote the fermionic (a) or bosonic (s) $N$-particle spaces by
\[
\Hil^N_{a/s}=L_{a/s}^2((\RR^d\times\{1,\ldots,q\})^N),
\]
where $L^2_a$ (resp. $L^2_s$) is the Hilbert space of totally antisymmetric (resp. symmetric) square-integrable complex-valued functions both equipped with the inner product
\begin{multline*}
\dua{\phi}{\psi}_{\Hil^N}
=\sum_{\sigma_1,\ldots,\sigma_N\in\{1,\ldots,q\}}
\int_{\RR^{dN}}
\ol{\phi(\vx_1,\sigma_1,\ldots,\vx_N,\sigma_N)}\\
{}\times\psi(\vx_1,\sigma_1,\ldots,\vx_N,\sigma_N)
\,\dd\vx_1\ldots\dd\vx_N.
\end{multline*}
Total antisymmetry (resp. symmetry) is understood with respect to simultaneous permutations of the spatial and spin variables.
In what follows, we will often aggregate the spatial variables as $\vX=(\vx_1,\ldots,\vx_N)$, and the spin variables as $\vsigma=(\sigma_1,\ldots,\sigma_N)$
and the spatial-spin variables as $\vZ=(\vz_1,\ldots,\vz_N)$, where $\vz_j=(\vx_j,\sigma_j)$. 
By abuse of notation, we write $\psi(\vX,\vsigma)=\psi(\vZ)$. Whenever necessary, we distinguish the number of particles $N$ in the aggregate 
vectors as $\vX^{(N)}$, $\vsigma^{(N)}$ and $\vZ^{(N)}$.
We equip $\RR^d\times\{1,\ldots,q\}$ with the product of Lebesgue and
counting measure and use the shorthand
\[
 \int_{\RR^d\times\{1,\ldots,q\}}h(\vz)\,\dd\vz
 :=\sum_{\sigma=1}^q\int_{\RR^d}h(\vx,\sigma)\,\dd\vx.
\]
Accordingly, for $\vz=(\vx,\sigma)$ and $\vz'=(\vx',\sigma')$,
\[
 \delta(\vz-\vz'):=\delta(\vx-\vx')\delta_{\sigma,\sigma'}.
\]

Electrons will be described by the antisymmetric space $\Hil_a^N$ with
fixed $N$, and photons by the bosonic Fock space built from the
symmetric spaces.

The bosonic Fock space is defined as the infinite direct sum
\[
\Fock_s=\Hil^0 \oplus \Hil^1 \oplus \Hil_s^2 \oplus \ldots,\quad\text{with the convention}\quad \Hil^0\equiv \CC,
\]
consisting of infinite tuples $\Psi=(\psi^0,\psi^1,\psi^2,\ldots)$, where $\psi^N\in \Hil^N_s$, such that 
\[
\|\Psi\|^2=\sum_{N\ge 0} \|\psi^N\|^2 <\infty.
\]
This space is separable, since $\Hil^1$ is.
Equipped with the corresponding inner product, i.e.,
\[
\dua{\Phi}{\Psi}_\Fock=\sum_{N\ge 0} \dua{\phi^N}{\psi^N}_{\Hil^N},
\]
the space $\Fock_s$ is a Hilbert space. Instead of the tuple notation, $\Psi=\psi^0\oplus\psi^1\oplus\psi^2\oplus\ldots$ is used sometimes. 
The canonically embedded subspaces $\Hil_s^N\subset\Fock_s$ are called the $N$-particle sectors.
For any $N\in\NN$, the truncated Fock spaces are given by
\[
\Fock_s^{\le N}=\Hil^0 \oplus \Hil^1 \oplus \Hil_s^2 \oplus \ldots \oplus \Hil_s^N.
\]
Moreover, the finite-particle vectors form the dense subspace
\[
\Fock_s^\fin=\bigcup_{N\ge 1} \Fock_s^{\le N}.
\]
We denote by $\Omega$ the canonical vacuum
vector $\Omega=1\oplus0\oplus0\oplus\ldots$ of $\Fock_s$.

It is possible to symmetrically compose two elements $\psi_1\in\Hil_s^{N_1}$ and
$\psi_2\in\Hil_s^{N_2}$ and obtain an element in $\Hil_s^{N_1+N_2}$,
namely the symmetric tensor product
\begin{multline*}
\psi_1\vee\psi_2=(N_1!N_2!(N_1+N_2)!)^{-1/2}
\sum_{\tau\in\SymG_{N_1+N_2}}
\psi_1(\vz_{\tau(1)},\ldots,\vz_{\tau(N_1)})\\
{}\times\psi_2(\vz_{\tau(N_1+1)},\ldots,
\vz_{\tau(N_1+N_2)}),
\end{multline*}
where $\SymG_N$ denotes the symmetric group on $\{1,\ldots,N\}$.

Next, we define the creation and annihilation operators
\cite[Section~5.2]{bratteli-robinson-2}. For any
$f\in\Hil:=\Hil^1$, define the creation operator
$a^\dag(f):\Fock_s^\fin\to\Fock_s^\fin$ sectorwise by
\[
 a^\dag(f)\psi=f\vee\psi
\]
for $\psi\in\Hil_s^N$, and extend the definition linearly to
$\Fock_s^\fin$. Thus $a^\dag(f)\Hil_s^N\subset\Hil_s^{N+1}$.

On the same finite-particle space, define the annihilation operator by
\[
 (a(f)\psi)(\vz_1,\ldots,\vz_{N-1})
 =\sqrt{N}\int_{\RR^d\times\{1,\ldots,q\}}
 \ol{f(\vz)}\psi(\vz,\vz_1,\ldots,\vz_{N-1})\,\dd\vz,
 \qquad\psi\in\Hil_s^N,
\]
and extend it linearly to $\Fock_s^\fin$. We set $a(f)\Omega=0$, while
$a^\dag(f)\Omega=f$. Moreover,
\[
 \dua{\Phi}{a^\dag(f)\Psi}_\Fock
 =\dua{a(f)\Phi}{\Psi}_\Fock,
 \qquad\Phi,\Psi\in\Fock_s^\fin.
\]
The canonical commutation relations (CCR)
\begin{align*}
[a(g),a^\dag(f)]&=\dua{g}{f}_\Hil \iden_{\Fock_s},\\
[a(g),a(f)]&=[a^\dag(g),a^\dag(f)]=0
\end{align*} 
hold for any $f,g\in\Hil$ on $\Fock_s^\fin$.
These operators are closable, and we use the same symbols for their
closures, which are mutual Hilbert-space adjoints.
The Segal field operator
\[
 \boxed{\varPhi(f)=a^\dag(f)+a(f)}
\]
is essentially self-adjoint on $\Fock_s^\fin$, and we denote its closure
by the same symbol.  The \emph{conjugate} Segal field operator, also
called the field-momentum operator, is
\begin{equation}\label{conjugate-segal-field-def}
 \boxed{\varPi(f)=\varPhi(\i f)=\i\bigl(a^\dag(f)-a(f)\bigr).}
\end{equation}
It is likewise essentially self-adjoint on $\Fock_s^\fin$, and the
same symbol denotes its closure.  For the vacuum,
$\|\varPhi(f)\Omega\|=\|\varPi(f)\Omega\|=\|f\|$.
For $f\ne0$, the creation and annihilation operators are unbounded on
$\Fock_s$, but their restrictions to
truncated Fock spaces are bounded. More precisely, for $N\in\NN$,
\begin{align*}
 a(f)&:\Fock_s^{\le N}\longrightarrow\Fock_s^{\le N-1},
 &\|a(f)\Psi\|&\leq\sqrt{N}\,\|f\|\|\Psi\|,\\
 a^\dag(f)&:\Fock_s^{\le N}\longrightarrow\Fock_s^{\le N+1},
 &\|a^\dag(f)\Psi\|&\leq\sqrt{N+1}\,\|f\|\|\Psi\|.
\end{align*}

In contrast, the symbols $a(\vz)$ and $a^\dag(\vz)$ are operator-valued
distributions, not operators defined at each fixed $\vz$. If
$\psi^N\in\Hil_s^N$, then $\vz\mapsto a(\vz)\psi^N$ denotes the
$\Hil_s^{N-1}$-valued $L^2$ function defined, for almost every $\vz$, by
\[
 (a(\vz)\psi^N)(\vz_1,\ldots,\vz_{N-1})
 =\sqrt{N}\psi^N(\vz_1,\ldots,\vz_{N-1},\vz).
\]
In particular,
\[
 \int_{\RR^d\times\{1,\ldots,q\}}
 \|a(\vz)\psi^N\|^2\,\dd\vz
 =N\|\psi^N\|^2.
\]
Then, the creation operator is defined by the weak adjoint relation
\[
 \dua{\Phi}{a^\dag(\vz)\Psi}_\Fock
 :=\dua{a(\vz)\Phi}{\Psi}_\Fock,
 \qquad \Phi,\Psi\in\Fock_s^\fin.
\]
Let $\SC_{N+1}$ denote the normalized symmetrization operator,
\[
 (\SC_{N+1}F)(\vz_1,\ldots,\vz_{N+1})
 =\frac{1}{(N+1)!}\sum_{\tau\in\SymG_{N+1}}
 F(\vz_{\tau(1)},\ldots,\vz_{\tau(N+1)}).
\]
Then, as an identity of distributions,
\[
 (a^\dag(\vz)\psi^N)(\vz_1,\ldots,\vz_{N+1})
 =\sqrt{N+1}\,\SC_{N+1}\left[
 \psi^N(\vz_1,\ldots,\vz_N)\delta(\vz_{N+1}-\vz)
 \right].
\]

The smeared identities
\begin{align*}
 a^\dag(f)&=\int_{\RR^d\times\{1,\ldots,q\}}
 f(\vz)a^\dag(\vz)\,\dd\vz,\\
 a(f)&=\int_{\RR^d\times\{1,\ldots,q\}}
 \ol{f(\vz)}a(\vz)\,\dd\vz
\end{align*}
hold as quadratic forms on $\Fock_s^\fin$.  In the same sense, without
smearing, the point-field CCR are
\begin{align*}
 [a(\vz),a^\dag(\vz')]&=\delta(\vz-\vz')\iden_{\Fock_s},\\
 [a^\dag(\vz),a^\dag(\vz')]&=[a(\vz),a(\vz')]=0.
\end{align*}
Smearing the first identity against $\ol{g(\vz)}f(\vz')$ recovers the
first CCR above.

\subsection{Partial traces and reduced states}
Let $\HC_1$ and $\HC_2$ be separable Hilbert spaces and let
$\Gamma\in\mathfrak S_1(\HC_1\otimes\HC_2)$ be trace class.
Here, $\mathfrak S_p$ is the Banach space of $p$-Schatten-class operators.
The partial trace of $\Gamma$ over the second factor is the unique trace-class
operator $\Tr_2\Gamma$ on $\HC_1$ characterized by
\begin{equation}\label{partial-trace-duality}
 \Tr_{\HC_1}\big((\Tr_2\Gamma)B\big)
 =\Tr_{\HC_1\otimes\HC_2}
 \big(\Gamma(B\otimes\iden_{\HC_2})\big),
 \qquad \text{for all}\; B\in\mathcal B(\HC_1).
\end{equation}
Here $\mathcal B(\HC_1)$ denotes the bounded operators on $\HC_1$.
Similarly, $\Tr_1\Gamma$ is characterized by testing against
$\iden_{\HC_1}\otimes C$, with $C\in\mathcal B(\HC_2)$.  If
$(e_j)_{j\geq1}$ is any orthonormal basis of $\HC_2$, then
\begin{equation*}
 \Tr_2\Gamma
 =\sum_{j\geq1}
 (\iden_{\HC_1}\otimes\langle e_j|)\,\Gamma\,
 (\iden_{\HC_1}\otimes|e_j\rangle),
\end{equation*}
where $\iden_{\HC_1}\otimes|e_j\rangle:\HC_1\to\HC_1\otimes\HC_2$ maps
$\phi$ to $\phi\otimes e_j$, and $\iden_{\HC_1}\otimes\langle e_j|$ is
its adjoint.  The sum converges in trace norm and is
independent of the chosen basis.  Partial trace preserves positivity and
the trace.  Thus, if
$\Gamma\geq0$ and $\Tr\Gamma=1$, then both partial traces are again density
matrices.

In the electron--photon system below, $\HC_1=\Hil_\el^N$
and $\HC_2=\Fock_\ph$.  We denote density matrices on the full
composite space by $\vGamma$, retaining unbolded symbols for reduced
density matrices.  The states
\[
 \Gamma_\el=\Tr_\ph\vGamma,\qquad
 \Gamma_\ph=\Tr_\el\vGamma
\]
are called the electronic and photon reduced states.  When the total state
is a wavefunction $\Psi$, the phrase \emph{trace out the photons} means first
forming the rank-one density matrix
$|\Psi\rangle\langle\Psi|$ and then taking
$\Tr_\ph|\Psi\rangle\langle\Psi|$.

For a normalized pure state, its Schmidt decomposition has the form
\begin{equation}\label{Schmidt-decomposition}
 \Psi=\sum_{j=1}^{r}\sqrt{p_j}\,\Phi_j\otimes\Xi_j,
 \qquad p_j>0,\qquad\sum_{j=1}^{r}p_j=1,
\end{equation}
where $r\in\NN\cup\{\infty\}$ is the Schmidt rank, with
$r\leq\min\{\dim\HC_1,\dim\HC_2\}$, and
$(\Phi_j)_{j=1}^{r}$ and $(\Xi_j)_{j=1}^{r}$ are orthonormal families
in $\HC_1$ and $\HC_2$, respectively.
The two reduced states are then
\begin{equation}\label{Schmidt-reduced-states}
 \Tr_2|\Psi\rangle\langle\Psi|
 =\sum_{j=1}^{r}p_j|\Phi_j\rangle\langle\Phi_j|,
 \qquad
 \Tr_1|\Psi\rangle\langle\Psi|
 =\sum_{j=1}^{r}p_j|\Xi_j\rangle\langle\Xi_j|.
\end{equation}
Thus they have the same nonzero eigenvalues, and either reduced state is
pure exactly when $r=1$, equivalently when $\Psi$ is a product vector.
More precisely, a pure marginal is itself a tensor factor of $\Psi$: for
a unit vector $\Xi\in\HC_2$,
\begin{equation}\label{pure-marginal-product}
 \Tr_1|\Psi\rangle\langle\Psi|=|\Xi\rangle\langle\Xi|
 \quad\Longrightarrow\quad
 \Psi=\Phi\otimes\Xi\ \text{ for a unit vector }\Phi\in\HC_1.
\end{equation}
Indeed, \cref{Schmidt-reduced-states} then forces $r=1$ and
$|\Xi_1\rangle\langle\Xi_1|=|\Xi\rangle\langle\Xi|$, so
$\Xi_1=e^{\i\theta}\Xi$ for some $\theta\in\RR$ and
$\Psi=\Phi_1\otimes\Xi_1=(e^{\i\theta}\Phi_1)\otimes\Xi$.

More generally, a normalized vector $\Psi\in\HC_1\otimes\HC_2$ is called
a \emph{purification} of a density matrix $\gamma$ on $\HC_1$ if
\[
 \Tr_2|\Psi\rangle\langle\Psi|=\gamma.
\]
Such a purification in the fixed auxiliary space $\HC_2$ exists exactly
when $\operatorname{rank}\gamma\leq\dim\HC_2$. Indeed, if
$\gamma=\sum_{j=1}^{r}p_j|\Phi_j\rangle\langle\Phi_j|$, one may choose
an orthonormal family $(\Xi_j)_{j=1}^{r}$ in $\HC_2$ and use \cref{Schmidt-decomposition}. Since $\Fock_\ph$ is infinite
dimensional, every electronic density matrix admits such an abstract
Hilbert-space purification. However, the purification might have infinite photon energy.

For a positive trace-class $\Gamma$, the defining identity
\cref{partial-trace-duality} also applies to positive unbounded
observables by monotone approximation.  In particular, 
\begin{equation}\label{partial-trace-positive-observables}
 \Tr\big((A\otimes\iden_{\HC_2})\Gamma\big)
 =\Tr\big(A\,\Tr_2\Gamma\big),\qquad A\geq0,
\end{equation}
where both sides are form traces. In particular, the left-hand side means
\[
 \Tr\!\left((\sqrt{A}\otimes\iden_{\HC_2})\Gamma
 (\sqrt{A}\otimes\iden_{\HC_2})\right).
\]
The analogous identity holds for an observable acting only on $\HC_2$.
This is the sense in which finite electronic kinetic energy and finite
photon energy pass to the corresponding reduced states.

The electronic density and paramagnetic current density used below depend only on
$\Gamma_\el$, while the free-field energy depends only on $\Gamma_\ph$.
By contrast, the field-induced current density contains a joint
electron-position--photon observable and is generally not determined by
either reduced state separately. The pair of reduced states retains all
expectations of observables acting on only one factor, but does not in
general determine the electron--photon correlations needed for the
field-induced current density.

\subsection{Second quantization of operators}\label{sec-second-quantization}

We briefly recall the second quantization of one-particle operators and
refer to~\cite[Section~7]{solovej2014many} for details.  Let $A$ be a
self-adjoint operator on $\Hil$.  Its $N$-particle lift $A^{(N)}$ is the
closure of $\sum_{j=1}^NA_{\vz_j}$, initially defined on the algebraic
symmetric tensor product of $N$ copies of $D(A)$, where $A_{\vz_j}$ acts
on the $j$th variable.  It is self-adjoint on $\Hil_s^N$, and we set
$A^{(0)}=0$.  The second quantization of $A$ is the self-adjoint operator
\[
 \dd\varGamma(A)=\bigoplus_{N\geq0}A^{(N)}
\]
on $\Fock_s$, with domain consisting of all $\Psi=\bigoplus_N\psi^N$ such
that $\psi^N\in D(A^{(N)})$ and $\sum_N\|A^{(N)}\psi^N\|^2<\infty$.  If
$(f_j)_{j\geq1}\subset D(A)$ is an orthonormal basis of $\Hil$, then
\[
 \dd\varGamma(A)=\sum_{j,k\geq1}\dua{f_j}{Af_k}\,a^\dag(f_j)a(f_k)
\]
in the sense of quadratic forms on finite-particle
vectors~\cite[Lemma~7.8]{solovej2014many}.  In particular, the number
operator $\NC=\dd\varGamma(\iden_\Hil)=\sum_{j\geq1}a^\dag(f_j)a(f_j)$ acts
as multiplication by $N$ on $\Hil_s^N$.  For a tuple
$\boldsymbol A=(A_1,\ldots,A_m)$ of strongly commuting self-adjoint
operators and $\boldsymbol\alpha\in\RR^m$, we write
$\dd\varGamma(\boldsymbol A)=(\dd\varGamma(A_1),\ldots,\dd\varGamma(A_m))$ and
$\boldsymbol\alpha\cdot\dd\varGamma(\boldsymbol A)
:=\dd\varGamma(\boldsymbol\alpha\cdot\boldsymbol A)$.

The multiplicative second quantization of a unitary $U$ on $\Hil$ is the
unitary operator $\varGamma(U)=\bigoplus_{N\geq0}U^{\otimes N}|_{\Hil_s^N}$
on $\Fock_s$, with $U^{\otimes0}=1$~\cite[Remark~7.10]{solovej2014many}.
It satisfies
\begin{gather*}
 \varGamma(U)a^\#(f)\varGamma(U)^\dag=a^\#(Uf),\qquad
 \varGamma(U)\dd\varGamma(A)\varGamma(U)^\dag=\dd\varGamma(UAU^\dag),\\
 \varGamma(e^{-\i tA})=e^{-\i t\dd\varGamma(A)},
\end{gather*}
where $a^\#$ stands for either $a^\dag$ or $a$.  When no confusion can
arise, we use the same symbol $U$ for $\varGamma(U)$.

\section{Nonrelativistic quantum electrodynamics}\label{sec-nrqed}

For the remainder of the paper we take $d=3$ and $q=1$ (spinless
electrons), and write
\[
 \Hil_\el^N:=\Hil_a^N,\qquad
 \Fock_\ph:=\Fock_s(\Hil_\ph),\qquad
 \HC_N:=\Hil_\el^N\otimes\Fock_\ph,
\]
where the one-photon space $\Hil_\ph$ is defined in
\cref{sec-one-photon-space} below.
The restriction to spinless electrons is for notational simplicity only,
and all results extend to the spinful case with the appropriate modifications.
Writing $\Hil_\ph^{\otimes_sM}$ for
the $M$-photon sector of $\Fock_\ph$, with
$\Hil_\ph^{\otimes_s0}=\CC$ and $\Hil_\ph^{\otimes_s1}=\Hil_\ph$,
every $\Psi\in\HC_N$ has the unique
decomposition into $M$-photon components
\begin{equation}\label{psifock}
 \Psi=\bigoplus_{M=0}^\infty\psi^{(M)},\qquad
 \psi^{(M)}\in\Hil_\el^N\otimes\Hil_\ph^{\otimes_sM},\qquad
 \sum_{M=0}^\infty\|\psi^{(M)}\|^2<\infty.
\end{equation}
We often view the elements of $\HC_N$ as Fock-space-valued functions of
the electronic positions.  The canonical unitary identification
$L^2(\RR^{3N})\otimes\Fock_\ph\cong L^2(\RR^{3N};\Fock_\ph)$ maps
$\HC_N$ onto the functions $\vX\mapsto\Psi(\vX)\in\Fock_\ph$ that are
antisymmetric under permutations of $\vx_1,\ldots,\vx_N$, and
\[
 \|\Psi\|^2=\int_{\RR^{3N}}\|\Psi(\vX)\|_{\Fock_\ph}^2\dd\vX.
\]
Derivatives in the electronic positions are weak derivatives of such
functions.  For $\Psi\in L^2(\RR^{3N};\Fock_\ph)$, $1\leq n\leq N$ and
$\mu=1,2,3$, we say that $\partial_{x_n^\mu}\Psi\in
L^2(\RR^{3N};\Fock_\ph)$ if there is an element
$\partial_{x_n^\mu}\Psi$ of $L^2(\RR^{3N};\Fock_\ph)$ such that
\[
 \int_{\RR^{3N}}\partial_{x_n^\mu}\Phi(\vX)\,\Psi(\vX)\dd\vX
 =-\int_{\RR^{3N}}\Phi(\vX)\,\partial_{x_n^\mu}\Psi(\vX)\dd\vX
 \qquad\text{for every }\Phi\in C_c^\infty(\RR^{3N}),
\]
where the integrals are Bochner integrals in $\Fock_\ph$.  Pairing with
a fixed $\Xi\in\Fock_\ph$ shows that this holds if and only if, for every
$\Xi\in\Fock_\ph$, the scalar function
$\vX\mapsto\dua{\Xi}{\Psi(\vX)}_{\Fock_\ph}$ has weak derivative
$\vX\mapsto\dua{\Xi}{\partial_{x_n^\mu}\Psi(\vX)}_{\Fock_\ph}$.  In terms of
\cref{psifock}, the $M$-photon component of $\partial_{x_n^\mu}\Psi$ is
the weak partial derivative of $\psi^{(M)}$, viewed as a function of
the electronic and photon variables, with respect to $x_n^\mu$.  We
write $\grad_{\vx_n}\Psi=(\partial_{x_n^1}\Psi,\partial_{x_n^2}\Psi,
\partial_{x_n^3}\Psi)$ and
$\|\grad_{\vx_n}\Psi\|^2=\sum_{\mu=1}^3\|\partial_{x_n^\mu}\Psi\|^2$.

\subsection{One-photon space}\label{sec-one-photon-space}

Photons have two physical polarizations.  One may either use the
polarization representation
\[
 \Hil_\ph=L^2(\RR^3\times\{1,2\})
\]
or, without choosing a polarization frame, the solenoidal representation
\[
 \Hil_\sol=\{\vf\in L^2(\RR^3,\CC^3):\vk\cdot\vf(\vk)=0
 \text{ for a.a.\ }\vk\}.
\]
Choose measurable, real, orthonormal polarization vectors
$\vepsilon_1,\vepsilon_2$ such that
$\vepsilon_\lambda(\vk)\cdot\vk=0$ and use the identity
\begin{equation}\label{polarization-projector}
 \sum_{\lambda=1}^2\vepsilon_\lambda(\vk)\otimes
 \vepsilon_\lambda(\vk)
 =\iden-\frac{\vk\otimes\vk}{|\vk|^2}=:P_\sol(\vk)
\end{equation}
to define the solenoidal projector $P_\sol$.
All identities involving $P_\sol$ are understood almost everywhere, and its
value at $\vk=0$ may be fixed arbitrarily. Pointwise multiplication by $P_\sol$ is
an orthogonal projection on $L^2(\RR^3,\CC^3)$, and
\[
 \Hil_\sol=\operatorname{Ran}P_\sol
\]
is therefore closed.  Under the inverse Fourier transform it is the space
of square-integrable vector fields that are divergence-free in the sense
of distributions.

More explicitly, the map
\begin{equation*}
 \iota_{\vepsilon}:\Hil_\ph\longrightarrow\Hil_\sol,
 \qquad
 \iota_{\vepsilon} f=\sum_{\lambda=1}^2f_\lambda\vepsilon_\lambda,
 \qquad
 (\iota_{\vepsilon}^\dag\vf)_\lambda
 =\vepsilon_\lambda\cdot\vf,
\end{equation*}
is unitary.  If $\vepsilon'$ is another measurable polarization frame,
then $\iota_{\vepsilon'}^\dag\iota_{\vepsilon}$ is unitary on $\Hil_\ph$,
and its second quantization satisfies
\[
 \varGamma(\iota_{\vepsilon'}^\dag\iota_{\vepsilon})\,
 a^\#(\iota_{\vepsilon}^\dag\vf)\,
 \varGamma(\iota_{\vepsilon'}^\dag\iota_{\vepsilon})^\dag
 =a^\#(\iota_{\vepsilon'}^\dag\vf),
 \qquad \vf\in\Hil_\sol.
\]
Thus the creation and annihilation operators of a solenoidal mode $\vf$,
and all operators built from them below, are unitarily equivalent in the two frames.

\subsection{The vector potential operator}
The photon \emph{dispersion relation}, or simply the
\emph{dispersion}, is $\omega(\vk)=|\vk|$.
The free-field Hamiltonian is $H_\rad=\dd\varGamma(\omega)$,
discussed further in \cref{sec-free-field}.
We fix a momentum-space coupling profile, or form factor, $\wh\chi$,
incorporating the ultraviolet cutoff, and assume that it satisfies
\begin{equation}\label{cutoff-assumption}
 \boxed{\wh\chi(\vk)=\ol{\wh\chi(-\vk)},\qquad
 \wh\chi\in L^2(\RR^3),\qquad
 \frac{\wh\chi}{\sqrt\omega}\in L^2(\RR^3).}
\end{equation}
Such a form factor is necessary to obtain a well-defined
Hamiltonian \cite[Chapter~11]{lieb2010stability}.
The $L^2$ condition controls, in particular, the vacuum term in the
creation-operator estimate.  The weighted condition gives relative
bounds with respect to $\sqrt{H_\rad}$ and also imposes an infrared
condition at $\vk=0$.
The hard ultraviolet cutoff used in the examples is
\begin{equation}\label{harduv}
 \wh\chi_\Lambda(\vk)=\frac{\iden_{|\vk|\leq\Lambda}(\vk)}{\sqrt{|\vk|}}.
\end{equation}
For $0\leq\Lambda<\infty$ this satisfies \cref{cutoff-assumption}, with
the value at $\vk=0$ again irrelevant, and
\begin{equation}\label{harduv-norms}
 \|\wh\chi_\Lambda\|^2=2\pi\Lambda^2,
 \qquad
 \left\|\frac{\wh\chi_\Lambda}{\sqrt\omega}\right\|^2
 =4\pi\Lambda.
\end{equation}
The assumptions and representability constructions below apply to a
general fixed $\wh\chi$.
For $\mu=1,2,3$, define the coupling function
\begin{equation}\label{Gx-def}
 G^\mu_{\vx}(\vk,\lambda)
 =\frac{\epsilon_\lambda^\mu(\vk)\wh\chi(-\vk)
 e^{-\i\vk\cdot\vx}}{\sqrt{2}}
\end{equation}
and, equivalently in the solenoidal representation,
\begin{equation}\label{Gx-solenoidal}
 \vG^\mu_{\vx}(\vk)
 =\frac{P_\sol(\vk)\ve_\mu\wh\chi(-\vk)
 e^{-\i\vk\cdot\vx}}{\sqrt{2}},
 \qquad
 \iota_{\vepsilon} G^\mu_{\vx}=\vG^\mu_{\vx},
\end{equation}
where $\ve_\mu$ is the $\mu$th coordinate vector.  The identity
for the solenoidal projector~\eqref{polarization-projector}
gives the exact, uniform relations
\begin{equation}\label{Gx-norms}
 \sum_{\mu=1}^3\|G^\mu_{\vx}\|^2
 =\|\wh\chi\|^2,
 \qquad
 \sum_{\mu=1}^3
 \left\|\frac{G^\mu_{\vx}}{\sqrt\omega}\right\|^2
 =\left\|\frac{\wh\chi}{\sqrt\omega}\right\|^2.
\end{equation}
Dominated convergence also shows that $\vx\mapsto G^\mu_{\vx}$ is
continuous both in $\Hil_\ph$ and in the norm
$f\mapsto\|f/\sqrt\omega\|$.

On the finite-particle core $\Fock_\ph^\fin$, define the quantized vector
potential, componentwise, by the Segal field
\begin{equation}\label{Afield-def}
 \boxed{\Af^\mu(\vx)=\varPhi(G^\mu_{\vx})=a^\dag(G^\mu_{\vx})+a(G^\mu_{\vx}).}
\end{equation}
It is symmetric and essentially self-adjoint there, and the same notation
will denote its self-adjoint closure.
We suppress the dependence of $G^\mu_{\vx}$ and $\Af^\mu(\vx)$ on the
fixed form factor $\wh\chi$.  A superscript $\chi$ on the representability
sets and universal functionals records this choice.
The field $\Af^\mu(\vx)$ maps each
$M$-photon sector into the direct sum of the $(M-1)$- and
$(M+1)$-photon sectors (with the negative sector omitted when $M=0$).

In the point-field notation of \cref{sec-fock-space}, with
$\vz=(\vk,\lambda)$, in the weak smeared sense on $\Fock_\ph^\fin$,
\begin{equation*}
 a^\dag(f)=\sum_{\lambda=1}^2\int_{\RR^3}
 f(\vk,\lambda)a^\dag(\vk,\lambda)\dd\vk,
 \qquad
 a(f)=\sum_{\lambda=1}^2\int_{\RR^3}
 \ol{f(\vk,\lambda)}a(\vk,\lambda)\dd\vk.
\end{equation*}
We may write \cref{Afield-def} as
\begin{align}
\boxed{ \AF(\vx)=\frac{1}{\sqrt{2}}
 \sum_{\lambda=1}^2\int_{\RR^3}\vepsilon_\lambda(\vk)
 \big[\wh\chi(-\vk)e^{-\i\vk\cdot\vx}a^\dag(\vk,\lambda)
 +\wh\chi(\vk)e^{\i\vk\cdot\vx}a(\vk,\lambda)\big]\dd\vk .}
 \label{Afield-momentum}
\end{align}
The integral in \cref{Afield-momentum} is an operator-valued distribution. On the electron--photon space the
notation $\Af^\mu(\vx_n)$ means operator-valued multiplication:
\begin{equation*}
 (\Af^\mu(\vx_n)\Psi)(\vX)
 =\Af^\mu(\vx_n)\Psi(\vX),
 \qquad \vX=(\vx_1,\ldots,\vx_N),
\end{equation*}
initially on the natural finite-particle form core.  The continuity and
uniform bounds above make this multiplication strongly measurable and
extend it to the domain used in the Pauli--Fierz form.
In the sense of operator-valued distributions we have the Coulomb gauge
\begin{equation*}
\boxed{\div\AF=0.}
\end{equation*}
More explicitly, for every $\eta\in C_c^\infty(\RR^3)$,
\begin{equation}\label{coulomb-gauge-weak}
 \sum_{\mu=1}^3\int_{\RR^3}
 (\partial_\mu\eta)(\vx)\Af^\mu(\vx)\dd\vx=0.
\end{equation}
This is first an identity on $\Fock_\ph^\fin$ and, by
\cref{ann-Hrad-bound,cre-Hrad-bound} below, also a form identity on $D(\sqrt{H_\rad})$.  It follows directly from
$\vk\cdot\vepsilon_\lambda(\vk)=0$.

\subsection{The free field and field bounds}\label{sec-free-field}

The free-field Hamiltonian is the second quantization of the dispersion,
\begin{equation*}
 H_\rad=\dd\varGamma(\omega)
 =\sum_{\lambda=1}^2\int_{\RR^3}|\vk|\,
 a^\dag(\vk,\lambda)a(\vk,\lambda)\dd\vk.
\end{equation*}
On the $M$-photon sector, $H_\rad$ is multiplication by
$\sum_{m=1}^M|\vk_m|$.  It is positive and self-adjoint,
\[
 \sigma(H_\rad)=[0,\infty),\qquad
 \ker H_\rad=\Span\{\Omega\},
\]
and its restriction to $\Omega^\perp$ has purely absolutely continuous
spectrum.

The standard creation--annihilation estimates imply, uniformly in $\vx$,
\begin{align}
 \|a(G^\mu_{\vx})\Psi\|&\leq
 \|G^\mu_{\vx}/\sqrt\omega\|\,\|\sqrt{H_\rad}\Psi\|,
 \label{ann-Hrad-bound}\\
 \|a^\dag(G^\mu_{\vx})\Psi\|&\leq
 \|G^\mu_{\vx}/\sqrt\omega\|\,\|\sqrt{H_\rad}\Psi\|
 +\|G^\mu_{\vx}\|\,\|\Psi\|.
 \label{cre-Hrad-bound}
\end{align}
The same estimates hold with $G^\mu_{\vx}$ replaced by any
$f\in\Hil_\ph$ such that $f/\sqrt\omega\in\Hil_\ph$, for every
$\Psi\in D(\sqrt{H_\rad})$.
Consequently, for a constant $C_\chi<\infty$ depending only on the form factor,
\begin{equation}\label{A-Hrad-bound}
 \|\AF(\vx)\Psi\|^2\leq C_\chi
 \big(\|\sqrt{H_\rad}\Psi\|^2+\|\Psi\|^2\big).
\end{equation}
These standard field estimates and the associated Pauli--Fierz form
construction are discussed, for example, in \cite{lieb2003existence}.

The transverse electric and magnetic fields are
\begin{equation}\label{EB-field-def}
 \EF_\perp(\vx)=-\i[H_\rad,\AF(\vx)],\qquad
 \BF(\vx)=\grad\times\AF(\vx),
\end{equation}
where the commutator and the curl are understood in the sense of
operator-valued distributions, that is, after smearing with test fields
as in \cref{coulomb-gauge-weak}.  The subscript records that $\EF_\perp$ is
only the divergence-free part of the electric field.  In Coulomb gauge
the longitudinal part is the instantaneous Coulomb field of the
charges, which manifests itself as the Coulomb interaction in the
Hamiltonian~\cite{spohn2004dynamics}.
More explicitly, let $\vu\in C_c^\infty(\RR^3;\RR^3)$ and
$g_\vu=\sum_{\mu=1}^3\int_{\RR^3}u^\mu(\vx)G^\mu_{\vx}\dd\vx$, so that
$\int_{\RR^3}\vu\cdot\AF=a^\dag(g_\vu)+a(g_\vu)$.  The Fourier
transform of $\vu$ decays rapidly, hence $\omega g_\vu\in\Hil_\ph$, and
\[
 \int_{\RR^3}\vu\cdot\EF_\perp
 =-\i\big(a^\dag(\omega g_\vu)-a(\omega g_\vu)\big),
 \qquad
 \int_{\RR^3}\vu\cdot\BF
 =\int_{\RR^3}(\grad\times\vu)\cdot\AF\,\dd\vx
\]
are self-adjoint Segal fields.  For a general form factor the fields need not
be defined pointwise.  If $\sqrt\omega\,\wh\chi\in L^2(\RR^3)$, the
estimates \cref{ann-Hrad-bound,cre-Hrad-bound} define $\EF_\perp(\vx)$ and
$\BF(\vx)$ as forms on $D(\sqrt{H_\rad})$ for every $\vx$.

Let us summarize the basic properties of these fields.  Identities between
fields at points are understood after smearing with test fields
$\vu,\vv\in C_c^\infty(\RR^3;\RR^3)$, as operator identities on
$\Fock_\ph^\fin$.  Both fields are transverse,
$\div\EF_\perp=\div\BF=0$. In fact, for $\EF_\perp$ the vector
$g_{\grad\eta}$ vanishes for every $\eta\in C_c^\infty(\RR^3)$, exactly as
in \cref{coulomb-gauge-weak}, and for $\BF$ this follows from
$\grad\times\grad\eta=0$.
The components of $\AF$ commute, as do those
of $\EF_\perp$, and the two fields form a canonical pair:
\begin{gather*}
 [\Af^\mu(\vx),\Af^\nu(\vy)]=[\Ef^\mu_\perp(\vx),\Ef^\nu_\perp(\vy)]=0,\\
 [\Af^\mu(\vx),\Ef^\nu_\perp(\vy)]
 =-\i\int_{\RR^3}\omega(\vk)|\wh\chi(\vk)|^2
   P_\sol(\vk)^{\mu\nu}e^{\i\vk\cdot(\vx-\vy)}\dd\vk.
\end{gather*}
Indeed, $[a^\dag(f)+a(f),a^\dag(h)+a(h)]=2\i\Im\dua{f}{h}$.  The scalar
products $\dua{g_\vu}{g_\vv}$ and $\dua{\omega g_\vu}{\omega g_\vv}$ are
real, since $\vu,\vv$ are real and $|\wh\chi|$, $\omega$ and $P_\sol$
are even in $\vk$, while
\[
 \Big[\int_{\RR^3}\vu\cdot\AF,\int_{\RR^3}\vv\cdot\EF_\perp\Big]
 =-2\i\Re\dua{g_\vu}{\omega g_\vv}.
\]
Commutators involving $\BF$ follow by taking the curl in the
corresponding variable.  In particular, the components of $\BF$
commute with those of $\AF$ and with each other.  For the hard cutoff \cref{harduv}, $\omega|\wh\chi_\Lambda|^2$
is the indicator function of the ball of radius $\Lambda$, so
\begin{equation}\label{hard-cutoff-commutator}
 [\Af^\mu(\vx),\Ef^\nu_\perp(\vy)]
 =-\i\int_{|\vk|\leq\Lambda}P_\sol(\vk)^{\mu\nu}
   e^{\i\vk\cdot(\vx-\vy)}\dd\vk.
\end{equation}
The right-hand side is $-\i(2\pi)^3$ times the transverse delta function
with momenta cut off at $\Lambda$.  The factor $(2\pi)^3$ reflects the
normalization of $\wh\chi$.

Since $\EF_\perp$ and $\BF$ carry the form factor, $H_\rad$ is not their
field energy.  Write ${:}\,\cdot\,{:}$ for normal ordering, which places
all creation operators to the left of all annihilation operators.
Inserting \cref{Afield-momentum}, the terms with two creation or two
annihilation operators cancel between the electric and magnetic parts,
and a formal computation gives
\[
 \frac12\int_{\RR^3}{:}\,|\EF_\perp|^2+|\BF|^2\,{:}
 =(2\pi)^3\dd\varGamma\big(\omega^2|\wh\chi|^2\big).
\]
This equals $H_\rad$ exactly when $|\wh\chi|=(2\pi)^{-3/2}\omega^{-1/2}$,
that is, without cutoff.

\subsection{The Pauli--Fierz Hamiltonian}\label{sec-PF-hamiltonian}

We first construct the internal Hamiltonian.  Put
\begin{equation}\label{PF-reference-form-domain}
 \mathcal Q_\PF:=\left\{\Psi\in\HC_N:
 \grad_{\vx_n}\Psi\in L^2(\RR^{3N};\Fock_\ph)^3\ (1\leq n\leq N),\quad
 \sqrt{H_\rad}\Psi\in\HC_N\right\}
\end{equation}
and equip it with the reference norm
\begin{equation}\label{PF-reference-form-norm}
 \|\Psi\|_{\mathcal Q_\PF}^2
 :=\|\Psi\|^2+\sum_{n=1}^N\|\grad_{\vx_n}\Psi\|^2
 +\|\sqrt{H_\rad}\Psi\|^2.
\end{equation}
Thus $\mathcal Q_\PF$ is the antisymmetric, Fock-space-valued electronic
$H^1$ space intersected with the form domain of the radiation energy.  The
algebraic tensor product
\begin{equation}\label{PF-algebraic-core}
 \mathscr C_\PF=
 \big(C_c^\infty(\RR^{3N})\cap\Hil_\el^N\big)
 \otimes_{\mathrm{alg}}
 \Fock_s^{\mathrm{alg}}\!\left(
 C_c^\infty\big((\RR^3\setminus\{0\})\times\{1,2\}\big)\right)
\end{equation}
is dense in $\mathcal Q_\PF$ for $\|\cdot\|_{\mathcal Q_\PF}$.  Here
$\Fock_s^{\mathrm{alg}}(\mathcal D)$ is the algebraic symmetric Fock space
over a space $\mathcal D$ of one-photon functions, that is, the finite
linear combinations of the vacuum and of symmetric tensor products of
finitely many elements of $\mathcal D$.
Fix $e\geq0$\footnote{The photon parity
$(-1)^\NC=\varGamma(-\iden)$ satisfies $(-1)^\NC a^\#(f)(-1)^\NC=-a^\#(f)$,
so it reverses the sign of $\AF(\vx)$, commutes with $H_\rad$ and acts
trivially on the electrons.  It therefore maps the Hamiltonians below at
coupling $e$ to those at $-e$ and preserves $\rho_\Psi$ and the combined
current density $\vjc_\Psi$ of \cref{jc-def}.  Hence the
representability sets and universal functionals are the same for $e$
and $-e$.} and set
\begin{equation*}
\boxed{ \vD_n=-\i\grad_{\vx_n}-e\AF(\vx_n).}
\end{equation*}
The internal Pauli--Fierz Hamiltonian is
\begin{equation}\label{q00-def}
\boxed{ H^{0,0}=\frac12\sum_{n=1}^N\vD_n^2
 +\sum_{1\leq n<m\leq N}\frac{1}{|\vx_n-\vx_m|}
 +H_\rad}
\end{equation}
on $\mathscr C_\PF$.  Expanding the square, the weak Coulomb gauge
\cref{coulomb-gauge-weak} gives
$\vD_n^2\Psi=-\Delta_{\vx_n}\Psi+2\i e\AF(\vx_n)\cdot\grad_{\vx_n}\Psi
+e^2|\AF(\vx_n)|^2\Psi$, so
\cref{q00-def} is a nonnegative symmetric operator on $\mathscr C_\PF$.
The field bound \cref{A-Hrad-bound} and
$-\i\grad_{\vx_n}=\vD_n+e\AF(\vx_n)$ control the ordinary and covariant
derivatives in terms of each other, and the Coulomb repulsion is
infinitesimally form bounded with respect to the electronic kinetic
energy~\cite[Theorem~X.16]{reed-simon-2}.  Hence there are constants $0<c\leq C<\infty$, depending on $N$,
$e$ and the cutoff, such that on $\mathscr C_\PF$,
\begin{equation}\label{PF-form-norm-equivalence}
 c\|\Psi\|_{\mathcal Q_\PF}^2
 \leq \dua{\Psi}{(H^{0,0}+1)\Psi}
 \leq C\|\Psi\|_{\mathcal Q_\PF}^2.
\end{equation}
From now on, $H^{0,0}$ denotes the Friedrichs extension
\cite[Chapter~3]{lewin2024spectral} of \cref{q00-def}.  Since $\mathscr C_\PF$ is dense in $\mathcal Q_\PF$,
\cref{PF-form-norm-equivalence} identifies its form domain:
\begin{equation}\label{PF-canonical-form-domain}
 Q(H^{0,0})=\mathcal Q_\PF.
\end{equation}
For a semibounded self-adjoint operator $H$ and $\Psi\in Q(H)$, we write
$\dua{\Psi}{H\Psi}$ for the value of its closed form at $\Psi$, even
if $\Psi\notin D(H)$.  Operator cores and operator domains are
studied in
\cite{hiroshima2002selfadjoint,hasler2008selfadjoint,matte2017paulifierz},
but only the form domain will be of relevance here.

For real external potentials satisfying
\begin{equation}\label{potential-spaces}
 v\in L^{3/2}(\RR^3)+L^\infty(\RR^3),\qquad
 \vA\in L^3(\RR^3,\RR^3)+L^\infty(\RR^3,\RR^3),
\end{equation}
the Hamiltonian is formally
\begin{equation*}
 \boxed{H^{v,\vA}
 =\sum_{n=1}^N\left[\frac12\big(\vD_n+\vA(\vx_n)\big)^2+v(\vx_n)\right]
 +\sum_{1\leq n<m\leq N}\frac{1}{|\vx_n-\vx_m|}+H_\rad.}
\end{equation*}
By the Sobolev inequality, $|\vA(\vx_n)|^2$ and $v(\vx_n)$
are infinitesimally form bounded with respect to $-\Delta_{\vx_n}$, hence
by \cref{PF-form-norm-equivalence} also with respect to $H^{0,0}$.
Cauchy--Schwarz controls the cross term
$\Re\dua{\vD_n\Psi}{\vA(\vx_n)\Psi}$.  The KLMN theorem~\cite[Theorem~X.17]{reed-simon-2} therefore defines
$H^{v,\vA}$ as a semibounded self-adjoint operator with
\begin{equation}\label{external-PF-form-domain}
 Q(H^{v,\vA})=\mathcal Q_\PF
\end{equation}
and shifted form norm equivalent to $\|\cdot\|_{\mathcal Q_\PF}$.

The bottom of the spectrum is
\begin{equation}\label{ground-energy-def}
 E(v,\vA)=\inf\sigma(H^{v,\vA})
 =\inf_{\substack{\Psi\in \mathcal Q_\PF\\ \|\Psi\|=1}} \dua{\Psi}{H^{v,\vA}\Psi},
\end{equation}
which is greater than $-\infty$ by the KLMN theorem.
We call $E(v,\vA)$ the ground-state energy even
when the infimum is not attained.  A ground state is a normalized minimizer,
equivalently an eigenvector at the bottom of the spectrum.

Mixed states of the electron--photon system are the density matrices on
$\HC_N$, that is, the operators $\vGamma\in\mathfrak S_1(\HC_N)$ with
$\vGamma\geq0$ and $\Tr\vGamma=1$.  Among these, the finite-energy ones
form the class
\begin{equation}\label{finite-energy-PF-ensembles}
 \dm=\left\{\vGamma\in\mathfrak S_1(\HC_N):
 \vGamma\geq0,\quad \Tr\vGamma=1,\quad
 \Tr(H^{0,0}\vGamma)<\infty\right\}.
\end{equation}
Here the energy trace is understood in the form sense: for a spectral
decomposition $\vGamma=\sum_rp_r|\Psi_r\rangle\langle\Psi_r|$ over the
positive eigenvalues we take
\[
 \Tr(H^{0,0}\vGamma)=\sum_rp_r\|\sqrt{H^{0,0}}\Psi_r\|^2\in[0,\infty],
\]
where the $r$th term is $+\infty$ unless $\Psi_r\in \mathcal Q_\PF$.  In
particular, every eigenvector of a $\vGamma\in\dm$ with positive
eigenvalue lies in $\mathcal Q_\PF$.  The elements of $\dm$ are the
finite-energy density matrices, often called ensembles in DFT.

Ground-state existence is a separate compactness and infrared question and
does not follow from self-adjointness.  For atoms and molecules satisfying
appropriate binding conditions, existence with massless photons, arbitrary
coupling, and a finite ultraviolet cutoff was established in
\cite{griesemer2001ground,lieb2003existence}.  Abstract existence criteria
for massless Pauli--Fierz Hamiltonians were given in
\cite{gerard2000ground}.  Conversely, translation-invariant models with
nonzero net charge can have no Fock-space ground state
\cite{hasler2008absence}.  Ionization thresholds and exponential
localization below them are analyzed in \cite{griesemer2004exponential}.
For particles bound by external potentials with a spatially cut off
interaction, ground states, localization, and resonance phenomena were
studied in \cite{bach1998confined}.  Finally, for abstract
massless, linearly coupled Pauli--Fierz models, local finiteness of the
point spectrum and a limiting absorption principle away from eigenvalues
are among the results of \cite{georgescu2004spectral}. 

\section{Main results}\label{sec-main-results}
In this section, we present our main results, which are proved
in \cref{sec-proofs}.

\subsection{The combined current density}

For $\Phi,\Psi\in\HC_N$, let $\dua{\Phi}{\Psi}_{\vx}$ denote the partial
inner product in every variable (including the photon inner product)
except the first electronic position $\vx=\vx_1$, and
$\|\Phi\|_{\vx}:=\dua{\Phi}{\Phi}_{\vx}^{1/2}$ the corresponding
partial norm.  For
$\Psi\in \mathcal Q_\PF$, define the \emph{electron density} $\rho_\Psi$, the
\emph{paramagnetic current density} $\vjp_\Psi$, the \emph{field-induced current
density} $\vjph_\Psi$, and the \emph{combined current density} $\vjc_\Psi$ by
\begin{align}
 \rho_\Psi(\vx)&=N\dua{\Psi}{\Psi}_{\vx},
 \label{density-def}\\
 \vjp_\Psi(\vx)&=N\Im\dua{\Psi}{\grad_{\vx}\Psi}_{\vx},
 \label{jp-def}\\
 \vjph_\Psi(\vx)&=-eN\Re\dua{\Psi}{\AF(\vx)\Psi}_{\vx},
 \label{jph-def}\\
 \vjc_\Psi(\vx)&=N\Re\dua{\Psi}{
 \big(-\i\grad_{\vx}-e\AF(\vx)\big)\Psi}_{\vx}
 =\vjp_\Psi(\vx)+\vjph_\Psi(\vx).
 \label{jc-def}
\end{align}
If $\Psi$ is normalized, then $\int_{\RR^3}\rho_\Psi=N$.
From the quadratic-form identity
\[
 \frac12(\vD_n+\vA)^2
 =\frac12\vD_n^2+\frac12(\vD_n\cdot\vA+\vA\cdot\vD_n)
 +\frac12|\vA|^2
\]
we obtain the decomposition
\begin{equation}\label{energy-decomposition}
\boxed{\dua{\Psi}{H^{v,\vA}\Psi}=\dua{\Psi}{H^{0,0}\Psi}
 +\dua{v+\tfrac12|\vA|^2}{\rho_\Psi}
 +\dua{\vA}{\vjc_\Psi}.}
\end{equation}
Equation~\eqref{energy-decomposition} shows that the external potentials couple to
the state through the density and combined current density.  We therefore
choose these as the basic variables of our density-functional formulation.

The following bounds show how finite energy controls the density and
current densities and determine the function spaces used below.  Recall
from \cref{sec-PF-hamiltonian} that
$\vD_n=-\i\grad_{\vx_n}-e\AF(\vx_n)$ is the covariant momentum of the
$n$th electron.  The constant $C_\chi$ is from the field bound
\cref{A-Hrad-bound}.

\begin{proposition}[A priori density--current bounds]
\label[proposition]{prop-current-bounds}
In the following estimates, quotients are defined to be
zero wherever the density vanishes.
For every normalized $\Psi\in \mathcal Q_\PF$, we have $\rho_\Psi\geq0$,
$\int_{\RR^3}\rho_\Psi=N$, $\sqrt{\rho_\Psi}\in H^1(\RR^3)$, and
\begin{equation}\label{joint-density-current-bound}
 \int_{\RR^3}\left(
 |\grad\sqrt{\rho_\Psi}|^2+
 \frac{|\vjc_\Psi|^2}{\rho_\Psi}\right)
 \leq\sum_{n=1}^N\|\vD_n\Psi\|^2.
\end{equation}
Moreover,
\begin{equation}\label{ph-current-bound}
 \int_{\RR^3}\frac{|\vjph_\Psi|^2}{\rho_\Psi}
 \leq e^2N\|\AF(\vx_1)\Psi\|^2
 \leq e^2NC_\chi\big(\|\sqrt{H_\rad}\Psi\|^2+1\big).
\end{equation}
In particular, $\rho_\Psi\in L^1(\RR^3)\cap L^3(\RR^3)$ and 
$\vjc_\Psi,\vjp_\Psi,\vjph_\Psi\in L^1(\RR^3)\cap L^{3/2}(\RR^3)$.
\end{proposition}
The same conclusions hold for every normalized finite-energy density
matrix $\vGamma=\sum_r p_r|\Psi_r\rangle\langle\Psi_r|\in\dm$, with
$\rho_\vGamma=\sum_r p_r\rho_{\Psi_r}$ and similarly for all three
current densities.  Each squared norm on the right-hand sides of the
estimates is replaced by the corresponding nonnegative form trace,
defined by spectral averaging, e.g.,
$\Tr(\vD_n^2\vGamma)=\sum_r p_r\|\vD_n\Psi_r\|^2$.

The Coulomb-gauge relation $\div\AF=0$ does \emph{not} imply that the
field-induced current density of an electron-photon entangled state is solenoidal.

\begin{proposition}[Divergence of the field-induced current density]
\label[proposition]{prop-div-jph}
For every $\Psi\in \mathcal Q_\PF$, in the sense of distributions,
\begin{equation}\label{div-jph}
 \div\vjph_\Psi(\vx)
 =-2eN\Re\sum_{\mu=1}^3
 \dua{\partial_\mu\Psi}{\Af^\mu(\vx)\Psi}_{\vx}.
\end{equation}
The right-hand side is integrable, and
\begin{equation}\label{div-jph-L1}
 \int_{\RR^3} \big|\div\vjph_\Psi\big|
 \leq 2eN\|\grad_{\vx_1}\Psi\|\,
 \|\AF(\vx_1)\Psi\|<\infty.
\end{equation}
Neither $\vjph_\Psi$ nor $\vjph_\Psi/\rho_\Psi$ is divergence-free in
general.
\end{proposition}

For a concrete one-electron example, let $\phi_0,\phi_1\in H^1(\RR^3)$,
let $\xi\in D(\sqrt\omega)$ be normalized, assume
$\|\phi_0\|^2+\|\phi_1\|^2=1$, and set $\Psi(\vx)=\phi_0(\vx)\Omega+\phi_1(\vx)a^\dag(\xi)\Omega$.
Then $\rho_\Psi=|\phi_0|^2+|\phi_1|^2$ and with $\vg_\xi(\vx)=\dua{\Omega}{\AF(\vx)a^\dag(\xi)\Omega}$ it holds
\begin{equation}\label{entangled-example}
 \vjph_\Psi=-2e\Re(\ol{\phi_0}\phi_1\vg_\xi).
\end{equation}
Although $\div\vg_\xi=0$, the right-hand side of
\cref{entangled-example} has nonzero divergence in general.

Recall the definition $\vj_\Psi=\vjc_\Psi+\rho_\Psi\vA$ of the physical current density.

\begin{proposition}[Fields in a ground state]
\label[proposition]{prop-ground-state-fields}
Let $\Psi$ be a ground state of $H^{v,\vA}$. Then
$\vj_\Psi\in L^1(\RR^3;\RR^3)\cap L^{3/2}(\RR^3;\RR^3)$, and
the following hold.
\begin{enumerate}[(i)]
\item The current density is divergence-free: $\div\vj_\Psi=0$ in the sense of
distributions.
\item The transverse electric field vanishes:
$\dua{\Psi}{\EF_\perp(\vx)\Psi}=0$.
\item The magnetic field satisfies the Amp\`ere law
\[
 \grad\times\dua{\Psi}{\BF(\vx)\Psi}
 =e\int_{\RR^3}\!\int_{\RR^3}\omega(\vk)|\wh\chi(\vk)|^2P_\sol(\vk)
   e^{\i\vk\cdot(\vx-\vy)}\vj_\Psi(\vy)\dd\vk\,\dd\vy.
\]
\item If $\vA=0$ and the ground state is nondegenerate, then $\vj_\Psi=0$
and $\dua{\Psi}{\AF(\vx)\Psi}=\dua{\Psi}{\BF(\vx)\Psi}=0$.
\end{enumerate}
\end{proposition}

Thus, the ground state carries the static field generated
by its own current density.  The transverse electric field vanishes, and the
magnetic field obeys the magnetostatic Amp\`ere law, with the
divergence-free current density $\vj_\Psi$ as its source and the kernel of the
commutator $[\AF,\EF_\perp]$ in place of the transverse delta function
(compare \cref{hard-cutoff-commutator}).

Items (i)--(iii) are well known at the formal level.  Since time
derivatives of expectation values vanish in an eigenstate, they are the
stationary forms of the continuity equation and of the Ehrenfest--Maxwell
equation for the vector potential, whose source is the current
density (see \cite[Eq.~(49), Sec.~IV.A and App.~C]{ruggenthaler2014quantum}).
Item (iv) is the usual time-reversal argument.  We record
\cref{prop-ground-state-fields}, with a variational proof under the
present hypotheses, for later comparison with the optimizers of the
constrained search in \cref{sec-optimizer-mean-fields}.

\subsection{\texorpdfstring{$N$-representability}{N-representability}}

A \emph{density--current pair} consists of a density and a current
density.  The main density variable is the pair $(\rho,\vjc)$ formed
with the combined current density, rather than a fixed
decomposition of $\vjc$ into electronic and photonic pieces.  The
pure-state $N$-representability class is then
\begin{equation*}
 \rcdens_N^{e,\chi}
 =\{(\rho_\Psi,\vjc_\Psi):
 \Psi\in \mathcal Q_\PF,\ \|\Psi\|=1\}.
\end{equation*}

The photon variables also turn a pure total state into a purification of a
mixed electronic state.  To make this precise, denote the finite-kinetic-energy
electronic mixed-state class by
\begin{equation}\label{electronic-mixed-class}
 \rcdens_{N,\mathrm{mix}}^{\el}
 =\left\{(\rho_\Gamma,\vjp_\Gamma):
 \begin{array}{l}
 \Gamma\in\mathfrak S_1(\Hil_\el^N),\;
 \Gamma\geq0,\;\Tr\Gamma=1,\;
 \Tr(-\lapl\Gamma)<\infty
 \end{array}\right\}.
\end{equation}
Here $\lapl$ is the Laplacian in all $N$ electronic coordinates.
Equivalently, by the mixed-state one-body density-matrix theorem~\cite{coleman1963structure}
(see \cite[Chapter~3]{lieb2010stability} for an elementary proof), a
pair $(\rho,\vj)$  is in
$\rcdens_{N,\mathrm{mix}}^{\el}$ if and only if there is a one-particle
density matrix $\gamma$ on $L^2(\RR^3)$ such that
\begin{equation}\label{electronic-1rdm-characterization}
 0\leq\gamma\leq1,\qquad \Tr\gamma=N,\qquad
 \Tr(-\Delta\gamma)<\infty,\qquad
 (\rho_\gamma,\vjp_\gamma)=(\rho,\vj).
\end{equation}
Equation \eqref{electronic-1rdm-characterization} is an exact operator
characterization, although it is not yet an intrinsic characterization
solely by elementary conditions on $(\rho,\vj)$.

\begin{theorem}[Photon purification and exact electronic projection]
\label{thm-photon-purification}
For every $e\geq0$,
\begin{equation}\label{electronic-projection}
 \left\{(\rho_\Psi,\vjp_\Psi):
 \Psi\in \mathcal Q_\PF,\ \|\Psi\|=1\right\}
 =\rcdens_{N,\mathrm{mix}}^{\el}.
\end{equation}
More sharply, the zero-field-current slice of the pure-state triple class
is exactly
\begin{equation}\label{zero-jph-slice}
 \left\{(\rho_\Psi,\vjp_\Psi,\vjph_\Psi):
 \Psi\in \mathcal Q_\PF,\ \|\Psi\|=1,\ \vjph_\Psi=0\right\}
 =\left\{(\rho,\vj,0):(\rho,\vj)
 \in\rcdens_{N,\mathrm{mix}}^{\el}\right\}.
\end{equation}
Thus the photon space allows one to purify every finite-energy electronic density matrix.
\end{theorem}

For the mixed-state statement below, recall the class $\dm$ of
finite-energy density matrices from \cref{finite-energy-PF-ensembles}.
For $\vGamma\in\dm$ with spectral
decomposition $\vGamma=\sum_rp_r|\Psi_r\rangle\langle\Psi_r|$ over
the positive eigenvalues, set
\begin{equation}\label{mixed-PF-density-current}
 \rho_\vGamma=\sum_rp_r\rho_{\Psi_r},\qquad
 \vjc_\vGamma=\sum_rp_r\vjc_{\Psi_r}.
\end{equation}

\begin{theorem}[Equivalence of pure and mixed $N$-representability]
\label{prop-spectator-purification}
Suppose that $\wh\chi=0$ almost everywhere on a measurable set
$S\subset\RR^3$ of positive measure.  Then
pure and mixed
$N$-representability coincide:
\begin{equation}\label{pure-mixed-PF-class}
 \rcdens_N^{e,\chi}
 =\left\{(\rho_\vGamma,\vjc_\vGamma):\vGamma\in\dm\right\}.
\end{equation}
In particular, $\rcdens_N^{e,\chi}$ is convex. 
\end{theorem}
Clearly, for the hard cutoff
\cref{harduv} with $0\leq\Lambda<\infty$, the hypothesis holds.

When an electronic representative can itself be chosen
pure, no photon purification is needed.  If $(\rho,\rho\vv)$ is
represented in ordinary paramagnetic CDFT by a finite-kinetic-energy
state $\Phi\in\Hil_\el^N$, then
$(\rho,\rho\vv)\in\rcdens_N^{e,\chi}$ for every $e\geq0$.
Indeed, we can simply take $\Psi=\Phi\otimes\Omega$.  Since
$\dua{\Omega}{\AF(\vx)\Omega}=0$, one has
$\rho_\Psi=\rho$, $\vjph_\Psi=0$, and
$\vjc_\Psi=\vjp_\Phi=\rho\vv$.  The field bound implies
$\dua{\Psi}{H^{0,0}\Psi}<\infty$, since the ultraviolet cutoff makes the
vacuum fluctuation in $|\AF|^2$ finite.

An intrinsic sufficient condition follows from a coherent displacement
that depends on the electronic positions.  The resulting
conditional field can be nontransverse even though the field operator is
transverse. 

\begin{theorem}[Photonic representability]
\label{thm-photonic-representability}
Let $N\in\NN$ and $e\geq0$, and suppose $\wh\chi\not\equiv0$.  Let
\[
 \rho\geq0,\qquad \int_{\RR^3}\rho=N,\qquad
 \sqrt\rho\in H^1(\RR^3),
\]
and let $\va,\vb\in H^1_{\mathrm{loc}}(\RR^3;\RR^3)$ and
$S\in H^1_{\mathrm{loc}}(\RR^3;\RR)$ satisfy
\begin{equation*}
 \int_{\RR^3}\rho\bigl(|\va|^2+|\grad\va|^2+|\vb|^2+|\grad\vb|^2
   +|\grad S|^2\bigr)<\infty,
\end{equation*}
where $\grad\va$ and $\grad\vb$ are weak Jacobians.  Then there is a
normalized pure state $\Psi\in \mathcal Q_\PF$ such that
\begin{equation}\label{photonic-representability-triple}
 \rho_\Psi=\rho,\qquad \vjp_\Psi=\rho(\grad S+\vb),\qquad
 \vjph_\Psi=-e\rho\va,
\end{equation}
and
\begin{equation}\label{photonic-representability-energy}
 \begin{aligned}
 &\sum_{n=1}^N\|\grad_{\vx_n}\Psi\|^2+\dua{\Psi}{H_\rad\Psi}\\
 &\quad\leq C_{N,\chi}\left(1+\int_{\RR^3}|\grad\sqrt\rho|^2
   +\int_{\RR^3}\rho\bigl(|\va|^2+|\grad\va|^2+|\vb|^2+|\grad\vb|^2
   +|\grad S|^2\bigr)\right),
 \end{aligned}
\end{equation}
with a constant independent of $\rho$, $\va$, $\vb$ and $S$.
\end{theorem}

In particular, every $\va\in W^{1,\infty}(\RR^3;\RR^3)$ is
allowed for every density in the theorem.  The paramagnetic velocity
$\vb$ is subject only to the same weighted Sobolev condition and need be
neither bounded nor curl-free.  For instance, the rigid rotation
$\vb=\vomega\times\vx$ is allowed whenever $\int\rho|\vx|^2<\infty$.
The local Sobolev condition
asks for a velocity defined on the whole space, but no positive lower
bound on $\rho$ is required.  Form equivalence
\cref{PF-form-norm-equivalence} also bounds the internal energy by the
right-hand side of \cref{photonic-representability-energy}, with a
constant depending additionally on $e$.

\begin{corollary}[A Sobolev class of combined current densities]
\label[corollary]{cor-photonic-combined-class}
Suppose $e>0$ and $\wh\chi\not\equiv0$.  For every $N\in\NN$,
\[
 \left\{(\rho,\rho\vv):
 \begin{array}{l}
 \rho\geq0,\quad\int_{\RR^3}\rho=N,\quad\sqrt\rho\in H^1,\\
 \vv\in H^1_{\mathrm{loc}}(\RR^3;\RR^3),\quad
 \displaystyle\int_{\RR^3}\rho\bigl(|\vv|^2+|\grad\vv|^2\bigr)<\infty
 \end{array}\right\}
 \subset\rcdens_N^{e,\chi}.
\]
The same conclusion holds for velocities $\grad S+\vw$ with
$S\in H^1_{\mathrm{loc}}(\RR^3;\RR)$,
$\vw\in H^1_{\mathrm{loc}}(\RR^3;\RR^3)$, and
\[
 \int_{\RR^3}\rho\bigl(|\grad S|^2+|\vw|^2+|\grad\vw|^2\bigr)<\infty.
\]
\end{corollary}
\begin{proof}
Apply \cref{thm-photonic-representability} with $\va=-\vv/e$, $S=0$
and $\vb=0$.  For the second assertion use $\va=-\vw/e$, the given
$S$, and $\vb=0$.
Then \cref{jc-def} gives the prescribed combined current density.
\end{proof}

\subsection{Universal density functionals}

By \cref{prop-current-bounds}, every $N$-representable pair $(\rho,\vjc)$ belongs to the
Banach space $X\times Y$, where
\begin{equation}\label{density-current-spaces}
 X=L^1(\RR^3)\cap L^3(\RR^3),\qquad
 Y=L^1(\RR^3,\RR^3)\cap L^{3/2}(\RR^3,\RR^3),
\end{equation}
with the max norms
$\|\rho\|_X=\max\{\|\rho\|_1,\|\rho\|_3\}$ and
$\|\vjc\|_Y=\max\{\|\vjc\|_1,\|\vjc\|_{3/2}\}$.
Their duals are $X^*=L^\infty(\RR^3)+L^{3/2}(\RR^3)$ and
$Y^*=L^\infty(\RR^3,\RR^3)+L^3(\RR^3,\RR^3)$, with dual norms given by
$\|v\|_{X^*}=\inf_{v=v_\infty+v_{3/2}}
\big(\|v_\infty\|_\infty+\|v_{3/2}\|_{3/2}\big)$ and
$\|\vA\|_{Y^*}=\inf_{\vA=\vA_\infty+\vA_3}
\big(\|\vA_\infty\|_\infty+\|\vA_3\|_3\big)$,
where the infima run over decompositions in the respective summand spaces.
The same spaces were considered for CDFT in \cite{MY-CDFTpaper2019}.

For $(\rho,\vjc)\in X\times Y$, define the \emph{Pauli--Fierz Levy--Lieb universal functional}
\begin{equation}\label{FLL-def}
\boxed{ F_\LL^{e,\chi}(\rho,\vjc)
 :=\inf_{\substack{\Psi\in \mathcal Q_\PF,\,\|\Psi\|=1\\
 \rho_\Psi=\rho\\ \vjc_\Psi=\vjc}}
 \dua{\Psi}{H^{0,0}\Psi}},
\end{equation}
with value $+\infty$ if the constraint set is empty.  The variational principle implies
\begin{equation}\label{density-variational-principle}
 E(v,\vA)=\inf_{(\rho,\vjc)\in X\times Y}
 \left[F_\LL^{e,\chi}(\rho,\vjc)
 +\dua{v+\tfrac12|\vA|^2}{\rho}+\dua{\vA}{\vjc}\right].
\end{equation}
The quadratic diamagnetic term $\tfrac12\dua{|\vA|^2}{\rho}$ means
that $E(v,\vA)$ need not be jointly concave in $(v,\vA)$.
To obtain a jointly concave energy, we follow \cite{tellgren2012choice}
and absorb this term into the scalar potential by setting
$u=v+\frac{1}{2}|\vA|^2$, and define
\begin{equation*}
 \bar E(u,\vA)=E(u-\frac{1}{2}|\vA|^2,\vA).
\end{equation*}
For each fixed $N$-representable pair $(\rho,\vjc)$, the expression inside
the infimum in \cref{density-variational-principle} is then affine in
$(u,\vA)$.  Thus $\bar E$ is a pointwise infimum of affine functions and
is jointly concave, providing the basis for the following convex-duality
construction.
The \emph{Pauli--Fierz Lieb universal functional} is then defined as the convex conjugate
\begin{equation}\label{FL-def}
 F_\LI^{e,\chi}(\rho,\vjc)
 =\sup_{(u,\vA)\in X^*\times Y^*}
 \big[\bar E(u,\vA)-\dua{u}{\rho}-\dua{\vA}{\vjc}\big].
\end{equation}
It is thus convex and lower semicontinuous in the norm topology of $X\times Y$,
and biconjugation gives
\begin{equation}\label{L-vs-LL}
 F_\LI^{e,\chi}=\ol{\operatorname{conv}}F_\LL^{e,\chi}
 \leq F_\LL^{e,\chi}.
\end{equation}
Here $\ol{\operatorname{conv}}$ denotes the norm-lower-semicontinuous convex
envelope.  Equivalently, it is the
weak-lower-semicontinuous convex envelope.
For comparison, recall the class $\dm$ of finite-energy density matrices from
\cref{finite-energy-PF-ensembles} and define the mixed-state functional
\begin{equation}\label{FLcs-def}
 \boxed{\wt{F}_\LI^{e,\chi}(\rho,\vjc)
 =\inf_{\substack{\vGamma\in\dm\\
 \rho_\vGamma=\rho\\ \vjc_\vGamma=\vjc}}
 \Tr(H^{0,0}\vGamma)}.
\end{equation}
Directly from the variational principle we have
\begin{equation}\label{L-cs-relations}
 F_\LI^{e,\chi}\leq \wt{F}_\LI^{e,\chi}
 \leq F_\LL^{e,\chi}.
\end{equation}
As in the case of ordinary DFT and CDFT, equality of the first two functionals
requires a separate lower-semicontinuity and compactness argument, supplied by
\cref{thm-unrestricted-attainment} below.
A closely related result is attainment of the defining infimum of $\wt{F}_\LI^{e,\chi}$,
also called expectation-valuedness of the functional.

\subsection{Existence and nonexistence of optimizers}
\label{sec-unrestricted-attainment}

We now establish attainment of the mixed-state constrained search and
identify it with the Lieb functional.  The main difficulty comes
from the possible occurrence of soft photons: an energy bound alone does not control photon number, since
a minimizing sequence may contain arbitrarily many photons of
arbitrarily small frequency.

\begin{theorem}[Existence of a PFL optimizer]
\label{thm-unrestricted-attainment}
Whenever finite, the mixed-state functional
$\wt{F}_\LI^{e,\chi}(\rho,\vjc)$ has an optimizer.  Every optimizer
$\vGamma$ has finite mean photon number, namely
\[
 \Tr(\NC\vGamma)\leq C_{N,e,\chi}
 \big(1+\wt{F}_\LI^{e,\chi}(\rho,\vjc)\big),
\]
with a constant $C_{N,e,\chi}$ depending only on $N$, $e$ and $\chi$.
The functional $\wt{F}_\LI^{e,\chi}(\rho,\vjc)$ is norm lower semicontinuous on $X\times Y$,
and
\begin{equation}\label{unrestricted-Lieb-equality}
 \boxed{\wt{F}_\LI^{e,\chi}(\rho,\vjc)=F_\LI^{e,\chi}(\rho,\vjc)}
\end{equation}
for all $(\rho,\vjc)\in X\times Y$.

\end{theorem}
The proof removes soft photons and restores the combined current density
exactly, at a controlled energy cost.  The electronic current shift
of Appendix~\ref{app-current-shift} preserves the density, fermionic
antisymmetry, and the remaining photon marginal.  Applied to a minimizing sequence, it yields a
uniform bound on mean photon number.  Photon localization and passage
to the limit then give a mixed-state minimizer.  The same estimates
establish norm lower semicontinuity and hence equality with the Lieb
functional.

However, pure-state attainment fails in general even for $N=1$.

\begin{proposition}[Nonexistence of a PFLL optimizer]
\label[proposition]{prop-pure-nonattainment}
Let $N=1$ and fix the form factor.
For every $e\geq0$, there is a density--current pair
$(\rho_e,\vj_e)\in\rcdens_1^{e,\chi}$ such that
$F_\LL^{e,\chi}(\rho_e,\vj_e)<\infty$ and its infimum is not
attained.  The pair may depend on both $e$ and the fixed form factor.
In particular, this holds for the standard hard ultraviolet cutoff
\cref{harduv} at every $e\geq0$ and every finite $\Lambda$.
\end{proposition}
The density $\rho_e$ is supported in a thin solid torus around a long,
narrow ellipse, and the velocity $\vj_e/\rho_e$ has circulation $\pi$ around
each loop of the torus, half the circulation quantum.  A single orbital
cannot carry this velocity, since the phase of a nonvanishing
wavefunction winds by a multiple of $2\pi$, but the equal mixture of two
orbitals with winding numbers $0$ and $1$ can.  Attaching orthogonal
soft-photon states to the two orbitals turns this mixture into a pure
state with the same density and combined current density, at an energy cost that can be made arbitrarily
small but not zero.  To exclude a pure minimizer, we compare a lower
bound for pure states with an upper bound from mixed states.  As the
ellipse is narrow, the field has small circulation around the loops, and
the two bounds force a pure minimizer to be almost photon-free.  Its
vacuum component is then bounded away from zero along some loop, so its
phase winds by a multiple of $2\pi$, which contradicts the circulation
$\pi$.

An obvious sufficient condition for pure-state attainment is the existence of
a pure ground state with the prescribed density-current pair ($v$-representability).
In fact, a pure ground state
of $H^{v,\vA}$, with potentials in \cref{potential-spaces}, minimizes
the PFLL constrained search at its own density--current pair: the
external terms in \cref{energy-decomposition} are the same for all
states with that pair, so lowering the internal energy would contradict
the ground-state property.  In particular, the PFLL infimum at the density--current pair of a
ground-state density matrix is attained whenever the corresponding ground state is
non-degenerate, since every ground-state density matrix
is then the rank-one projector onto that ground state.

\subsection{Soft-photon purification}
\label{sec-many-electron-soft-purification}

Arbitrarily soft photons allow a mixture to be represented by a pure
electron--photon state at arbitrarily small energy cost, while the density
and combined current density remain unchanged.

\begin{theorem}[Soft-photon purification and equality of functionals]
\label{thm-many-electron-soft-purification}
Fix $N\in\NN$ and $e\geq0$.  For every $\vGamma\in\dm$ and every
$\delta>0$, there is a normalized $\Psi\in \mathcal Q_\PF$ with finite
mean photon number such that
\begin{equation}\label{many-electron-pure-approximation}
 (\rho_\Psi,\vjc_\Psi)=(\rho_\vGamma,\vjc_\vGamma),\qquad
 \dua{\Psi}{H^{0,0}\Psi}\leq\Tr(H^{0,0}\vGamma)+\delta.
\end{equation}
Consequently,
\begin{equation}\label{many-electron-pure-ensemble-equality}
 \boxed{\quad F_\LL^{e,\chi}(\rho,\vjc)
 =F_\LI^{e,\chi}(\rho,\vjc)\quad}
\end{equation}
for all $(\rho,\vjc)\in X\times Y$.
If this value is finite at $(\rho,\vjc)$, the pure-state infimum is
attained if and only if the total Pauli--Fierz mixed-state minimization
has a rank-one minimizer.
\end{theorem}

This does not contradict Levy's and Lieb's nonconvexity result for the electronic
Levy--Lieb functional \cite{levy1982,Lieb1983}, since a pure electron--photon
state can have a mixed electronic marginal and, already at zero
coupling, arbitrarily soft photons purify electronic mixed states at
arbitrarily small energy cost.

\begin{corollary}[Convexity, $N$-representability, and pure minimizing sequences]
\label[corollary]{cor-many-electron-pure-functional}
For every $N\in\NN$ and $e\geq0$,
$F_\LL^{e,\chi}$ is proper, convex, and norm lower semicontinuous
on $X\times Y$.  Moreover,
\begin{equation}\label{unrestricted-pure-mixed-PF-class}
 \boxed{\rcdens_N^{e,\chi}
 =\{(\rho_\vGamma,\vjc_\vGamma):\vGamma\in\dm\}
 =\operatorname{dom}F_\LI^{e,\chi}.}
\end{equation}
In particular, the pure-state $N$-representability class is convex.
Let $(\rho,\vjc)\in X\times Y$ with
$F_\LI^{e,\chi}(\rho,\vjc)<\infty$, and let $\vGamma_*\in\dm$ be a
minimizer in \cref{FLcs-def}, so that
$(\rho_{\vGamma_*},\vjc_{\vGamma_*})=(\rho,\vjc)$.  Then there are
normalized $\Psi_j\in\mathcal Q_\PF$ with
\[
 (\rho_{\Psi_j},\vjc_{\Psi_j})=(\rho,\vjc),\qquad
 \dua{\Psi_j}{H^{0,0}\Psi_j}\longrightarrow F_\LL^{e,\chi}(\rho,\vjc),
\]
and
\begin{equation}\label{pure-minimizing-sequence-number-bound}
 \sup_j\dua{\Psi_j}{\NC\Psi_j}
 \leq 1+\Tr(\NC\vGamma_*)
 \leq 1+C_{N,e,\chi}
       \big(1+F_\LI^{e,\chi}(\rho,\vjc)\big).
\end{equation}
\end{corollary}

This result extends \cref{prop-spectator-purification} to general form factors.
Equality of the infima and the bound
\cref{pure-minimizing-sequence-number-bound} do not imply pure-state
attainment: the soft photons can still prevent the necessary compactness.
The construction uses the infinitely many continuum modes of
arbitrarily small positive frequency.  In contrast to \cref{prop-spectator-purification}, 
it requires no region on which the coupling vanishes, so it also strengthens
the representability conclusion of that previous result.

At zero combined current density, both functionals reduce to searches
constrained by the density alone.

\begin{proposition}[Vanishing combined current density]
\label[proposition]{prop-zero-current-density-functional}
For $\sqrt{\rho}\in H^1(\RR^3)$ with $\rho\ge0$ and $\int_{\RR^3}\rho=N$, the following relations hold true:
\begin{align}
 F_\LL^{e,\chi}(\rho,0)
 &=\inf_{\substack{\Psi\in\mathcal Q_\PF,\ \|\Psi\|=1\\
                   \rho_\Psi=\rho}}
       \dua{\Psi}{H^{0,0}\Psi},
 \label{zero-current-pure-density-search}\\
 F_\LI^{e,\chi}(\rho,0)
 &=\min_{\substack{\vGamma\in\dm\\\rho_\vGamma=\rho}}
       \Tr(H^{0,0}\vGamma)
 \label{zero-current-density-search}\\
 &=\sup_{v\in X^*}\big[E(v,0)-\dua{v}{\rho}\big].
 \label{zero-current-scalar-dual}
\end{align}
\end{proposition}
Using \cref{thm-many-electron-soft-purification}, the above two zero-current functionals coincide.
Note that the current constraint can be
dropped from both constrained searches.

These functionals still include the quantized field and its coupling to the
electrons.  At nonzero coupling, their common value generally differs
from the purely electronic Lieb functional $F_{\mathrm{L}}(\rho)=F_\VR(\rho,0)$ recalled below
in \cref{sec-zero-coupling}.  The difference is a nontrivial variational
contribution arising from electronic and photonic relaxation and their
correlations at fixed density.  The vacuum shift
$e^2N\|\wh\chi\|^2/2$ gives an upper bound on this difference
(\cref{thm-zero-coupling-convergence}).

\subsection{Properties of optimizers}
\label{sec-optimizer-properties}

By \cref{thm-unrestricted-attainment}, the constrained search has an
optimizer whenever its value is finite.  In this section we describe
such optimizers: their fields, their symmetries and dark modes,
and how many soft photons they contain.  By dark modes we mean photon
modes on which $\wh\chi$ vanishes; these do not couple to the
electrons.  In each proof we compare an optimizer with other states
that have the same density and combined current density, so we never
assume that the optimizer is a ground state of some $H^{v,\vA}$.

\subsubsection{Fields of constrained optimizers}
\label{sec-optimizer-mean-fields}

We now ask which conclusions of \cref{prop-ground-state-fields} require
a ground state, and which follow from constrained minimization alone.
Here an optimizer means any $\vGamma\in\dm$ satisfying
\[
 \rho_\vGamma=\rho,\qquad \vjc_\vGamma=\vjc,\qquad
 \Tr(H^{0,0}\vGamma)=F_\LI^{e,\chi}(\rho,\vjc)<\infty.
\]
There need not be supporting external potentials (no $v$-representability required).
In their absence, the prescribed current is $\vjc$, whereas a physical current
$\vj=\vjc+\rho\vA$ requires a specified external vector potential.
All traces below are understood in the form sense, with the fields
smeared as in Section~\ref{sec-free-field}.

\begin{proposition}[Fields of optimizers]
\label[proposition]{prop-optimizer-mean-fields}
Let $\vGamma$ be a PFL optimizer at $(\rho,\vjc)$.  The following properties hold true.
\begin{enumerate}[(i)]
\item For real $\vu_1,\ldots,\vu_n\in C_c^\infty(\RR^3;\RR^3)$ and every
real $\phi\in\mathcal S(\RR^n)$,
\begin{equation}\label{optimizer-field-statistics}
 \sum_{j=1}^n\Re\Tr\Big(\Big[(\partial_j\phi)\Big(\int_{\RR^3}\vu_1\cdot\AF,\ldots,
   \int_{\RR^3}\vu_n\cdot\AF\Big)\Big]\Big[\int_{\RR^3}\vu_j\cdot\EF_\perp\Big]\vGamma\Big)=0.
\end{equation}
\item In particular, the transverse electric field vanishes,
$\Tr(\EF_\perp\vGamma)=0$ in the sense of distributions. This means that
\begin{equation}\label{optimizer-mean-electric-zero}
 \Tr\Big(\Big[\int_{\RR^3}\vu\cdot\EF_\perp\Big]\vGamma\Big)=0
 \qquad\text{for every }\vu\in C_c^\infty(\RR^3;\RR^3).
\end{equation}
\item If the prescribed current has the form
\[
 \vjc=\rho\grad S,\qquad
 S\in W^{1,1}_{\rm loc}(\RR^3;\RR),\qquad
 \grad S\in L^3(\RR^3;\RR^3)+L^\infty(\RR^3;\RR^3),
\]
then every coherent photon amplitude vanishes:
$\Tr(a(h)\vGamma)=0$ for all $h\in\Hil_\ph$ with
$h/\sqrt\omega\in\Hil_\ph$.  In particular, the vector potential
vanishes pointwise and the magnetic field vanishes in the sense
of distributions:
\begin{equation*}
 \Tr(\AF(\vx)\vGamma)=0\quad(\vx\in\RR^3),
 \qquad \Tr(\BF\vGamma)=0.
\end{equation*}
\end{enumerate}
\end{proposition}

Item (i) extends \cref{optimizer-mean-electric-zero}, its limiting
linear case, from the expectation value to all equal-time statistics of the smeared
transverse field.  The kinetic, Coulomb, and external
terms of $H^{v,\vA}$ commute with
$\phi(\int\vu_1\cdot\AF,\ldots,\int\vu_n\cdot\AF)$.  Formally, the
left-hand side of \cref{optimizer-field-statistics} is therefore minus
the time derivative at $t=0$ of
$\Tr\big(\phi(\int\vu_1\cdot\AF,\ldots,\int\vu_n\cdot\AF)\,
e^{-\i tH^{v,\vA}}\vGamma e^{\i tH^{v,\vA}}\big)$: under the full
dynamics, every such expectation value is stationary to first order at an
optimizer.  In a Schr\"odinger representation of the field, in which the
smeared fields act as coordinates (Appendix~\ref{app-soft-gaussian}
with $\mathfrak h=\Hil_\ph$), \cref{optimizer-field-statistics}
says that the probability current in field configuration space is
weakly divergence-free.  It is thus the field-space counterpart of
\cref{prop-ground-state-fields}(i).

The field variations in the proof of \cref{prop-optimizer-mean-fields}
preserve both constraints on the density and the combined current density.
By contrast, an electronic phase changes $\vjc$ by $\rho\grad S$,
and a general photon displacement changes it by $-e\rho\vc_f$,
with $\vc_f$ as in \cref{weyl-classical-field}.
Consequently, the variations proving continuity and the Amp\`ere law
for ground states are not generally available at a fixed pair.
The next proposition shows that a constrained optimizer can have a
combined current density with nonzero divergence at positive coupling.
It also shows that even a unique optimizer need not be a ground state
for any external potentials in \cref{potential-spaces}.

\begin{proposition}[Ground-state conclusions that fail for optimizers]
\label[proposition]{prop-optimizer-counterexamples}
Let $N=1$, and set
\[
 \rho(\vx)=\pi^{-3/2}e^{-|\vx|^2},\qquad
 \vjc(\vx)=\rho(\vx)\vq,\qquad 0\ne\vq\in\RR^3.
\]
\begin{enumerate}[(i)]
\item Let $e>0$, and let $\wh\chi$ satisfy \cref{cutoff-assumption}.
The pair $(\rho,\vjc)$ has a PFL optimizer, and every such optimizer
satisfies
\[
 \Tr(\AF\vGamma)
 =\Tr(\BF\vGamma)
 =\Tr(\EF_\perp\vGamma)=0,
 \qquad
 \div\vjc=-2(\vq\cdot\vx)\rho\ne0.
\]
The vector-potential identity holds pointwise, and the magnetic and
electric field expectation values hold in the sense of distributions.
\item Let $e=0$.  The unique PFL optimizer at $(\rho,\vjc)$ is
\begin{equation}\label{unique-gaussian-optimizer}
 \begin{gathered}
 \vGamma_\vq=|\Psi_\vq\rangle\langle\Psi_\vq|,
 \qquad
 \Psi_\vq(\vx)=e^{\i\vq\cdot\vx}\pi^{-3/4}e^{-|\vx|^2/2}\Omega,
 \\
 F_\LI^{0,\chi}(\rho,\rho\vq)
       =\frac34+\frac{|\vq|^2}{2}.
 \end{gathered}
\end{equation}
It is not a ground state, or even an eigenstate, of any
$H^{v,\vA}$ with potentials in \cref{potential-spaces}.
\end{enumerate}
\end{proposition}

Part (i) shows that continuity and Amp\`ere's law are not consequences
of constrained optimality. Indeed, since it involves the physical current $\vjc + \rho \vA$, Amp\`ere's law is most natural in a situation where the external potential $\vA$ is given. By part (ii), uniqueness of a constrained optimizer cannot replace the ground-state assumptions in \cref{prop-ground-state-fields}(iv): the unique constrained optimizer can have nonzero current.

As is made clearer in the proof below (in \cref{proof:prop-optimizer-counterexamples}), the restriction on the potentials in \cref{potential-spaces} plays an essential
role. For example, in the case $e=0$, the given density-current pair does come from the ground state of an $H^{v,\vA}$ with $v(\vx) = \frac{1}{2} |\vx|^2 \not\in L^{3/2}(\RR^3) + L^{\infty}(\RR^3)$ and $\vA(\vx) = -\vq$.

\subsubsection{Symmetry and dark modes}
\label{sec-optimizer-symmetry}

For $\vR\in{\rm O}(3)$, define in the solenoidal representation of the
one-photon space the rotation $(\Upsilon_\vR\vf)(\vk)=\vR\vf(\vR^\top\vk)$, and put
\[
 (U_\vR\Psi)(\vX)
 =\Upsilon_\vR\Psi(\vR^\top\vx_1,\ldots,\vR^\top\vx_N)
\]
on $\HC_N$, as in \cref{prop-orthogonal-covariance}.  Let $\Theta$ be
the time reversal \cref{time-reversal-def}.  For
$g=(\vR,s)\in{\rm O}(3)\times\{1,-1\}$, put
\[
 V_g=\begin{cases}
 U_\vR, & s=1,\\
 U_\vR\Theta, & s=-1.
 \end{cases}
\]

\begin{proposition}[Symmetry and dark modes of optimizers]
\label[proposition]{prop-optimizer-symmetry}
Let $F_\LI^{e,\chi}(\rho,\vjc)<\infty$.
\begin{enumerate}[(i)]
\item Use the hard cutoff \cref{harduv}, and let
$G\subset{\rm O}(3)\times\{1,-1\}$ be a closed subgroup such that
\[
 \rho(\vR^\top\vx)=\rho(\vx),\qquad
 s\vR\vjc(\vR^\top\vx)=\vjc(\vx)
 \qquad((\vR,s)\in G).
\]
Then $\vGamma\mapsto V_g\vGamma V_g^{-1}$, $g\in G$, maps the set of
PFL optimizers at $(\rho,\vjc)$ onto itself, and this set contains a
$G$-invariant optimizer.
\item Suppose that $\wh\chi=0$ almost everywhere on a measurable set
$S\subset\RR^3$.  Let $\Hil_S\subset\Hil_\ph$ consist of the
one-photon states supported in $S$, and factor
$\Fock_\ph\simeq\Fock_s(\Hil_S^\perp)\otimes\Fock_s(\Hil_S)$.  Then
every PFL optimizer at $(\rho,\vjc)$ has the form
$\vGamma'\otimes|\Omega_S\rangle\langle\Omega_S|$, where $\Omega_S$ is
the vacuum of the second factor.
\end{enumerate}
\end{proposition}

For instance, if $\rho$ is radial and $\vjc=0$, part (i) with
$G={\rm O}(3)\times\{1,-1\}$ gives an optimizer invariant under all
rotations, reflections, and time reversal.  The group
$G=\{(\iden,\pm1)\}$ requires only $\vjc=0$, with no symmetry of $\rho$.
Its Haar average is $\frac12(\vGamma+\Theta\vGamma\Theta^{-1})$, so
every optimizer $\vGamma$ at $(\rho,0)$ yields a time-reversal-invariant
one.  If $\rho$ and $\vjc$ are invariant under
rotations about an axis, as for a circulating current, there is an
optimizer invariant under these rotations even though time reversal
is broken.  For the hard cutoff, part (ii) applies with
$S=\{|\vk|>\Lambda\}$: optimizers carry no photons above the cutoff.

\subsubsection{Soft-photon energy distributions and infrared laws}
\label{sec-soft-photon-spectrum}

We describe photon energy distributions of constrained optimizers and
their infrared behavior.  Here photon energy and
frequency are identified by $\omega(\vk)=|\vk|$, in the units
$\hbar=c=1$ used above.

Fix a density--current pair $(\rho,\vjc)$ with $F_\LI^{e,\chi}(\rho,\vjc)<\infty$, and let
$\vGamma_*$ be any mixed-state optimizer supplied by
\cref{thm-unrestricted-attainment}.  For a Borel set
$B\subset(0,\infty)$, define the photon-number and photon-energy
measures
\begin{align*}
 \nu_*(B)&=\Tr\!\left(\dd\varGamma(\iden_{\{|\vk|\in B\}})
                         \vGamma_*\right),
 \\
 \mu_*(B)&=\Tr\!\left(\dd\varGamma(|\vk|\iden_{\{|\vk|\in B\}})
                         \vGamma_*\right)
          =\int_B\varepsilon\,\nu_*(\dd\varepsilon).
\end{align*}
Both are finite.  They describe expected photon counts and expected
energy in frequency bands, but neither is a probability measure in general.
The total mass of $\mu_*$ is the expected total radiation energy,
$\mu_*((0,\infty))=\Tr(H_\rad\vGamma_*)$.
By contrast, $\Tr(\iden_B(H_\rad)\vGamma_*)$ is the probability that
the \emph{total photon energy} lies in $B$.

There is also a density formula.  Let $\gamma_*^{(1)}$ be the
one-photon reduced density operator, characterized by
\[
 \Tr(T\gamma_*^{(1)})=\Tr(\dd\varGamma(T)\vGamma_*)
\]
for bounded one-photon operators $T$.  It is positive semidefinite and
since its trace gives the mean photon number, it is also trace class.  Write
$\gamma_*^{(1)}=\sum_j s_j|f_j\rangle\langle f_j|$, with
$s_j\geq0$ and $(f_j)$ orthonormal.  The momentum-space photon-number
density is defined almost everywhere by
\[
 n_*(\vk,\lambda)=\sum_j s_j|f_j(\vk,\lambda)|^2.
\]
Polar coordinates give absolutely continuous measures
\begin{equation*}
 \begin{aligned}
 \nu_*(\dd\varepsilon)&=\mathcal N_*(\varepsilon)\dd\varepsilon,
 &\mathcal N_*(\varepsilon)
 &=\varepsilon^2\sum_{\lambda=1}^2
       \int_{|\vtheta|=1}n_*(\varepsilon\vtheta,\lambda)\dd\vtheta,\\
 \mu_*(\dd\varepsilon)&=\mathcal U_*(\varepsilon)\dd\varepsilon,
 &\mathcal U_*(\varepsilon)&=\varepsilon\mathcal N_*(\varepsilon).
 \end{aligned}
\end{equation*}
Summing over polarizations makes
these distributions independent of the polarization frame.

\begin{proposition}[Infrared distribution bounds for optimizers]
\label[proposition]{prop-optimizer-soft-spectrum}
There are $C<\infty$ and
$0<\kappa_0\leq1$, depending only on $N,e,\chi$, such that every
PFL optimizer satisfies, for $0<\kappa\leq\kappa_0$,
\begin{align}
 E_*(\kappa):=\int_0^\kappa\mathcal U_*(\varepsilon)\dd\varepsilon
 &\leq C(\kappa^2+b_\kappa^2),
 \label{optimizer-soft-energy-bound}\\
 N_*(\kappa):=\int_0^\kappa\mathcal N_*(\varepsilon)\dd\varepsilon
 &\leq C(\kappa+a_\kappa^2),
 \label{optimizer-soft-number-bound}
\end{align}
where
\begin{equation}\label{infrared-kernel-errors}
 a_\kappa=\left\|
 \frac{\iden_{\{|\vk|<\kappa\}}\wh\chi}{\sqrt{|\vk|}}\right\|,
 \qquad b_\kappa=\|\iden_{\{|\vk|<\kappa\}}\wh\chi\|.
\end{equation}
For the hard cutoff
\cref{harduv}, with $\Lambda>0$ and
$\kappa\leq\min\{\kappa_0,\Lambda\}$, these reduce to
\begin{equation}\label{optimizer-hard-cutoff-spectrum}
 E_*(\kappa)\leq C\kappa^2,\qquad
 N_*(\kappa)\leq C\kappa.
\end{equation}
\end{proposition}

By evaluating the constants
$a_\kappa$ and $b_\kappa$ explicitly, the hard cutoff gives the
frequency-bin estimates
\begin{equation*}
 \frac{2}{\kappa}\int_{\kappa/2}^\kappa
       \mathcal U_*(\varepsilon)\dd\varepsilon\leq C\kappa,
 \qquad
 \frac{2}{\kappa}\int_{\kappa/2}^\kappa
       \mathcal N_*(\varepsilon)\dd\varepsilon\leq C.
\end{equation*}
One cannot differentiate \cref{optimizer-hard-cutoff-spectrum} to
deduce the corresponding pointwise inequalities, so these must be seen
as cumulative bounds, not as an assured soft-photon spectrum or a
proved leading asymptotic.  In particular,
neither a Planck law nor a temperature follows only from these estimates.
If $P_{<\kappa}^{\mathrm{vac}}$ is the projection onto the vacuum
in the soft Fock factor, tensored with the identity on the other
factors, then
\[
 \Tr((\iden-P_{<\kappa}^{\mathrm{vac}})\vGamma_*)
 \leq N_*(\kappa)\longrightarrow0.
\]
Thus the probability of any photon below $\kappa$ tends to zero in
an exact optimizer, with rate $O(\kappa)$ for the hard cutoff.

\subsection{Zero coupling and soft-photon relaxation}
\label{sec-zero-coupling}

At $e=0$, the combined current density reduces to the paramagnetic current density and the
internal Pauli--Fierz Hamiltonian separates into electronic and photon parts:
\begin{align*}
 \vjc_\Psi&=\vjp_\Psi,\\
 \left.H^{0,0}\right|_{e=0}
 &=H_\el^0\otimes\iden_{\Fock_\ph}
 +\iden_{\Hil_\el^N}\otimes H_\rad,\\
 H_\el^0&:=\frac12\sum_{n=1}^N(-\Delta_{\vx_n})
 +\sum_{1\leq n<m\leq N}\frac{1}{|\vx_n-\vx_m|}.
\end{align*}
Thus the density and current constraints depend only on the electronic
reduced state.  However, the photon factor can be used to purify a mixed electronic
state, and the massless dispersion permits the orthogonal photon states in
that purification to have arbitrarily small, though nonzero, energy.

We define the mixed-state Vignale--Rasolt functional of CDFT by
\[
 F_\VR(\rho,\vjp)
 :=\inf_{\substack{
 \Gamma\in\mathfrak S_1(\Hil_\el^N)\\ \Gamma\geq0,\ \Tr\Gamma=1\\
 \rho_\Gamma=\rho\\ \vjp_\Gamma=\vjp}}
 \Tr(H_\el^0\Gamma),
\]
with the value $+\infty$ when the constraint set is empty.  As elsewhere,
the trace involving $H_\el^0$ is a form trace.

\begin{theorem}[Soft-photon relaxation at zero coupling]
\label{thm-soft-photon-relaxation}
For every $(\rho,\vjp)\in\rcdens_{N,\mathrm{mix}}^{\el}$,
\begin{equation}\label{soft-photon-equality}
 \boxed{F_\LL^{0,\chi}(\rho,\vjp)
 =F_\LI^{0,\chi}(\rho,\vjp)=F_\VR(\rho,\vjp).}
\end{equation}
The infimum defining $F_\LL^{0,\chi}(\rho,\vjp)$ is attained if and
only if the electronic mixed-state minimization has a rank-one
minimizer.  Equivalently, attainment holds if and only if there is a
normalized $\Phi\in Q(H_\el^0)$ such that
\[
 (\rho_\Phi,\vjp_\Phi)=(\rho,\vjp),\qquad
 \dua{\Phi}{H_\el^0\Phi}=F_\VR(\rho,\vjp).
\]
\end{theorem}

The proof purifies a fixed mixed electronic state using orthogonal
one-photon states whose energies tend to zero.  These purifications have
no strongly convergent subsequence: any limit would be a product with
the photon vacuum and therefore have a pure electronic reduced state,
contradicting the fixed mixed one.

The next theorem treats the limit $e\downarrow0$ at fixed form
factor.\footnote{Let $F_k,F\colon V\to(-\infty,\infty]$ be functionals
on a Banach space $V$.  A \emph{recovery sequence} for $u\in V$ is a
sequence $u_k\to u$ in norm with $\limsup_{k\to\infty}F_k(u_k)\leq
F(u)$.  The $F_k$ converge to $F$ in the \emph{norm-Gamma} sense if
$F(u)\leq\liminf_{k\to\infty}F_k(u_k)$ whenever $u_k\to u$ in norm, and
every $u\in V$ has a recovery sequence.  They converge in the
\emph{Mosco} sense if, in addition, this $\liminf$ inequality holds
whenever $u_k\wconv u$ weakly.  Mosco convergence therefore implies
norm-Gamma convergence.  We use the sequential versions of these
notions, and a limit $e\downarrow0$ means convergence along every
sequence $e_k\downarrow0$.}

\begin{theorem}[Zero-coupling limit]
\label{thm-zero-coupling-convergence}
As extended-real
functionals on $X\times Y$, the pure Pauli--Fierz universal functionals
have the following sequential limit as $e\downarrow0$.
\begin{enumerate}[(i)]
\item If $e_k>0$, $e_k\to0$, and
$(\rho_k,\vj_k)\wconv(\rho,\vj)$ weakly in $X\times Y$, then
\[
 F_\VR(\rho,\vj)
 \leq\liminf_{k\to\infty}
 F_\LL^{e_k,\chi}(\rho_k,\vj_k).
\]
\item If $F_\VR(\rho,\vj)<\infty$, then the constant sequence
of constraints is a recovery sequence.  More precisely,
\[
 F_\LL^{e,\chi}(\rho,\vj)
 \leq F_\VR(\rho,\vj)
 +\frac{N}{2}\|\wh\chi\|^2e^2,\qquad e\geq0.
\]
Moreover, $(\rho,\vj)\in\rcdens_N^{e,\chi}$ for every $e\geq0$.
\end{enumerate}
Consequently, for every $(\rho,\vj)\in X\times Y$,
\[
 \lim_{e\downarrow0}F_\LL^{e,\chi}(\rho,\vj)
 =F_\LL^{0,\chi}(\rho,\vj)
 =F_\VR(\rho,\vj).
\]
In particular, the convergence holds in the sequential Mosco sense on
$X\times Y$, hence also in the norm-Gamma sense, and every point in the
effective domain has a constant recovery sequence.
\end{theorem}

Part~(i) holds along arbitrary weakly convergent constraints, while
part~(ii) uses only the constant recovery sequence, which suffices for
Mosco convergence.  An upper bound along every norm-convergent sequence
$(\rho_k,\vj_k)\to(\rho,\vj)$ would fail in general, since norm
convergence in $X\times Y$ does not control $\int_{\RR^3}|\vj_k|^2/\rho_k$.

\subsection{The long-wavelength (dipole) limit}
\label{sec-dipole-limit}

We next keep $N$, $e$, the form factor, and the photon dispersion fixed.
For $0\leq\varepsilon\leq1$, introduce a parameter in the spatial
dependence of the quantized field:
\begin{equation*}
 \Af^\mu(\varepsilon\vx)=\varPhi(G^\mu_{\varepsilon\vx}),\qquad
 \vD_{n,\varepsilon}:=-\i\grad_{\vx_n}-e\AF(\varepsilon\vx_n).
\end{equation*}
The original model corresponds to $\varepsilon=1$.  At $\varepsilon=0$,
the field is the spatially uniform operator $\AF(0)$, giving the
velocity-gauge dipole approximation.  

A photon dilation shows that $\varepsilon\downarrow0$ is a
long-wavelength limit at fixed system size: rescaling
$\vx\mapsto\varepsilon\vx$ is equivalent to keeping the electrons
fixed while stretching the relevant photon wavelengths by a factor
$1/\varepsilon$. For $\varepsilon>0$, the
scaling $\varepsilon^{-3/2}\vf(\cdot/\varepsilon)$ is unitary
on $\Hil_\sol$.  As in \cref{prop-dilation-covariance}, conjugation by
its second quantization maps
$\AF(\varepsilon\vx)$ to the field at $\vx$ built with
$\varepsilon^{-3/2}\wh\chi(\cdot/\varepsilon)$ in place of $\wh\chi$,
and maps $H_\rad$ to $\varepsilon^{-1}H_\rad$.  The model at
$\varepsilon$ is therefore unitarily equivalent to one with photon
dispersion $|\vk|/\varepsilon$, that is, with the \emph{speed of light}
formally set to $1/\varepsilon$ while the system size is kept fixed.
In that picture the frequencies and
$\|\wh\chi\|^2$, the vacuum expectation of $|\AF(0)|^2$, are unchanged,
while for the hard cutoff the coupled wave numbers satisfy
$|\vk|\leq\varepsilon\Lambda$, so all coupled wavelengths diverge as
$\varepsilon\downarrow0$, becoming large compared to any fixed
system size. Correspondingly, with the fields
\cref{EB-field-def}, the electrons in the model at $\varepsilon$ feel
the transverse electric field $-\i[H_\rad,\AF(\varepsilon\vx)]=\EF_\perp(\varepsilon\vx)$
at unchanged strength, whereas the magnetic field
$\grad_\vx\times\AF(\varepsilon\vx)=\varepsilon\BF(\varepsilon\vx)$
carries a factor $\varepsilon$.

Let $H_\varepsilon^{0,0}$ be the internal Hamiltonian obtained from
$H^{0,0}$ by replacing $\vD_n$ with $\vD_{n,\varepsilon}$.
As in \cref{sec-PF-hamiltonian}, $\dua{\Psi}{H_\varepsilon^{0,0}\Psi}$
denotes its form value for $\Psi\in\mathcal Q_\PF$, and energies of
density matrices are written as $\Tr(H_\varepsilon^{0,0}\vGamma)$ in the
form sense.
The combined current density changes with the field:
\begin{equation*}
 \vjcmodel{\varepsilon}_{\Psi}(\vx)
 :=N\Re\dua{\Psi}{\vD_{1,\varepsilon}\Psi}_{\vx}
 =\vjp_\Psi(\vx)
   -eN\Re\dua{\Psi}{\AF(\varepsilon\vx)\Psi}_{\vx}.
\end{equation*}
The mixed-state current density is defined by affine extension.  The bounds in
\cref{prop-current-bounds} hold uniformly in $\varepsilon$ on sets of
bounded form energy.

Using the same spaces $X\times Y$ and external potentials as in
\cref{density-current-spaces}, set
\begin{align*}
 \bar E_\varepsilon(u,\vA)
 &:=\inf_{\substack{\Psi\in\mathcal Q_\PF\\\|\Psi\|=1}}
 \big[\dua{\Psi}{H_\varepsilon^{0,0}\Psi}+\dua{u}{\rho_\Psi}
                  +\dua{\vA}{\vjcmodel{\varepsilon}_{\Psi}}\big],
 \\
 F_{\LI,\varepsilon}(\rho,\vjc)
 &:=\sup_{(u,\vA)\in X^*\times Y^*}
 \big[\bar E_\varepsilon(u,\vA)-\dua{u}{\rho}-\dua{\vA}{\vjc}\big].
\end{align*}
Let $F_{\LL,\varepsilon}$ be the pure-state constrained search for
$H_\varepsilon^{0,0}$, with
$(\rho_\Psi,\vjcmodel{\varepsilon}_{\Psi})=(\rho,\vjc)$.  As in
\cref{L-vs-LL},
\begin{equation}\label{dipole-convex-envelope}
 F_{\LI,\varepsilon}
 =\ol{\operatorname{conv}}F_{\LL,\varepsilon}.
\end{equation}
The fixed parameters $N,e,\chi$ are suppressed in this notation.
We use the label $\dip$ for the endpoint $\varepsilon=0$ throughout,
keeping the coupling $e$ fixed, and write
$F_\dip:=\left.F_{\LI,\varepsilon}\right|_{\varepsilon=0}$ and
$\vjcdip_\Psi:=\left.\vjcmodel{\varepsilon}_{\Psi}\right|_{\varepsilon=0}$.
The external classical vector potential $\vA(\vx)$ is unchanged.

\begin{theorem}[Dipole limit of the PFL functional]
\label{thm-dipole-limit}
Fix $N\in\NN$, $e\geq0$, and the form factor.
The functionals $F_{\LI,\varepsilon}$ converge
to $F_\dip$ in the sequential Mosco sense on $X\times Y$:
\begin{enumerate}[(i)]
\item If $\varepsilon_k>0$, $\varepsilon_k\to0$, and
$(\rho_k,\vjc_k)\wconv(\rho,\vjc)$ weakly in $X\times Y$, then
\begin{equation}\label{dipole-functional-liminf}
 F_\dip(\rho,\vjc)
 \leq\liminf_{k\to\infty}F_{\LI,\varepsilon_k}(\rho_k,\vjc_k).
\end{equation}
\item For every $(\rho,\vjc)\in X\times Y$, there are pairs
$(\rho_\varepsilon,\vjc_\varepsilon)\to(\rho,\vjc)$ in the norm of
$X\times Y$ such that
\begin{equation}\label{dipole-functional-recovery}
 \limsup_{\varepsilon\downarrow0}
 F_{\LI,\varepsilon}(\rho_\varepsilon,\vjc_\varepsilon)
 \leq F_\dip(\rho,\vjc).
\end{equation}
\end{enumerate}
\end{theorem}

The dipole functional can be compared directly with the electronic
mixed-state functional defined above.  In this comparison, the same current density
$\vjcdip$ is used as the paramagnetic-current argument of $F_\VR$.
\begin{proposition}[Electronic comparison for the dipole functional]
\label[proposition]{prop-dipole-electronic-comparison}
For every $N\in\NN$, $e\geq0$, and
$(\rho,\vjcdip)\in X\times Y$,
\begin{equation}\label{dipole-electronic-comparison}
 F_\VR(\rho,\vjcdip)
 \leq F_\dip(\rho,\vjcdip)
 \leq F_\VR(\rho,\vjcdip)+\frac{Ne^2}{2}\|\wh\chi\|^2.
\end{equation}
Consequently,
\begin{equation}\label{dipole-effective-domain}
 \operatorname{dom}F_\dip=\operatorname{dom}F_\VR.
\end{equation}
For the standard hard cutoff \cref{harduv}, the upper correction in
\cref{dipole-electronic-comparison} is $N\pi e^2\Lambda^2$.
\end{proposition}

At the dipole endpoint $\varepsilon=0$, with $e$ fixed, the internal
Hamiltonian is
\begin{equation}\label{dipole-internal-hamiltonian}
 H_\dip
 =\frac12\sum_{n=1}^N
      \big(-\i\grad_{\vx_n}-e\AF(0)\big)^2
  +\sum_{1\leq n<m\leq N}\frac1{|\vx_n-\vx_m|}
  +H_\rad,
\end{equation}
understood as the positive self-adjoint operator associated with the
corresponding closed form on $\mathcal Q_\PF$.  The \emph{dipole PFLL
functional} is its pure-state constrained search, that is,
$F_{\LL,\dip}:=\left.F_{\LL,\varepsilon}\right|_{\varepsilon=0}$,
\begin{equation}\label{dipole-pure-search}
 F_{\LL,\dip}(\rho,\vjcdip)
 :=\inf_{\substack{\Psi\in \mathcal Q_\PF,\,\|\Psi\|=1\\
 \rho_\Psi=\rho\\ \vjcdip_\Psi=\vjcdip}}
 \dua{\Psi}{H_\dip\Psi},
\end{equation}
and its mixed-state constrained search is
\begin{equation}\label{dipole-ensemble-search}
 \wt{F}_\dip(\rho,\vjcdip)
 :=\inf_{\substack{
 \vGamma\in\mathfrak S_1(\HC_N),\ \vGamma\geq0,\ \Tr\vGamma=1\\
 \Tr(H_\dip\vGamma)<\infty\\
 \rho_\vGamma=\rho,\ \vjcdip_{\vGamma}=\vjcdip}}
 \Tr(H_\dip\vGamma).
\end{equation}
The expectation and the trace are taken in the form sense, and an empty
admissible set has infimum $+\infty$.  The following theorem consistently identifies the pure-state and
mixed-state searches with the dipole PFL functional for every finite $N$.

\begin{theorem}[Equality of dipole functionals]
\label{thm-dipole-ensemble-realization}
We have
\begin{equation}\label{dipole-search-equality}
\boxed{ F_{\LL,\dip}(\rho,\vjcdip)=\wt F_\dip(\rho,\vjcdip)=F_\dip(\rho,\vjcdip)}
\end{equation}
for all $(\rho,\vjcdip)\in X\times Y$.
Every finite mixed-state infimum is attained by a density matrix of
finite mean photon number.
\end{theorem}

As a preparation for passing to the length gauge, put $\vR=\sum_{n=1}^N\vx_n$, so that $e\vR$
is the dipole moment of the electrons.  If $\sqrt\omega\,\wh\chi\in L^2$,
define the real symmetric positive semidefinite $3\times3$ matrix $\vM=(M^{\mu\nu})_{\mu,\nu=1}^3$ by
\begin{equation}\label{dipole-polarization-matrix}
 M^{\mu\nu}
 =\Re\dua{\sqrt\omega\,G^\mu_{0}}{\sqrt\omega\,G^\nu_{0}}
 =\frac12\int_{\RR^3}\omega(\vk)|\wh\chi(\vk)|^2P_\sol(\vk)^{\mu\nu}\dd\vk;
\end{equation}
for the hard cutoff \cref{harduv}, $\vM=\frac{4\pi\Lambda^3}{9}\iden$
(Appendix~\ref{app-dipole-length-gauge}).
Under the same assumption, the transverse electric field
\cref{EB-field-def} at the origin is a form on $D(\sqrt{H_\rad})$
(\cref{sec-free-field}), namely
\begin{equation}\label{dipole-electric-field-origin}
 \Ef^\mu_\perp(0)=-\varPi(\omega G^\mu_{0}),
\end{equation}
where $\omega G^\mu_{0}/\sqrt\omega\in\Hil_\ph$ and $\varPi$ is the
extension of the conjugate Segal field \cref{conjugate-segal-field-def}
to such arguments (\cref{prop-dipole-length-gauge}(iv)).

\begin{corollary}[Length-gauge form of the dipole functional]
\label[corollary]{cor-dipole-length-gauge}
Suppose that $\sqrt\omega\,\wh\chi\in L^2$.
For all $(\rho,\vjcdip)\in X\times Y$ with $\int_{\RR^3}|\vx|^2\rho<\infty$,
\begin{equation}\label{dipole-length-hamiltonian-functional}
 \boxed{F_\dip(\rho,\vjcdip)
 =\inf_{\substack{\vGamma\in\dm\\
 \rho_\vGamma=\rho,\ \vjp_\vGamma=\vjcdip}}
 \Tr\Big(H_\el^0+H_\rad-e\vR\cdot\EF_\perp(0)
 +e^2\vR^\top\vM\vR\Big)\vGamma.}
\end{equation}
Here $\dm$ is the class \cref{finite-energy-PF-ensembles} of
finite-energy density matrices, the trace is taken in the form sense
and converges absolutely, and every finite infimum is attained.
\end{corollary}

This is the mixed-state constrained search for the usual length-gauge
dipole Hamiltonian \cite{rokaj2018lightmatter}.  By
\cref{dipole-search-equality}, the same identity holds for the
pure-state search over normalized $\Psi\in\mathcal Q_\PF$, but that
infimum need not be attained: at $e=0$, where the dipole model coincides
with the Pauli--Fierz model, \cref{prop-pure-nonattainment} provides an
example.  The constant field has disappeared from the kinetic energy and
from the constraint, which becomes the paramagnetic current density as
in $F_\VR$.  It enters only through the last three terms, which add up
to the radiation energy
$\Tr(H_\rad U_{\mathrm d}\vGamma U_{\mathrm d}^\dag)\geq0$ of the
transformed density matrix, with $U_{\mathrm d}$ the unitary
\cref{dipole-length-unitary}.  Dropping them
gives the lower bound in \cref{dipole-electronic-comparison}.  

\clearpage
\section{Proofs}\label{sec-proofs}

The rest of the paper is devoted to proofs.

\subsection{A priori bounds}

\phantomsection
\label{proof:prop-current-bounds}
\begin{proof}[Proof of \cref{prop-current-bounds}]
For $\mu=1,2,3$, write $\partial_\mu$ for differentiation
with respect to the $\mu$th component of $\vx=\vx_1$, and $D_1^\mu$ for
the corresponding component of $\vD_1$.
By \cref{PF-canonical-form-domain}, the state $\Psi$ has square-integrable
weak electronic derivatives and finite photon form energy.  The field
bound \cref{A-Hrad-bound} also shows that
$\|\AF(\vx_n)\Psi\|$ and $\|\vD_n\Psi\|$ are finite for every $n$.
We may therefore regard $\vx\mapsto\Psi(\vx,\cdot)$ as an $H^1$
function taking values in the Hilbert space of the remaining electronic
and photon variables.  In terms of the partial norm $\|\cdot\|_{\vx}$,
the definition of the density gives
\[
 \|\Psi\|_{\vx}
 =\sqrt{\rho_\Psi(\vx)/N}.
\]
In particular, $\rho_\Psi\geq0$, and Fubini's theorem and normalization
give
\[
 \int_{\RR^3}\rho_\Psi(\vx)\dd\vx
 =N\|\Psi\|^2=N.
\]

We next prove the joint density--current bound
\cref{joint-density-current-bound}.
The Sobolev chain rule for the Hilbert-space norm gives
$\sqrt{\rho_\Psi}\in H^1$ and, almost everywhere on $\{\rho_\Psi>0\}$,
\[
 \partial_\mu\sqrt{\rho_\Psi}
 =\frac{N\Re\dua{\Psi}{\partial_\mu\Psi}_{\vx}}{\sqrt{\rho_\Psi}}.
\]
One can obtain this formula by first differentiating
$(\|\Psi\|_{\vx}^2+\epsilon^2)^{1/2}-\epsilon$ and then
letting $\epsilon\downarrow0$.  The derivatives are dominated by
$\|\partial_\mu\Psi\|_{\vx}$.  On $\{\rho_\Psi=0\}$, the weak
gradient of $\sqrt{\rho_\Psi}$ vanishes almost everywhere.
Since $\Af^\mu(\vx)$ is self-adjoint, the partial inner
product $\dua{\Psi}{\Af^\mu(\vx)\Psi}_{\vx}$ is real.
Set $\zeta^\mu(\vx):=\dua{\Psi}{D_1^\mu\Psi}_{\vx}$.
The identity $\partial_\mu\Psi=\i D_1^\mu\Psi+\i e\Af^\mu(\vx)\Psi$
and the definition of the combined current density then give
\[
 \Im\zeta^\mu=-\frac{\sqrt{\rho_\Psi}\,\partial_\mu\sqrt{\rho_\Psi}}{N},
 \qquad
 \Re\zeta^\mu=\frac{(\vjc_\Psi)^\mu}{N}.
\]
These identities express the gradient of $\sqrt{\rho_\Psi}$
and the combined current through the imaginary and real parts of the same
partial inner product.
Cauchy--Schwarz in the remaining variables gives, on $\{\rho_\Psi>0\}$,
\begin{align*}
 |\grad\sqrt{\rho_\Psi}|^2+
 \frac{|\vjc_\Psi|^2}{\rho_\Psi}
 &=\frac{N^2}{\rho_\Psi}\sum_{\mu=1}^3
 \big(|\Im\zeta^\mu|^2+|\Re\zeta^\mu|^2\big)\\
 &=\frac{N^2}{\rho_\Psi}\sum_{\mu=1}^3|\zeta^\mu|^2
 \leq N\|\vD_1\Psi\|_{\vx}^2.
\end{align*}
On $\{\rho_\Psi=0\}$ we have $\|\Psi\|_{\vx}=0$, so all three
current densities vanish there.  With the stated convention
for the quotient, both terms on the left vanish almost everywhere on
this set, and the inequality holds there as well.
Exchanging electronic coordinates sends
$\vD_1$ to the corresponding $\vD_n$, and the sign from antisymmetry
disappears upon taking squared norms.  Consequently,
\[
 N\|\vD_1\Psi\|^2=\sum_{n=1}^N\|\vD_n\Psi\|^2.
\]
Integrating the pointwise joint bound proves
\cref{joint-density-current-bound}.

We now prove the bound \cref{ph-current-bound} on the field-induced
current density.  Applying conditional Cauchy--Schwarz to the definition of
$\vjph_\Psi$ and then integrating gives
\[
 |\vjph_\Psi(\vx)|^2
 \leq e^2N\rho_\Psi(\vx)\|\AF(\vx)\Psi\|_{\vx}^2,
 \qquad
 \int_{\RR^3}\frac{|\vjph_\Psi|^2}{\rho_\Psi}
 \leq e^2N\|\AF(\vx_1)\Psi\|^2.
\]
For almost every electronic configuration $\vX$, apply
\cref{A-Hrad-bound} to the Fock-space vector $\Psi(\vX)$ at the
position $\vx_1$.  Its constant is independent of that position, so
integration in $\vX$ gives
\[
 \|\AF(\vx_1)\Psi\|^2
 \leq C_\chi\big(\|\sqrt{H_\rad}\Psi\|^2+\|\Psi\|^2\big)
 =C_\chi\big(\|\sqrt{H_\rad}\Psi\|^2+1\big).
\]
This proves \cref{ph-current-bound}.

It remains to prove the integrability of the density and the current densities.
The following integrability argument extends the estimates for electronic
paramagnetic currents in~\cite{laestadius2014density,MY-CDFTpaper2019}
to the combined and field-induced current densities, using the bounds
established above.
Since $\sqrt{\rho_\Psi}\in H^1(\RR^3)$, the Sobolev embedding gives
$\sqrt{\rho_\Psi}\in L^2\cap L^6$, hence
$\rho_\Psi\in L^1\cap L^3$.
For either $\vj=\vjc_\Psi$ or $\vj=\vjph_\Psi$, the bounds just proved
show that $\vj/\sqrt{\rho_\Psi}\in L^2$, with value zero where the
density vanishes.  Factoring
$\vj=\sqrt{\rho_\Psi}\,(\vj/\sqrt{\rho_\Psi})$ and applying
H\"older's inequality gives
\[
 \|\vj\|_{L^1}
 \leq\|\sqrt{\rho_\Psi}\|_{L^2}
 \left\|\frac{\vj}{\sqrt{\rho_\Psi}}\right\|_{L^2},
 \qquad
 \|\vj\|_{L^{3/2}}
 \leq\|\sqrt{\rho_\Psi}\|_{L^6}
 \left\|\frac{\vj}{\sqrt{\rho_\Psi}}\right\|_{L^2}.
\]
Finally, $\vjp_\Psi=\vjc_\Psi-\vjph_\Psi$ belongs to the same two
spaces.

For a density matrix $\vGamma=\sum_r p_r|\Psi_r\rangle\langle\Psi_r|\in\dm$,
apply the preceding argument to the auxiliary vector
$(\sqrt{p_r}\Psi_r)_r$, taking the Hilbert direct sum over $r$ as part of
the space of remaining variables.  This vector is normalized and has
finite form energy.  Its partial density and currents are the ensemble
averages, and the operators act componentwise in $r$.  Their squared
norms are therefore the corresponding form traces, which proves the
mixed-state extension.
\end{proof}

\phantomsection
\label{proof:prop-div-jph}
\begin{proof}[Proof of \cref{prop-div-jph}]
Let $\Psi\in \mathcal Q_\PF$, write $\partial_\mu$ for the derivative with
respect to the $\mu$th component of $\vx=\vx_1$, and denote the
right-hand side of \cref{div-jph} by
\[
 f_\Psi(\vx):=-2eN\Re\sum_{\mu=1}^3
 \dua{\partial_\mu\Psi}{\Af^\mu(\vx)\Psi}_{\vx}.
\]
Let $\|\cdot\|_{\mathcal Q_\PF}$ be the
reference norm \cref{PF-reference-form-norm}.  As in the proof of
\cref{prop-current-bounds}, applying
\cref{A-Hrad-bound} to $\Phi(\vX)$ for almost every $\vX$ gives
\begin{equation}\label{div-jph-field-bound}
 \|\AF(\vx_1)\Phi\|\leq C_\chi^{1/2}\|\Phi\|_{\mathcal Q_\PF},
 \qquad \Phi\in\mathcal Q_\PF.
\end{equation}

We first show that $f_\Psi$ is integrable.  The conditional
Cauchy--Schwarz inequality
$|\dua{\partial_\mu\Psi}{\Af^\mu(\vx)\Psi}_{\vx}|
\leq\|\partial_\mu\Psi\|_{\vx}\|\Af^\mu(\vx)\Psi\|_{\vx}$, followed by
Cauchy--Schwarz in $\vx$ and in $\mu$, gives
\begin{equation}\label{div-jph-f-bound}
 \int_{\RR^3}|f_\Psi|
 \leq2eN\sum_{\mu=1}^3\|\partial_\mu\Psi\|\,\|\Af^\mu(\vx_1)\Psi\|
 \leq2eN\|\grad_{\vx_1}\Psi\|\,\|\AF(\vx_1)\Psi\|.
\end{equation}
Both factors are finite by \cref{PF-reference-form-domain} and
\cref{div-jph-field-bound}.

We now prove \cref{div-jph} in the sense of distributions.  Let
$v\in C_c^\infty(\RR^3)$ be real.  For $\Phi,\Phi'\in\mathcal Q_\PF$ put
\[
 \begin{aligned}
 s_v(\Phi,\Phi')
 :=\sum_{\mu=1}^3\Big(
 &\dua{\Phi}{(\partial_\mu v)(\vx_1)\Af^\mu(\vx_1)\Phi'}
 +\dua{\partial_\mu\Phi}{v(\vx_1)\Af^\mu(\vx_1)\Phi'}\\
 &+\dua{v(\vx_1)\Af^\mu(\vx_1)\Phi}{\partial_\mu\Phi'}\Big).
 \end{aligned}
\]
Since $\dua{\Psi}{\Af^\mu(\vx)\Psi}_{\vx}$ is real, Fubini's theorem and
the definitions give
\[
 \sum_{\mu=1}^3\int_{\RR^3}(\partial_\mu v)(\vjph_\Psi)^\mu
 +\int_{\RR^3}vf_\Psi
 =-eN s_v(\Psi,\Psi),
\]
so \cref{div-jph} holds in the sense of distributions once
$s_v(\Psi,\Psi)=0$.  The form $s_v$ is sesquilinear, and Cauchy--Schwarz
together with \cref{div-jph-field-bound} gives
\[
 |s_v(\Phi,\Phi')|
 \leq C_\chi^{1/2}\big(\sqrt3\|\grad v\|_\infty+2\|v\|_\infty\big)
 \|\Phi\|_{\mathcal Q_\PF}\|\Phi'\|_{\mathcal Q_\PF}.
\]
By \cref{PF-form-norm-equivalence,PF-canonical-form-domain}, the
algebraic core $\mathscr C_\PF$ of \cref{PF-algebraic-core} is dense in
$\mathcal Q_\PF$ for $\|\cdot\|_{\mathcal Q_\PF}$.  It therefore suffices
to show $s_v(\Phi,\Phi')=0$ for the product vectors
$\Phi=\phi\otimes\Sigma$ and $\Phi'=\phi'\otimes\Sigma'$ spanning
$\mathscr C_\PF$, where $\phi,\phi'\in C_c^\infty(\RR^{3N})\cap\Hil_\el^N$
and $\Sigma,\Sigma'$ are finite-particle vectors in $D(\sqrt{H_\rad})$.  Since
$\Af^\mu(\vx)$ is symmetric on $D(\sqrt{H_\rad})$, the three terms of
$s_v$ combine by the product rule into
\[
 s_v(\Phi,\Phi')
 =\sum_{\mu=1}^3\int_{\RR^{3N}}
 \partial_\mu\big(v(\vx_1)\,\ol{\phi(\vX)}\,\phi'(\vX)\big)\,
 \dua{\Sigma}{\Af^\mu(\vx_1)\Sigma'}\dd\vX.
\]
The function $\vx\mapsto\dua{\Sigma}{\Af^\mu(\vx)\Sigma'}$ is bounded and
continuous by \cref{A-Hrad-bound} and the continuity of
$\vx\mapsto G^\mu_{\vx}$.  For fixed $\vx_2,\ldots,\vx_N$, the function
$\vx_1\mapsto v(\vx_1)\ol{\phi(\vX)}\phi'(\vX)$ lies in
$C_c^\infty(\RR^3)$.  By Fubini's theorem we may first integrate in
$\vx_1$, and that integral vanishes by the weak Coulomb gauge
\cref{coulomb-gauge-weak}, applied to the real and imaginary parts of
this test function.  Hence $s_v=0$ on
$\mathcal Q_\PF$, which proves \cref{div-jph}.  Now
$\div\vjph_\Psi=f_\Psi$, and \cref{div-jph-f-bound} is
\cref{div-jph-L1}.

Finally, the right-hand side of \cref{div-jph} contains no derivative of
the field and need not vanish.  In the one-electron example following
\cref{prop-div-jph}, with $\vjph_\Psi$ given by \cref{entangled-example},
it equals
$-2e\Re\big(\vg_\xi\cdot\grad(\ol{\phi_0}\phi_1)\big)$, which is nonzero
for $e>0$ and generic $\phi_0,\phi_1$.  Choosing $\phi_0,\phi_1$ such that
$\rho_\Psi=|\phi_0|^2+|\phi_1|^2$ is constant on an open set where this
expression does not vanish shows that $\vjph_\Psi/\rho_\Psi$ is not
divergence-free either.
\end{proof}

\subsection{Proof of \texorpdfstring{\cref{prop-ground-state-fields}}{Proposition \getrefnumber{prop-ground-state-fields}}}
\label{proof:prop-ground-state-fields}

Let $\Psi$ be a ground state of $H^{v,\vA}$.
Items (i)--(iii) are first-order conditions of the variational
principle \cref{ground-energy-def}: a one-parameter family of trial
states through $\Psi$ has energy at least $E(v,\vA)$, with equality at
$t=0$.  Item (iv) follows from time-reversal symmetry.
By \cref{prop-current-bounds}, $\vjc_\Psi\in L^1\cap L^{3/2}$ and
$\rho_\Psi\in L^1\cap L^3$, so H\"older's inequality gives
$\rho_\Psi\vA\in L^1\cap L^{3/2}$ for $\vA\in L^3+L^\infty$.

We first prove (i).  For real $\varphi\in C_c^\infty(\RR^3)$ and
$t\in\RR$, the gauge unitary $U_{t\varphi}$ of \cref{gauge-unitary}
preserves $\mathcal Q_\PF$, and by \cref{H-gauge,energy-decomposition}
\[
 \dua{U_{t\varphi}\Psi}{H^{v,\vA}U_{t\varphi}\Psi}
 =E(v,\vA)+t\int_{\RR^3}\grad\varphi\cdot\vj_\Psi
  +\frac{t^2}{2}\int_{\RR^3}|\grad\varphi|^2\rho_\Psi.
\]
The minimum at $t=0$ gives $\int\grad\varphi\cdot\vj_\Psi=0$, which is (i).

We next derive an identity for the mean photon amplitude, with
\[
 g_{\vj_\Psi}=\sum_{\mu=1}^3\int_{\RR^3}j_\Psi^\mu(\vx)G^\mu_{\vx}\dd\vx.
\]
Let $f\in D(\omega)$.  By \cref{cor-weyl-PF-shifts}, applied to
$\vGamma=|\Psi\rangle\langle\Psi|$, $W(tf)$ preserves $\mathcal Q_\PF$,
and the energy of $W(tf)\Psi$ is the quadratic polynomial
\cref{weyl-PF-energy-shift} in $t$.  Its linear coefficient is
\[
 2\Re\dua{\Psi}{a(\omega f)\Psi}-e\int\vc_f\cdot\vj_\Psi
 =2\Re\dua{\Psi}{a(\omega f)\Psi}-2e\Re\dua{f}{g_{\vj_\Psi}},
\]
with $\vc_f$ as in \cref{weyl-classical-field}.
This coefficient vanishes, and applying this also to $\i f$ in place of
$f$ gives
\[
 \dua{\Psi}{a(\omega f)\Psi}=e\dua{f}{g_{\vj_\Psi}}
 \qquad\text{for all }f\in D(\omega).
\]

We next prove (ii) and (iii).  For test fields
$\vu,\vv\in C_c^\infty(\RR^3;\RR^3)$, the kernels of $\int\vu\cdot\EF_\perp$
and of $\int(\grad\times\vv)\cdot\BF=\int(\grad\times\grad\times\vv)\cdot\AF$
are $-\i\omega g_\vu$ and $g_{-\Delta\vv}=\omega^2g_\vv$, since
$\grad\times\grad\times\vv=-\Delta\vv+\grad\div\vv$ and gradients have
zero kernel.  The choices $f=g_\vu$ and $f=\omega g_\vv$ in this
identity therefore give
\begin{gather*}
 \Big\langle \Psi, \Big(\int\vu\cdot\EF_\perp\Big)\Psi\Big\rangle=-2e\Im\dua{g_\vu}{g_{\vj_\Psi}} =0,\\
 \Big\langle \Psi, \Big(\int(\grad\times\vv)\cdot\BF\Big)\Psi \Big\rangle
 =2e\Re\dua{\omega g_\vv}{g_{\vj_\Psi}}.
\end{gather*}
The first scalar product is real as in Section~\ref{sec-free-field},
which proves (ii).  Since
\[
 2\Re\dua{\omega G^\mu_{\vx}}{G^\nu_{\vy}}
 =\int_{\RR^3}\omega(\vk)|\wh\chi(\vk)|^2P_\sol(\vk)^{\mu\nu}
   e^{\i\vk\cdot(\vx-\vy)}\dd\vk,
\]
the second identity is (iii) paired with $\vv$.

We finally prove (iv).  By \cref{prop-time-reversal}, the time
reversal $\Theta$ of \cref{time-reversal-def} reverses $\AF$, $\BF$,
and $\vD_n$ and commutes with $H^{v,0}$.  Hence $\Theta\Psi$ is a
ground state, and by nondegeneracy a multiple of $\Psi$.  On the other
hand, $\vjc_{\Theta\Psi}=-\vjc_\Psi$ and
$\dua{\Theta\Psi}{\AF\Theta\Psi}=-\dua{\Psi}{\AF\Psi}$, and likewise
for $\BF$.  All three quantities therefore vanish.

\subsection{Representability results}

\phantomsection
\label{proof:thm-photon-purification}
\begin{proof}[Proof of \cref{thm-photon-purification}]
Let
$\Gamma=\sum_{j\geq1}p_j|\Phi_j\rangle\langle\Phi_j|$ be an electronic
mixed state in \cref{electronic-mixed-class}, written in its spectral
decomposition.  For $\epsilon>0$, choose an
orthonormal family
\[
 f_j\in L^2(\{0<|\vk|<\epsilon\}\times\{1,2\})
\]
and put $\Xi_j=a^\dag(f_j)\Omega$.  All $\Xi_j$ are normalized and lie in
the one-photon sector.  The vector
\begin{equation}\label{one-photon-purification}
 \Psi_\epsilon=\sum_{j\geq1}\sqrt{p_j}\,
 \Phi_j\otimes\Xi_j
\end{equation}
is normalized and
$\Tr_\ph|\Psi_\epsilon\rangle\langle\Psi_\epsilon|=\Gamma$.  It therefore
has density $\rho_\Gamma$ and paramagnetic current density $\vjp_\Gamma$.
Moreover, $\AF(\vx)$ changes photon number by exactly one, whence
\[
 \dua{\Xi_j}{\AF(\vx)\Xi_k}=0
 \quad\text{for all }j,k,
\]
and consequently $\vjph_{\Psi_\epsilon}=0$.

It remains to check that $\Psi_\epsilon\in\mathcal Q_\PF$.  Orthonormality
of the photon vectors $\Xi_j$ eliminates the cross terms in the electronic
kinetic energy, so
\[
 \sum_{n=1}^N\|\grad_{\vx_n}\Psi_\epsilon\|^2
 =\sum_j p_j\sum_{n=1}^N\|\grad_{\vx_n}\Phi_j\|^2
 =\Tr(-\lapl\Gamma)<\infty.
\]
Likewise, orthonormality of the electronic vectors $\Phi_j$ eliminates
the cross terms in the photon energy, and $H_\rad$ acts as multiplication
by $\omega(\vk)=|\vk|$ on the one-photon sector.  Hence
\begin{align*}
 \dua{\Psi_\epsilon}{H_\rad\Psi_\epsilon}
 &=\sum_j p_j\dua{\Xi_j}{H_\rad\Xi_j}\\
 &=\sum_j p_j\|\sqrt{\omega}f_j\|^2
 \leq\epsilon\sum_j p_j\|f_j\|^2=\epsilon,
\end{align*}
where the inequality uses the support condition $|\vk|<\epsilon$,
and the last equality uses $\|f_j\|=1$ and $\sum_j p_j=1$.
Thus $\Psi_\epsilon\in\mathcal Q_\PF$, proving the reverse inclusion in
\cref{electronic-projection} as well as the reverse inclusion in
\cref{zero-jph-slice}.

Conversely, let $\Psi\in \mathcal Q_\PF$ be normalized and define its
electronic reduced state by
\[
 \Gamma:=\Tr_\ph|\Psi\rangle\langle\Psi|.
\]
Partial trace preserves positivity and the trace, so $\Gamma\geq0$ and
$\Tr\Gamma=1$.  We must check that this electronic density matrix has
finite kinetic energy and reproduces the pair
$(\rho_\Psi,\vjp_\Psi)$.

The positive-observable
partial-trace identity \cref{partial-trace-positive-observables} gives,
in the quadratic-form sense,
\begin{align*}
 \Tr(-\lapl\Gamma)
 &=\sum_{n=1}^N\|\grad_{\vx_n}\Psi\|^2\\
 &\leq 2\sum_{n=1}^N\|\vD_n\Psi\|^2
 +2e^2\sum_{n=1}^N\|\AF(\vx_n)\Psi\|^2<\infty.
\end{align*}
The equality follows by applying the definition
\cref{partial-trace-duality} of the partial trace first to bounded
spectral truncations of $-\lapl$ and then using monotone convergence.
The inequality uses
$-\i\grad_{\vx_n}\Psi=\vD_n\Psi+e\AF(\vx_n)\Psi$.
Both sums are finite by \cref{PF-canonical-form-domain,A-Hrad-bound}.
Thus $\Gamma$ satisfies all the state conditions in
\cref{electronic-mixed-class}.

To see explicitly why the electronic density and paramagnetic current density
are preserved, we use a Schmidt decomposition as in
\cref{Schmidt-decomposition,Schmidt-reduced-states},
\[
 \Psi=\sum_j\sqrt{p_j}\,\Phi_j\otimes\Xi_j,
 \qquad
 \Gamma=\sum_j p_j|\Phi_j\rangle\langle\Phi_j|,
 \qquad \sum_j p_j=1.
\]
The electronic vectors $\Phi_j$ and the photon vectors $\Xi_j$ form
orthonormal families.  The finite kinetic trace implies that each
$\Phi_j$ with $p_j>0$ has finite kinetic energy and that the Schmidt
series also converges in the electronic $H^1$ norm.  Inserting this
series into \cref{density-def,jp-def}, the photon inner products
$\dua{\Xi_j}{\Xi_k}=\delta_{jk}$ remove all cross terms, giving
\[
 \rho_\Psi=\sum_j p_j\rho_{\Phi_j}=\rho_\Gamma,
 \qquad
 \vjp_\Psi=\sum_j p_j\vjp_{\Phi_j}=\vjp_\Gamma.
\]
The density series converges in $L^1$ by normalization, and the current
series by Cauchy--Schwarz and the finite kinetic
trace.  These identities therefore hold almost everywhere, and prove
the forward inclusion in \cref{electronic-projection}.

If the original state additionally satisfies $\vjph_\Psi=0$, its triple
is $(\rho_\Gamma,\vjp_\Gamma,0)$ with
$(\rho_\Gamma,\vjp_\Gamma)\in\rcdens_{N,\mathrm{mix}}^{\el}$.
This proves the forward inclusion in \cref{zero-jph-slice}.  The reverse
inclusion was supplied by the one-photon purification above.
\end{proof}

\phantomsection
\label{proof:prop-spectator-purification}
\begin{proof}[Proof of \cref{prop-spectator-purification}]
\noindent\textbf{Step 1. Splitting off the uncoupled photon modes.}
Replacing $S$ by $S\cap B_r$ for sufficiently large $r$, we may assume
that $S$ is bounded.  By the symmetry $\wh\chi(\vk)=\ol{\wh\chi(-\vk)}$ in
\cref{cutoff-assumption}, $\wh\chi(-\vk)=0$ for almost every
$\vk\in S$, so by \cref{Gx-def} every $G^\mu_{\vx}$ vanishes almost
everywhere on $S\times\{1,2\}$.  The orthogonal decomposition
\[
 \Hil_\ph=\mathfrak h_R\oplus\mathfrak h_S,\qquad
 \mathfrak h_R=L^2(S^c\times\{1,2\}),\qquad
 \mathfrak h_S=L^2(S\times\{1,2\}),
\]
induces the canonical factorization
$\Fock_s(\Hil_\ph)\simeq\Fock_s(\mathfrak h_R)\otimes\Fock_s(\mathfrak h_S)$,
and hence
\[
 \HC_N\simeq\mathcal K_R\otimes\mathcal F_S,\qquad
 \mathcal K_R:=\Hil_\el^N\otimes\Fock_s(\mathfrak h_R),\qquad
 \mathcal F_S:=\Fock_s(\mathfrak h_S).
\]
Under this identification
$\AF(\vx)=\AF_R(\vx)\otimes\iden_{\mathcal F_S}$, where $\AF_R$ is the
field on $\Fock_s(\mathfrak h_R)$ with the same kernels $G^\mu_{\vx}$,
and
$H_\rad=H_{\rad,R}\otimes\iden_{\mathcal F_S}
+\iden\otimes H_{\rad,S}$, where $H_{\rad,R}$ and $H_{\rad,S}$ are the
second quantizations of the restrictions of $\omega$ to $S^c$ and $S$.
For real $v\in C_c^\infty(\RR^3)$ and $\vA\in C_c^\infty(\RR^3;\RR^3)$,
let $H_R^{v,\vA}$ be the Hamiltonian on $\mathcal K_R$ defined like
$H^{v,\vA}$, with $\AF_R$ and $H_{\rad,R}$ in place of $\AF$ and
$H_\rad$.  Then, as closed forms,
\begin{equation}\label{spectator-Hamiltonian-splitting}
 H^{v,\vA}=H_R^{v,\vA}\otimes\iden_{\mathcal F_S}
 +\iden_{\mathcal K_R}\otimes H_{\rad,S},
\end{equation}
and in particular
$Q(H^{0,0})=Q(H_R^{0,0}\otimes\iden_{\mathcal F_S})
\cap Q(\iden_{\mathcal K_R}\otimes H_{\rad,S})$.

\noindent\textbf{Step 2. The density and current density depend only on the
coupled marginal.}
Let $\vGamma\in\dm$, and put $\Gamma_R=\Tr_{\mathcal F_S}\vGamma$ and
$\Gamma_S=\Tr_{\mathcal K_R}\vGamma$.  Both terms in
\cref{spectator-Hamiltonian-splitting} are nonnegative after
$N\|v\|_\infty$ is added to the first.  The positive-observable identity
\cref{partial-trace-positive-observables} therefore gives
\begin{equation}\label{spectator-energy-trace-splitting}
 \Tr(H^{v,\vA}\vGamma)
 =\Tr(H_R^{v,\vA}\Gamma_R)+\Tr(H_{\rad,S}\Gamma_S),
\end{equation}
where all terms are finite by \cref{external-PF-form-domain}.
Subtracting the case $v=0$, $\vA=0$ removes the last term.  Applying
\cref{energy-decomposition} to the spectral decomposition of
$\vGamma$, we obtain
\[
 \int_{\RR^3}\big(v+\tfrac12|\vA|^2\big)\rho_\vGamma
 +\int_{\RR^3}\vA\cdot\vjc_\vGamma
 =\Tr(H_R^{v,\vA}\Gamma_R)-\Tr(H_R^{0,0}\Gamma_R).
\]
The right-hand side depends on $\vGamma$ only through $\Gamma_R$.
Taking first $\vA=0$ and then general $\vA$ shows that two density matrices in
$\dm$ with the same marginal on $\mathcal K_R$ have the same density
and the same combined current density.

\noindent\textbf{Step 3. Purification by uncoupled one-photon states.}
Fix $\vGamma\in\dm$ and write
$\Gamma_R=\sum_rp_r|\Phi_r\rangle\langle\Phi_r|$ with $p_r>0$ and
orthonormal $\Phi_r$, the index $r$ ranging over a finite or countable
set.  Since $S$ has positive measure, $\mathfrak h_S$ is infinite
dimensional and contains an orthonormal family $(f_r)$ indexed by the
same set.  Put $\Xi_r=a_S^\dag(f_r)\Omega_S$ and
$R_S=\operatorname*{ess\,sup}_{\vk\in S}|\vk|<\infty$.  The $\Xi_r$
are orthonormal, so by \cref{Schmidt-decomposition,Schmidt-reduced-states}
the unit vector
\[
 \Psi=\sum_r\sqrt{p_r}\,\Phi_r\otimes\Xi_r
\]
satisfies
\[
 \Tr_{\mathcal F_S}|\Psi\rangle\langle\Psi|=\Gamma_R,
 \qquad
 \Tr_{\mathcal K_R}|\Psi\rangle\langle\Psi|
 =\sum_rp_r|\Xi_r\rangle\langle\Xi_r|.
\]
Applying \cref{partial-trace-positive-observables} to the two terms of
\cref{spectator-Hamiltonian-splitting} with $v=0$ and $\vA=0$, and
using \cref{spectator-energy-trace-splitting} for $\vGamma$, we get
\begin{align*}
 \|(H_R^{0,0}\otimes\iden_{\mathcal F_S})^{1/2}\Psi\|^2
 &=\Tr(H_R^{0,0}\Gamma_R)\leq\Tr(H^{0,0}\vGamma),\\
 \|(\iden_{\mathcal K_R}\otimes H_{\rad,S})^{1/2}\Psi\|^2
 &=\sum_rp_r\dua{f_r}{\omega f_r}\leq R_S.
\end{align*}
Thus $\Psi\in \mathcal Q_\PF$, and $|\Psi\rangle\langle\Psi|\in\dm$ has
the same marginal $\Gamma_R$ on $\mathcal K_R$ as $\vGamma$.  By Step 2,
$(\rho_\Psi,\vjc_\Psi)=(\rho_\vGamma,\vjc_\vGamma)$.  This proves that
the right-hand side of \cref{pure-mixed-PF-class} is contained in the
left-hand side.

Conversely, if $\Psi\in \mathcal Q_\PF$ is normalized, then
$|\Psi\rangle\langle\Psi|\in\dm$, which proves the reverse inclusion.
Finally, $\dm$ is convex, its form energy is affine, and the map
$\vGamma\mapsto(\rho_\vGamma,\vjc_\vGamma)$ is affine by the weak trace
interpretation of \cref{mixed-PF-density-current}.  Its image is
therefore convex, and the equality \cref{pure-mixed-PF-class} proves the
last assertion.
\end{proof}

\subsection{Proof of \texorpdfstring{\cref{thm-photonic-representability}}{Theorem \getrefnumber{thm-photonic-representability}}}
\label{proof:thm-photonic-representability}

The idea is to distribute the electrons independently, with the real
amplitude $\prod_{n=1}^N\sqrt{\rho(\vx_n)/N}$, and to let the photon
state depend on their positions.  The electronic amplitude gives
density $\rho$ and no paramagnetic current density, but it is symmetric, so
antisymmetry is supplied by a real unit vector built from the pair
directions $\vx_i-\vx_j$ and placed in a few auxiliary photon modes.
This factor contributes neither current density nor field expectation.  The
field-induced current density comes from a coherent state in a separate band of
modes, displaced along the coupling functions at the electron positions
with weights $\vu(\vx_n)$.  Its conditional field expectation at $\vx$ is
$(\vT\vu)(\vx)$ for a positive operator $\vT$, and solving $\vT\vu=\va$ gives
$\vjph_\Psi=-e\rho\va$.  All photon factors are real, so no paramagnetic
current density is created.  The paramagnetic current density $\rho(\grad S+\vb)$ is
produced by a further factor in the tag: a phase times a coherent state
whose Berry connection is $\grad S+\vb$ (\cref{lem-coherent-tag}), made
permutation invariant by the lift of \cref{lem-symmetric-tensor-lift}.

\noindent\textbf{Step 1. An antisymmetric unit vector.}
Let $m=N(N-1)/2$ and let $V=\bigvee^m\RR^3$ be the
$m$th symmetric power of $\RR^3$, with its Hilbert tensor norm and
dimension $d=\binom{m+2}{2}$.  If $N=1$, then $V=\RR$ and we take
$\eta=1$.  For $N>1$, let
\[
 \mathcal O_N:=\big\{\vX=(\vx_1,\ldots,\vx_N)\in\RR^{3N}:
 \vx_i\neq\vx_j\text{ for all }i\neq j\big\}
\]
be the complement of the collision diagonals, and define
$\tau,\eta:\mathcal O_N\to V$ by
\[
 \tau(\vX)=\bigvee_{i<j}\vepsilon_{ij}(\vX),\qquad
 \eta(\vX)=\frac{\tau(\vX)}{\|\tau(\vX)\|_V},\qquad
 \vepsilon_{ij}(\vX)=\frac{\vx_i-\vx_j}{|\vx_i-\vx_j|}\in S^2,
\]
where $\vee$ denotes the symmetric tensor product.  For
$v_1,\ldots,v_m\in S^2$ the product $v_1\vee\cdots\vee v_m$ is nonzero:
its pairing with $y^{\otimes m}$ is a positive multiple of
$\prod_k(v_k\cdot y)$, which does not vanish for $y\in\RR^3$ outside
the finitely many planes $v_k^\perp$.  By continuity and compactness of
$(S^2)^m$, $\|\tau\|_V\geq c_N>0$ on $\mathcal O_N$, so $\eta$ is well
defined.  Since
$\grad\eta=\|\tau\|_V^{-1}(1-|\eta\rangle\langle\eta|)\grad\tau$,
$\tau$ is multilinear in the $\vepsilon_{ij}$, and
$|\grad_{\vx_n}\vepsilon_{ij}|\leq C|\vx_i-\vx_j|^{-1}$ for
$n\in\{i,j\}$, we obtain
\begin{equation}\label{photonic-antisymmetric-tag-bound}
 \|\eta(\vX)\|_V=1,\qquad
 \sum_{n=1}^N\|\grad_{\vx_n}\eta(\vX)\|_V^2
 \leq C_N\sum_{i<j}\frac1{|\vx_i-\vx_j|^2}.
\end{equation}
Permuting the electronic positions permutes the unordered pair factors
and reverses their orientations once for every inversion.  Thus
$\eta(\pi\vX)=\operatorname{sgn}(\pi)\eta(\vX)$.
Its coordinates in a real orthonormal basis of $V$ are real
antisymmetric functions.\footnote{A vector-valued tag is necessary.
For $N\geq2$ the configuration space of distinct points in $\RR^3$ is
connected, so a continuous real antisymmetric function must vanish
somewhere.  Rotating a pair $\vx_i,\vx_j$ about its midpoint traces a
sphere of directions $\vepsilon_{ij}$ on which the transposition $(ij)$
acts antipodally, so by the Borsuk--Ulam theorem a complex-valued one
must vanish as well.  For $N=2$, the tag $\eta=\vepsilon_{12}$ with
$\dim V=3$ is therefore of minimal dimension.}

The singularities in \cref{photonic-antisymmetric-tag-bound} are locally
integrable in three relative dimensions, and it is easy to see that
$\eta\in H^1_{\mathrm{loc}}(\RR^{3N};V)$.

\noindent\textbf{Step 2. Photon modes and their real structure.}
Work in the solenoidal one-photon representation
\cref{Gx-solenoidal}, using the unitary identification with the original
Fock space.  Define the conjugation by
$(J\vf)(\vk)=\ol{\vf(-\vk)}$.  A vector $\vf$ is called $J$-real if $J\vf=\vf$.
Every coupling function satisfies $J\vG^\mu_{\vx}=\vG^\mu_{\vx}$ for all
$\vx$ and $\mu$.  Indeed,
$P_\sol(-\vk)=P_\sol(\vk)$ is real and
$\ol{\wh\chi(\vk)}=\wh\chi(-\vk)$ by \cref{cutoff-assumption}, so
\[
 (J\vG^\mu_{\vx})(\vk)
 =\frac{P_\sol(\vk)\ve_\mu\ol{\wh\chi(\vk)}e^{-\i\vk\cdot\vx}}{\sqrt2}
 =\vG^\mu_{\vx}(\vk).
\]
Since the form factor is
nonzero, there is a bounded annulus
$A=\{\vk:r<|\vk|<R\}$, with $0<r<R<\infty$, such that
$\int_A|\wh\chi|^2>0$.  Define the translation-invariant matrix kernel
$\vK=(K^{\mu\nu})_{\mu,\nu=1}^3$ by
\[
 K^{\mu\nu}(\vx,\vy)
     =2\Re\dua{\iden_A\vG^\mu_{\vx}}{\iden_A\vG^\nu_{\vy}},\qquad
 \vM=\vK(\vx,\vx).
\]
The function $\iden_A\vG^\mu_{\vx}$ is $J$-real.  All its spatial
derivatives are
uniformly bounded both in the one-photon norm and after multiplication
by $\sqrt\omega$, since their momentum supports are bounded.
The real matrix $\vM$ is independent of $\vx$, and
\begin{equation*}
 \vc^{\mathsf T}\vM\vc
 =\int_A|\wh\chi(\vk)|^2|P_\sol(\vk)\vc|^2\dd\vk>0
 \qquad(\vc\in\RR^3\setminus\{0\}).
\end{equation*}
Indeed, $P_\sol(\vk)\vc=0$ only when $\vk$ is parallel to $\vc$, which has measure zero.  Thus $\vM$ is positive definite.  The kernel $\vK$ is
smooth with bounded derivatives and
$\vK(\vy,\vx)=\vK(\vx,\vy)^{\mathsf T}$.

Write $\Hil_A=\iden_A\Hil_\sol$ and
$\Hil_{\RR^3\setminus A}=\iden_{\RR^3\setminus A}\Hil_\sol$.  Under
the Fock factorization
\[
 \Fock_s(\Hil_\sol)
 \simeq\Fock_s(\Hil_A)\otimes\Fock_s(\Hil_{\RR^3\setminus A}),
\]
choose orthonormal $\xi_1,\ldots,\xi_d\in\Hil_{\RR^3\setminus A}$,
supported in the annulus $\{\vk:r/2<|\vk|<r\}$, which is disjoint
from $A$, and put
$\Xi_\alpha=a^\dag(\xi_\alpha)\Omega_{\RR^3\setminus A}$, where
$\Omega_{\RR^3\setminus A}$ is the vacuum of the second factor.  The free photon energy is bounded on
the span of the $\Xi_\alpha$.  Writing $\eta_\alpha$ for the
coordinates of $\eta$ in a real orthonormal basis of $V$, set
\[
 \zeta(\vX)=\sum_{\alpha=1}^d\eta_\alpha(\vX)\Xi_\alpha
 =a^\dag\Big(\sum_{\alpha=1}^d\eta_\alpha(\vX)\xi_\alpha\Big)\Omega_{\RR^3\setminus A}.
\]
This is a unit antisymmetric vector and, for $n=1,\ldots,N$,
\begin{equation}\label{photonic-tag-zero-current}
 \dua{\zeta}{\grad_{\vx_n}\zeta}=0,\qquad
 \dua{\zeta}{\AF_{\RR^3\setminus A}(\vx)\zeta}=0.
\end{equation}
The first equality follows from
$\sum_\alpha\eta_\alpha\grad_{\vx_n}\eta_\alpha=\frac12\grad_{\vx_n}\|\eta\|_V^2=0$, and the
second because all $\Xi_\alpha$ have the same photon number and
the field changes photon number by one.
Here $\AF_{\RR^3\setminus A}$ denotes the field with coupling functions
projected onto $\Hil_{\RR^3\setminus A}$.  The auxiliary modes need not be
decoupled from the electrons, since $\wh\chi$ may be nonzero on their
support; the second identity ensures that they nevertheless contribute
nothing to the field expectation.

\noindent\textbf{Step 3. The displacement profile.}
On the real Hilbert space $L^2(\rho\,\dd\vx;\RR^3)$ (weighted $L^2$ space),
define
\[
 (\vK_\rho\vu)(\vx)=\int_{\RR^3} \vK(\vx,\vy)\vu(\vy)\rho(\vy)\dd\vy,
 \qquad \vT=\vM+\frac{N-1}{N}\vK_\rho.
\]
The integral operator $\vK_\rho$ is bounded and self-adjoint.  It is positive,
since
\begin{equation*}
 \dua{\vu}{\vK_\rho\vu}_{L^2(\rho\,\dd\vx;\RR^3)}
 =2\left\|\int_{\RR^3}\rho(\vy)\sum_{\mu=1}^3
                      u^\mu(\vy)\iden_A(\cdot)\vG^\mu_{\vy}(\cdot)\dd\vy\right\|^2
 \geq0,
\end{equation*}
where the norm is that of the one-photon space $\Hil_\sol$.
If $\lambda>0$ is the smallest eigenvalue of $\vM$, then
$\vT\geq\lambda\iden$.  Hence for $\vu=\vT^{-1}\va$, 
\begin{equation}\label{photonic-field-equation}
 \|\sqrt\rho\,\vu\|\leq\lambda^{-1}\|\sqrt\rho\,\va\|.
\end{equation}
Because $\vK$ has bounded derivatives and
$\int_{\RR^3}\rho|\vu|\leq\sqrt N\|\sqrt\rho\,\vu\|$, the function
$\vK_\rho\vu$ is smooth with bounded derivatives.  The solution $\vu$ is
determined only $\rho$-almost everywhere.  The equation $\vT\vu=\va$ is equivalent to
$\vM\vu=\va-\frac{N-1}N\vK_\rho\vu$ $\rho$-almost everywhere, and the
right-hand side is defined almost everywhere on $\RR^3$, we fix $\vu$ on
all of $\RR^3$ by
\[
 \vu(\vx)=\vM^{-1}\Bigl(\va(\vx)-\frac{N-1}{N}(\vK_\rho\vu)(\vx)\Bigr).
\]
Then $\vu$ is in $H^1_{\mathrm{loc}}$ and satisfies
\begin{equation}\label{photonic-t-Sobolev-bound}
 \int_{\RR^3}\rho\bigl(|\vu|^2+|\grad\vu|^2\bigr)
 \leq C_{N,\chi}\int_{\RR^3}\rho\bigl(|\va|^2+|\grad\va|^2\bigr).
\end{equation}
Indeed, the zeroth-order part is \cref{photonic-field-equation}.  For
the gradient, Cauchy--Schwarz with $\int_{\RR^3}\rho=N$ gives
\[
 \sup_{\vx}|\grad(\vK_\rho\vu)(\vx)|
 \leq\|\grad\vK\|_\infty\int_{\RR^3}\rho|\vu|
 \leq\sqrt N\lambda^{-1}\|\grad\vK\|_\infty\|\sqrt\rho\,\va\|,
\]
and $\grad\vu=\vM^{-1}(\grad\va-\frac{N-1}N\grad\vK_\rho\vu)$.
Only the bounded first derivatives of $\vK$ are used here, and no derivative
of the density is taken.

\noindent\textbf{Step 4. The representing vector and its energy.}
Let $W(\vf)$ be the Weyl operator \cref{weyl-definitions} on
$\Fock_s(\Hil_A)$, and let $\Omega_A$ be the vacuum of this factor.
Let $B(\vX)=\prod_{n=1}^N\sqrt{\rho(\vx_n)/N}$ be the electronic
amplitude.  It has norm one and
$\sum_n\|\grad_{\vx_n}B\|^2=\|\grad\sqrt\rho\|^2$.  Define the
displacement
\[
 \vf_{\vX}=\sum_{n=1}^N\sum_{\mu=1}^3
                 u^\mu(\vx_n)\iden_A\vG^\mu_{\vx_n}
\]
and the state
\begin{equation}\label{photonic-conditional-state}
\boxed{\Psi(\vX)=B(\vX)W(\vf_{\vX})\Omega_A\otimes\zeta(\vX).}
\end{equation}
The displacement $\vf_\vX$ is symmetric in the electronic positions and
$J$-real.  Thus $\Psi$ is antisymmetric and
\[
 \|\Psi(\vX)\|^2=B(\vX)^2=\prod_{n=1}^N\frac{\rho(\vx_n)}N.
\]
Here $\|\Psi(\vX)\|$ is the norm in the photon Fock space.
The state $\Psi$ is normalized and has density $\rho$.
Since $\vX\mapsto\vf_\vX$ is $J$-real and lies in
$H^1_{\rm loc}(\RR^{3N};\Hil_A)$, \cref{prop-weyl-coherent-derivatives}(iv)
shows that $W(\vf_\vX)\Omega_A$ is a unit $H^1_{\rm loc}$ vector and
gives, for $n=1,\ldots,N$,
\begin{equation}\label{photonic-real-coherent-derivatives}
 \begin{gathered}
 \dua{W(\vf_\vX)\Omega_A}{\grad_{\vx_n}\bigl(W(\vf_\vX)\Omega_A\bigr)}
     =0,\\
 \|\grad_{\vx_n}\bigl(W(\vf_\vX)\Omega_A\bigr)\|^2
     =\|\grad_{\vx_n}\vf_\vX\|^2.
 \end{gathered}
\end{equation}
Together with \cref{photonic-tag-zero-current}, these identities make
the three terms obtained by differentiating
\cref{photonic-conditional-state} pairwise orthogonal.  Hence
\begin{align*}
 \sum_n\|\grad_{\vx_n}\Psi\|^2
 &=\sum_n\|\grad_{\vx_n}B\|^2
   +\int_{\RR^{3N}} B(\vX)^2\sum_n\|\grad_{\vx_n}\zeta\|^2\dd\vX\\
 &\quad+\int_{\RR^{3N}} B(\vX)^2\sum_n\|\grad_{\vx_n}\vf_{\vX}\|^2\dd\vX\\
 &\leq C_N\|\grad\sqrt\rho\|^2
       +C_{N,\chi}\int_{\RR^3}\rho\bigl(|\va|^2+|\grad\va|^2\bigr).
\end{align*}
To justify the last inequality, first consider the term involving
$\grad_{\vx_n}\zeta$.  The identity
$\|\grad_{\vx_n}\zeta\|=\|\grad_{\vx_n}\eta\|_V$ and
\cref{photonic-antisymmetric-tag-bound} reduce it to
$C_N\int B^2\sum_{i<j}|\vx_i-\vx_j|^{-2}$.  For each pair, integrating
out the other positions and applying Hardy's inequality
\[
 \int_{\RR^3}\frac{\rho(\vx)}{|\vx-\vy|^2}\,\dd\vx
 \leq4\|\grad\sqrt\rho\|^2,\qquad\vy\in\RR^3,
\]
gives
\begin{equation}\label{photonic-tag-Hardy-bound}
 \int_{\RR^{3N}}\frac{B(\vX)^2}{|\vx_i-\vx_j|^2}\dd\vX
 =\frac1{N^2}\int_{\RR^3}\int_{\RR^3}
     \frac{\rho(\vx)\rho(\vy)}{|\vx-\vy|^2}\dd\vx\dd\vy
 \leq\frac4N\|\grad\sqrt\rho\|^2.
\end{equation}
For the last term, differentiation in $\vx_n$ acts only on the $n$th
summand of $\vf_{\vX}$, so the uniform bounds on
$\iden_A\vG^\mu_{\vx}$ and its derivatives and \cref{photonic-t-Sobolev-bound} apply.

The auxiliary modes $\xi_\alpha$ in Step 2 are supported where
$|\vk|<r<R$.  On the one-photon sector the free
photon energy acts as multiplication by $\omega(\vk)=|\vk|$, so
\begin{align*}
 \dua{\zeta(\vX)}{H_{\rad,\RR^3\setminus A}\zeta(\vX)}
 &=\left\|\sqrt\omega\sum_{\alpha=1}^d
             \eta_\alpha(\vX)\xi_\alpha\right\|^2\leq R\sum_{\alpha=1}^d|\eta_\alpha(\vX)|^2
 =R.
\end{align*}
Here we used orthonormality of the $\xi_\alpha$ and $\|\eta(\vX)\|_V=1$.
Since $\int_{\RR^{3N}}B(\vX)^2\dd\vX=1$, the integrated contribution
of this term is also at most $R$.
The factorization \cref{weyl-factorization} and the coherent-vector
energy \cref{weyl-coherent-energy} give
\begin{align*}
 \|\sqrt{H_\rad}\Psi\|^2
 &=\int_{\RR^{3N}} B(\vX)^2\left(\|\sqrt\omega\vf_{\vX}\|^2
          +\dua{\zeta(\vX)}{H_{\rad,\RR^3\setminus A}\zeta(\vX)}\right)\dd\vX\\
 &\leq N\int_{\RR^3}\rho(\vx)
       \left\|\sqrt\omega\sum_{\mu=1}^3
       u^\mu(\vx)\iden_A\vG^\mu_{\vx}\right\|^2\dd\vx+R\\
 &\leq C_{N,\chi}\int_{\RR^3}\rho|\vu|^2+R\\
 &\leq C_{N,\chi}\lambda^{-2}\int_{\RR^3}\rho|\va|^2+R\\
 &\leq C_{N,\chi}\int_{\RR^3}\rho|\va|^2+C_{N,\chi}.
\end{align*}
The first inequality uses $\|\sum_ng_n\|^2\leq N\sum_n\|g_n\|^2$,
the one-particle marginal $\rho/N$ of $B(\vX)^2$, and the photon energy
bound above.  The second uses the uniform bounds on
$\sqrt\omega\iden_A\vG^\mu_{\vx}$, and the third is
\cref{photonic-field-equation} squared.
Fixing $A$ once for the given form factor allows $\lambda^{-2}$ and
$R$ to be absorbed into $C_{N,\chi}$ in the last inequality,
independently of $\rho$ and $\va$.
This proves \cref{photonic-representability-energy} for $S=0$
and $\vb=0$.  By \cref{PF-canonical-form-domain}, $\Psi$ lies in the Pauli--Fierz
form domain, including the Coulomb form.

For completeness, these derivative computations are valid for the
stated weak regularity.  Since $W(\vf_\vX)\Omega_A$ is a unit
$H^1_{\rm loc}$ vector, so is
\[
U(\vX)=W(\vf_{\vX})\Omega_A\otimes\zeta(\vX).
\]
  The estimates above give
$B\grad U\in L^2$.  Truncating $B$ at height $L$, applying the local
Sobolev product rule, and passing to the limit proves
$\grad(BU)=(\grad B)U+B\grad U$ in $L^2$.  No boundedness or strict positivity
of $\rho$ is required.

\noindent\textbf{Step 5. The prescribed current densities.}
By \cref{photonic-tag-zero-current,photonic-real-coherent-derivatives},
the photon scalar product satisfies
$\dua{U(\vX)}{\grad_{\vx_n}U(\vX)}=0$ for almost every electronic
configuration $\vX$.  Hence
$\dua{\Psi(\vX)}{\grad_{\vx_n}\Psi(\vX)}=B(\vX)\grad_{\vx_n}B(\vX)$,
which is real, so $\vjp_\Psi=0$.

To compute the field-induced current density, we first evaluate the
field expectation in the photon state $U(\vX)$ at fixed electronic
configuration $\vX$.  By \cref{weyl-field-shift}, the $\mu$th component of the field
expectation in the coherent state
$W(\vf_{\vX})\Omega_A$ of the first photon factor is
$2\Re\dua{\vG^\mu_{\vx}}{\vf_{\vX}}$, and the second
factor contributes zero.  Expanding $\vf_{\vX}\in\Hil_A$ and using the
definition of $\vK$ therefore gives
\[
 \dua{U(\vX)}{\AF(\vx)U(\vX)}
 =\sum_{n=1}^N\vK(\vx,\vx_n)\vu(\vx_n).
\]
For $\vx_1=\vx$, the $n=1$ term in this field expectation is
$\vK(\vx,\vx)\vu(\vx)=\vM\vu(\vx)$.
Moreover, for $\rho$-almost every $\vx$,
\[
 \frac{N B(\vx,\vx_2,\ldots,\vx_N)^2}{\rho(\vx)}
 =\prod_{n=2}^N\frac{\rho(\vx_n)}N.
\]
Thus, conditional on $\vx_1=\vx$, the other positions have independent
density $\rho/N$.  Averaging leaves the $n=1$ contribution unchanged,
while each of the remaining $N-1$ contributions gives the same integral.
Consequently,
\begin{align*}
 \frac{N\Re\dua{\Psi}{\AF(\vx)\Psi}_{\vx}}{\rho(\vx)}
 &=\vM\vu(\vx)+\frac{N-1}{N}\int_{\RR^3}
     \vK(\vx,\vy)\vu(\vy)\rho(\vy)\dd\vy\\
 &=(\vT\vu)(\vx)=\va(\vx),
\end{align*}
and \cref{jph-def} gives $\vjph_\Psi=-e\rho\va$.  Both current densities vanish on $\{\rho=0\}$, which proves
\cref{photonic-representability-triple} for $S=0$ and $\vb=0$.

\noindent\textbf{Step 6. The paramagnetic current density.}
Let $t$ be the tag of \cref{lem-coherent-tag} with $\vw=\vb$ and
$\lambda=1$, with values in $\Fock_s(\CC^3)$.  Thus $\|t\|=1$,
$\Im\dua{t}{\grad t}=\grad S+\vb$, and
\[
 \|\grad t\|^2=|\grad S+\vb|^2+\frac12|\grad\vb|^2+\frac32 .
\]
Let $L$ be its symmetric lift from \cref{lem-symmetric-tensor-lift},
fix an orthonormal basis $(e_\beta)$ of the separable Hilbert space
$\mathcal K_N$ in which $L$ takes its values, and put
$L_\beta=\dua{e_\beta}{L}$.  By
\cref{symmetric-tensor-lift-connection,symmetric-tensor-lift-metric},
\begin{gather*}
 \sum_\beta|L_\beta|^2=1,\qquad
 \sum_\beta\ol L_\beta\grad_{\vx_n}L_\beta=\i(\grad S+\vb)(\vx_n),\\
 \sum_{n,\beta}|\grad_{\vx_n}L_\beta|^2
 \leq\sum_n\|\grad t(\vx_n)\|^2+b_N\sum_{i<j}\frac1{|\vx_i-\vx_j|^2}.
\end{gather*}
Choose orthonormal $\xi_{\alpha\beta}\in\Hil_{\RR^3\setminus A}$,
$1\leq\alpha\leq d$, $\beta\geq1$, supported in the annulus of Step 2,
and replace $\zeta$ by
\[
 \tilde{\zeta}(\vX)=\sum_{\alpha,\beta}\eta_\alpha(\vX)L_\beta(\vX)\,
   a^\dag(\xi_{\alpha\beta})\Omega_{\RR^3\setminus A}.
\]
By orthonormality of the $\xi_{\alpha\beta}$, $\tilde{\zeta}(\vX)$ has unit norm,
\[
 \|\tilde{\zeta}(\vX)\|^2
 =\sum_{\alpha,\beta}|\eta_\alpha(\vX)|^2|L_\beta(\vX)|^2=1.
\]
It is again an antisymmetric one-photon vector, so
$\dua{\tilde{\zeta}}{\AF_{\RR^3\setminus A}(\vx)\tilde{\zeta}}=0$, and its photon
energy is bounded as before.  Since
$\sum_\alpha\eta_\alpha\grad\eta_\alpha=\frac12\grad\|\eta\|_V^2=0$,
\[
 \dua{\tilde{\zeta}}{\grad_{\vx_n}\tilde{\zeta}}=\i(\grad S+\vb)(\vx_n),\qquad
 \|\grad_{\vx_n}\tilde{\zeta}\|^2
 =\|\grad_{\vx_n}\eta\|_V^2+\sum_\beta|\grad_{\vx_n}L_\beta|^2.
\]
In Step 4, each new cross term contains either the real factor
$B\grad_{\vx_n}B$ or the vanishing coherent-state pairing in
\cref{photonic-real-coherent-derivatives}, so it does not contribute.
Under $B^2$ each $\vx_n$ has density $\rho/N$, so by
\cref{photonic-tag-Hardy-bound} the kinetic energy increases by at most
\begin{align*}
 &\int_{\RR^3}\rho\|\grad t\|^2+2(N-1)b_N\|\grad\sqrt\rho\|^2\\
 &\quad\leq\int_{\RR^3}\rho\Big(2|\grad S|^2+2|\vb|^2
   +\frac12|\grad\vb|^2\Big)+\frac32N+2(N-1)b_N\|\grad\sqrt\rho\|^2.
\end{align*}
Since $L\in H^1_{\mathrm{loc}}$, the regularity argument of Step 4
applies verbatim.  Step 5 now gives $\vjp_\Psi=\rho(\grad S+\vb)$ and
still $\vjph_\Psi=-e\rho\va$.  This proves
\cref{photonic-representability-triple,photonic-representability-energy}.

\subsection{Proof of \texorpdfstring{\cref{prop-pure-nonattainment}}{Proposition \getrefnumber{prop-pure-nonattainment}}}
\label{proof:prop-pure-nonattainment}

Fix $e\geq0$ and the form factor.  On thin elliptic loops we prescribe a
velocity with circulation $\pi$, half the circulation quantum of a pure
electronic state.  A pure minimizer has few soft photons by
\cref{prop-optimizer-soft-spectrum} and, by matching energy bounds, few
hard photons.  The phase of its vacuum component would then wind by
$\pi$ around a loop, which is impossible.

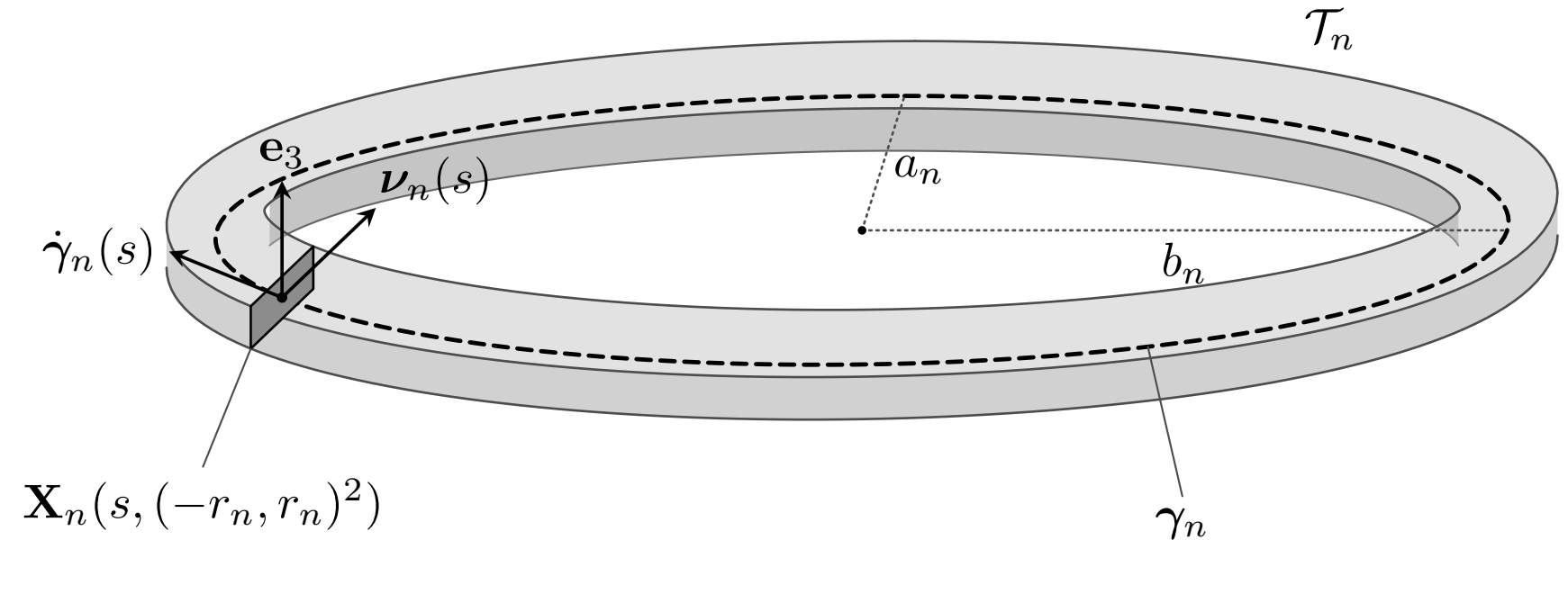
\begin{figure}[!htbp]
\centering
\begin{tikzpicture}[
 x={(0.39cm,1.235cm)},
 y={(1.755cm,0cm)},
 z={(0cm,0.78cm)},
 line cap=round,
 line join=round,
 declare function={
  NN(\p)=sqrt((2.7*cos(\p))^2+(0.8*sin(\p))^2);
  ox(\p,\t)=(0.8-\t*2.7/NN(\p))*cos(\p);
  oy(\p,\t)=(2.7-\t*0.8/NN(\p))*sin(\p);
 }
]
 \def\R{0.2}
 \fill[black!45,fill opacity=.5]
   plot[domain=-80:80,samples=60,variable=\p]
     ({ox(\p,\R)},{oy(\p,\R)},\R)
   -- plot[domain=80:-80,samples=60,variable=\p]
     ({ox(\p,\R)},{oy(\p,\R)},-\R) -- cycle;
 \draw[black!60,line width=.4pt]
   plot[domain=-80:80,samples=60,variable=\p]
     ({ox(\p,\R)},{oy(\p,\R)},-\R);
 \fill[black!25,fill opacity=.45]
   plot[domain=0:360,samples=160,variable=\p]
     ({ox(\p,-\R)},{oy(\p,-\R)},\R)
   -- plot[domain=360:0,samples=160,variable=\p]
     ({ox(\p,\R)},{oy(\p,\R)},\R) -- cycle;
 \fill[black!40,fill opacity=.45]
   plot[domain=86.23:266.23,samples=100,variable=\p]
     ({ox(\p,-\R)},{oy(\p,-\R)},\R)
   -- plot[domain=266.23:86.23,samples=100,variable=\p]
     ({ox(\p,-\R)},{oy(\p,-\R)},-\R) -- cycle;
 \draw[black!70,line width=.5pt]
   plot[domain=0:360,samples=160,variable=\p]
     ({ox(\p,-\R)},{oy(\p,-\R)},\R);
 \draw[black!70,line width=.5pt]
   plot[domain=0:360,samples=160,variable=\p]
     ({ox(\p,\R)},{oy(\p,\R)},\R);
 \draw[black!70,line width=.5pt]
   plot[domain=86.23:266.23,samples=100,variable=\p]
     ({ox(\p,-\R)},{oy(\p,-\R)},-\R);
 \draw[black,densely dashed,line width=.9pt]
   plot[domain=0:360,samples=160,variable=\p]
     ({0.8*cos(\p)},{2.7*sin(\p)},0);
 \draw[black!70,densely dotted,line width=.5pt] (0,0,0) -- (0,2.7,0)
   node[pos=.5,below=-1pt,black] {$b_n$};
 \draw[black!70,densely dotted,line width=.5pt] (0,0,0) -- (0.8,0,0)
   node[pos=.45,right=-1pt,black] {$a_n$};
 \fill[black] (0,0,0) circle (.9pt);
 \pgfmathsetmacro{\pz}{240}
 \pgfmathsetmacro{\gx}{0.8*cos(\pz)}
 \pgfmathsetmacro{\gy}{2.7*sin(\pz)}
 \pgfmathsetmacro{\nx}{-2.7*cos(\pz)/NN(\pz)}
 \pgfmathsetmacro{\ny}{-0.8*sin(\pz)/NN(\pz)}
 \pgfmathsetmacro{\tx}{-0.8*sin(\pz)/NN(\pz)}
 \pgfmathsetmacro{\ty}{2.7*cos(\pz)/NN(\pz)}
 \filldraw[fill=black!45,draw=black,line width=.5pt]
   (\gx-\R*\nx,\gy-\R*\ny,-\R) -- (\gx+\R*\nx,\gy+\R*\ny,-\R)
   -- (\gx+\R*\nx,\gy+\R*\ny,\R) -- (\gx-\R*\nx,\gy-\R*\ny,\R) -- cycle;
 \draw[-stealth,line width=.7pt] (\gx,\gy,0) -- ++(0.6*\tx,0.6*\ty,0)
   node[left=-1pt] {$\dot{\vgamma}_n(s)$};
 \draw[-stealth,line width=.7pt] (\gx,\gy,0) -- ++(0.6*\nx,0.6*\ny,0)
   node[above right=-3pt] {$\vnu_n(s)$};
 \draw[-stealth,line width=.7pt] (\gx,\gy,0) -- ++(0,0,1.1)
   node[above=-1pt] {$\ve_3$};
 \fill[black] (\gx,\gy,0) circle (1.1pt);
 \draw[black!70,line width=.4pt] (\gx-\R*\nx,\gy-\R*\ny,-\R)
   -- ++(-0.45,-0.1,-0.4)
   node[below=-1pt,black] {$\vX_n(s,(-r_n,r_n)^2)$};
 \draw[black!70,line width=.4pt] ({0.8*cos(150)},{2.7*sin(150)},0)
   -- ++(-0.7,0.3,-0.3) node[below=-1pt,black] {$\vgamma_n$};
 \node[black,above right=0pt] at ({ox(35,-\R)},{oy(35,-\R)},\R)
   {$\mathcal T_n$};
\end{tikzpicture}
\caption{The solid torus $\mathcal T_n$ around the ellipse $\vgamma_n$
(dashed), with the orthonormal frame
$(\dot{\vgamma}_n,\vnu_n,\ve_3)$ and the square cross-section
$\vX_n(s,(-r_n,r_n)^2)$ at $\vgamma_n(s)$.  Not to scale: in the proof,
$a_n\leq b_n^3$ and $r_n=a_n^2b_n^3/10$.}
\label{fig:thin-loop}
\end{figure}

\noindent\textbf{Step 1. The pairs $(\rho_n,\vj_n)$.}
Let $0<a_n<b_n$ and $0<r_n<a_n^2/b_n$ be scales to be chosen
later.  Let $\vgamma_n:\RR/L_n\ZZ\to\RR^3$ parametrize the ellipse
$\{x_1^2/a_n^2+x_2^2/b_n^2=1,\ x_3=0\}$ by arclength, with inward unit
normal $\vnu_n$ and curvature $\kappa_n$.  Then
$4b_n\leq L_n\leq2\pi b_n$, $\kappa_n\leq\tilde{\kappa}_n=b_n/a_n^2$, and
$\int_0^{L_n}\kappa_n\dd s=2\pi$.  The coordinates
\[
 \vX_n(s,\vt)=\vgamma_n(s)+t_1\vnu_n(s)+t_2\ve_3,
 \qquad \vt\in(-r_n,r_n)^2,
\]
parametrize an open solid torus $\mathcal T_n$ (\cref{fig:thin-loop}),
since the ellipse is convex and $r_n<1/\tilde{\kappa}_n$.  By the Frenet equations,
$\partial_s\vX_n=J_n\dot{\vgamma}_n$ with $J_n=1-\kappa_nt_1$, and
$(\dot{\vgamma}_n,\vnu_n,\ve_3)$ is orthonormal.  Hence
$\dd\vx=J_n\dd s\dd\vt$, $\grad s=J_n^{-1}\dot{\vgamma}_n$, and
\begin{equation*}
 |J_n-1|\leq\epsilon_n:=r_n\tilde{\kappa}_n<1.
\end{equation*}
With $w(\vz)=\cos^2(\pi z_1/2)\cos^2(\pi z_2/2)$ on $(-1,1)^2$, define, on
$\mathcal T_n$ and by zero outside,
\begin{equation}\label{arbitrary-coupling-nonattainment-pair}
 \rho_n(\vX_n(s,\vt))=
 \frac{w(\vt/r_n)}{L_nr_n^2J_n(s,\vt)},\qquad
 \vj_n=\frac{\pi}{L_n}\rho_n\grad s.
\end{equation}
In the variables $\theta=s/L_n\in\mathbb T:=\RR/\ZZ$ and
$\vz=\vt/r_n\in(-1,1)^2$,
\begin{equation}\label{thin-loop-measure}
 \rho_n\dd\vx=w(\vz)\dd\theta\dd\vz.
\end{equation}
Hence $\int\rho_n=1$, and $\sqrt{\rho_n}$ is Lipschitz with compact
support, as $\sqrt w$ is Lipschitz and vanishes on the boundary of
$(-1,1)^2$.  The equal mixture of the normalized orbitals $\phi_{n,0}=\sqrt{\rho_n}$ and
$\phi_{n,1}=\sqrt{\rho_n}\,e^{2\pi\i s/L_n}$ in $H^1(\RR^3)$ has density
$\rho_n$ and paramagnetic current density $\vj_n$.  A single orbital
cannot do the same, because its velocity has quantized circulation.

\noindent\textbf{Claim.}  Let $\psi\in H^1(\RR^3)$ with
$|\psi|^2=\rho_n$.  Then, for almost every $\vt\in(-r_n,r_n)^2$, the
velocity $\Im(\ol\psi\grad\psi)/|\psi|^2$ has circulation
\[
 \oint_{\mathcal C_{n,\vt}}\frac{\Im(\ol\psi\grad\psi)}{|\psi|^2}\cdot\dd\bm{\ell}
 =\int_0^{L_n}\frac{\Im(\ol\psi\grad\psi)}{\rho_n}\big(\vX_n(s,\vt)\big)
 \cdot\partial_s\vX_n(s,\vt)\dd s\in2\pi\ZZ
\]
around the loop $\mathcal C_{n,\vt}$ parametrized by
$s\mapsto\vX_n(s,\vt)$.

\noindent\textit{Proof of the claim.}
On $\mathcal T_n$, where $\rho_n$ is smooth and positive,
$u=\psi/\sqrt{\rho_n}\in H^1_{\mathrm{loc}}(\mathcal T_n)$ satisfies
$|u|=1$.  As
$\ol\psi\grad\psi=\sqrt{\rho_n}\grad\sqrt{\rho_n}+\rho_n\ol u\grad u$,
the velocity is $\Im(\ol u\grad u)$.  For almost every $\vt$, the
function $u_\vt(s)=u(\vX_n(s,\vt))$ is absolutely continuous on
$\RR/L_n\ZZ$ with values in the unit circle.  Hence
$u_\vt=e^{\i\vartheta}$ for an absolutely continuous
$\vartheta:[0,L_n]\to\RR$ with
$\vartheta(L_n)-\vartheta(0)\in2\pi\ZZ$.  The integrand above equals
$\Im(\ol{u_\vt}\,\partial_su_\vt)=\partial_s\vartheta$, so the
circulation is
$\vartheta(L_n)-\vartheta(0)$.  This proves the claim.

Since $\grad s\cdot\partial_s\vX_n=1$, the velocity
$\vj_n/\rho_n=(\pi/L_n)\grad s$ has circulation
\[
 \oint_{\mathcal C_{n,\vt}}\frac{\vj_n}{\rho_n}\cdot\dd\bm{\ell}
 =\frac{\pi}{L_n}\int_0^{L_n}\grad s\cdot\partial_s\vX_n\dd s=\pi
\]
around every loop $\mathcal C_{n,\vt}$, half the circulation quantum
$2\pi$.  Hence no
$\psi\in H^1(\RR^3)$ satisfies $|\psi|^2=\rho_n$ and
$\Im(\ol\psi\grad\psi)=\vj_n$.  Consequently, for every $e\geq0$, no
photon-free state $\psi\otimes\Omega$ has the pair $(\rho_n,\vj_n)$,
since $\dua{\Omega}{\AF(\vx)\Omega}=0$ makes its combined current
density equal to $\Im(\ol\psi\grad\psi)$.

The claim does not yet prove the proposition, because a minimizer need
not be photon-free.  Photons can carry the missing half quantum: by
\cref{thm-many-electron-soft-purification}, there are pure states with
the pair $(\rho_n,\vj_n)$ whose energies approach the infimum.  For
$e>0$, the field also enters the combined current, and its circulation
around a loop could make up the difference.  The remaining steps rule
this out for a minimizer.  Step~2 shows that the field has small
circulation around thin loops.  Steps~3 and~4 give matching lower and
upper energy bounds, from which Step~5 deduces that a pure minimizer is
almost photon-free.  Step~6 then applies the winding argument of the
claim to its vacuum component.

\noindent\textbf{Step 2. The flux through the loops.}
For $\vt\in(-r_n,r_n)^2$, let
\[
 g_{n,s,\vt}=\sum_{\mu=1}^3G^\mu_{\vX_n(s,\vt)}\,
                 \partial_sX_n^\mu(s,\vt),\qquad
 \mathcal F_{n,\vt}=\int_0^{L_n}g_{n,s,\vt}\dd s.
\]
Thus $\varPhi(g_{n,s,\vt})=\AF(\vX_n)\cdot\partial_s\vX_n$, and
$\varPhi(\mathcal F_{n,\vt})$ is the circulation of the field around the loop
$s\mapsto\vX_n(s,\vt)$.

For $\nu=0,1$, put $M_\nu=\|\omega^{-\nu/2}\wh\chi\|$ and
$d_\nu(\ell)=\|\min\{1,\ell\omega\}\,\omega^{-\nu/2}\wh\chi\|$, with the
norm of $L^2(\RR^3)$.  By \cref{cutoff-assumption}, $M_\nu<\infty$ and
$d_\nu(\ell)\downarrow0$ as $\ell\downarrow0$.  By \cref{Gx-norms} and
$|e^{\i t}-1|\leq2\min\{1,|t|\}$,
\begin{equation}\label{thin-loop-kernel-continuity}
 \sum_{\mu=1}^3\|\omega^{-\nu/2}G^\mu_{\vx}\|^2=M_\nu^2,\qquad
 \sum_{\mu=1}^3\|\omega^{-\nu/2}(G^\mu_{\vx}-G^\mu_{\vy})\|^2
 \leq4d_\nu(|\vx-\vy|)^2.
\end{equation}
We now choose $a_n$.  As $d_\nu(\ell)\downarrow0$, we may take $a_n$ so
small that
\begin{equation}\label{thin-loop-scales}
 a_n\leq b_n^3,\qquad d_0(2a_n)+d_1(2a_n)\leq b_n^2.
\end{equation}
We claim that, with $C$ depending only on the form factor,
\begin{equation}\label{thin-loop-flux-kernel}
 \sup_{\vt\in(-r_n,r_n)^2}\|\omega^{-\nu/2}\mathcal F_{n,\vt}\|\leq Cb_nL_n^2,
 \qquad\nu=0,1.
\end{equation}
All norms in this step are $\|\omega^{-\nu/2}\cdot\|$ on $\Hil_\ph$.
Parametrize the central ellipse by
$\phi\mapsto(a_n\cos\phi,b_n\sin\phi,0)$.  Since $x_1$ has total
variation $4a_n$, the $\mu=1$ part of $\mathcal F_{n,0}$ is at most $4a_nM_\nu$.
In the $\mu=2$ part, $\phi\mapsto\pi-\phi$ pairs points with equal $x_2$,
opposite orientation, and distance at most $2a_n$, so
\cref{thin-loop-kernel-continuity} bounds it by $4b_nd_\nu(2a_n)$.  Off
the central ellipse, $\partial_s\vX_n=(1-\kappa_nt_1)\dot{\vgamma}_n$
with $|\kappa_nt_1|\leq\kappa_nr_n$, and
$|\vX_n(s,\vt)-\vgamma_n(s)|\leq\sqrt2r_n$.  Since
$\int_0^{L_n}\kappa_n\dd s=2\pi$, this gives
$\|\mathcal F_{n,\vt}-\mathcal F_{n,0}\|\leq2L_nd_\nu(2r_n)+2\pi r_nM_\nu$.  By
\cref{thin-loop-scales} and $r_n<a_n^2/b_n<a_n$, all four terms are at most
$C(1+M_\nu)b_n^3\leq C(1+M_\nu)b_nL_n^2$.

\noindent\textbf{Step 3. A lower bound for pure representatives.}
Let $\Psi\in \mathcal Q_\PF$ be normalized with
$(\rho_\Psi,\vjc_\Psi)=(\rho_n,\vj_n)$.  For $N=1$,
$\rho_\Psi(\vx)=\|\Psi(\vx)\|^2$ in the norm of $\Fock_\ph$.  Hence
$\Psi=0$ outside $\mathcal T_n$.  On $\mathcal T_n$, where $\rho_n$ is
smooth and positive, $\Psi=\sqrt{\rho_n}\,u$ with
$u\in H^1_{\mathrm{loc}}(\mathcal T_n;\Fock_\ph)$, $\|u\|=1$, and
$u(\vx)\in D(\sqrt{H_\rad})$ for almost every $\vx$.  Since
$\Re\dua{u}{\grad u}=\frac12\grad\|u\|^2=0$ and $\dua{u}{\Af^\mu u}$ is
real, differentiating $\sqrt{\rho_n}$ produces no cross terms, and
\begin{gather*}
 \vjc_\Psi=\rho_n\Re\dua{u}{(-\i\grad-e\AF)u},\\
 \dua{\Psi}{H^{0,0}\Psi}=\frac12\int_{\RR^3}|\grad\sqrt{\rho_n}|^2
 +\frac12\int_{\mathcal T_n}\rho_n\|(-\i\grad-e\AF)u\|^2\dd\vx
 +\dua{\Psi}{H_\rad\Psi}.
\end{gather*}
In the coordinates $(s,\vt)$ of Step~1, put
$\mathcal A_n=\varPhi(g_{n,s,\vt})=\AF(\vX_n)\cdot\partial_s\vX_n$ and
$D_s=-\i\partial_s-e\mathcal A_n$.  By the chain rule,
$(-\i\grad-e\AF)u\cdot\partial_s\vX_n=D_su$.  Taking the dot product of
$\vjc_\Psi=\vj_n$ with $\partial_s\vX_n$ and using
$\grad s\cdot\partial_s\vX_n=1$, we get
\[
 \rho_n\Re\dua{u}{D_su}
 =\vjc_\Psi\cdot\partial_s\vX_n
 =\vj_n\cdot\partial_s\vX_n
 =\frac{\pi}{L_n}\rho_n\,\grad s\cdot\partial_s\vX_n
 =\frac{\pi}{L_n}\rho_n.
\]
Dividing by $\rho_n>0$ gives
\begin{equation}\label{thin-loop-circulation}
 \Re\dua{u}{D_su}=\frac{\pi}{L_n}
 \qquad\text{almost everywhere on }\mathcal T_n.
\end{equation}
As $\|u\|=1$, this gives $\|D_su\|^2=\|(D_s-\pi/L_n)u\|^2+\pi^2/L_n^2$.
Keeping only the $\dot{\vgamma}_n$-component of the covariant gradient and
using $J_n\leq1+\epsilon_n$, we obtain
\begin{equation}\label{thin-loop-energy-split}
 \begin{aligned}
 &\dua{\Psi}{H^{0,0}\Psi}-\frac12\int_{\RR^3}|\grad\sqrt{\rho_n}|^2\\
 &\qquad\geq\frac{1}{2(1+\epsilon_n)^2}\left[\frac{\pi^2}{L_n^2}
 +\int_{\mathcal T_n}\rho_n\|(D_s-\pi/L_n)u\|^2\dd\vx\right]
 +\dua{\Psi}{H_\rad\Psi}.
 \end{aligned}
\end{equation}

To bound the last integral, we represent $\Fock_\ph$ by the
Wiener--It\^o--Segal isomorphism of Appendix~\ref{app-soft-gaussian}
with $\mathfrak h=\Hil_\ph$.  Then $\mathcal A_n$ becomes multiplication
by a real function $\mathcal A_n(s,\vt;q)$ of the Gaussian point $q$,
square integrable in $s$ for almost every $q$, and $\varPhi(\mathcal F_{n,\vt})$
becomes multiplication by its integral over $s$.  On $\RR/L\ZZ$, a real
potential $\beta\in L^2$ is gauge equivalent to its mean, so Fourier
series give
$(-\i\partial_s+\beta)^2\geq L^{-2}
\operatorname{dist}(\int_0^L\beta,2\pi\ZZ)^2$ as forms on
$H^1(\RR/L\ZZ)$.  For almost every $\vt$, the map
$s\mapsto u(\vX_n(s,\vt))$ lies in $H^1(\RR/L_n\ZZ;\Fock_\ph)$.  For such
$\vt$ and almost every $q$, apply this with
$\beta=-\pi/L_n-e\mathcal A_n(\cdot,\vt;q)$, and use
$\operatorname{dist}(\pi+x,2\pi\ZZ)^2\geq\pi^2-2\pi|x|$.  Integrating
over $q$, and then over $\vt$ by \cref{thin-loop-measure}, gives
\[
 \int_{\mathcal T_n}\rho_n\|(D_s-\pi/L_n)u\|^2\dd\vx
 \geq\frac{\pi^2}{L_n^2}
 -\frac{2\pi e}{L_n^2}\int_{\mathcal T_n}\rho_n
   \dua{u}{|\varPhi(\mathcal F_{n,\vt})|u}\dd\vx.
\]
By \cref{ann-Hrad-bound,cre-Hrad-bound},
$\dua{u}{|\varPhi(\mathcal F_{n,\vt})|u}\leq\|\varPhi(\mathcal F_{n,\vt})u\|
\leq2\|\mathcal F_{n,\vt}/\sqrt\omega\|\,\|\sqrt{H_\rad}u\|
+\|\mathcal F_{n,\vt}\|$, and
$\int\rho_n\|\sqrt{H_\rad}u\|\leq\|\sqrt{H_\rad}\Psi\|$.  Inserting these bounds and
\cref{thin-loop-flux-kernel} into \cref{thin-loop-energy-split} proves
\begin{equation}\label{thin-loop-lower-bound}
 \begin{aligned}
 &\dua{\Psi}{H^{0,0}\Psi}-\frac12\int_{\RR^3}|\grad\sqrt{\rho_n}|^2\\
 &\qquad\geq\frac{\pi^2}{(1+\epsilon_n)^2L_n^2}+\dua{\Psi}{H_\rad\Psi}
 -\delta_n\big(1+\|\sqrt{H_\rad}\Psi\|\big),
 \end{aligned}
\end{equation}
where $\delta_n=2\pi eCb_n$.

\noindent\textbf{Step 4. A mixed-state upper bound.}
For $K>0$, let $\AF_{<K}$ be the field with coupling functions
$\iden_{\{\omega<K\}}G^\mu_{\vx}$, and put
\[
 U_K(\vx)=e^{\i e\vx\cdot\AF_{<K}(0)}=W(\i f_\vx),\qquad
 f_\vx=e\sum_{\mu=1}^3x^\mu\iden_{\{\omega<K\}}G^\mu_{0},
\]
where $W$ is the Weyl displacement operator \cref{weyl-definitions} of
Appendix~\ref{app-weyl-displacements}.  By the symmetry of the form factor in
\cref{cutoff-assumption}, all scalar products among the coupling functions
$G^\mu_{\vx}$ and $\iden_{\{\omega<K\}}G^\nu_0$, for $\vx\in\RR^3$ and
$\mu,\nu=1,2,3$, are real.  By \cref{prop-weyl-algebra}(iii), the field
components $\Af^\mu(\vx)$ and $\Af^\nu_{<K}(0)$ therefore strongly
commute with each other.  Hence, for $\phi=\phi_{n,0},\phi_{n,1}$,
\[
 (-\i\grad-e\AF(\vx))U_K(\phi\otimes\Omega)
 =U_K\big[(-\i\grad\phi)\,\Omega
   -e\phi\,(\AF(\vx)-\AF_{<K}(0))\Omega\big].
\]
In the bracket, a zero-photon vector is added to a one-photon vector,
since
$(\Af^\mu(\vx)-\Af^\mu_{<K}(0))\Omega
=a^\dag(G^\mu_{\vx}-\iden_{\{\omega<K\}}G^\mu_{0})\Omega$.  Hence the
equal mixture $\vGamma_{n,K}$ of the two states $U_K(\phi\otimes\Omega)$
has the pair $(\rho_n,\vj_n)$.  By $|\grad s|=J_n^{-1}$ and the
coherent-state radiation energy \cref{weyl-coherent-energy},
\[
 \begin{aligned}
 \Tr(H^{0,0}\vGamma_{n,K})-\frac12\int_{\RR^3}|\grad\sqrt{\rho_n}|^2
 &=\frac{\pi^2}{L_n^2}\int\frac{\rho_n}{J_n^2}\dd\vx
 +\int\rho_n(\vx)\|\sqrt\omega f_\vx\|^2\dd\vx\\
 &\quad+\frac{e^2}{2}\int\rho_n(\vx)\sum_{\mu=1}^3
   \|G^\mu_{\vx}-\iden_{\{\omega<K\}}G^\mu_{0}\|^2\dd\vx.
 \end{aligned}
\]
Take $K=b_n^{-1}$.  Since $|\vx|\leq b_n+\sqrt2r_n<3b_n$ on
$\supp\rho_n$ and
$\|\sqrt\omega f_\vx\|^2\leq e^2K|\vx|^2\|\wh\chi\|^2$,
\cref{thin-loop-kernel-continuity} bounds the last two integrals by
\[
 \tau_n=e^2\big(4d_0(3b_n)^2
   +\|\iden_{\{\omega\geq b_n^{-1}\}}\wh\chi\|^2
   +9b_n\|\wh\chi\|^2\big).
\]
By \cref{thm-unrestricted-attainment,thm-many-electron-soft-purification},
$F_\LL^{e,\chi}=\wt F_\LI^{e,\chi}$.  Together with
$J_n\geq1-\epsilon_n$, this proves
\begin{equation}\label{thin-loop-upper-bound}
 F_\LL^{e,\chi}(\rho_n,\vj_n)-\frac12\int_{\RR^3}|\grad\sqrt{\rho_n}|^2
 \leq\frac{\pi^2}{(1-\epsilon_n)^2L_n^2}+\tau_n.
\end{equation}
In particular, $F_\LL^{e,\chi}(\rho_n,\vj_n)<\infty$, so
$(\rho_n,\vj_n)\in\rcdens_1^{e,\chi}$.

\noindent\textbf{Step 5. Pure minimizers are almost photon-free.}
We now choose the remaining scales.  Let $b_n\downarrow0$ with $b_n<1$,
and take $r_n=a_n^2b_n^3/10$, so that $r_n<a_n^2/b_n$ as required in
Step~1.  Then $\epsilon_n=b_n^4/10$, and $L_n\geq4b_n$ gives
$\epsilon_n/L_n^2\leq b_n^2/160$.  Since moreover $d_0(\ell)\downarrow0$
as $\ell\downarrow0$ and $\wh\chi\in L^2(\RR^3)$,
\begin{equation}\label{thin-loop-limits}
 \frac{\epsilon_n}{L_n^2}\longrightarrow0,\qquad
 \delta_n\longrightarrow0,\qquad
 \tau_n\longrightarrow0.
\end{equation}
Suppose, for contradiction, that pure minimizers $\Psi_n$ exist for
infinitely many $n$.
Pass to that subsequence, and let $u_n$ be as in Step~3.  By
\cref{thin-loop-lower-bound,thin-loop-upper-bound,thin-loop-limits},
\[
 \|\sqrt{H_\rad}\Psi_n\|^2-\delta_n(1+\|\sqrt{H_\rad}\Psi_n\|)
 \leq\frac{\pi^2}{L_n^2}
   \big((1-\epsilon_n)^{-2}-(1+\epsilon_n)^{-2}\big)+\tau_n
 \longrightarrow0,
\]
so $\|\sqrt{H_\rad}\Psi_n\|\to0$.  By
\cref{thm-unrestricted-attainment,thm-many-electron-soft-purification},
$|\Psi_n\rangle\langle\Psi_n|$ is a mixed-state optimizer.  For
$0<\kappa\leq\kappa_0$, \cref{prop-optimizer-soft-spectrum} therefore
bounds its mean number of photons of frequency below $\kappa$ by the
right-hand side $c_\kappa$ of \cref{optimizer-soft-number-bound},
independently of $n$.  The mean number of photons of frequency at least
$\kappa$ is at most $\kappa^{-1}\dua{\Psi_n}{H_\rad\Psi_n}$.  With
$P_\Omega=|\Omega\rangle\langle\Omega|$ on $\Fock_\ph$ and
$\iden-P_\Omega\leq\NC$,
\[
 \int_{\mathcal T_n}\rho_n\|(\iden-P_\Omega)u_n\|^2\dd\vx
 =\|(\iden-P_\Omega)\Psi_n\|^2
 \leq\dua{\Psi_n}{\NC\Psi_n}\leq c_\kappa+\frac{\dua{\Psi_n}{H_\rad\Psi_n}}{\kappa}.
\]
Since $c_\kappa\to0$ as $\kappa\downarrow0$, letting $n\to\infty$ and
then $\kappa\downarrow0$ proves
\begin{equation}\label{thin-loop-vacuum-limit}
 \int_{\mathcal T_n}\rho_n\|(\iden-P_\Omega)u_n\|^2\dd\vx
 \longrightarrow0.
\end{equation}
Moreover, $\|\mathcal A_nu\|\leq J_n\|\AF(\vX_n)u\|$ with $J_n\leq2$,
so \cref{A-Hrad-bound} gives
$\int\rho_n\|\mathcal A_nu_n\|^2\leq4C_\chi(\dua{\Psi_n}{H_\rad\Psi_n}+1)$, while
\cref{thin-loop-energy-split,thin-loop-upper-bound} bound
$\int\rho_n\|(D_s-\pi/L_n)u_n\|^2$ by $C/L_n^2$.  Writing
$\partial_su_n=\i(D_s-\pi/L_n)u_n+\i(\pi/L_n)u_n+\i e\mathcal A_nu_n$,
we obtain
\begin{equation}\label{thin-loop-longitudinal-bound}
 \int_{\mathcal T_n}\rho_n\|\partial_su_n\|^2\dd\vx\leq\frac{C}{L_n^2},
 \qquad
 eL_n\int_{\mathcal T_n}\rho_n\|\mathcal A_nu_n\|\dd\vx\longrightarrow0.
\end{equation}

\noindent\textbf{Step 6. A pure minimizer cannot carry circulation
$\pi$.}
We keep the assumption of Step~5 and the notation introduced there.
For $\vz\in(-1,1)^2$, let $u_\vz(\theta)=u_n(\vX_n(L_n\theta,r_n\vz))$ on
$\mathbb T$.  Let $\|\cdot\|$ be the norm of $\Fock_\ph$ and
$\|\cdot\|_{L^2}$ that of $L^2(\mathbb T;\Fock_\ph)$, and put
\[
 k_n(\vz)=\|\partial_\theta u_\vz\|_{L^2}^2,\qquad
 p_n(\vz)=\|(\iden-P_\Omega)u_\vz\|_{L^2}^2
   +eL_n\int_{\mathbb T}\|\mathcal A_nu_\vz\|\dd\theta.
\]
By \cref{thin-loop-measure},
\cref{thin-loop-vacuum-limit,thin-loop-longitudinal-bound} give
$\int_{(-1,1)^2}wk_n\leq C$ and $\bar p_n:=\int_{(-1,1)^2}wp_n\to0$.  Since
$w\dd\vz$ is a probability measure, Markov's inequality gives a $\vz$
with $k_n(\vz)\leq3C$ and $p_n(\vz)\leq3\bar p_n$ for which
$u=u_\vz\in H^1(\mathbb T;\Fock_\ph)$, $\|u\|=1$, and
\cref{thin-loop-circulation} holds along $u$.  Multiplying
\cref{thin-loop-circulation} by $L_n$ and integrating over $\mathbb T$
gives
\[
 \Big|\int_{\mathbb T}\Im\dua{u}{\partial_\theta u}\dd\theta-\pi\Big|
 =eL_n\Big|\int_{\mathbb T}\dua{u}{\mathcal A_nu}\dd\theta\Big|
 \leq3\bar p_n.
\]
Let $\alpha=\dua{\Omega}{u}$ and $u_\perp=(\iden-P_\Omega)u$.  Since
$\|\partial_\theta u_\perp\|\leq\|\partial_\theta u\|$, the
one-dimensional Sobolev inequality gives
\[
 \sup_{\mathbb T}\big(1-|\alpha|^2\big)=\sup_{\mathbb T}\|u_\perp\|^2
 \leq\|u_\perp\|_{L^2}^2+2\|u_\perp\|_{L^2}\|\partial_\theta u\|_{L^2}
 \leq3\bar p_n+6\sqrt{C\bar p_n}.
\]
For large $n$ the right-hand side is at most $\frac12$.  Then
$|\alpha|^2\geq\frac12$ on $\mathbb T$, and the winding number
$m=(2\pi)^{-1}\int_{\mathbb T}\Im(\partial_\theta\alpha/\alpha)\dd\theta$
is an integer.  As $u_\perp$ and $\partial_\theta u_\perp$ are
orthogonal to $\Omega$,
\[
 \Im\dua{u}{\partial_\theta u}
 =|\alpha|^2\Im(\partial_\theta\alpha/\alpha)
 +\Im\dua{u_\perp}{\partial_\theta u_\perp}.
\]
Integrating over $\mathbb T$ and using
$|\partial_\theta\alpha|\leq\|\partial_\theta u\|$ and the two
preceding bounds, we find that
$\int_{\mathbb T}\Im\dua{u}{\partial_\theta u}\dd\theta$ differs from
$2\pi m$ by at most
\[
 \sqrt2\sup_{\mathbb T}\big(1-|\alpha|^2\big)\|\partial_\theta u\|_{L^2}
 +\|u_\perp\|_{L^2}\|\partial_\theta u\|_{L^2}\longrightarrow0,
\]
and from $\pi$ by at most $3\bar p_n\to0$.  This contradicts
$|\pi-2\pi m|\geq\pi$ for large $n$, so the assumption of Step~5 is
false.  Hence pure minimizers exist for at most finitely many $n$, and
any larger $n$ in
\cref{arbitrary-coupling-nonattainment-pair} gives the asserted pair at
the fixed coupling and cutoff.

\subsection{Lower semicontinuity with photon loss}

The pure-state infimum need not be attained, as the explicit
one-electron examples above show.  Along a minimizing sequence, photons
may escape to infinity, so a limit state must be allowed to lose
photons.  We implement photon loss by localizing the photons at a scale
$R$ and tracing out the rest.  For every $R$, localization followed by
partial trace preserves the normalization and the electronic reduced
state exactly, and hence also the electron density and the
paramagnetic current density.  It does not preserve the combined
current density at fixed $R$, because the field-induced current density
depends on the field expectation, which loses the contribution of the
traced-out photons.  The combined current density is preserved only in
the limit $R\to\infty$: the proof of \cref{thm-continuum-compactness}
bounds the change in $\int\vzeta\cdot\vjc$, for each smooth compactly
supported $\vzeta$, by a constant depending on the internal energy times
a factor that vanishes as $R\to\infty$.  The partial trace can also turn
a pure state into a mixed state, which motivates the mixed-state
compactness arguments here.
We first record an abstract trace-norm bound and its compactness
consequence.  We then prove lower semicontinuity of the energy and
passage to the constraints under trace-norm convergence.  Photon
localization gives compactness under a condition on photon number.

\begin{lemma}
\label[lemma]{lem-trace-compression}
Let $\mathfrak K$ be a Hilbert space.  A positive trace-class operator
$\Gamma$ on $\mathfrak K$ with $\Tr\Gamma\leq1$ and an orthogonal
projection $S$ on $\mathfrak K$ satisfy
\begin{equation}\label{density-matrix-compression}
 \|\Gamma-S\Gamma S\|_{\mathfrak S_1}
 \leq2\big(\Tr((1-S)\Gamma)\big)^{1/2}.
\end{equation}
Moreover, a sequence $(\Gamma_j)$ of positive trace-class operators
on $\mathfrak K$, with $\Tr\Gamma_j\leq1$ for every $j$, is
precompact in trace norm if, for every $\varepsilon>0$, there is a finite-rank orthogonal
projection $S$ with $\sup_j\Tr((1-S)\Gamma_j)<\varepsilon$.
\end{lemma}

\begin{proof}
Writing
$\Gamma-S\Gamma S=(1-S)\Gamma+S\Gamma(1-S)$ and factoring through
$\sqrt{\Gamma}$, the Hilbert--Schmidt ideal inequality gives
\[
 \|\Gamma-S\Gamma S\|_{\mathfrak S_1}
 \leq2\|\sqrt{\Gamma}\|_{\mathfrak S_2}
          \|(1-S)\sqrt{\Gamma}\|_{\mathfrak S_2}
 \leq2\big(\Tr((1-S)\Gamma)\big)^{1/2}.
\]
To prove precompactness, let $\delta>0$ and choose a finite-rank orthogonal projection $S$ with
$\sup_j\Tr((1-S)\Gamma_j)<\delta^2/16$.
The bound \cref{density-matrix-compression} then gives the uniform
approximation
\[
 \sup_j\|\Gamma_j-S\Gamma_jS\|_{\mathfrak S_1}<\frac\delta2.
\]
Each compression is positive and satisfies
\[
 \Tr(S\Gamma_jS)=\Tr(S\Gamma_j)\leq\Tr\Gamma_j\leq1.
\]
Thus all the compressions belong to
\[
 \mathcal K_S
 :=\{A\in\mathfrak S_1(\mathfrak K): A=SAS,\ A\geq0,\ \Tr A\leq1\}.
\]
This set is closed in trace norm: the map $A\mapsto SAS$ and the trace
are continuous, and positivity is preserved under trace-norm limits.
It is bounded because
$\|A\|_{\mathfrak S_1}=\Tr A\leq1$ for $A\in\mathcal K_S$.  Moreover, the condition
$A=SAS$ identifies these operators with operators on the
finite-dimensional space $\ran S$.  Hence $\mathcal K_S$ is compact
in trace norm.

Choose a finite $\delta/2$-net $A_1,\ldots,A_m$ for $\mathcal K_S$.
For each $j$, some $k\in\{1,\ldots,m\}$ therefore satisfies
$\|S\Gamma_jS-A_k\|_{\mathfrak S_1}<\delta/2$, and the triangle inequality gives
\[
 \|\Gamma_j-A_k\|_{\mathfrak S_1}
 \leq\|\Gamma_j-S\Gamma_jS\|_{\mathfrak S_1}+\|S\Gamma_jS-A_k\|_{\mathfrak S_1}
 <\delta.
\]
The sequence is thus covered by finitely many trace-norm balls of
radius $\delta$.  Since $\delta>0$ was arbitrary, the set
$\{\Gamma_j:j\geq1\}$ is totally bounded.  The trace class
$\mathfrak S_1(\mathfrak K)$ is complete, so the closure of this set is
compact, which proves precompactness in trace norm.
\end{proof}

For a positive trace-class operator $\vGamma$ on $\HC_N$ with
$\Tr\vGamma\leq1$, not necessarily equal to one, define the internal
energy $\Tr(H^{0,0}\vGamma)\in[0,\infty]$ by the spectral sum following
\cref{finite-energy-PF-ensembles}, and similarly the trace of $\vGamma$
against any nonnegative quadratic form in place of $H^{0,0}$.  When the
internal energy is finite, define the density $\rho_\vGamma$ and the
combined current density $\vjc_\vGamma$ by
\cref{mixed-PF-density-current}.  These definitions are linear in
$\vGamma$ and agree with the earlier ones on $\dm$.

\begin{lemma}[Lower semicontinuity and passage to the constraints]
\label[lemma]{lem-trace-compactness-and-limits}
Let $\vGamma_j$ be positive trace-class operators on $\HC_N$ with
$\Tr\vGamma_j\leq1$ and uniformly bounded internal energies,
\[
 \sup_j\Tr(H^{0,0}\vGamma_j)<\infty,
\]
and suppose that $\vGamma_j\to\vGamma$ in trace norm.
Then $\vGamma$ is positive, $\Tr\vGamma=\lim_j\Tr\vGamma_j$, and
\begin{equation}\label{mixed-energy-lsc}
 \Tr(H^{0,0}\vGamma)
 \leq\liminf_j\Tr(H^{0,0}\vGamma_j).
\end{equation}
In particular, $\vGamma$ has finite internal energy, so $\rho_\vGamma$
and $\vjc_\vGamma$ are defined.
The same liminf inequality holds for every nonnegative quadratic form
on $Q(H^{0,0})$ bounded relative to $H^{0,0}+1$.
For bounded scalar $f$ and real
$\vzeta\in C_c^\infty(\RR^3,\RR^3)$,
\begin{align}
 \int_{\RR^3} f\rho_{\vGamma_j}&\longrightarrow\int_{\RR^3} f\rho_\vGamma,
 \label{mixed-density-passage}\\
 \int_{\RR^3}\vzeta\cdot\vjc_{\vGamma_j}
   &\longrightarrow\int_{\RR^3}\vzeta\cdot\vjc_\vGamma.
 \label{mixed-current-passage}
\end{align}
\end{lemma}
\begin{proof}
Positivity and convergence of the traces follow
from trace-norm convergence.  The Powers--St\o rmer inequality gives
\[
 \|\sqrt{\vGamma_j}-\sqrt{\vGamma}\|_{\mathfrak S_2}^2
 \leq\|\vGamma_j-\vGamma\|_{\mathfrak S_1}\longrightarrow0.
\]
Equip the form domain $Q(H^{0,0})$ with the energy norm
$u\mapsto\dua{u}{(H^{0,0}+1)u}^{1/2}$, making it a Hilbert
space continuously embedded in $\HC_N$.  Let
$\mathfrak S_2(\HC_N,Q(H^{0,0}))$ be the Hilbert--Schmidt operators from
$\HC_N$ into this space.  Thus an operator $T$ on $\HC_N$ belongs to it
when its range lies in the form domain and $(H^{0,0}+1)^{1/2}T$ is
Hilbert--Schmidt, and then
$\|T\|_{\mathfrak S_2(\HC_N,Q(H^{0,0}))}
 =\|(H^{0,0}+1)^{1/2}T\|_{\mathfrak S_2}$.
Evaluating this norm on an eigenbasis of $\vGamma_j$ gives
\[
 \|\sqrt{\vGamma_j}\|_{\mathfrak S_2(\HC_N,Q(H^{0,0}))}^2
 =\Tr\vGamma_j+\Tr(H^{0,0}\vGamma_j),
\]
which is uniformly bounded.  The continuous embedding into $\mathfrak S_2$
and the strong convergence there identify every weak subsequential
limit in this stronger Hilbert space with $\sqrt{\vGamma}$.
Thus $\sqrt{\vGamma}$ belongs to that space and
$\sqrt{\vGamma_j}\rightharpoonup\sqrt{\vGamma}$ there.
Weak lower semicontinuity proves \cref{mixed-energy-lsc}.  A nonnegative
form bounded relative to $H^{0,0}+1$ is $u\mapsto\|Bu\|^2$ for some
operator $B$ bounded from $Q(H^{0,0})$ into a Hilbert space.  Composition
with $B$ preserves weak Hilbert--Schmidt convergence, so weak lower
semicontinuity applies to it as well.

For the density, test the trace-norm convergence
$\vGamma_j\to\vGamma$ against the multiplication operator
$M_f:=\sum_{i=1}^N f(\vx_i)$, acting as the identity on the photon
factor.  It preserves electronic antisymmetry and satisfies
$\|M_f\|_{\mathrm{op}}\leq N\|f\|_\infty$.  Hence
\begin{align*}
 \left|\int_{\RR^3}f(\rho_{\vGamma_j}-\rho_\vGamma)\right|
 &=\left|\Tr\big(M_f(\vGamma_j-\vGamma)\big)\right|\\
 &\leq N\|f\|_\infty
        \|\vGamma_j-\vGamma\|_{\mathfrak S_1}\longrightarrow0,
\end{align*}
which proves \cref{mixed-density-passage}.

For the current density, recall from \cref{sec-PF-hamiltonian} that
$\vD_1=-\i\grad_{\vx_1}-e\AF(\vx_1)$ is the covariant momentum of the
first electron.  Every term in \cref{q00-def} is nonnegative, so
$\|\vD_1u\|^2\leq2\dua{u}{H^{0,0}u}$ for $u\in Q(H^{0,0})$.  Hence
$\vD_1$ is bounded from $Q(H^{0,0})$, with its energy norm, to the ambient
vector-valued electron--photon $L^2$ space.  We therefore also have
$\vD_1\sqrt{\vGamma_j}\rightharpoonup\vD_1\sqrt{\vGamma}$
in the corresponding Hilbert--Schmidt class.
As operators into this vector-valued space, the factors
$\vzeta(\vx_1)\sqrt{\vGamma_j}$ and
$\vD_1\sqrt{\vGamma_j}$ converge strongly and weakly in
$\mathfrak S_2$, respectively.  The product of the adjoint of the first
factor with the second is trace class, and
\begin{align*}
 \int_{\RR^3}\vzeta\cdot\vjc_{\vGamma_j}
 &=N\Re\Tr\Big[
       \big(\vzeta(\vx_1)\sqrt{\vGamma_j}\big)^\dag
       \vD_1\sqrt{\vGamma_j}\Big]\\
 &\longrightarrow
 N\Re\Tr\Big[
       \big(\vzeta(\vx_1)\sqrt{\vGamma}\big)^\dag
       \vD_1\sqrt{\vGamma}\Big]
 =\int_{\RR^3}\vzeta\cdot\vjc_\vGamma.
\end{align*}
This proves \cref{mixed-current-passage}.
\end{proof}

\begin{remark}[Continuum compactness obstruction]
\label{rem-continuum-obstruction}
Uniform energy bounds and a fixed electronic density do not ensure
trace-norm compactness in the continuum.  Choose an orthonormal
sequence $\xi_n\in\Hil_\ph$ in a bounded momentum shell and a
normalized smooth compactly supported electronic state $\Phi$.
Then $\Psi_n=\Phi\otimes a^\dag(\xi_n)\Omega$ has fixed density,
one photon, and bounded energy, while
$|\Psi_n\rangle\langle\Psi_n|$ has no trace-norm convergent subsequence.
Photon loss must therefore be allowed in constructing limits.
\Cref{thm-continuum-compactness} implements this using localization
and partial trace, and \cref{sec-unrestricted-attainment} supplies its photon-number
condition by modifying minimizing sequences at fixed density and
combined current density.
\end{remark}

Photon loss can be measured by localization, which preserves
normalization but can turn a pure state into a mixed state.  We use
geometric localization in Fock space, as developed in
\cite[Section~3]{lewin2011geometric}.

We first recall the notation of \cref{sec-fock-space,sec-second-quantization},
specialized to the photon Fock space $\Fock_\ph=\Fock_s(\Hil_\ph)$.  For
$f\in\Hil_\ph$, the Segal field $\varPhi(f)$ is the self-adjoint closure of
$a^\dag(f)+a(f)$, initially defined on the finite-particle space
$\Fock_\ph^\fin$.  It depends real-linearly on $f$, the conjugate field is
$\varPi(f)=\varPhi(\i f)$, and the vector potential \cref{Afield-def} is
$\Af^\mu(\vx)=\varPhi(G^\mu_{\vx})$.  In the vacuum,
$\langle\Omega,\varPhi(f)\Omega\rangle=0$ and
$\|\varPhi(f)\Omega\|=\|f\|$.  For a self-adjoint operator $b$ on
$\Hil_\ph$, its second quantization
$\dd\varGamma(b)=\bigoplus_{M\geq0}b^{(M)}$ acts on the $M$-photon sector
as the closure of $\sum_{j=1}^Mb_j$, where $b_j$ acts on the $j$th photon
variable.  For nonnegative $b$, the same symbol denotes the corresponding
nonnegative quadratic form.  For a unitary $T$ on $\Hil_\ph$, or on
$\Hil_\ph\oplus\Hil_\ph$ as below, the multiplicative second quantization
$\varGamma(T)=\bigoplus_{M\geq0}T^{\otimes M}$, restricted to symmetric
tensors, is unitary on the corresponding bosonic Fock space and satisfies
$\varGamma(T)\varPhi(f)\varGamma(T)^\dag=\varPhi(Tf)$.  Finally, the Weyl
operator \cref{weyl-definitions} is $W(f)=\exp(a^\dag(f)-a(f))$, so that
$W(\i f)=e^{\i\varPhi(f)}$.

Fix the polarization representation
$L^2(\RR^3\times\{1,2\})$ and Fourier transform each photon momentum
to a position variable $\vy$.  Then
$H_\rad=\dd\varGamma(\sqrt{-\Delta})$, and
$\NC=\dd\varGamma(1)$ denotes photon number.
Here $\sqrt{-\Delta}$ is the Fourier multiplier $|\vk|$.
The variable $\vy$ is a technical device rather than a physical photon
position.\footnote{It is the Fourier dual of $\vk$ in the fixed
polarization frame $\vepsilon_\lambda$, which cannot be continuous on
spheres $\{|\vk|=r\}$ by the hairy ball theorem, so the localization
below depends on this frame.  We use the polarization representation
because its one-photon space is all of $L^2(\RR^3;\CC^2)$, so
multiplication by functions of $\vy$ is available; in the solenoidal
representation it would not preserve the divergence-free constraint.
Only the photon-number hypothesis
\cref{continuum-local-sector-tightness} of
\cref{thm-continuum-compactness} refers to the frame; its conclusion
involves only the density, the combined current density, and the
internal energy.}
Fix smooth real functions $\theta_1,\phi_1$ on $\RR^3$ with
$0\leq\theta_1,\phi_1\leq1$, $\theta_1^2+\phi_1^2=1$, $\theta_1=1$ on
$B_1$, and $\theta_1=0$ outside $B_2$.  For $R>0$, set
$\theta_R(\vy):=\theta_1(\vy/R)$ and $\phi_R(\vy):=\phi_1(\vy/R)$.  Thus
$\theta_R^2+\phi_R^2=1$, $\theta_R=1$ on $B_R$, and $\theta_R=0$
outside $B_{2R}$.  The one-photon isometry
$f\mapsto\theta_Rf\oplus\phi_Rf$ induces a Fock-space isometry
$\mathcal U_R$.  With the electronic identity suppressed, set
\[
 \mathcal L_R(\vGamma)
 :=\Tr_{\mathrm{out}}(\mathcal U_R\vGamma\mathcal U_R^\dag).
\]
Thus $\mathcal L_R$ retains nearby photons and traces out the second
photon factor.

\begin{lemma}[Photon localization and field identities]
\label[lemma]{lem-photon-localization}
For every $R>0$, the map $\mathcal L_R$ is positive and trace
preserving, and leaves the electronic reduced state unchanged.
For $f\in\Hil_\ph$, its adjoint with respect to the trace pairing
satisfies
\begin{align}
 \mathcal L_R^\dag(\varPhi(f))&=\varPhi(\theta_Rf),
 \label{localized-linear-field}\\
 \mathcal L_R^\dag(\varPhi(f)^2)&=\varPhi(\theta_Rf)^2+\|\phi_Rf\|^2\iden.
 \label{localized-quadratic-field}
\end{align}
These identities are understood as quadratic-form identities on
$D(\varPhi(\theta_Rf))$, with the electronic identity suppressed.
\end{lemma}

\begin{proof}
Since $\mathcal U_R^\dag\mathcal U_R=\iden$, conjugation by
$\mathcal U_R$ followed by partial trace is positive and trace
preserving.  For every bounded electronic operator $A$,
\[
 \mathcal U_R^\dag(A\otimes\iden\otimes\iden)\mathcal U_R
 =A\otimes\iden.
\]
Thus the electronic reduced state is unchanged.  By monotone spectral
truncation, this also preserves the expectations of nonnegative
unbounded electronic observables, including the ordinary kinetic
energy and the Coulomb interaction.

To prove the field identities, complete the one-photon isometry
$j_Ru=(\theta_Ru,\phi_Ru)$ to the unitary
\[
 T_R=\begin{pmatrix}\theta_R&-\phi_R\\
                         \phi_R&\theta_R\end{pmatrix}
 \quad\text{on }\Hil_\ph\oplus\Hil_\ph.
\]
Unitarity follows from $\theta_R^2+\phi_R^2=1$ and commutation of
these multiplication operators.  Under the canonical identification
$\Fock_s(\Hil_\ph\oplus\Hil_\ph)
 \simeq\Fock_\ph\otimes\Fock_\ph$, write
$V_R=\varGamma(T_R)$ for its second quantization.  Then
\[
 \mathcal U_R\Psi=V_R(\Psi\otimes\Omega),\qquad
 V_R^\dag(\varPhi(f)\otimes\iden)V_R
 =\varPhi(\theta_Rf)\otimes\iden
   -\iden\otimes\varPhi(\phi_Rf).
\]
For the second identity, \cref{weyl-second-quantization,weyl-factorization}
give, for $s\in\RR$,
\[
 V_R^\dag\big(e^{\i s\varPhi(f)}\otimes\iden\big)V_R
 =W\big(\i sT_R^\dag(f\oplus0)\big)
 =e^{\i s\varPhi(\theta_Rf)}\otimes e^{-\i s\varPhi(\phi_Rf)},
\]
and the generators of the two sides agree.  Since
\[
 \langle\Omega,\varPhi(\phi_Rf)\Omega\rangle=0,\qquad
 \|\varPhi(\phi_Rf)\Omega\|^2=\|\phi_Rf\|^2,
\]
vacuum compression gives \cref{localized-linear-field} and
\[
 \|(\varPhi(f)\otimes\iden)\mathcal U_R\Psi\|^2
 =\|\varPhi(\theta_Rf)\Psi\|^2
   +\|\phi_Rf\|^2\|\Psi\|^2.
\]
This proves \cref{localized-quadratic-field} by polarization.
The two fields on the separate factors strongly commute.  Their joint
spectral calculus, and the finite second moment of the vacuum in the
second factor, show that the norm on the left is finite exactly when
$\Psi\in D(\varPhi(\theta_Rf))$.  Thus these are identities on the
stated form domain, without a mean-photon-number assumption.  The
extra scalar term is the vacuum fluctuation in the discarded modes.
\end{proof}

\begin{lemma}[Localization of one-photon forms]
\label[lemma]{lem-photon-localization-forms}
For every $R>0$ and nonnegative self-adjoint one-photon operator $b$,
the corresponding nonnegative forms obey
\begin{equation}\label{localized-second-quantization}
 \mathcal L_R^\dag(\dd\varGamma(b))
 =\dd\varGamma(\theta_R b\theta_R),
\end{equation}
where $\theta_R b\theta_R$ denotes the form
$u\mapsto\|\sqrt{b}\theta_Ru\|^2$.
Moreover, with $C$ depending only on $\theta_1$,
\begin{equation}\label{localized-one-photon-bounds}
 \theta_R^2\leq C R\sqrt{-\Delta},\qquad
 \theta_R\sqrt{-\Delta}\theta_R\leq C\sqrt{-\Delta}
\end{equation}
as forms on $H^{1/2}(\RR^3;\CC^2)$, uniformly for $R>0$.
\end{lemma}

\begin{proof}
Write $j_Ru=(\theta_Ru,\phi_Ru)$ for the one-photon localization
isometry.  For \cref{localized-second-quantization}, observe that
$\dd\varGamma(b)\otimes\iden$ corresponds to $\dd\varGamma(b\oplus0)$
and
\[
 j_R^\dag(b\oplus0)j_R=\theta_R b\theta_R.
\]
Since $\mathcal U_R$ is induced by $j_R$, it preserves the total photon
number and acts on the $m$-photon sector as $j_R^{\otimes m}$, followed
by the canonical identification.  On this sector, write $B_k$ for the
operator $b\oplus0$ acting on the $k$th photon variable, so that
$\dd\varGamma(b\oplus0)=\sum_{k=1}^mB_k$ there.  Each term factors over
the $m$ photon variables, and since $j_R^\dag j_R=\iden$,
\[
 (j_R^{\otimes m})^\dag B_k\,j_R^{\otimes m}
 =j_R^\dag j_R\otimes\cdots\otimes j_R^\dag(b\oplus0)j_R
  \otimes\cdots\otimes j_R^\dag j_R
 =\iden\otimes\cdots\otimes\theta_Rb\theta_R\otimes\cdots\otimes\iden,
\]
with the nontrivial factor in the $k$th position.  Summing over $k$
gives $\dd\varGamma(\theta_Rb\theta_R)$ on the $m$-photon sector, and
summing over $m$ gives the asserted identity.  The argument first
applies to bounded $b$ on finitely many sectors.  Spectral truncation
of $b$ and monotone convergence over the sectors give the identity
for arbitrary nonnegative $b$, with possibly infinite form values.

We next prove \cref{localized-one-photon-bounds}.  Work first with
$u\in C_c^\infty(\RR^3;\CC^2)$, with the two components corresponding
to the photon polarizations.
The fractional Hardy inequality, applied to each component, gives
\[
 \int_{\RR^3}\frac{|u(\vy)|^2}{|\vy|}\dd\vy
 \leq C_{\mathrm H}\|(-\Delta)^{1/4}u\|^2;
\]
see \cite[Section~2]{frank2008hardy}.
Since $0\leq\theta_R\leq1$ and $\theta_R=0$ outside $B_{2R}$,
we have $\theta_R(\vy)^2\leq2R/|\vy|$ almost everywhere.  Thus
\[
 \|\theta_Ru\|^2
 \leq2R\int_{\RR^3}\frac{|u(\vy)|^2}{|\vy|}\dd\vy
 \leq2C_{\mathrm H}R\|(-\Delta)^{1/4}u\|^2,
\]
which proves the first inequality.

For the second inequality, use the integral representation
\[
 \|(-\Delta)^{1/4}u\|^2
 =c_3\iint_{\RR^3\times\RR^3}
   \frac{|u(\vy)-u(\vz)|^2}{|\vy-\vz|^4}\dd\vy\dd\vz,
 \qquad c_3>0.
\]
Expanding the difference as
\[
 \theta_R(\vy)u(\vy)-\theta_R(\vz)u(\vz)
 =\theta_R(\vy)\big(u(\vy)-u(\vz)\big)
  +\big(\theta_R(\vy)-\theta_R(\vz)\big)u(\vz)
\]
and using $|a+b|^2\leq2|a|^2+2|b|^2$ yields
\begin{gather*}
 \|(-\Delta)^{1/4}(\theta_Ru)\|^2
 \leq2\|(-\Delta)^{1/4}u\|^2
      +2c_3\int_{\RR^3} K_R(\vz)|u(\vz)|^2\dd\vz,\\
 K_R(\vz):=\int_{\RR^3}
 \frac{|\theta_R(\vy)-\theta_R(\vz)|^2}{|\vy-\vz|^4}\dd\vy.
\end{gather*}
We claim that $K_R(\vz)\leq C/|\vz|$ for $\vz\neq0$.
Indeed, since $\theta_R(\vy)=\theta_1(\vy/R)$, smoothness and
boundedness of $\theta_1$ give
\[
 |\theta_R(\vy)-\theta_R(\vz)|
 \leq C\min\{|\vy-\vz|/R,1\}.
\]
Splitting the integral at $|\vy-\vz|=R$ and using spherical
coordinates therefore gives, for every $\vz$,
\[
 K_R(\vz)
 \leq C\left(R^{-2}\int_0^R\dd r
                 +\int_R^\infty r^{-2}\dd r\right)
 \leq\frac{C}{R}.
\]
For $0<|\vz|\leq4R$ this is at most $4C/|\vz|$.
If $|\vz|>4R$, then $\theta_R(\vz)=0$, and on the support of
$\theta_R$ we have $|\vy-\vz|\geq|\vz|/2$.  Hence
\[
 K_R(\vz)
 =\int_{B_{2R}}\frac{\theta_R(\vy)^2}{|\vy-\vz|^4}\dd\vy
 \leq\frac{C R^3}{|\vz|^4}
 \leq\frac{C}{|\vz|},
\]
which proves the claim.  Applying the Hardy inequality once more,
we obtain
\begin{align*}
 \|(-\Delta)^{1/4}(\theta_Ru)\|^2
 &\leq2\|(-\Delta)^{1/4}u\|^2
      +C\int_{\RR^3}\frac{|u(\vz)|^2}{|\vz|}\dd\vz\\
 &\leq C\|(-\Delta)^{1/4}u\|^2.
\end{align*}
By density of $C_c^\infty(\RR^3;\CC^2)$ in
$H^{1/2}(\RR^3;\CC^2)$ and boundedness of multiplication
by $\theta_R$ on $L^2$, these estimates extend to all of $H^{1/2}(\RR^3;\CC^2)$.
In particular, $\theta_R$ preserves this form domain.  This proves
both form inequalities with constants depending only on
$\theta_1$.
\end{proof}

\begin{lemma}[Photon-number and energy bounds]
\label[lemma]{lem-photon-localization-energy}
There is a constant $C$, depending only on $\theta_1$,
such that for every $R>0$ and every positive trace-class operator
$\vGamma$,
\begin{align}
 \Tr(\NC\mathcal L_R\vGamma)
 &=\Tr\big(\dd\varGamma(\theta_R^2)\vGamma\big)
 \leq\Tr(\NC\vGamma),
 \label{localized-photon-number}\\
 \Tr(\NC\mathcal L_R\vGamma)
 &\leq C R\Tr(H_\rad\vGamma),
 \label{localized-number-energy-bound}\\
 \Tr(H_\rad\mathcal L_R\vGamma)
 &\leq C\Tr(H_\rad\vGamma).
 \label{localized-radiation-bound}
\end{align}
The nonnegative traces may take the value $+\infty$.
In particular, $\mathcal L_R$ maps $\dm$ into $\dm$, and
\begin{equation}\label{localized-internal-energy-bound}
 \Tr(H^{0,0}\mathcal L_R\vGamma)
 \leq C\big(\Tr(H^{0,0}\vGamma)+\Tr\vGamma\big)
\end{equation}
for positive $\vGamma$ of finite internal energy, with $C$ independent
of $R$ (but also depending on $N,e$, and the form factor).
\end{lemma}

\begin{proof}
Second quantization preserves nonnegative form inequalities by summing
them over the photons in each sector.  Taking $b=1$ in
\cref{localized-second-quantization} and using $0\leq\theta_R^2\leq1$
therefore proves \cref{localized-photon-number}.  Taking
$b=1,\sqrt{-\Delta}$ and applying \cref{localized-one-photon-bounds}
gives \cref{localized-number-energy-bound,localized-radiation-bound}.
The trace assertions follow by summing the vector identities and
inequalities over a spectral decomposition of $\vGamma$, using
monotone convergence.  In particular, the localized mean photon
number is finite whenever the original radiation energy is finite,
even if $\Tr(\NC\vGamma)=\infty$.

By \cref{lem-photon-localization}, the ordinary electronic kinetic
energy is unchanged, while
\cref{localized-radiation-bound} bounds the radiation energy uniformly
in $R$.  Applying the form-norm equivalence
\cref{PF-form-norm-equivalence} before and after localization proves
\cref{localized-internal-energy-bound}.  Positivity and preservation
of the trace then show that $\mathcal L_R(\dm)\subset\dm$.
\end{proof}

\begin{lemma}[Trace-norm bound for photon localization]
\label[lemma]{lem-localization-trace-distance}
Let $0\leq\eta\leq1$ be an operator on $\Hil_\ph$, and let
$\mathcal U_\eta$ be the Fock-space localization isometry induced by
$u\mapsto\eta u\oplus\sqrt{1-\eta^2}\,u$.  With the electronic
identity suppressed, define
\[
 \mathcal L_\eta(\tau)
 :=\Tr_{\mathrm{out}}(\mathcal U_\eta\tau\mathcal U_\eta^\dag).
\]
Then
\begin{align*}
 \|\tau-\mathcal L_\eta\tau\|_{\mathfrak S_1}&\leq2\sqrt t+t,\\
 t:=\Tr\big((1-\varGamma(\eta^2))\tau\big)
 &\leq\Tr\big(\dd\varGamma(1-\eta^2)\tau\big)
\end{align*}
for every positive trace-one operator $\tau$ on $\HC_N$. 
\end{lemma}
Here
\[
 \varGamma(\eta)
 :=\bigoplus_{m\geq0}\eta^{\otimes m}\big|_{\Hil_\ph^{\otimes_s m}},
 \qquad \eta^{\otimes0}=1,
\]
is the multiplicative second quantization, extending the earlier
$\varGamma(U)$ construction from unitaries to contractions.
The additive lift $\dd\varGamma(b)$ acts on the $m$-photon sector as
$\sum_{j=1}^m b_j$ and vanishes on the vacuum.
\begin{proof}
Projecting the discarded
factor onto its vacuum gives
\[
 K_0:=(\iden\otimes\langle\Omega|)\mathcal U_\eta
 =\varGamma(\eta).
\]
Indeed, on the $m$-photon sector this projection retains precisely
the term $\eta^{\otimes m}$.  Splitting the partial trace into the
discarded vacuum and its orthogonal complement yields
\[
 \mathcal L_\eta\tau=K_0\tau K_0+\sigma,\qquad
 \sigma\geq0,\qquad
 \Tr\sigma=\Tr\big((1-K_0^2)\tau\big)=t,
\]
where $K_0^2=\varGamma(\eta^2)$.  Thus $t\in[0,1]$ is the
probability that the discarded factor contains at least one photon.
Since $K_0$ is a positive contraction and
$(1-K_0)^2\leq1-K_0^2$, Hilbert--Schmidt factorization gives
$\|\tau-K_0\tau K_0\|_{\mathfrak S_1}\leq2\sqrt t$.
The triangle inequality and $\|\sigma\|_{\mathfrak S_1}=t$ prove the
trace-norm bound.  Finally, on each photon sector the commuting
contractions $(\eta^2)_j$ satisfy
$1-\prod_j(\eta^2)_j\leq\sum_j(1-(\eta^2)_j)$.
Summing over sectors gives
$1-\varGamma(\eta^2)\leq\dd\varGamma(1-\eta^2)$ and hence the bound on $t$.
\end{proof}

We now combine \cref{lem-trace-compactness-and-limits} with the
localization bounds of
\cref{lem-photon-localization,lem-photon-localization-forms,lem-photon-localization-energy,lem-localization-trace-distance}
to construct a limit state that has the limiting density and current,
and whose internal energy is at most the liminf of the internal
energies along the sequence.  Photons are localized at the scales
$R=3^m$, and those that escape to infinity are discarded.

\begin{lemma}[Lower semicontinuity with photon loss]
\label[lemma]{thm-continuum-compactness}
Let $(\vGamma_j)\subset\dm$ have uniformly
bounded internal energies and uniformly tight electronic densities:
\begin{align*}
 &\sup_j\Tr(H^{0,0}\vGamma_j)<\infty,
 \\
 &\lim_{R\to\infty}\sup_j\int_{|\vx|>R}\rho_{\vGamma_j}(\vx)\dd\vx=0.
\end{align*}
Suppose also that the photon-number distributions of the localized
states $\mathcal L_{3^m}\vGamma_j$ are asymptotically tight, uniformly
in $m$:
\begin{equation}\label{continuum-local-sector-tightness}
 \lim_{M\to\infty}\sup_{m\geq1}\limsup_{j\to\infty}
 \Tr\big(\iden_{\{\NC>M\}}\mathcal L_{3^m}\vGamma_j\big)=0.
\end{equation}
If
$(\rho_{\vGamma_j},\vjc_{\vGamma_j})\to(\rho,\vjc)$ in the sense of distributions,
then there is a $\vGamma\in\dm$ such that
\begin{equation}\label{continuum-retained-limit}
 \rho_\vGamma=\rho,\qquad \vjc_\vGamma=\vjc,\qquad
 \Tr(H^{0,0}\vGamma)
 \leq\liminf_{j\to\infty}\Tr(H^{0,0}\vGamma_j).
\end{equation}
A sufficient condition for \cref{continuum-local-sector-tightness} is
\[
 \lim_{M\to\infty}\sup_j\Tr(\iden_{\{\NC>M\}}\vGamma_j)=0;
\]
in particular, a uniform bound on $\Tr(\NC\vGamma_j)$ suffices.
\end{lemma}
Note that neither a bound on mean photon number nor photon-position tightness is
assumed, and the original sequence need not converge in trace norm.

The order of limits in \cref{continuum-local-sector-tightness} matters.
For each fixed localization scale $3^m$, the limit superior as
$j\to\infty$ is taken first, so only the eventual behavior of the
localized states at that scale enters.  The resulting photon-number
tails, one for each $m$, are then required to vanish uniformly in $m$
as $M\to\infty$.  Explicitly, for every $\varepsilon>0$ there is an $M$
such that, for every $m\geq1$, some $j_0$ satisfies
\[
 \Tr\big(\iden_{\{\NC>M\}}\mathcal L_{3^m}\vGamma_j\big)<\varepsilon
 \qquad\text{for all }j\geq j_0.
\]
The threshold $j_0$ may depend on $m$, as well as on $\varepsilon$.  A
uniform bound on the original mean photon number is much stronger.  By
\cref{localized-photon-number} and $\iden_{\{\NC>M\}}\leq\NC/M$, it
gives
\[
 \Tr\big(\iden_{\{\NC>M\}}\mathcal L_{3^m}\vGamma_j\big)
 \leq\frac1M\sup_i\Tr(\NC\vGamma_i)
\]
for all $j$ and $m$ at once, with no threshold.  In contrast,
\cref{continuum-local-sector-tightness} constrains only the photons
retained at each fixed scale, and only for large $j$.  This permits
the removal of arbitrarily many soft photons. 

For example, let $\xi$ be
a normalized smooth one-photon function supported in $\{0<|\vk|<1\}$,
put $\xi_j(\vk,\lambda)=j^3\xi(j^2\vk,\lambda)$, and set
$\Xi_j=(j!)^{-1/2}a^\dag(\xi_j)^j\Omega$.  Then
\[
 \dua{\Xi_j}{\NC\Xi_j}=j,\qquad
 \dua{\Xi_j}{H_\rad\Xi_j}=j^{-1}\dua{\xi}{\omega\xi}.
\]
Write $\check\xi$ for the inverse Fourier transform of $\xi$ in $\vk$,
taken separately for each polarization $\lambda$, so that $\check\xi$
is $\xi$ in the photon position variable $\vy$.  Then
$\check\xi_j(\vy)=j^{-3}\check\xi(\vy/j^2)$, so for each fixed $R$,
\[
 \Tr\big(\NC\mathcal L_R(|\Xi_j\rangle\langle\Xi_j|)\big)
 =j\|\theta_R\check\xi_j\|^2\leq C R^3j^{-5}\longrightarrow0.
\]
Thus the localized states converge to the vacuum and satisfy
\cref{continuum-local-sector-tightness}, although their original
photon-number distributions are not uniformly tight.

Now fix a normalized $\Phi\in Q(H_\el^0)$ and set
$\Psi_j=\Phi\otimes\Xi_j$.  We compare $\Psi_j$ with the vacuum
product $\Phi\otimes\Omega$.  Since $\Xi_j$ is normalized,
$\rho_{\Psi_j}=\rho_\Phi$ and $\vjp_{\Psi_j}=\vjp_\Phi$.  Since
$\Xi_j$ has exactly $j$ photons and $\Af^\mu(\vx)$ changes the photon
number by one, $\dua{\Xi_j}{\Af^\mu(\vx)\Xi_j}=0$.  Hence
$\vjph_{\Psi_j}=0$, and $\Psi_j$ has the same density and combined
current density $\vjc_{\Psi_j}=\vjp_\Phi$ as $\Phi\otimes\Omega$.

The internal energies differ only through the radiation energy and the
field fluctuations.  The cross terms in $\|\vD_n\Psi_j\|^2$ vanish
because the field expectation in $\Xi_j$ is zero, and the canonical
commutation relations give
\[
 \|\Af^\mu(\vx)\Xi_j\|^2
 =\|G^\mu_{\vx}\|^2+2j|\dua{G^\mu_{\vx}}{\xi_j}|^2,
\]
whose first term is the vacuum value.  The Cauchy--Schwarz inequality,
\cref{Gx-norms}, and the identity
$\|\sqrt\omega\xi_j\|^2=j^{-2}\dua{\xi}{\omega\xi}$
bound the sum over $\mu$ of the second term by
$2j^{-1}\|\wh\chi/\sqrt\omega\|^2\dua{\xi}{\omega\xi}$.  Integrating
against $|\Phi|^2$, summing over the electrons, and using
$\int_{\RR^3}\rho_\Phi=N$, we find that the additional internal energy
obeys
\[
 0\leq\dua{\Psi_j}{H^{0,0}\Psi_j}
   -\dua{\Phi\otimes\Omega}{H^{0,0}(\Phi\otimes\Omega)}
 \leq\frac1j\left(1+e^2N\left\|\frac{\wh\chi}{\sqrt\omega}\right\|^2\right)
   \dua{\xi}{\omega\xi}.
\]
Thus the states $|\Psi_j\rangle\langle\Psi_j|$ satisfy all hypotheses
of \cref{thm-continuum-compactness}, with constant density and current,
although $\dua{\Psi_j}{\NC\Psi_j}=j\to\infty$.  By the photon-number
computation above, their localized states converge in trace norm to
$|\Phi\otimes\Omega\rangle\langle\Phi\otimes\Omega|$, so all $j$
photons are lost in the limit.  The limit state
$\vGamma=|\Phi\otimes\Omega\rangle\langle\Phi\otimes\Omega|$ satisfies
\cref{continuum-retained-limit} with equality.

\begin{proof}[Proof of \cref{thm-continuum-compactness}]
Write $H=H^{0,0}$ and first pass to a subsequence realizing the energy
liminf.  The form-norm equivalence controls the ordinary electronic
kinetic energy and $H_\rad$ uniformly.

By \cref{lem-photon-localization,lem-photon-localization-energy}, the
localized states have the same electronic reduced state, their radiation
energies are uniformly bounded, and
\[
 \Tr(\NC\mathcal L_R\vGamma_j)
 \leq C R\Tr(H_\rad\vGamma_j).
\]
Their full internal energies are uniformly bounded in both $j$ and
$R>0$ by \cref{localized-internal-energy-bound}.

\noindent\textbf{Step 1. Local trace-norm compactness.}
Fix $R>0$.  Set
\[
 -\lapl:=\sum_{n=1}^N(-\Delta_{\vx_n}),\qquad
 E:=1+\sup_j\Tr(H\vGamma_j)<\infty.
\]
The form-norm equivalence and the preceding estimates give
\[
 \sup_j\Tr\big((1-\lapl+H_\rad)\mathcal L_R\vGamma_j\big)\leq CE,
 \qquad
 \sup_j\Tr(\NC\mathcal L_R\vGamma_j)\leq CRE.
\]
All constants in the estimates below are independent of
$j,R,M,L,s$.
We construct finite-rank orthogonal projections whose complements
have uniformly small expectation in $\mathcal L_R\vGamma_j$.

First, denote the photon number cutoff by 
\[
P_M=\iden_{\{\NC\leq M\}},
\]
 for integers $M\geq0$.
The electron--photon interaction terms in the internal Hamiltonian
$H=H^{0,0}$ can change photon number, so $P_M$ need not commute with
$H$.  We deal with these terms later.

Since $1-P_M\leq\NC/(M+1)$,
\[
 \sup_j\Tr\big((1-P_M)\mathcal L_R\vGamma_j\big)
 \leq\frac1{M+1}\sup_j\Tr(\NC\mathcal L_R\vGamma_j)
 \leq\frac{CRE}{M+1}.
\]
Next choose a fixed $\zeta\in C_c^\infty(\RR^3)$ with
$0\leq\zeta\leq1$, equal to one on $B_1$ and zero outside $B_2$,
and define the symmetric electronic cutoff
\[
 \zeta_L(\vx_1,\ldots,\vx_N)
 :=\prod_{n=1}^N\zeta(\vx_n/L),\qquad L\geq1.
\]
Thus $0\leq\zeta_L\leq1$, $\zeta_L=1$ on $(B_L)^N$, and
$\supp\zeta_L\subset(\overline{B_{2L}})^N$: the cutoff localizes all
$N$ electrons in the ball $B_{2L}$.
Electronic tightness is used through the quantity
\[
 \epsilon_\el(L):=\sup_j\int_{|\vx|>L}\rho_{\vGamma_j}(\vx)\dd\vx,
\]
which tends to zero as $L\to\infty$ by the assumed uniform tightness of
the electronic densities.
Using the elementary fact that
\[
 (1-\zeta_L)^2\leq\iden_{\{\max_n|\vx_n|>L\}}
 \leq\sum_{n=1}^N\iden_{\{|\vx_n|>L\}}
\]
and taking expectations and suprema of both sides yields
\[
\sup_j\Tr((1-\zeta_L)^2\mathcal L_R\vGamma_j)
\leq\sup_j\int_{|\vx|>L}\rho_{\mathcal L_R\vGamma_j}(\vx)\dd\vx
=\epsilon_\el(L).
\]
Indeed, photon localization leaves the electronic reduced state, and
hence the density, unchanged: $\rho_{\mathcal L_R\vGamma_j}=\rho_{\vGamma_j}$
by \cref{lem-photon-localization}. 

Further, the positive contractions $P_M$ and $\zeta_L$ commute.  Thus
\[
 (1-\zeta_LP_M)^2=(1-P_M)+P_M(1-\zeta_L)^2,
\]
and hence
\begin{align}
 \sup_j\|(1-\zeta_LP_M)\sqrt{\mathcal L_R\vGamma_j}\|_{\mathfrak S_2}^2
 &\leq\sup_j\Tr\big((1-P_M)\mathcal L_R\vGamma_j\big)
   +\sup_j\Tr\big((1-\zeta_L)^2\mathcal L_R\vGamma_j\big)\nonumber\\
 &\leq\frac{CRE}{M+1}+\epsilon_\el(L).
 \label{cutoff-error}
\end{align}
The cutoff also preserves the required regularity with an explicit
bound. We have
\[
 \sum_{n=1}^N|\grad_{\vx_n}\zeta_L|^2
 \leq\frac1{L^2}\sum_{n=1}^N\big|(\grad\zeta)(\vx_n/L)\big|^2
 \leq\frac{N\|\grad\zeta\|_\infty^2}{L^2}.
\]
Let $\psi\in Q(-\lapl)$ and put $\varphi=P_M\psi$.  Since $P_M$ acts
only on the photon factor, it commutes with the electronic
derivatives, so $\|\grad_{\vx_n}\varphi\|\leq\|\grad_{\vx_n}\psi\|$
and $\|\varphi\|\leq\|\psi\|$.  The product rule
and $0\leq\zeta_L\leq1$
give
\begin{align*}
 \sum_{n=1}^N\|\grad_{\vx_n}(\zeta_LP_M\psi)\|^2
 &\leq2\sum_{n=1}^N\|\grad_{\vx_n}\varphi\|^2
   +2\Big\langle \varphi, \sum_{n=1}^N|\grad_{\vx_n}\zeta_L|^2\varphi \Big\rangle\\
 &\leq2\sum_{n=1}^N\|\grad_{\vx_n}\psi\|^2+\frac{CN}{L^2}\|\psi\|^2.
\end{align*}
Applying this to the vectors in a spectral decomposition of
$\mathcal L_R\vGamma_j$ and summing, with
$\Tr\mathcal L_R\vGamma_j=1$, gives
\[
 \Tr\big((-\lapl)\zeta_LP_M(\mathcal L_R\vGamma_j)P_M\zeta_L\big)
 \leq2\Tr\big((-\lapl)\mathcal L_R\vGamma_j\big)+\frac{CN}{L^2}.
\]
Moreover,
$\zeta_LP_M$ commutes with $H_\rad$ and is a contraction, so
\begin{align*}
 \Tr\big(H_\rad\zeta_LP_M(\mathcal L_R\vGamma_j)P_M\zeta_L\big)
 &\leq\Tr(H_\rad\mathcal L_R\vGamma_j),\\
 \Tr\big((1-\lapl+H_\rad)\zeta_LP_M(\mathcal L_R\vGamma_j)P_M\zeta_L\big)
 &\leq C(E+N/L^2).
\end{align*}

We now apply Rellich compactness after both cutoffs.  In the
$m$-photon sector of $\mathcal L_R\vGamma_j$, every retained photon is supported
in $B_{2R}$, because its localization factor is $\theta_R$.
Consequently, $\zeta_LP_M\sqrt{\mathcal L_R\vGamma_j}$ takes values in the closed
subspace
\begin{align*}
 \mathscr H_{L,R,M}
 :=\bigoplus_{m=0}^M\big\{&\psi^{(m)}\in\Hil_\el^N\otimes L_s^2((\RR^3\times\{1,2\})^m):\psi^{(m)}\equiv 0\text{ a.e.\ outside }\Omega_{L,R,m}\big\},
\end{align*}
where
\[
 \Omega_{L,R,m}:=(B_{2L})^N\times(B_{2R})^m.
\]
Here the photon variables $(\vy_l,\lambda_l)$ are in the position
representation, and the support condition is imposed for every value
of the polarization indices $\lambda_l$.  On this subspace we restrict the
closed form of $1-\lapl+H_\rad$, computing all derivatives after
extension by zero to the full configuration space.  This gives the form
\[
 k_{L,R,M}[\psi]:=\sum_{m=0}^Mk_m[\psi^{(m)}],\qquad
 \psi=\bigoplus_{m=0}^M\psi^{(m)}\in\mathscr H_{L,R,M},
\]
whose value in the $m$-photon sector is
\[
 k_m[\psi^{(m)}]=\|\psi^{(m)}\|^2+\sum_{n=1}^N\|\grad_{\vx_n}\psi^{(m)}\|^2
 +\sum_{l=1}^m\big\|(-\Delta_{\vy_l})^{1/4}\psi^{(m)}\big\|^2.
\]
Its domain $Q(k_{L,R,M})$ consists of those $\psi\in\mathscr H_{L,R,M}$
for which every $k_m[\psi^{(m)}]$ is finite.  The form $k_{L,R,M}$ is
closed because $\mathscr H_{L,R,M}$ is closed in $\HC_N$.  It is densely defined because smooth functions compactly
supported in $\Omega_{L,R,m}$, with the prescribed symmetries, are
dense in the corresponding $L^2$ space.
The form $k_m[\cdot]$ controls the $H^{1/2}$ norm in all $3(N+m)$ variables
jointly.  Indeed, by Plancherel's theorem, with $\vxi_n$ dual to $\vx_n$
and $\vk_l$ dual to $\vy_l$, the Fourier symbol of $k_m[\cdot]$ is
$1+\sum_n|\vxi_n|^2+\sum_l|\vk_l|$.  Using
$\sum_l|\vk_l|^2\leq\big(\sum_l|\vk_l|\big)^2$,
$\sqrt{s+t}\leq\sqrt s+\sqrt t$ for $s,t\geq0$, and
$\sqrt{1+s}\leq1+s$, we obtain
\begin{align*}
 \left(1+\sum_{n=1}^N|\vxi_n|^2
          +\sum_{l=1}^m|\vk_l|^2\right)^{1/2}
 &\leq\left(1+\sum_{n=1}^N|\vxi_n|^2\right)^{1/2}
      +\sum_{l=1}^m|\vk_l|\\
 &\leq1+\sum_{n=1}^N|\vxi_n|^2+\sum_{l=1}^m|\vk_l|,
\end{align*}
and hence $\|u\|_{H^{1/2}(\RR^{3(N+m)})}^2\leq k_m[u]$.
Since $\Omega_{L,R,m}$ is a bounded convex domain and hence a Lipschitz domain, the
Rellich--Kondrashov theorem for $H^{1/2}$ gives a compact embedding of
the $m$-photon component of $Q(k_{L,R,M})$ into $L^2(\Omega_{L,R,m})$
\cite[Theorem~7.1]{dinezza2012hitchhiker}: the Gagliardo seminorm on
$\Omega_{L,R,m}$ is bounded by the $H^{1/2}(\RR^{3(N+m)})$ norm
\cite[Proposition~3.4]{dinezza2012hitchhiker}, and Lipschitz domains
are extension domains \cite[Theorem~5.4]{dinezza2012hitchhiker}.
There are only finitely many sectors and finitely many polarization
components in each.  Therefore $Q(k_{L,R,M})$, normed by
$k_{L,R,M}[\cdot]^{1/2}$, embeds compactly into $\mathscr H_{L,R,M}$.

Let $K_{L,R,M}\geq1$ be the self-adjoint operator on
$\mathscr H_{L,R,M}$ associated with the closed form $k_{L,R,M}[\cdot]$.  It has compact resolvent, so for each $s\geq1$ its spectral
projection
\[
 S:=\iden_{[0,s]}(K_{L,R,M}),
\]
extended by zero on $\mathscr H_{L,R,M}^\perp$, is a finite-rank
orthogonal projection on $\HC_N$.  The spectral theorem gives
\begin{align*}
 \|(1-S)\zeta_LP_M\sqrt{\mathcal L_R\vGamma_j}\|_{\mathfrak S_2}^2
 &=\Tr\big((1-S)\zeta_LP_M(\mathcal L_R\vGamma_j)P_M\zeta_L\big)\\
 &\leq s^{-1}\Tr\big(K_{L,R,M}\zeta_LP_M(\mathcal L_R\vGamma_j)P_M\zeta_L\big)
 \leq\frac{C(E+N/L^2)}{s}.
\end{align*}
Combining this with the preceding cutoff error \cref{cutoff-error}, we obtain
\begin{align*}
 \Tr\big((1-S)\mathcal L_R\vGamma_j\big)
 &=\|(1-S)\sqrt{\mathcal L_R\vGamma_j}\|_{\mathfrak S_2}^2\\
 &\leq2\|(1-\zeta_LP_M)\sqrt{\mathcal L_R\vGamma_j}\|_{\mathfrak S_2}^2
      +2\|(1-S)\zeta_LP_M\sqrt{\mathcal L_R\vGamma_j}\|_{\mathfrak S_2}^2\\
 &\leq\frac{2CRE}{M+1}+2\epsilon_\el(L)
      +\frac{C(E+N/L^2)}{s}.
\end{align*}
For fixed $R$, first choose $M$, then $L$, and finally $s$ sufficiently
large to make this bound arbitrarily small, uniformly in $j$.
The states $\mathcal L_R\vGamma_j$ are positive with unit trace, so the
compactness part of \cref{lem-trace-compression} shows that the sequence
$(\mathcal L_R\vGamma_j)_j$ is precompact in trace norm.

\noindent\textbf{Step 2. A trace-norm limit of the retained states.}
A diagonal extraction gives a subsequence, still denoted
$(\vGamma_j)$, such that
\[
 \mathcal L_{3^m}\vGamma_j\longrightarrow\vGamma^{(m)}
 \quad\text{in trace norm as }j\to\infty,\text{ for every }m\geq1.
\]
Localization maps compose according to multiplication of their
one-photon contractions.  For $l>m$, the function $\theta_{3^m}$
vanishes outside $B_{2\cdot3^m}$, while $\theta_{3^l}=1$ on
$B_{3^l}\supset B_{3^{m+1}}\supset B_{2\cdot3^m}$.  Hence
$\theta_{3^m}\theta_{3^l}=\theta_{3^m}$, and therefore
\[
 \mathcal L_{3^m}\vGamma^{(l)}=\vGamma^{(m)}.
\]
Since photon-sector projections are bounded, the hypothesis implies
\[
 \delta(M):=\sup_m\Tr(\iden_{\{\NC>M\}}\vGamma^{(m)})
 \longrightarrow0.
\]

We next show that $(\vGamma^{(m)})$ is Cauchy in trace norm, using
\cref{lem-localization-trace-distance} with $\tau=\vGamma^{(l)}$ and
$\eta=\theta_{3^m}$ for $l>m$.  For this, the probability that passing
from $\vGamma^{(l)}$ to $\vGamma^{(m)}$ discards at least one photon
must be small.  The bound on this probability in
\cref{lem-localization-trace-distance} involves the mean photon number,
which need not be finite here.  We therefore track instead the
expectation of the bounded function $n\mapsto2^{-n}$ of the photon
number,
\[
 b_m:=\Tr(2^{-\NC}\vGamma^{(m)})\in[0,1],
\]
which is finite for every state.  We show below that discarding photons
can only increase this expectation, so $(b_m)$ is nonincreasing and
converges.  Its decrease $b_m-b_l$, together with the photon-number
tightness $\delta(M)$, then controls the probability of discarding a
photon.

For $l>m$, the localization isometry $\mathcal U_{3^m}$ underlying
$\mathcal L_{3^m}$ takes values in
$\Hil_\el^N\otimes\Fock_\ph^{\mathrm{in}}\otimes\Fock_\ph^{\mathrm{out}}$,
where the two Fock factors contain the retained and discarded photons,
respectively.  In photon position, the retained photons carry the
factor $\theta_{3^m}$ and are supported in the ball $B_{2\cdot3^m}$,
while the discarded photons carry the factor $\phi_{3^m}$, which
vanishes on $B_{3^m}$.  Define their number operators by
\[
 \NC_{\mathrm{in}}
 :=\iden_{\Hil_\el^N}\otimes\NC\otimes\iden_{\Fock_\ph^{\mathrm{out}}},
 \qquad
 \NC_{\mathrm{out}}
 :=\iden_{\Hil_\el^N}\otimes\iden_{\Fock_\ph^{\mathrm{in}}}\otimes\NC.
\]
Thus on the sector with $p$ retained and $q$ discarded photons they
act as multiplication by $p$ and $q$, respectively, and strongly
commute.  Recall that $\mathcal U_{3^m}=\varGamma(j_{3^m})$ is the
second quantization of the one-photon isometry
$j_{3^m}f=(\theta_{3^m}f,\phi_{3^m}f)$, followed by the identification
$\Fock_s(\Hil_\ph\oplus\Hil_\ph)
\simeq\Fock_\ph^{\mathrm{in}}\otimes\Fock_\ph^{\mathrm{out}}$, with the
electronic identity suppressed, and that
$\mathcal L_{3^m}(\vGamma)
=\Tr_{\mathrm{out}}(\mathcal U_{3^m}\vGamma\mathcal U_{3^m}^\dag)$.
In the joint state
$\mathcal U_{3^m}\vGamma^{(l)}\mathcal U_{3^m}^\dag$, their sum has the
number distribution of $\vGamma^{(l)}$, and the first has that of
$\vGamma^{(m)}$.  Indeed, $\mathcal U_{3^m}$ maps the $n$-photon sector
into the sum of the sectors with $p+q=n$, so
$\mathcal U_{3^m}^\dag f(\NC_{\mathrm{in}}+\NC_{\mathrm{out}})
\mathcal U_{3^m}=f(\NC)$ for every bounded function $f$ on
$\{0,1,2,\ldots\}$.  Moreover, $f(\NC_{\mathrm{in}})$ acts trivially on
the discarded factor, and the partial trace over that factor is
$\mathcal L_{3^m}\vGamma^{(l)}=\vGamma^{(m)}$.  Hence
\begin{align*}
 \Tr\!\left(f(\NC_{\mathrm{in}}+\NC_{\mathrm{out}})\,
   \mathcal U_{3^m}\vGamma^{(l)}\mathcal U_{3^m}^\dag\right)
 &=\Tr\big(f(\NC)\vGamma^{(l)}\big),\\
 \Tr\!\left(f(\NC_{\mathrm{in}})\,
   \mathcal U_{3^m}\vGamma^{(l)}\mathcal U_{3^m}^\dag\right)
 &=\Tr\big(f(\NC)\vGamma^{(m)}\big).
\end{align*}
Take $f(n)=2^{-n}$.  The operators $\NC_{\mathrm{in}}$ and
$\NC_{\mathrm{out}}$ strongly commute, so
$2^{-(\NC_{\mathrm{in}}+\NC_{\mathrm{out}})}
=2^{-\NC_{\mathrm{in}}}2^{-\NC_{\mathrm{out}}}$, and the two identities
give
\begin{align*}
 b_l&=\Tr\!\left(2^{-\NC_{\mathrm{in}}}2^{-\NC_{\mathrm{out}}}
   \mathcal U_{3^m}\vGamma^{(l)}\mathcal U_{3^m}^\dag\right),\\
 b_m&=\Tr\!\left(2^{-\NC_{\mathrm{in}}}
   \mathcal U_{3^m}\vGamma^{(l)}\mathcal U_{3^m}^\dag\right).
\end{align*}
Subtracting, we obtain
\[
 b_m-b_l
 =\Tr\!\left(2^{-\NC_{\mathrm{in}}}
 (1-2^{-\NC_{\mathrm{out}}})
 \mathcal U_{3^m}\vGamma^{(l)}\mathcal U_{3^m}^\dag\right)\geq0,
\]
where the inequality holds because $2^{-p}(1-2^{-q})\geq0$ for all
integers $p,q\geq0$.  In particular, the sequence $(b_m)$ is
nonincreasing.
Let
\[
 t_{m,l}:=\Tr\!\left(\iden_{\{\NC_{\mathrm{out}}\geq1\}}
      \mathcal U_{3^m}\vGamma^{(l)}\mathcal U_{3^m}^\dag\right)
\]
denote the probability that at least one photon is discarded.
Splitting according to $\NC_{\mathrm{in}}>M$ and
$\NC_{\mathrm{in}}\leq M$ gives
\begin{align*}
 t_{m,l}
 &\leq\Tr\!\left(\iden_{\{\NC_{\mathrm{in}}>M\}}
      \mathcal U_{3^m}\vGamma^{(l)}\mathcal U_{3^m}^\dag\right)\\
 &\quad+\Tr\!\left(\iden_{\{\NC_{\mathrm{in}}\leq M\}}
      \iden_{\{\NC_{\mathrm{out}}\geq1\}}
      \mathcal U_{3^m}\vGamma^{(l)}\mathcal U_{3^m}^\dag\right)\\
 &\leq\delta(M)+2^{M+1}(b_m-b_l).
\end{align*}
For the first term, we take $f=\iden_{(M,\infty)}$ in the second trace
identity above.  This shows that the first term equals
$\Tr(\iden_{\{\NC>M\}}\vGamma^{(m)})$, which is at most $\delta(M)$.
For the second
term, the operator $2^{-\NC_{\mathrm{in}}}(1-2^{-\NC_{\mathrm{out}}})$
is at least $2^{-M-1}$ on the joint spectral subspace
$\NC_{\mathrm{in}}\leq M$, $\NC_{\mathrm{out}}\geq1$, so the formula
for $b_m-b_l$ bounds this term by $2^{M+1}(b_m-b_l)$.  Finally,
$t_{m,l}$ is the quantity $t$ of \cref{lem-localization-trace-distance}
for $\tau=\vGamma^{(l)}$ and $\eta=\theta_{3^m}$, and
$\mathcal L_{3^m}\vGamma^{(l)}=\vGamma^{(m)}$, so
\[
 \|\vGamma^{(l)}-\vGamma^{(m)}\|_{\mathfrak S_1}
 \leq2\sqrt{t_{m,l}}+t_{m,l}.
\]
The nonincreasing sequence $b_m$ has a limit.  First choose $M$ so that
$\delta(M)$ is small, then $m$ so that $b_m-b_l$ is small uniformly
for $l>m$.  Hence $\vGamma^{(m)}$ is Cauchy in trace norm and converges to a
density matrix $\vGamma$ on the original space $\HC_N$, possibly with
infinite mean photon number.  The rough
energy bound \cref{localized-internal-energy-bound} gives
\[
 \sup_{j,m}\Tr(H\mathcal L_{3^m}\vGamma_j)\leq CE.
\]
Applying \cref{lem-trace-compactness-and-limits} first to
$\mathcal L_{3^m}\vGamma_j\to\vGamma^{(m)}$ and then to
$\vGamma^{(m)}\to\vGamma$ yields
\[
 \sup_m\Tr(H\vGamma^{(m)})\leq CE,\qquad
 \Tr(H\vGamma)\leq CE<\infty,
\]
and hence $\vGamma\in\dm$.  We next prove the sharp energy inequality.

\noindent\textbf{Step 3. The energy inequality.}
The argument involves an electronic cutoff at scale $L$, the photon
localization scale $R=3^m$, and a truncation of $H_\rad$ by second
quantizations of one-photon operators of rank $K$, all introduced below.
We pass to the limits in a fixed order: first $j\to\infty$, then
$R=3^m\to\infty$ with $L$ and $K$ fixed, then $K\to\infty$, and finally
$L\to\infty$.  The order matters because the bound on the localized
truncated radiation energy below carries an error $\delta_K(R)$ that is
established only for fixed $K$, and tends to zero as $R\to\infty$ but not
uniformly in $K$.

To control the interaction error, we first choose a symmetric electronic
cutoff $\eta_L$, equal to one when $\max_i|\vx_i|\leq L$ and zero
when $\max_i|\vx_i|\geq2L$, together with a smooth complementary
cutoff $\wt\eta_L$.  For example, using the electronic cutoff
$\zeta_L$ constructed above, take
\[
 \eta_L=\sin\!\left(\frac\pi2\zeta_L\right),\qquad
 \wt\eta_L=\cos\!\left(\frac\pi2\zeta_L\right).
\]
Then $\eta_L^2+\wt\eta_L^2=1$ and
\[
 \sum_{i=1}^N\left(|\grad_{\vx_i}\eta_L|^2
                    +|\grad_{\vx_i}\wt\eta_L|^2\right)
 =\frac{\pi^2}{4}\sum_{i=1}^N|\grad_{\vx_i}\zeta_L|^2
 \leq\frac{C}{L^2}.
\]
Both cutoffs preserve $Q(H)$: their bounded gradients preserve the
electronic $H^1$ regularity, and they commute with $\sqrt{H_\rad}$.

For either cutoff $f\in\{\eta_L,\wt\eta_L\}$, the covariant derivative obeys
\[
 \vD_i(f\Psi)=f\vD_i\Psi-\i(\grad_{\vx_i}f)\Psi,
 \qquad \vD_i=-\i\grad_{\vx_i}-e\AF(\vx_i),
\]
because $f$ commutes with the field operator $\AF(\vx_i)$.
Since $\eta_L$ and $\wt\eta_L$ are real, expanding gives
\begin{align*}
 &\|\vD_i(\eta_L\Psi)\|^2+\|\vD_i(\wt\eta_L\Psi)\|^2\\
 &\quad=\dua{\vD_i\Psi}{(\eta_L^2+\wt\eta_L^2)\vD_i\Psi}
   +\|(\grad_{\vx_i}\eta_L)\Psi\|^2
   +\|(\grad_{\vx_i}\wt\eta_L)\Psi\|^2\\
 &\qquad+2\Re\dua{\vD_i\Psi}{-\i\big(\eta_L\grad_{\vx_i}\eta_L
   +\wt\eta_L\grad_{\vx_i}\wt\eta_L\big)\Psi}\\
 &\quad=\|\vD_i\Psi\|^2+\|(\grad_{\vx_i}\eta_L)\Psi\|^2
   +\|(\grad_{\vx_i}\wt\eta_L)\Psi\|^2.
\end{align*}
In the last step we used $\eta_L^2+\wt\eta_L^2=1$, and the cross terms
cancel because
\[
 \eta_L\grad_{\vx_i}\eta_L
 +\wt\eta_L\grad_{\vx_i}\wt\eta_L
 =\tfrac12\grad_{\vx_i}(\eta_L^2+\wt\eta_L^2)=0.
\]
The Coulomb interaction is an electronic multiplication operator, and
$H_\rad$ acts only on the photon factor.  Their localized forms
therefore sum exactly to their original forms, using
$\eta_L^2+\wt\eta_L^2=1$.  With the factor $1/2$ in the kinetic
energy, the resulting covariant IMS identity is
\begin{align*}
 &\dua{\eta_L\Psi}{H\eta_L\Psi}
   +\dua{\wt\eta_L\Psi}{H\wt\eta_L\Psi}\\
 &\quad=\dua{\Psi}{H\Psi}
   +\frac12\sum_{i=1}^N\left(
      \|(\grad_{\vx_i}\eta_L)\Psi\|^2
      +\|(\grad_{\vx_i}\wt\eta_L)\Psi\|^2\right)\\
 &\quad\leq\dua{\Psi}{H\Psi}+\frac{C}{L^2}\|\Psi\|^2.
\end{align*}
The computation first holds on the form core and extends to $Q(H)$
by continuity.  Applying it to the vectors in a spectral decomposition
of $\vGamma_j$ and summing, with $\Tr\vGamma_j=1$, gives
\[
 \Tr(H\eta_L\vGamma_j\eta_L)
 +\Tr(H\wt\eta_L\vGamma_j\wt\eta_L)
 \leq\Tr(H\vGamma_j)+\frac{C}{L^2}.
\]
Since $H\geq0$, the second term on the left is nonnegative and can
be dropped.  Thus
\[
 \Tr(H\eta_L\vGamma_j\eta_L)\leq\Tr(H\vGamma_j)+C/L^2.
\]

Let $\vGamma$ be a positive trace-class operator with
$\Tr(H\vGamma)<\infty$.  Every term in \cref{q00-def} is nonnegative,
so $\Tr(H_\rad\vGamma)\leq\Tr(H\vGamma)<\infty$, and we may define
\[
 \mathcal E(\vGamma):=\Tr(H\vGamma)-\Tr(H_\rad\vGamma),
\]
the internal energy with the free-photon contribution removed.  It
still contains the electron--photon coupling through the covariant
kinetic energy, as well as the Coulomb repulsion.
The field identities \cref{localized-linear-field,localized-quadratic-field},
the annihilation--creation bounds
\cref{ann-Hrad-bound,cre-Hrad-bound} relative to $\sqrt{H_\rad}$, and
Cauchy--Schwarz give
\[
 \mathcal E\big(\mathcal L_R(\eta_L\vGamma_j\eta_L)\big)
 \leq\mathcal E(\eta_L\vGamma_j\eta_L)
 +C\epsilon_L(R)+C\epsilon_L(R)^2,
\]
where
\[
 \epsilon_L(R):=\max_\mu\sup_{|\vx|\leq2L}
 \big(\|(1-\theta_R)G^\mu_{\vx}\|
 +\|(-\Delta)^{-1/4}(1-\theta_R)G^\mu_{\vx}\|
 +\|\phi_RG^\mu_{\vx}\|\big).
\]
Here the negative Sobolev norm is
$\|f\|_{\dot H^{-1/2}}=\|(-\Delta)^{-1/4}f\|$ in photon position
representation.
The Coulomb term is unchanged and the quadratic-field identity includes
the vacuum fluctuations.  The constants are uniform in $j$ for fixed
$L$.

We now show that $\epsilon_L(R)\to0$ as $R\to\infty$ for fixed $L$.
Since $0\leq1-\theta_R,\phi_R\leq1$ and both vanish on $B_R$, dominated
convergence gives $(1-\theta_R)f\to0$ and $\phi_Rf\to0$ in $\Hil_\ph$
for every $f\in\Hil_\ph$.
Smooth multiplication by $\theta_R$ is uniformly bounded on
$\dot H^{1/2}$, and therefore by duality on $\dot H^{-1/2}$.  Density of
compactly supported smooth functions shows that $\theta_R\to1$ strongly in
both spaces.  The map $\vx\mapsto G^\mu_{\vx}$ is continuous into
$\Hil_\ph$ and into $\dot H^{-1/2}$, as noted after \cref{Gx-norms}.
Hence the form factors $G^\mu_{\vx}$ with $|\vx|\leq2L$ and
$\mu=1,2,3$ form a compact family in both norms.  The operators
$1-\theta_R$ on both spaces and $\phi_R$ on $\Hil_\ph$ are uniformly
bounded in $R$ and converge strongly to zero, and strong convergence of uniformly bounded
operators is uniform on compact sets.  The supremum over $|\vx|\leq2L$
in the definition of $\epsilon_L(R)$ is therefore harmless, and
\[
 \lim_{R\to\infty}\epsilon_L(R)=0.
\]

The rough bound \cref{localized-radiation-bound} only controls the
localized radiation energy up to a constant factor.  To obtain it with
coefficient one, we truncate $H_\rad$ from below by second
quantizations of finite-rank one-photon operators: choose an
orthonormal basis $(e_k)$ whose finite span is a core for
$(-\Delta)^{1/4}$, and set
\begin{gather*}
 f_k=(-\Delta)^{1/4}e_k,\qquad
 t_K=\sum_{k=1}^K|f_k\rangle\langle f_k|\leq\sqrt{-\Delta},\\
 T_K=\dd\varGamma(t_K)=\sum_{k=1}^Ka^\dag(f_k)a(f_k).
\end{gather*}
Then $T_K\uparrow H_\rad$ in the form sense.

\noindent\textbf{Claim.}  For fixed $K$ there are numbers
$\delta_K(R)\to0$ as $R\to\infty$ such that
\[
 \theta_Rt_K\theta_R\leq\big(1+\delta_K(R)\big)\sqrt{-\Delta}
\]
in the sense of quadratic forms.

\noindent\textit{Proof of the claim.}
Since $\|(-\Delta)^{-1/4}f_k\|=\|e_k\|=1$, each $f_k$ lies in
$\dot H^{-1/2}$, and the strong convergence $\theta_R\to1$ on
$\dot H^{-1/2}$ established above gives
\[
 e_{k,R}:=(-\Delta)^{-1/4}\theta_Rf_k\longrightarrow e_k
 \quad\text{in }\Hil_\ph,\qquad k=1,\ldots,K.
\]
Let $\Pi_K$ be the orthogonal projection onto the span of
$e_1,\ldots,e_K$, and put
\[
 \delta_K(R):=\Big\|\sum_{k=1}^K|e_{k,R}\rangle\langle e_{k,R}|-\Pi_K\Big\|
 \leq\sum_{k=1}^K\big(\|e_{k,R}\|+1\big)\|e_{k,R}-e_k\|
 \longrightarrow0.
\]
Let $\psi$ be in the form domain of $\sqrt{-\Delta}$.  Since $\theta_R$
is real, moving $(-\Delta)^{\pm1/4}$ across the inner product gives
\begin{align*}
 \dua{\psi}{\theta_Rt_K\theta_R\psi}
 &=\sum_{k=1}^K|\dua{\theta_Rf_k}{\psi}|^2
 =\sum_{k=1}^K\big|\dua{e_{k,R}}{(-\Delta)^{1/4}\psi}\big|^2\\
 &=\Big\langle(-\Delta)^{1/4}\psi,
   \sum_{k=1}^K|e_{k,R}\rangle\langle e_{k,R}|\,(-\Delta)^{1/4}\psi\Big\rangle\\
 &\leq\big(\|\Pi_K\|+\delta_K(R)\big)\|(-\Delta)^{1/4}\psi\|^2
 =\big(1+\delta_K(R)\big)\dua{\psi}{\sqrt{-\Delta}\,\psi}.
\end{align*}
This proves the claim.

The convergence $\delta_K(R)\to0$ is not uniform in $K$.  Since second
quantization preserves form inequalities,
\cref{localized-second-quantization} and the claim give
\[
 \Tr\big(T_K\mathcal L_R(\eta_L\vGamma_j\eta_L)\big)
 =\Tr\big(\dd\varGamma(\theta_Rt_K\theta_R)\eta_L\vGamma_j\eta_L\big)
 \leq\big(1+\delta_K(R)\big)\Tr(H_\rad\eta_L\vGamma_j\eta_L).
\]
Combining the last two estimates before taking any limits yields
\begin{equation}\label{localized-combined-energy-bound}
 \mathcal E\big(\mathcal L_R(\eta_L\vGamma_j\eta_L)\big)
 +\Tr\big(T_K\mathcal L_R(\eta_L\vGamma_j\eta_L)\big)
 \leq\Tr(H\vGamma_j)+C/L^2+o_R(1).
\end{equation}
Here $o_R(1)$ denotes an error that tends to zero as $R\to\infty$,
uniformly in $j$, for fixed $L$ and $K$.  It collects the term
$C\epsilon_L(R)+C\epsilon_L(R)^2$ from the first estimate and the term
$\delta_K(R)\Tr(H_\rad\eta_L\vGamma_j\eta_L)\leq\delta_K(R)(E+C/L^2)$
from the second, and both are bounded independently of $j$.
We justify the two limit passages for fixed $L,K$ in detail.  Since
photon localization commutes with the electronic multiplier $\eta_L$,
the positive, possibly subnormalized operators
\[
 \sigma_{j,m}:=\mathcal L_{3^m}(\eta_L\vGamma_j\eta_L)
 =\eta_L(\mathcal L_{3^m}\vGamma_j)\eta_L,\qquad
 \sigma_m:=\eta_L\vGamma^{(m)}\eta_L,\qquad
 \sigma:=\eta_L\vGamma\eta_L
\]
satisfy $\sigma_{j,m}\to\sigma_m$ in trace norm as $j\to\infty$
for each $m$, and $\sigma_m\to\sigma$ in trace norm as $m\to\infty$.
Indeed, multiplication on both sides by $\eta_L$ is a contraction
in trace norm.  The rough energy bound
\cref{localized-internal-energy-bound} for
$\mathcal L_{3^m}\vGamma_j$, followed by the covariant IMS estimate,
gives
\[
 \sup_{j,m}\big(\Tr\sigma_{j,m}+\Tr(H\sigma_{j,m})\big)
 \leq C_L<\infty.
\]
By \cref{lem-trace-compactness-and-limits}, the same bound holds for $\sigma_m$,
uniformly in $m$, and for $\sigma$.

Write $W=\sum_{n<m}|\vx_n-\vx_m|^{-1}$.  Since $T_K\leq H_\rad$,
for $\Psi\in Q(H)$
\[
 \frac12\sum_{i=1}^N\|\vD_i\Psi\|^2+\|\sqrt{W}\Psi\|^2+\|\sqrt{T_K}\Psi\|^2
 \leq\|\sqrt{H}\Psi\|^2.
\]
Thus $\mathcal E(\cdot)+\Tr(T_K\cdot)$ is the trace of a
nonnegative form bounded relative to $H+1$, and the final assertion
of \cref{lem-trace-compactness-and-limits} applies to it along
trace-norm convergent sequences with uniformly bounded energy.
Apply it first to $\sigma_{j,m}\to\sigma_m$ with $m$
fixed.  The combined energy estimate \cref{localized-combined-energy-bound} yields
\begin{align*}
 \mathcal E(\sigma_m)+\Tr(T_K\sigma_m)
 &\leq\liminf_j\big[\mathcal E(\sigma_{j,m})
                       +\Tr(T_K\sigma_{j,m})\big]\\
 &\leq\liminf_j\Tr(H\vGamma_j)+C/L^2+o_m(1),
\end{align*}
where $o_m(1)$ is the error $o_R(1)$ at $R=3^m$.  Applying the same assertion
to $\sigma_m\to\sigma$ and then letting $m\to\infty$ gives
\[
 \mathcal E(\eta_L\vGamma\eta_L)
 +\Tr(T_K\eta_L\vGamma\eta_L)
 \leq\liminf_j\Tr(H\vGamma_j)+C/L^2.
\]
For fixed $L$, the nonnegative forms $T_K$ increase to $H_\rad$.
Monotone convergence of their traces and the definition of
$\mathcal E$ therefore give
\[
 \Tr(H\eta_L\vGamma\eta_L)
 \leq\liminf_j\Tr(H\vGamma_j)+C/L^2.
\]
Next, $\zeta_L\to1$ pointwise as $L\to\infty$, so $\eta_L\to1$
pointwise with $0\leq\eta_L\leq1$.  By dominated convergence,
$\eta_L\to1$ strongly as multiplication operators on $\HC_N$.  Since
$\sqrt{\vGamma}$ is Hilbert--Schmidt, dominated convergence over an
orthonormal basis gives
$\|(1-\eta_L)\sqrt{\vGamma}\|_{\mathfrak S_2}\to0$.  Factoring
through $\sqrt{\vGamma}$ and using $\Tr\vGamma=1$, we obtain
\[
 \|\eta_L\vGamma\eta_L-\vGamma\|_{\mathfrak S_1}
 \leq2\|(1-\eta_L)\sqrt{\vGamma}\|_{\mathfrak S_2}
 \longrightarrow0.
\]
The preceding energy bound allows us to apply
\cref{lem-trace-compactness-and-limits} to this trace-norm convergence.  Hence
\[
 \Tr(H\vGamma)
 \leq\liminf_{L\to\infty}\Tr(H\eta_L\vGamma\eta_L)
 \leq\liminf_j\Tr(H\vGamma_j).
\]

\noindent\textbf{Step 4. The constraints.}
We first pass the density through the two successive limits.
For $f\in C_c^\infty(\RR^3)$ and fixed $m$, invariance of the
electronic reduced state under photon localization and
\cref{mixed-density-passage} give
\[
 \int_{\RR^3} f\rho_{\vGamma^{(m)}}
 =\lim_{j\to\infty}\int_{\RR^3} f\rho_{\mathcal L_{3^m}\vGamma_j}
 =\lim_{j\to\infty}\int_{\RR^3} f\rho_{\vGamma_j}
 =\int_{\RR^3} f\rho.
\]
Since $\vGamma^{(m)}\to\vGamma$ in trace norm, a second application of
\cref{mixed-density-passage} gives $\int f\rho_\vGamma=\int f\rho$.
Thus $\rho_\vGamma=\rho$ in distributions.

For the current density, fix real
$\vzeta\in C_c^\infty(\RR^3,\RR^3)$ and choose $L$ with
$\supp\vzeta\subset B_{2L}$.  Put
\[
 h^\mu_{R,\vx}:=(1-\theta_R)G^\mu_{\vx},\qquad
 A_{R,\vzeta}:=e\sum_{i=1}^N\sum_{\mu=1}^3
     \zeta^\mu(\vx_i)\varPhi(h^\mu_{R,\vx_i}).
\]
Photon localization preserves the paramagnetic current density, since it
depends only on the electronic reduced state.  Using
$\vjc=\vjp+\vjph$ and \cref{localized-linear-field}, we consequently
have the form identity
\[
 \int_{\RR^3}\vzeta\cdot
 (\vjc_{\mathcal L_R\vGamma_j}-\vjc_{\vGamma_j})
 =\Tr(A_{R,\vzeta}\vGamma_j).
\]
The definition of $\epsilon_L(R)$ and the field bounds
\cref{ann-Hrad-bound,cre-Hrad-bound} imply
\[
 \|A_{R,\vzeta}\Psi\|
 \leq C_\vzeta\epsilon_L(R)
       \big(\|\sqrt{H_\rad}\Psi\|+\|\Psi\|\big),
 \qquad \Psi\in Q(H).
\]
Thus the trace pairing is well defined without a mean-photon-number
bound.  Hilbert--Schmidt Cauchy--Schwarz yields
\begin{align*}
 \left|\Tr(A_{R,\vzeta}\vGamma_j)\right|
 &\leq\|\sqrt{\vGamma_j}\|_{\mathfrak S_2}
        \|A_{R,\vzeta}\sqrt{\vGamma_j}\|_{\mathfrak S_2}\\
 &\leq C_\vzeta\epsilon_L(R)
       \big(1+\Tr(H_\rad\vGamma_j)\big)^{1/2}
 \leq C_\vzeta\epsilon_L(R)E^{1/2},
\end{align*}
where $E=1+\sup_j\Tr(H\vGamma_j)$ was fixed above.

Now fix $m$ and take $R=3^m$.  The states
$\mathcal L_{3^m}\vGamma_j$ converge in trace norm to $\vGamma^{(m)}$
and have uniformly bounded internal energies.  Hence
\cref{mixed-current-passage} passes their tested current densities to the
limit as $j\to\infty$.  Combining this with the assumed
distributional convergence of $\vjc_{\vGamma_j}$ gives
\[
 \left|\int_{\RR^3}\vzeta\cdot
 (\vjc_{\vGamma^{(m)}}-\vjc)\right|
 \leq C_\vzeta\epsilon_L(3^m)E^{1/2}.
\]
The energies of $\vGamma^{(m)}$ are bounded uniformly in $m$, so
\cref{mixed-current-passage} also applies to
$\vGamma^{(m)}\to\vGamma$.  Letting $m\to\infty$ in the last inequality,
and using $\epsilon_L(3^m)\to0$ for this fixed $L$, proves
$\int\vzeta\cdot\vjc_\vGamma=\int\vzeta\cdot\vjc$.
This identifies the combined current density in distributions.
Since $\vGamma\in\dm$, \cref{prop-current-bounds} and the
mixed-state series \cref{mixed-PF-density-current}, which converge in
$L^1$ by normalization and finite energy, show that both
$\rho_\vGamma$ and $\vjc_\vGamma$ lie in $L^1$.  Hence these
distributional identities hold almost everywhere, and in particular
$\rho$ and $\vjc$ are integrable functions.  This completes
\cref{continuum-retained-limit}.

\noindent\textbf{Step 5. Sufficient photon-number conditions.}
On the input $k$-photon sector, $\mathcal U_R$ takes values in the
sum of retained/discarded sectors with $p+q=k$.  Since $p\leq k$,
\[
 \mathcal U_R^\dag
 (\iden_{\{\NC>M\}}\otimes\iden)\mathcal U_R
 \leq\iden_{\{\NC>M\}}.
\]
Taking the trace against $\vGamma_j$ and then the indicated limits
gives
\[
 \sup_{m\geq1}\limsup_{j\to\infty}
 \Tr\big(\iden_{\{\NC>M\}}\mathcal L_{3^m}\vGamma_j\big)
 \leq\sup_j\Tr\big(\iden_{\{\NC>M\}}\vGamma_j\big).
\]
Uniform tails on the right therefore imply
\cref{continuum-local-sector-tightness}.  If the mean photon numbers
are uniformly bounded, the spectral inequality
$\iden_{\{\NC>M\}}\leq\NC/(M+1)$ for integers $M\geq0$ gives
\[
 \sup_j\Tr\big(\iden_{\{\NC>M\}}\vGamma_j\big)
 \leq\frac{\sup_j\Tr(\NC\vGamma_j)}{M+1}
 \longrightarrow0,
\]
which proves the final assertion.
\end{proof}

\subsection{Soft photon removal lemma}

In this section, we prove \cref{lem-exact-soft-removal},
 which is a central technical ingredient in the proofs of mixed-state attainment, 
soft-photon purification, and infrared bounds on optimizers.

First, we split the photon space into soft- and hard-photon parts. For $\kappa>0$, let
\[
 \Hil_{\geq\kappa}=\iden_{\{\omega\geq\kappa\}}\Hil_\ph,\qquad
 \Hil_{<\kappa}=\iden_{\{\omega<\kappa\}}\Hil_\ph
\]
be the hard and soft one-photon spaces, so that
$\Hil_\ph=\Hil_{\geq\kappa}\oplus\Hil_{<\kappa}$, and let
$\Fock_{\geq\kappa}=\Fock_s(\Hil_{\geq\kappa})$ and
$\Fock_{<\kappa}=\Fock_s(\Hil_{<\kappa})$.  The exponential law for
Fock spaces gives the factorization
\[
 \HC_N\simeq\Hil_\el^N\otimes\Fock_{\geq\kappa}
                         \otimes\Fock_{<\kappa}.
\]
Accordingly, $\Af^\mu(\vx)=\Af^\mu_{\geq\kappa}(\vx)+\Af^\mu_{<\kappa}(\vx)$,
where $\Af^\mu_{\geq\kappa}(\vx)=\varPhi(\iden_{\{\omega\geq\kappa\}}G^\mu_{\vx})$
acts on $\Fock_{\geq\kappa}$ and
$\Af^\mu_{<\kappa}(\vx)=\varPhi(\iden_{\{\omega<\kappa\}}G^\mu_{\vx})$ acts on
$\Fock_{<\kappa}$.  Similarly, $H_\rad=H_{\rad,\geq\kappa}+H_{\rad,<\kappa}$
with
\[
 H_{\rad,\geq\kappa}=\dd\varGamma(\omega\iden_{\{\omega\geq\kappa\}}),\qquad
 H_{\rad,<\kappa}=\dd\varGamma(\omega\iden_{\{\omega<\kappa\}}).
\]
The \emph{hard model} is the internal Pauli--Fierz Hamiltonian
$H^{0,0}_{\geq\kappa}$ on $\Hil_\el^N\otimes\Fock_{\geq\kappa}$ with field
$\AF_{\geq\kappa}$ and radiation energy $H_{\rad,\geq\kappa}$.  The
restricted form factor still satisfies \cref{cutoff-assumption}, so the form
construction above applies to it.
The notation
$\Tr_{\el,<\kappa}\vGamma$ denotes the partial trace over the
electronic and soft-photon factors, leaving the reduced state on
$\Fock_{\geq\kappa}$.  Similarly, $\Tr_{<\kappa}$ denotes the partial
trace over $\Fock_{<\kappa}$ alone, leaving an operator on
$\Hil_\el^N\otimes\Fock_{\geq\kappa}$.  We write $\Omega_{<\kappa}$ for the normalized
vacuum in $\Fock_{<\kappa}$, so that
$|\Omega_{<\kappa}\rangle\langle\Omega_{<\kappa}|$ is the state with
no photons in the modes $\omega<\kappa$.

We are now in a position to state and prove the main result of this section.

\begin{lemma}[Exact soft-photon removal]
\label[lemma]{lem-exact-soft-removal}
Fix $N\in\NN$.  For every
$\vGamma\in\dm$ and $0<\kappa\leq1$,
there is a $\vGamma_\kappa\in\dm$ with
$\rho_{\vGamma_\kappa}=\rho_\vGamma$ and
$\vjc_{\vGamma_\kappa}=\vjc_\vGamma$, and with photon marginal
\begin{equation}\label{exact-soft-removal-marginal}
 \Tr_\el\vGamma_\kappa
 =\big(\Tr_{\el,<\kappa}\vGamma\big)
       \otimes|\Omega_{<\kappa}\rangle\langle\Omega_{<\kappa}|.
\end{equation}
Put
$E_< =\Tr(H_{\rad,<\kappa}\vGamma)$.  There exists a constant $C$,
independent of $\vGamma$ and $\kappa$ (but allowed to depend on $N$),
such that
\begin{equation}\label{exact-soft-removal-cost}
 \begin{split}
 \Tr(H^{0,0}\vGamma_\kappa)
 &\leq\Tr(H^{0,0}\vGamma)-E_<\\
 &+C\Big(\kappa^2+e^2b_\kappa^2
       +e\kappa\sqrt{a_\kappa^2E_<+b_\kappa^2}+e^2\kappa^2\big(a_\kappa^2E_<+b_\kappa^2\big)\Big).
 \end{split}
\end{equation}
In particular,
$\Tr(\NC\vGamma_\kappa)\leq\kappa^{-1}\Tr(H_\rad\vGamma)$.
\end{lemma}

The quantities $a_\kappa,b_\kappa$ defined in
\cref{infrared-kernel-errors} both tend to zero as $\kappa\downarrow0$.

\phantomsection
\label{proof:lem-exact-soft-removal}
\begin{proof}
Replacing the soft photons by the vacuum removes their radiation energy
$E_<$, but it also removes the contribution of the soft field to the
combined current density, so by itself it would violate the current constraint.
We compensate for this loss by converting the field-induced
current density carried by the soft photons into an equal paramagnetic current density.
The components of the soft field commute, so we can condition on their
values $u$.  For each $u$, the soft field becomes a classical vector
potential $\va_u$ that varies on the scale $\kappa^{-1}$.  After the
electrons are localized into clusters of diameter $O(R)$, the field
$\va_u$ is nearly constant on each cluster.  Its constant part is
removed exactly by a suitable gauge transformation in the electronic variables.
The small remainder is removed by the current
shift of \cref{lem-fermionic-current-repair}, at a kinetic cost
controlled by $\kappa R$ and the infrared quantities $a_\kappa,b_\kappa$.

\noindent\textbf{Step 1. Electronic localization.}
Fix a cluster scale $1\leq R\leq\kappa^{-1}$, to be chosen in Step~6.
Choose a locally finite smooth partition $(\eta_a)_{a\in\NN}$ on
$\RR^3$ with $\sum_a\eta_a^2=1$, $\supp\eta_a\subset B(c_a,c_0R)$, and
$\sum_a|\grad\eta_a|^2\leq C/R^2$.
For each multi-index $\alpha=(\alpha_1,\ldots,\alpha_N)\in\NN^N$ with
$\alpha_1\leq\cdots\leq\alpha_N$, let $\mathscr S_N\alpha$ be the set of
its distinct rearrangements $(\alpha_{\pi(1)},\ldots,\alpha_{\pi(N)})$,
$\pi\in\mathscr S_N$, and put
\[
 \zeta_\alpha(\vX)=
 \left(\sum_{(a_1,\ldots,a_N)\in\mathscr S_N\alpha}
                    \prod_{i=1}^N\eta_{a_i}(\vx_i)^2\right)^{1/2}.
\]
These are real symmetric $W^{1,\infty}$ multipliers.  Since $\NN^N$
is the disjoint union of the sets $\mathscr S_N\alpha$, expanding
$\prod_{i=1}^N\sum_a\eta_a(\vx_i)^2=1$ gives
$\sum_\alpha\zeta_\alpha^2=1$.  Moreover, $\zeta_\alpha$ is the Euclidean
norm of the vector
$\big(\prod_i\eta_{a_i}(\vx_i)\big)_{(a_1,\ldots,a_N)\in\mathscr S_N\alpha}$,
and the gradient of a norm is bounded by the norm of the gradient.
Hence
\[
 \sum_\alpha\sum_{i=1}^N|\grad_{\vx_i}\zeta_\alpha|^2
 \leq\sum_{i=1}^N\sum_{(a_1,\ldots,a_N)\in\NN^N}
   \Big|\grad_{\vx_i}\prod_{j=1}^N\eta_{a_j}(\vx_j)\Big|^2
 =\sum_{i=1}^N\sum_a|\grad\eta_a(\vx_i)|^2
 \leq\frac{C_N}{R^2}.
\]
Set
\[
 \vGamma_\alpha=\zeta_\alpha\vGamma\zeta_\alpha,\qquad
 E_\alpha=\Tr(H_{\rad,<\kappa}\vGamma_\alpha).
\]
All these operators act within the antisymmetric space.
Real multiplication and the covariant IMS identity give
\begin{align}
 \sum_\alpha(\rho_{\vGamma_\alpha},\vjc_{\vGamma_\alpha})
   &=(\rho_\vGamma,\vjc_\vGamma),\qquad
 \sum_\alpha\Tr\vGamma_\alpha=1,\qquad \sum_\alpha E_\alpha=E_<,\notag\\
 \sum_\alpha\Tr(H^{0,0}\vGamma_\alpha)&\leq\Tr(H^{0,0}\vGamma)+C/R^2.
 \label{current-preserving-electronic-IMS}
\end{align}
Here densities, current densities, and form traces are extended linearly
to positive trace-class operators of arbitrary finite trace.  Whenever
current densities are used, the operator is assumed to have finite
energy in the relevant model.

\noindent\textbf{Step 2. Conditioning on the soft field.}
We use the Gaussian representation and the unnormalized conditional operators
constructed in Appendix~\ref{app-soft-gaussian}, with the soft space
$\mathfrak h=\Hil_{<\kappa}$ and
$\mathcal K=\Hil_\el^N\otimes\Fock_{\geq\kappa}$.
Under $\iden\otimes\mathcal G$, the soft field $\AF_{<\kappa}(\vx)$
becomes multiplication by the real random field $\va_u(\vx)$ with
components
\[
 a_u^\mu(\vx)=\sum_{n\geq1}
 \dua{\xi_n}{\iden_{\{\omega<\kappa\}}G^\mu_{\vx}}\,u_n,
\]
where $(\xi_n)_{n\geq1}$ is a $J$-real orthonormal basis of
$\Hil_{<\kappa}$ and $u=(u_n)_{n\geq1}\in\mathcal Q=\RR^{\NN}$ carries
independent standard Gaussian coordinates.  Let $\Theta_\alpha(u)$ be
the family $\Theta(u)$ of \cref{lem-soft-disintegration} for
$\vGamma_\alpha$.  These are unnormalized conditional operators
for the electrons and hard photons at the soft-field value $u$.  They satisfy
\[
 \int_{\mathcal Q}\Tr\Theta_\alpha(u)\dd\mu(u)
 =\Tr\vGamma_\alpha\leq1,
\]
but their individual traces can exceed one.
More precisely, for every bounded
operator $B$ on $\Hil_\el^N\otimes\Fock_{\geq\kappa}$ and every bounded
measurable function $f$ on $\mathcal Q$,
\begin{equation}\label{conditioned-disintegration}
 \Tr\big(\vGamma_\alpha\,(B\otimes\mathcal G^{-1}M_f\mathcal G)\big)
 =\int_{\mathcal Q}f(u)\Tr\big(B\Theta_\alpha(u)\big)\dd\mu(u),
\end{equation}
where $M_f$ is multiplication by $f$ on $L^2(\mathcal Q,\mu)$.
Explicitly, if $(\Psi_j)_{j\geq1}$ is an orthonormal basis of $\HC_N$,
then
\[
 \Theta_\alpha(u)=\sum_{j\geq1}
 \big|\big((\iden\otimes\mathcal G)\sqrt{\vGamma_\alpha}\Psi_j\big)(u)\big\rangle
 \big\langle\big((\iden\otimes\mathcal G)\sqrt{\vGamma_\alpha}\Psi_j\big)(u)\big|,
\]
where $(\iden\otimes\mathcal G)\sqrt{\vGamma_\alpha}\Psi_j$ is regarded as
an element of $L^2(\mathcal Q,\mu;\Hil_\el^N\otimes\Fock_{\geq\kappa})$.
The series converges in trace norm for $\mu$-almost every $u$, and by
\cref{lem-soft-disintegration}(iii) the result does not depend on the
choice of the basis, up to a $\mu$-null set.

The electronic coordinates remain in the union of support balls
selected by $\zeta_\alpha$, without assigning fixed particle labels.
The unitary $\iden\otimes\mathcal G$ acts only on the soft factor.  It
therefore commutes with electronic differentiation and with hard-photon
operators, and it maps $\AF_{<\kappa}(\vx_i)$ to multiplication by
$\va_u(\vx_i)$.  Writing
\[
 \vD_{i,\geq\kappa}=-\i\grad_{\vx_i}-e\AF_{\geq\kappa}(\vx_i),\qquad
 \vD_i=\vD_{i,\geq\kappa}-e\AF_{<\kappa}(\vx_i),
\]
we obtain, for $\mu$-almost every $u$,
\[
 \big((\iden\otimes\mathcal G)\vD_i\Psi\big)(u)
 =\big(\vD_{i,\geq\kappa}-e\va_u(\vx_i)\big)
  \big((\iden\otimes\mathcal G)\Psi\big)(u),
\]
first for $\Psi$ in the finite-particle core and then, by the field
bounds, for every $\Psi$ in the form domain.  Apply this to
$\Psi=\sqrt{\vGamma_\alpha}\Psi_j$, where $(\Psi_j)_{j\geq1}$ is the
orthonormal basis used in the series representation of $\Theta_\alpha$
above.  For
$\Theta=\sum_j|v_j\rangle\langle v_j|$ one has
$\sum_j\|Xv_j\|^2=\|X\sqrt{\Theta}\|_{\mathfrak S_2}^2$, so Tonelli's
theorem shows that the kinetic energy of $\vGamma_\alpha$ is exactly
\begin{equation}\label{conditioned-soft-kinetic}
 \begin{aligned}
 \frac12\sum_{i=1}^N\|\vD_i\sqrt{\vGamma_\alpha}\|_{\mathfrak S_2}^2
 &=\frac12\sum_{i=1}^N\sum_{j\geq1}\|\vD_i\sqrt{\vGamma_\alpha}\Psi_j\|^2\\
 &=\frac12\int_{\mathcal Q}\sum_{i=1}^N
 \|(\vD_{i,\geq\kappa}-e\va_u(\vx_i))\sqrt{\Theta_\alpha(u)}\|_{\mathfrak S_2}^2
 \dd\mu(u).
 \end{aligned}
\end{equation}
We next derive the corresponding formula for the current density.
For a positive trace-class operator $\Theta$ on
$\Hil_\el^N\otimes\Fock_{\geq\kappa}$ with
$\Tr(H^{0,0}_{\geq\kappa}\Theta)<\infty$, let
$\vjc_{\Theta,\geq\kappa}$ denote the combined current density of $\Theta$ in the
hard model, defined by \cref{jc-def} with $\AF$ replaced by
$\AF_{\geq\kappa}$.  Fix $\vzeta\in C_c^\infty(\RR^3,\RR^3)$, let $\Psi$
be in the form domain, and write $\psi_u=((\iden\otimes\mathcal G)\Psi)(u)$
for its fibers.  Integrating \cref{jc-def} against $\vzeta$ turns the
partial inner product into a full one, and
$-\i\grad_{\vx_1}-e\AF(\vx_1)=\vD_{1,\geq\kappa}-e\AF_{<\kappa}(\vx_1)$.
Hence
\begin{align*}
 \int_{\RR^3}\vzeta(\vx)\cdot\vjc_\Psi(\vx)\dd\vx
 &=N\Re\dua{\Psi}{\vzeta(\vx_1)\cdot
     \big(\vD_{1,\geq\kappa}-e\AF_{<\kappa}(\vx_1)\big)\Psi}\\
 &=N\Re\int_{\mathcal Q}\dua{\psi_u}{\vzeta(\vx_1)\cdot
     \big(\vD_{1,\geq\kappa}-e\va_u(\vx_1)\big)\psi_u}\dd\mu(u)\\
 &=\int_{\mathcal Q}\int_{\RR^3}\vzeta(\vx)\cdot
     \big(\vjc_{\psi_u,\geq\kappa}(\vx)
          -e\va_u(\vx)\rho_{\psi_u}(\vx)\big)\dd\vx\,\dd\mu(u).
\end{align*}
The second line uses the unitarity of $\iden\otimes\mathcal G$, which
commutes with $\vzeta(\vx_1)$, and the fiberwise identity for $\vD_1$
above.  In the third line, the hard part is the hard-model current density of
$\psi_u$ tested against $\vzeta$.  For the soft part, $\va_u(\vx_1)$ is
real, so
$N\Re\dua{\psi_u}{\vzeta(\vx_1)\cdot\va_u(\vx_1)\psi_u}
=\int_{\RR^3}\vzeta\cdot\va_u\,\rho_{\psi_u}$ by \cref{density-def}.
Densities and current densities are linear in the state.  Apply the last
identity to $\Psi=\sqrt{\vGamma_\alpha}\Psi_j$ and sum over $j$.  By
\cref{A-Hrad-bound}, applied to $\AF_{\geq\kappa}$ and $\AF_{<\kappa}$,
the resulting integrand
$\vjc_{\Theta_\alpha(u),\geq\kappa}(\vx)
-e\rho_{\Theta_\alpha(u)}(\vx)\va_u(\vx)$ is integrable over
$\RR^3\times\mathcal Q$, which justifies exchanging the sum with the
integrals.  Since $\vzeta$ is arbitrary, Fubini's theorem and the
fundamental lemma of the calculus of variations give
\begin{equation}\label{conditioned-soft-current}
 \vjc_{\vGamma_\alpha}(\vx)
 =\int_{\mathcal Q}\big(\vjc_{\Theta_\alpha(u),\geq\kappa}(\vx)
        -e\rho_{\Theta_\alpha(u)}(\vx)\va_u(\vx)\big)\dd\mu(u)
 \qquad\text{for a.e. }\vx\in\RR^3.
\end{equation}
Since $\iden_{\{\omega<\kappa\}}G^\mu_{\vx}$ has norms at most those of
$G^\mu_{\vx}$, the estimates \cref{ann-Hrad-bound,cre-Hrad-bound} and
hence \cref{A-Hrad-bound} also hold for the removed field
$\AF_{<\kappa}(\vx)$.  As $\vD_{i,\geq\kappa}=\vD_i+e\AF_{<\kappa}(\vx_i)$
and $\Tr(H^{0,0}\vGamma_\alpha)<\infty$, this gives
\[
 \sum_{i=1}^N\|\vD_{i,\geq\kappa}\sqrt{\vGamma_\alpha}\|_{\mathfrak S_2}^2
 \leq2\sum_{i=1}^N\|\vD_i\sqrt{\vGamma_\alpha}\|_{\mathfrak S_2}^2
  +2e^2NC_\chi\big(\Tr(H_\rad\vGamma_\alpha)+\Tr\vGamma_\alpha\big)<\infty.
\]
The computation leading to \cref{conditioned-soft-kinetic}, with
$\vD_i$ replaced by $\vD_{i,\geq\kappa}$, and the fact that
$H_{\rad,\geq\kappa}$ acts only on $\Fock_{\geq\kappa}$ therefore give
\[
 \int_{\mathcal Q}\Big(\sum_{i=1}^N\|\vD_{i,\geq\kappa}\sqrt{\Theta_\alpha(u)}\|_{\mathfrak S_2}^2
       +\Tr(H_{\rad,\geq\kappa}\Theta_\alpha(u))\Big)\dd\mu(u)<\infty.
\]
The Coulomb interaction $W=\sum_{n<m}|\vx_n-\vx_m|^{-1}$ acts only on
the electronic variables and hence commutes with $\iden\otimes\mathcal G$.
The argument leading to \cref{conditioned-soft-kinetic} therefore gives
\[
 \Tr(W\vGamma_\alpha)=\int_{\mathcal Q}\Tr\big(W\Theta_\alpha(u)\big)\dd\mu(u).
\]
Together, these estimates show that
$\Tr(H^{0,0}_{\geq\kappa}\Theta_\alpha(u))<\infty$ for almost every $u$.
For $\Tr\Theta_\alpha(u)>0$, the density matrix
$\Theta_\alpha(u)/\Tr\Theta_\alpha(u)$ has finite hard-model energy.
The operations applied to $\Theta_\alpha(u)$ in Step 4, namely
multiplication by a phase of modulus one and the current shift of
\cref{lem-fermionic-current-repair}, leave $\Tr(W\Theta_\alpha(u))$
unchanged.  Hence only the kinetic and radiation energies change in the
remaining steps.

\noindent\textbf{Step 3. Local bounds on the measured field.}
Fix a multi-index $\beta$ and $\vx\in\RR^3$.  The differentiated soft
field $\partial_{\vx}^\beta\Af^\mu_{<\kappa}(\vx)$ is the Segal field with
kernel
$\partial_{\vx}^\beta(\iden_{\{\omega<\kappa\}}G^\mu_{\vx})
=(-\i\vk)^\beta\iden_{\{\omega<\kappa\}}G^\mu_{\vx}$.  Since $|\vk|<\kappa$
on the soft region, the projector identity behind \cref{Gx-norms} gives
\[
 \sum_{\mu=1}^3\Big\|\frac{(-\i\vk)^\beta\iden_{\{\omega<\kappa\}}
   G^\mu_{\vx}}{\sqrt\omega}\Big\|^2
 \leq\kappa^{2|\beta|}a_\kappa^2,\qquad
 \sum_{\mu=1}^3\big\|(-\i\vk)^\beta\iden_{\{\omega<\kappa\}}
   G^\mu_{\vx}\big\|^2
 \leq\kappa^{2|\beta|}b_\kappa^2,
\]
with $a_\kappa,b_\kappa$ as in \cref{infrared-kernel-errors}.  These
kernels are soft, that is, they vanish for $\omega\geq\kappa$, so
\cref{ann-Hrad-bound,cre-Hrad-bound} hold with
$H_{\rad,<\kappa}$ in place of $H_\rad$ and give
\[
 \sum_{\mu=1}^3\|\partial_{\vx}^\beta\Af^\mu_{<\kappa}(\vx)\Psi\|^2
 \leq C\kappa^{2|\beta|}
 \big(a_\kappa^2\|\sqrt{H_{\rad,<\kappa}}\Psi\|^2+b_\kappa^2\|\Psi\|^2\big).
\]
Apply this to $\Psi=\sqrt{\vGamma_\alpha}\Psi_j$ and sum over $j$.
Then use \cref{conditioned-disintegration} with $B=\iden$ and the unbounded function
$f(u)=|\partial^\beta\va_u(\vx)|^2$, to which the
identity extends by monotone convergence.  As
$E_\alpha=\Tr(H_{\rad,<\kappa}\vGamma_\alpha)$, we obtain
\begin{equation}\label{conditioned-soft-field-moments}
 \int_{\mathcal Q} \Tr\Theta_\alpha(u)|\partial^\beta\va_u(\vx)|^2\dd\mu(u)
 \leq C_\beta\kappa^{2|\beta|}\big(a_\kappa^2E_\alpha+b_\kappa^2\Tr\vGamma_\alpha\big),
\end{equation}
uniformly in $\vx$ and $\alpha$.  The fields have smooth versions
locally: this follows from these Gaussian derivative moments and local
Sobolev embedding, and remains true under the measures
$\Tr\Theta_\alpha(u)\dd\mu(u)$.

For a fixed $\alpha$, every electron in $\vGamma_\alpha$ lies in one of
the balls $B(c_{\alpha_i},c_0R)$.  Consider the graph whose vertices are
the distinct centers among $c_{\alpha_1},\ldots,c_{\alpha_N}$, two centers
being joined by an edge if their distance is at most $4c_0R$.  We call
the vertex sets of its connected components the \emph{clusters} of
$\alpha$ and index them by $s$.  A cluster has at most $N$ centers, so
its diameter is at most $4(N-1)c_0R\leq C_NR$, while centers in
different clusters are more than $4c_0R$ apart.  The measured soft field
varies on the scale $\kappa^{-1}\geq R$, so in Step 4 it will be
replaced on each cluster by a constant.  Choose one center of each
cluster as its anchor $c_{\alpha,s}$.  The construction is illustrated in
\cref{fig:soft-clusters}.
Let $\theta_a=1$ on a neighborhood of $B(c_a,c_0R)$, with
$\supp\theta_a\subset B(c_a,3c_0R/2)$ and
$|\grad\theta_a|\leq C/R$, and define the cluster cutoff
\[
 \theta_{\alpha,s}=1-\prod_{a\text{ in cluster }s}(1-\theta_a).
\]
Then
\begin{gather*}
 \theta_{\alpha,s}=1\ \text{ on }\bigcup_{a\text{ in cluster }s}B(c_a,c_0R),
 \qquad
 \supp\theta_{\alpha,s}\subset
 \bigcup_{a\text{ in cluster }s}B(c_a,3c_0R/2),\\
 |\grad\theta_{\alpha,s}|\leq\frac{C_N}{R}.
\end{gather*}
Since centers in different clusters are more than $4c_0R$ apart, the
supports of $\theta_{\alpha,s}$ for different $s$ are disjoint.
Let $\mathcal B_{\alpha,s}$ be a ball centered at $c_{\alpha,s}$ of radius $C_NR$
containing $\supp\theta_{\alpha,s}$.
All constants are independent of the separation between clusters.
Put $Z_\alpha(u)=\max_s\sup_{\mathcal B_{\alpha,s}}|\grad\va_u|$.
For a ball $B\subset\RR^3$ of radius $r$ and $f\in H^2(B)$, the
Sobolev inequality
\[
 \sup_B|f|^2\leq C\sum_{|\beta|\leq2}r^{2|\beta|-3}
 \int_B|\partial^\beta f|^2\dd\vx
\]
holds with $C$ independent of $r$.
Apply it with $r=C_NR$ to the components of $\grad\va_u$ on each
$\mathcal B_{\alpha,s}$, multiply by $\Tr\Theta_\alpha(u)$, and integrate
over $u$.  By \cref{conditioned-soft-field-moments} for multi-indices of
order $|\beta|+1$, and since $|\mathcal B_{\alpha,s}|\leq CR^3$, this
yields
\begin{align*}
 \int_{\mathcal Q}\Tr\Theta_\alpha(u)Z_\alpha(u)^2\dd\mu(u)
 &\leq C\sum_s\sum_{|\beta|\leq2}R^{2|\beta|-3}
       \int_{\mathcal B_{\alpha,s}}\int_{\mathcal Q}\Tr\Theta_\alpha(u)
          |\partial^\beta\grad\va_u|^2\dd\mu(u)\dd\vx\\
 &\leq C\kappa^2\big(a_\kappa^2E_\alpha+b_\kappa^2\Tr\vGamma_\alpha\big),
\end{align*}
because $\kappa R\leq1$ and there are at most $N$ clusters
(the constant may depend on $N$).
No global supremum of the random field is used.

\begin{figure}[!htbp]
\centering
\begin{tikzpicture}[
 x=.82cm,
 y=.82cm,
 line cap=round,
 line join=round
]
 \fill[black!7]
   (1.4,2.2) circle (.75) (3.0,2.9) circle (.75) (3.2,1.2) circle (.75)
   (9.2,2.4) circle (.75) (10.6,1.6) circle (.75)
   (6.4,4.7) circle (.75);
 \foreach \c in {(1.4,2.2),(3.0,2.9),(3.2,1.2),(9.2,2.4),(10.6,1.6),(6.4,4.7)}{
   \filldraw[fill=black!18,draw=black!55,line width=.4pt] \c circle (.5);
 }

 \draw[black!80,dashed,line width=.6pt] (3.0,2.9) circle (2.55);
 \draw[black!80,dashed,line width=.6pt] (9.2,2.4) circle (2.4);
 \draw[black!80,dashed,line width=.6pt] (6.4,4.7) circle (1.0);

 \draw[black,line width=.9pt] (1.4,2.2) -- (3.0,2.9) -- (3.2,1.2);
 \draw[black,line width=.9pt] (9.2,2.4) -- (10.6,1.6);

 \foreach \c in {(1.4,2.2),(3.0,2.9),(3.2,1.2),(9.2,2.4),(10.6,1.6),(6.4,4.7)}{
   \fill[black] \c circle (1.5pt);
 }
 \foreach \c in {(3.0,2.9),(9.2,2.4),(6.4,4.7)}{
   \draw[black,line width=.8pt] \c circle (3.6pt);
 }

 \foreach \x/\y in {1.2/2.38,1.62/2.04,3.32/2.72,2.98/1.0,
                    9.02/2.72,10.82/1.42,6.68/4.46}{
   \filldraw[fill=white,draw=black,line width=.6pt] (\x,\y) circle (.13);
   \node[font=\bfseries\tiny,inner sep=0pt] at (\x,\y) {$-$};
 }
\end{tikzpicture}
\caption{The clusters of a multi-index $\alpha$ ($N=7$,
$\alpha_1=\alpha_2$).  Centers at distance at most $4c_0R$ are joined.
Anchors are ringed and the balls $\mathcal B_{\alpha,s}$ are dashed.}
\label{fig:soft-clusters}
\end{figure}
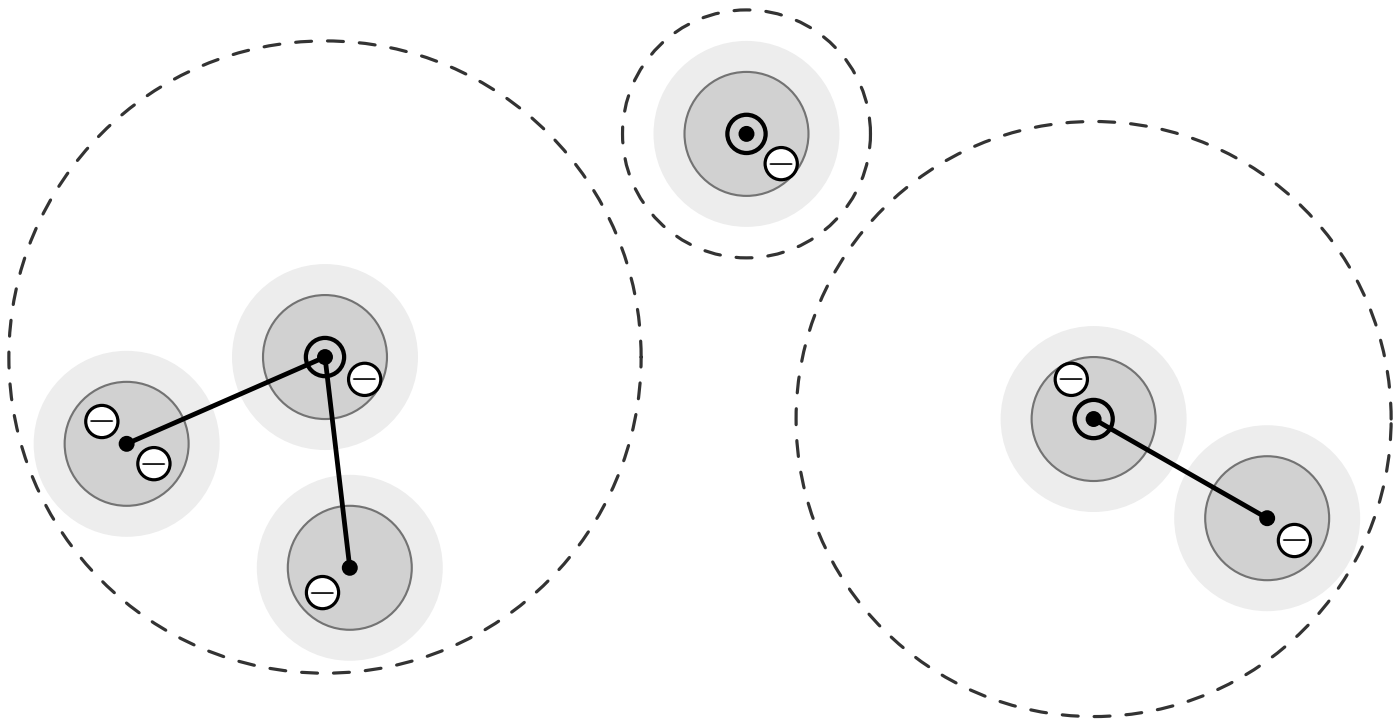

\noindent\textbf{Step 4. Current shift on each conditional state.}
For almost every $u$, define the one-body phase and residual velocity
\[
 \begin{split}
 \varphi_{\alpha,u}(\vx)
 &=-e\sum_s\theta_{\alpha,s}(\vx)\va_u(c_{\alpha,s})\cdot(\vx-c_{\alpha,s}),\\
 \vv_{\alpha,u}(\vx)
 &=-e\sum_s\theta_{\alpha,s}(\vx)
                \big(\va_u(\vx)-\va_u(c_{\alpha,s})\big).
 \end{split}
\]
In the electronic variables, let $U_{\alpha,u}$ be multiplication by
the permutation-symmetric phase
\[
 \exp\Big(\i\sum_{i=1}^N\varphi_{\alpha,u}(\vx_i)\Big).
\]
First replace $\Theta_\alpha(u)$ by
$U_{\alpha,u}\Theta_\alpha(u)U_{\alpha,u}^\dag$.  This leaves the density
and the photon marginal unchanged and adds
$\rho_{\Theta_\alpha(u)}\grad\varphi_{\alpha,u}$ to the current density.  Where
$\theta_{\alpha,s}=1$, the velocity
$\grad\varphi_{\alpha,u}=-e\va_u(c_{\alpha,s})$ is constant, and
$\grad\varphi_{\alpha,u}+\vv_{\alpha,u}=-e\va_u$.  Next apply
\cref{lem-fermionic-current-repair} with velocity $\vv_{\alpha,u}$ to the
hard model $H^{0,0}_{\geq\kappa}$ instead of $H^{0,0}$.  Its proof applies
verbatim, since the multipliers act only on the electronic variables.
If $\Tr\Theta_\alpha(u)=0$, set $\widetilde\Theta_\alpha(u)=0$.
Positivity gives $\Theta_\alpha(u)=0$ in this case.
If $\Tr\Theta_\alpha(u)>0$, apply it to the normalized state
$U_{\alpha,u}\Theta_\alpha(u)U_{\alpha,u}^\dag/\Tr\Theta_\alpha(u)$,
multiply the result by $\Tr\Theta_\alpha(u)$, and call the resulting
operator $\widetilde\Theta_\alpha(u)$.  Its current density is
$\vjc_{\Theta_\alpha(u),\geq\kappa}-e\rho_{\Theta_\alpha(u)}\va_u$, the
integrand in \cref{conditioned-soft-current}, while
\[
 \|\vv_{\alpha,u}\|_\infty\leq CeR Z_\alpha(u),\qquad
 \|\grad\vv_{\alpha,u}\|_\infty\leq CeZ_\alpha(u).
\]
The wavenumber chosen in the proof of
\cref{lem-fermionic-current-repair} depends measurably on
$\vv_{\alpha,u}$.  When the velocity vanishes, use the identity channel.
The tensor lift is fixed independently of the velocity, so this
defines a measurable family of channels.  The kinetic remainder
on an unnormalized conditional operator is bounded
by
\[
C\Tr\Theta_\alpha(u)(\|\vv_{\alpha,u}\|_\infty^2+\|\grad\vv_{\alpha,u}\|_\infty).
\]

We now compare the hard-model kinetic energy of the output with the
integrand of \cref{conditioned-soft-kinetic}, that is, we estimate
\[
 \delta_\alpha(u)
 =\frac12\sum_{i=1}^N
   \big\|\vD_{i,\geq\kappa}\sqrt{\widetilde\Theta_\alpha(u)}\big\|_{\mathfrak S_2}^2
 -\frac12\sum_{i=1}^N
   \big\|(\vD_{i,\geq\kappa}-e\va_u(\vx_i))\sqrt{\Theta_\alpha(u)}\big\|_{\mathfrak S_2}^2.
\]
The phase is exact: since
$\vD_{i,\geq\kappa}U_{\alpha,u}
=U_{\alpha,u}(\vD_{i,\geq\kappa}+\grad\varphi_{\alpha,u}(\vx_i))$, it
replaces $\vD_{i,\geq\kappa}$ by $\vD_{i,\geq\kappa}+\grad\varphi_{\alpha,u}(\vx_i)$.
By \cref{fermionic-repair-cost}, the current shift then adds
\[
 \int_{\RR^3}\vjc_{U_{\alpha,u}\Theta_\alpha(u)U_{\alpha,u}^\dag,\geq\kappa}
 \cdot\vv_{\alpha,u},
\]
the hard-model current density of $U_{\alpha,u}\Theta_\alpha(u)U_{\alpha,u}^\dag$
paired with $\vv_{\alpha,u}$, plus the kinetic remainder.
For the square completion, abbreviate
$\Theta=\Theta_\alpha(u)$, $\vg=\grad\varphi_{\alpha,u}$,
$\vv=\vv_{\alpha,u}$, and $\vD_i^{\mathrm h}=\vD_{i,\geq\kappa}$.
The phase-transformed current is
$\vjc_{\Theta,\geq\kappa}+\rho_\Theta\vg$, so
\[
 \begin{aligned}
 &\frac12\sum_{i=1}^N
   \big\|(\vD_i^{\mathrm h}+\vg(\vx_i))\sqrt\Theta\big\|_{\mathfrak S_2}^2
   +\int_{\RR^3}(\vjc_{\Theta,\geq\kappa}+\rho_\Theta\vg)\cdot\vv\\
 &\quad=\frac12\sum_{i=1}^N
   \big\|(\vD_i^{\mathrm h}+\vg(\vx_i)+\vv(\vx_i))\sqrt\Theta\big\|_{\mathfrak S_2}^2
   -\frac12\int_{\RR^3}\rho_\Theta|\vv|^2\\
 &\quad=\frac12\sum_{i=1}^N
   \big\|(\vD_i^{\mathrm h}-e\va_u(\vx_i))\sqrt\Theta\big\|_{\mathfrak S_2}^2
   -\frac12\int_{\RR^3}\rho_\Theta|\vv|^2.
 \end{aligned}
\]
The last equality uses $\vg+\vv=-e\va_u$ on the electronic support.
Dropping this nonpositive term and using the bounds on $\vv_{\alpha,u}$,
\[
 \delta_\alpha(u)
 \leq C\Tr\Theta_\alpha(u)\big(\|\vv_{\alpha,u}\|_\infty^2
      +\|\grad\vv_{\alpha,u}\|_\infty\big)
 \leq C\Tr\Theta_\alpha(u)\big(e^2R^2Z_\alpha(u)^2+eZ_\alpha(u)\big).
\]
Integrate over $u$.  The first term is controlled by the Step 3 bound
on the integral of $\Tr\Theta_\alpha(u)Z_\alpha(u)^2$.  For the second,
the Cauchy--Schwarz inequality, together with
$\int_{\mathcal Q}\Tr\Theta_\alpha\dd\mu=\Tr\vGamma_\alpha$, gives the
same bound under a square root.  Hence
\[
 \int_{\mathcal Q}\delta_\alpha(u)\dd\mu(u)
 \leq Ce^2\kappa^2R^2\big(a_\kappa^2E_\alpha+b_\kappa^2\Tr\vGamma_\alpha\big)
 +Ce\kappa\sqrt{\big(a_\kappa^2E_\alpha+b_\kappa^2\Tr\vGamma_\alpha\big)
                \Tr\vGamma_\alpha}.
\]
Thus constant measured fields on each cluster are compensated without
a kinetic remainder depending on their magnitude.

\noindent\textbf{Step 5. Assembly.}
Using $\HC_N\simeq\Hil_\el^N\otimes\Fock_{\geq\kappa}\otimes\Fock_{<\kappa}$,
define
\[
 \vGamma_{\kappa,R}
 =\sum_\alpha\int_{\mathcal Q}\widetilde\Theta_\alpha(u)\otimes
   |\Omega_{<\kappa}\rangle\langle\Omega_{<\kappa}|\dd\mu(u).
\]
The map $u\mapsto\widetilde\Theta_\alpha(u)$ is measurable by Step 4, and
the integral converges in trace norm because the conditional operations
preserve traces, so that
$\Tr\widetilde\Theta_\alpha(u)=\Tr\Theta_\alpha(u)$.  They also preserve
photon marginals.  By \cref{conditioned-disintegration} with $f=1$,
$\int_{\mathcal Q}\Theta_\alpha\dd\mu=\Tr_{<\kappa}\vGamma_\alpha$, and
$\vGamma_{\kappa,R}$ is therefore a positive trace-class operator with
\[
 \Tr\vGamma_{\kappa,R}
 =\sum_\alpha\Tr\big(\Tr_{<\kappa}\vGamma_\alpha\big)
 =\sum_\alpha\Tr\vGamma_\alpha=1,
\]
and its photon marginal is
\[
 \sum_\alpha\Tr_\el\big(\Tr_{<\kappa}\vGamma_\alpha\big)\otimes
 |\Omega_{<\kappa}\rangle\langle\Omega_{<\kappa}|
 =\big(\Tr_{\el,<\kappa}\vGamma\big)\otimes
 |\Omega_{<\kappa}\rangle\langle\Omega_{<\kappa}|,
\]
since $\sum_\alpha\zeta_\alpha^2=1$.  This is \cref{exact-soft-removal-marginal}.
Preparing the soft vacuum adds only its covariant-kinetic
contribution $e^2N b_\kappa^2/2$, and the soft radiation energy $E_<$
is removed.
For the constraints, let $\Theta$ be a positive trace-class operator on
$\Hil_\el^N\otimes\Fock_{\geq\kappa}$ with finite hard-model energy,
$\Tr(H^{0,0}_{\geq\kappa}\Theta)<\infty$.  Since
$\dua{\Omega_{<\kappa}}{\Af^\mu_{<\kappa}(\vx)\Omega_{<\kappa}}=0$,
the definitions \cref{density-def,jc-def} give
\[
 \rho_{\Theta\otimes|\Omega_{<\kappa}\rangle\langle\Omega_{<\kappa}|}
 =\rho_\Theta,\qquad
 \vjc_{\Theta\otimes|\Omega_{<\kappa}\rangle\langle\Omega_{<\kappa}|}
 =\vjc_{\Theta,\geq\kappa}.
\]
By linearity in the state, Step 4, \cref{conditioned-soft-current}, and
\cref{current-preserving-electronic-IMS}, for almost every $\vx$,
\begin{align*}
 \vjc_{\vGamma_{\kappa,R}}(\vx)
 &=\sum_\alpha\int_{\mathcal Q}
   \vjc_{\widetilde\Theta_\alpha(u),\geq\kappa}(\vx)\dd\mu(u)\\
 &=\sum_\alpha\int_{\mathcal Q}
   \big(\vjc_{\Theta_\alpha(u),\geq\kappa}(\vx)
        -e\rho_{\Theta_\alpha(u)}(\vx)\va_u(\vx)\big)\dd\mu(u)\\
 &=\sum_\alpha\vjc_{\vGamma_\alpha}(\vx)
  =\vjc_\vGamma(\vx).
\end{align*}
The phase and the current shift leave the density unchanged, and the
density depends only on the electronic reduced state.  Hence
\cref{conditioned-disintegration} with $f=1$ and
\cref{current-preserving-electronic-IMS} give
\[
 \rho_{\vGamma_{\kappa,R}}
 =\sum_\alpha\int_{\mathcal Q}\rho_{\Theta_\alpha(u)}\dd\mu(u)
 =\sum_\alpha\rho_{\vGamma_\alpha}
 =\rho_\vGamma.
\]
Since
$\sum_\alpha(a_\kappa^2E_\alpha+b_\kappa^2\Tr\vGamma_\alpha)
=a_\kappa^2E_<+b_\kappa^2$ and, by the Cauchy--Schwarz inequality,
\[
 \sum_\alpha\sqrt{\big(a_\kappa^2E_\alpha+b_\kappa^2\Tr\vGamma_\alpha\big)
                  \Tr\vGamma_\alpha}
 \leq\sqrt{a_\kappa^2E_<+b_\kappa^2},
\]
we obtain
\begin{equation}\label{exact-soft-removal-cost-R}
 \begin{split}
 \Tr(H^{0,0}\vGamma_{\kappa,R})
 &\leq\Tr(H^{0,0}\vGamma)-E_<
 +C\Big(R^{-2}+e^2b_\kappa^2
       +e^2\kappa^2R^2\big(a_\kappa^2E_<+b_\kappa^2\big)\\
 &\qquad+e\kappa\sqrt{a_\kappa^2E_<+b_\kappa^2}\Big),
 \end{split}
\end{equation}
with $C$ independent of $\vGamma$, $\kappa$ and $R$.
It also proves finite form energy of the output.
Only the output is asserted to have finite full form energy, so no
soft-radiation regularity of an intermediate measured state is needed.
The marginal identity proves the photon-number bound.

\noindent\textbf{Step 6. Choice of the cluster scale.}
Only the terms $R^{-2}$ and
$e^2\kappa^2R^2(a_\kappa^2E_<+b_\kappa^2)$ in
\cref{exact-soft-removal-cost-R} depend on $R$.
Put $q=e\kappa\sqrt{a_\kappa^2E_<+b_\kappa^2}\geq0$.
If $\kappa^2\leq q\leq1$, then $q>0$, and we choose
$R=q^{-1/2}\in[1,\kappa^{-1}]$.  The two terms then balance and
both equal $q$.  If $q<\kappa^2$, including $q=0$, take
$R=\kappa^{-1}$.  Then $e^2(a_\kappa^2E_<+b_\kappa^2)<\kappa^2$, so the
two terms add up to less than $2\kappa^2$.  If $q>1$, take $R=1$.
Then the two terms add up to less than
$2q^2=2e^2\kappa^2(a_\kappa^2E_<+b_\kappa^2)$.
In each case, $\vGamma_\kappa=\vGamma_{\kappa,R}$ satisfies
\cref{exact-soft-removal-cost} after changing $C$.  The constant $C$ in
\cref{exact-soft-removal-cost-R} does not depend on $\vGamma$, so $R$
may depend on $E_<$.
\end{proof}

\subsection{The proof of \texorpdfstring{\cref{thm-unrestricted-attainment}}{Theorem \getrefnumber{thm-unrestricted-attainment}}}
\label{proof:thm-unrestricted-attainment}
We remove the soft photons of near-minimizers using
\cref{lem-exact-soft-removal}, which keeps $\rho$ and $\vjc$ fixed at
a small energy cost and bounds the mean photon number.
\Cref{thm-continuum-compactness} then gives a minimizer and lower
semicontinuity, and convex conjugation identifies the functional with
$F_\LI^{e,\chi}$.

\noindent\textbf{Step 1. Soft energy of near-minimizers.}
Apply \cref{lem-exact-soft-removal}, with
$H_{\rad,<\kappa}$ as defined above.
The resulting state $\vGamma_\kappa$ replaces photons
with $\omega<\kappa$ by the vacuum and adjusts the electronic current density
so that its density and combined current density agree exactly with those of
$\vGamma$.
The estimate \cref{exact-soft-removal-cost}, together with
$\sqrt{s+t}\leq\sqrt s+\sqrt t$ and $\kappa\leq1$, gives
\begin{align*}
 &\Tr(H^{0,0}\vGamma_\kappa)
 -\Tr(H^{0,0}\vGamma)+\Tr(H_{\rad,<\kappa}\vGamma)\\
 &\quad\leq C\Big[\kappa^2+e^2b_\kappa^2+
       e^2\kappa^2a_\kappa^2\Tr(H_{\rad,<\kappa}\vGamma)
       +e\kappa a_\kappa
          \sqrt{\Tr(H_{\rad,<\kappa}\vGamma)}
       +e\kappa b_\kappa\Big].
\end{align*}
Since $\kappa a_\kappa\to0$, choose $\kappa_0\leq1$, depending only on
$N,e$ and $\chi$, so the term proportional to
$\kappa^2a_\kappa^2\Tr(H_{\rad,<\kappa}\vGamma)$ can be absorbed by the term
$\Tr(H_{\rad,<\kappa}\vGamma)$ on the left-hand side for
$0<\kappa\leq\kappa_0$.
Young's inequality absorbs the square-root term as well, giving
\begin{equation}\label{unrestricted-soft-energy-gain}
 \Tr(H^{0,0}\vGamma_\kappa)
 \leq\Tr(H^{0,0}\vGamma)
       -\tfrac12\Tr(H_{\rad,<\kappa}\vGamma)
                   +C_{N,e,\chi}(\kappa^2+b_\kappa^2).
\end{equation}
For an exact feasible $\vGamma$, set
$d_\vGamma=\Tr(H^{0,0}\vGamma)
-\wt{F}_\LI^{e,\chi}(\rho,\vjc)$.
The competitor $\vGamma_\kappa$ has exactly the same constraints.  Consequently
\begin{equation}\label{unrestricted-uniform-soft-energy}
 \Tr(H_{\rad,<\kappa}\vGamma)\leq
 2d_\vGamma+C_{N,e,\chi}(\kappa^2+b_\kappa^2),
 \qquad 0<\kappa\leq\kappa_0.
\end{equation}

\noindent\textbf{Step 2. A uniformly number-bounded minimizing sequence.}
For $0<\kappa<\kappa_0$, the pointwise inequality
\[
 \iden_{\{\omega\geq\kappa\}}
 \leq\frac{\omega}{\kappa_0}
       +\omega\int_\kappa^{\kappa_0}
                    \frac{\iden_{\{\omega<t\}}}{t^2}\dd t
\]
and \cref{exact-soft-removal-marginal} imply
\[
 \Tr(\NC\vGamma_\kappa)
 \leq\frac{\Tr(H^{0,0}\vGamma)}{\kappa_0}
             +\int_\kappa^{\kappa_0}
                 \frac{\Tr(H_{\rad,<t}\vGamma)}{t^2}\dd t.
\]
The weighted condition $\wh\chi/\sqrt\omega\in L^2$ in
\cref{cutoff-assumption} gives, by Tonelli's theorem,
\[
 \int_0^{\kappa_0}\frac{b_t^2}{t^2}\dd t
 =\int_{\omega<\kappa_0}|\wh\chi(\vk)|^2
       \left(\frac1{\omega(\vk)}-\frac1{\kappa_0}\right)\dd\vk
 \leq a_{\kappa_0}^2<\infty.
\]
Thus \cref{unrestricted-uniform-soft-energy} yields
\begin{equation}\label{unrestricted-near-minimizer-number}
 \Tr(\NC\vGamma_\kappa)
 \leq \frac{\Tr(H^{0,0}\vGamma)}{\kappa_0}
              +\frac{2d_\vGamma}{\kappa}+C_{N,e,\chi}.
\end{equation}

Now take any minimizing sequence $\vGamma_n\in\dm$ with
$(\rho_{\vGamma_n},\vjc_{\vGamma_n})=(\rho,\vjc)$, with excess energies $d_n\to0$, and choose $\kappa_n\to0$ with
$d_n/\kappa_n\to0$, for example
$\kappa_n=\max\{\sqrt{d_n},1/n\}$ for all sufficiently large $n$.
Set $\widetilde\vGamma_n=(\vGamma_n)_{\kappa_n}$, using the output of
\cref{lem-exact-soft-removal} with the cluster scale chosen in Step~6
of its proof.
These states satisfy
$(\rho_{\widetilde\vGamma_n},\vjc_{\widetilde\vGamma_n})=(\rho,\vjc)$
by \cref{lem-exact-soft-removal}, so
$\Tr(H^{0,0}\widetilde\vGamma_n)\geq\wt{F}_\LI^{e,\chi}(\rho,\vjc)$.
For $n$ large enough that $\kappa_n\leq\kappa_0$, dropping the
nonpositive term in \cref{unrestricted-soft-energy-gain} gives
\[
 \Tr(H^{0,0}\widetilde\vGamma_n)
 \leq\wt{F}_\LI^{e,\chi}(\rho,\vjc)+d_n
   +C_{N,e,\chi}\big(\kappa_n^2+b_{\kappa_n}^2\big).
\]
Since $d_n\to0$, $\kappa_n\to0$, and $b_\kappa\to0$ as
$\kappa\downarrow0$, their energies converge to
$\wt{F}_\LI^{e,\chi}(\rho,\vjc)$.  Moreover, since $d_n/\kappa_n\to0$
and the energies $\Tr(H^{0,0}\vGamma_n)$ are bounded,
\cref{unrestricted-near-minimizer-number} gives a uniform bound on
the mean photon numbers of $\widetilde\vGamma_n$.
The fixed density supplies electronic tightness.
Hence \cref{thm-continuum-compactness} applies and produces a minimizer
$\vGamma\in\dm$ with $(\rho_\vGamma,\vjc_\vGamma)=(\rho,\vjc)$.

Now let $\vGamma$ be any minimizer at $(\rho,\vjc)$, not only the one
just constructed.  Then $d_\vGamma=0$.
Using the same pointwise inequality and then letting $\kappa\downarrow0$
in \cref{unrestricted-uniform-soft-energy} gives
\begin{equation}\label{unrestricted-uniform-number}
 \Tr(\NC\vGamma)\leq C_{N,e,\chi}
       \big(1+\wt{F}_\LI^{e,\chi}(\rho,\vjc)\big).
\end{equation}
In particular, every optimizer has finite mean photon number.

\noindent\textbf{Step 3. Lower semicontinuity and identification of the functional.}
Take pairs $(\rho_k,\vjc_k)$ converging in $X\times Y$ along a
subsequence on which $\wt{F}_\LI^{e,\chi}(\rho_k,\vjc_k)$ has finite
liminf, and choose the minimizers just constructed.  Their energies and mean photon numbers are uniformly
bounded by \cref{unrestricted-uniform-number}.  Convergence of their
densities in $L^1$ gives uniform electronic tightness.
A further application of \cref{thm-continuum-compactness} preserves
the limiting pair and proves the energy liminf inequality.
Finally, we identify the functional.  Fix $(u,\vA)\in X^*\times Y^*$
and put $v=u-\tfrac12|\vA|^2$.  By linearity in the state,
\cref{energy-decomposition} extends to density matrices:
\[
 \Tr(H^{v,\vA}\vGamma)=\Tr(H^{0,0}\vGamma)+\dua{u}{\rho_\vGamma}
 +\dua{\vA}{\vjc_\vGamma},\qquad\vGamma\in\dm.
\]
Minimizing first over $\vGamma$ at fixed $(\rho,\vjc)$ and then over
$(\rho,\vjc)$ gives
\[
 \inf_{(\rho,\vjc)\in X\times Y}
 \Big[\wt{F}_\LI^{e,\chi}(\rho,\vjc)+\dua{u}{\rho}+\dua{\vA}{\vjc}\Big]
 =\inf_{\vGamma\in\dm}\Tr(H^{v,\vA}\vGamma)
 =\bar E(u,\vA).
\]
The last equality holds because density matrices and pure states have
the same spectral infimum.  Indeed, pure states are density matrices,
and for a spectral decomposition
$\vGamma=\sum_rp_r|\Psi_r\rangle\langle\Psi_r|$,
\[
 \Tr(H^{v,\vA}\vGamma)=\sum_rp_r\dua{\Psi_r}{H^{v,\vA}\Psi_r}
 \geq E(v,\vA).
\]
Let
$(\wt{F}_\LI^{e,\chi})^*$ denote the convex conjugate with respect to
the pairing of $X\times Y$ with $X^*\times Y^*$.  The identity for the
infimum over $X\times Y$ states that $(\wt{F}_\LI^{e,\chi})^*(-u,-\vA)=-\bar E(u,\vA)$, so the
definition \cref{FL-def} reads
$F_\LI^{e,\chi}=(\wt{F}_\LI^{e,\chi})^{**}$.  The functional
$\wt{F}_\LI^{e,\chi}$ is proper, since it is nonnegative and finite at
every $N$-representable pair.  It is convex, since $\dm$ is convex and
$\rho_\vGamma$, $\vjc_\vGamma$, and $\Tr(H^{0,0}\vGamma)$ are linear
in $\vGamma$.  The Fenchel--Moreau theorem therefore gives, for any
$N$,
\begin{equation*}
 F_\LI^{e,\chi}
 =\overline{\wt{F}_\LI^{e,\chi}}^{\,\|\cdot\|_{X\times Y}},
\end{equation*}
the norm-lower-semicontinuous hull of $\wt{F}_\LI^{e,\chi}$; for convex
functions it coincides with the weakly lower semicontinuous hull.
Since $\wt{F}_\LI^{e,\chi}$ is norm lower semicontinuous by the first
part of this step, the hull equals $\wt{F}_\LI^{e,\chi}$, which
establishes \cref{unrestricted-Lieb-equality}.

\subsection{Soft-photon purification}

\phantomsection
\label{proof:thm-many-electron-soft-purification}
\begin{proof}[Proof of \cref{thm-many-electron-soft-purification}]
Fix $\vGamma\in\dm$ and $0<\kappa\leq1$.  Apply
\cref{lem-exact-soft-removal} and let $\vGamma_\kappa$ be its output.
This state has exactly the same density and combined current density as $\vGamma$.  Since
$a_\kappa,b_\kappa\to0$ and $E_<\leq\Tr(H^{0,0}\vGamma)$,
\cref{exact-soft-removal-cost} gives
\begin{equation}\label{many-electron-soft-removal-limit}
 \limsup_{\kappa\downarrow0}\Tr(H^{0,0}\vGamma_\kappa)
 \leq\Tr(H^{0,0}\vGamma).
\end{equation}

Use the Fock factorization into modes with $\omega\geq\kappa$ and
$\omega<\kappa$, and let
$Q=1\otimes\big(1-|\Omega_{<\kappa}\rangle\langle\Omega_{<\kappa}|\big)$
be the projection onto states with at least one soft photon.  By
\cref{exact-soft-removal-marginal}, the soft marginal of
$\vGamma_\kappa$ is $|\Omega_{<\kappa}\rangle\langle\Omega_{<\kappa}|$,
so $\Tr(Q\vGamma_\kappa Q)=\Tr(\vGamma_\kappa Q)=0$.  Since
$\vGamma_\kappa\geq0$, the operator $Q\vGamma_\kappa Q$ is positive
with zero trace, hence zero.  Thus $\vGamma_\kappa^{1/2}Q=0$, so
$\vGamma_\kappa Q=Q\vGamma_\kappa=0$ and
$\vGamma_\kappa=(1-Q)\vGamma_\kappa(1-Q)$.  This is the product
decomposition
\[
 \vGamma_\kappa
 =\gamma_\kappa\otimes
       |\Omega_{<\kappa}\rangle\langle\Omega_{<\kappa}|,
 \qquad
 \gamma_\kappa=\sum_rp_r|\Phi_r\rangle\langle\Phi_r|,
 \qquad p_r>0,\quad\sum_rp_r=1.
\]
Here $(\Phi_r)$ is orthonormal in
$\Hil_\el^N\otimes\Fock_{\geq\kappa}$, and each $\Phi_r$ is
antisymmetric in the electronic coordinates.
Let $H_{\geq\kappa}^{0,0}$ be the hard model introduced before
\cref{lem-exact-soft-removal}, and put
$\vD_{n,\geq\kappa}=-\i\grad_{\vx_n}
                         -e\AF_{\geq\kappa}(\vx_n)$.  The soft-vacuum expectation and
\cref{Gx-norms} give the exact identity
\begin{equation}\label{many-electron-soft-vacuum-energy}
 \Tr(H^{0,0}\vGamma_\kappa)
 =\sum_rp_r\dua{\Phi_r}{H_{\geq\kappa}^{0,0}\Phi_r}
       +\frac{Ne^2}{2}b_\kappa^2.
\end{equation}
In particular, every $\Phi_r$ is in the hard-model form domain and
the displayed sum of nonnegative form energies is finite.

The soft one-photon space $\Hil_{<\kappa}$ is
infinite dimensional.  Choose an orthonormal family $(f_r)$ in it,
write $\eta_r=a^\dag(f_r)\Omega_{<\kappa}$, and consider
\begin{equation}\label{many-electron-soft-tag-state}
 \Psi_\kappa=\sum_r\sqrt{p_r}\,\Phi_r\otimes\eta_r.
\end{equation}
The vectors $\Phi_r\otimes\eta_r$ are orthonormal, so the series
\cref{many-electron-soft-tag-state} converges in norm to a unit
vector.  We show that it also converges in the form norm of
$H^{0,0}$, so that $\Psi_\kappa$ lies in its form domain
$\mathcal Q_\PF$.

Each term of the series has exactly one soft photon.  Identify
$\Hil_{<\kappa}$ with the one-photon sector of $\Fock_{<\kappa}$, and
let $\Psi\in\Hil_\el^N\otimes\Fock_{\geq\kappa}\otimes\Hil_{<\kappa}$.
For almost every $\vX=(\vx_1,\dots,\vx_N)$, the value $\Psi(\vX)$ lies in
$\Fock_{\geq\kappa}\otimes\Hil_{<\kappa}$, and
$\Af^\mu_{<\kappa}(\vx_n)$ acts on it as $a^\dag(g)+a(g)$ on the
soft factor, with $g=\iden_{\{\omega<\kappa\}}G^\mu_{\vx_n}$.  The
vector $a^\dag(g)\Psi(\vX)$ has two soft photons and $a(g)\Psi(\vX)$ has
none, so these two vectors are orthogonal.  Together with the CCR
$\|a^\dag(g)\Psi(\vX)\|^2=\|g\|^2\|\Psi(\vX)\|^2+\|a(g)\Psi(\vX)\|^2$
and $\|a(g)\Psi(\vX)\|\leq\|g\|\|\Psi(\vX)\|$, this gives
\[
 \|[a^\dag(g)+a(g)]\Psi(\vX)\|^2
 =\|g\|^2\|\Psi(\vX)\|^2+2\|a(g)\Psi(\vX)\|^2
 \leq3\|g\|^2\|\Psi(\vX)\|^2.
\]
By \cref{Gx-norms} restricted to $\{\omega<\kappa\}$, the sum of
$\|g\|^2$ over $\mu$ is $b_\kappa^2$.  Summing over $\mu$ and $n$ and
integrating over $\vX$ therefore gives
\begin{equation}\label{many-electron-soft-tag-variance}
 \sum_{n=1}^N\|\AF_{<\kappa}(\vx_n)\Psi\|^2
 \leq3Nb_\kappa^2\|\Psi\|^2.
\end{equation}
Consider a finite partial sum
$\Psi_I=\sum_{r\in I}\sqrt{p_r}\,\Phi_r\otimes\eta_r$.  Both $\Psi_I$
and $\vD_{n,\geq\kappa}\Psi_I$ have exactly one soft photon, while
$\AF_{<\kappa}(\vx_n)\Psi_I$ has zero or two.  Since
$\vD_n=\vD_{n,\geq\kappa}-e\AF_{<\kappa}(\vx_n)$, it follows that
\[
 \dua{\vD_{n,\geq\kappa}\Psi_I}{\AF_{<\kappa}(\vx_n)\Psi_I}=0,
 \qquad
 \|\vD_n\Psi_I\|^2
 =\|\vD_{n,\geq\kappa}\Psi_I\|^2+e^2\|\AF_{<\kappa}(\vx_n)\Psi_I\|^2.
\]
Since the tags $\eta_r$ are orthonormal, tracing out the soft photons
gives
\begin{equation}\label{many-electron-soft-tag-hard-state}
 \Tr_{<\kappa}|\Psi_I\rangle\langle\Psi_I|
 =\sum_{r\in I}p_r|\Phi_r\rangle\langle\Phi_r|.
\end{equation}
The hard model $H^{0,0}_{\geq\kappa}$ acts only on
$\Hil_\el^N\otimes\Fock_{\geq\kappa}$, so
\[
 \dua{\Psi_I}{H^{0,0}_{\geq\kappa}\Psi_I}
 =\sum_{r\in I}p_r\dua{\Phi_r}{H^{0,0}_{\geq\kappa}\Phi_r}.
\]
By the
identity for $\|\vD_n\Psi_I\|^2$, the rest of
$\dua{\Psi_I}{H^{0,0}\Psi_I}$ is
$\frac{e^2}{2}\sum_n\|\AF_{<\kappa}(\vx_n)\Psi_I\|^2$, which
\cref{many-electron-soft-tag-variance} bounds, plus the soft
radiation energy.  Since $\omega<\kappa$ on $\Hil_{<\kappa}$, the
latter satisfies
$\dua{\Psi_I}{H_{\rad,<\kappa}\Psi_I}\leq\kappa\|\Psi_I\|^2
=\kappa\sum_{r\in I}p_r$.  Hence
\begin{equation}\label{many-electron-soft-tag-partial-energy}
 \dua{\Psi_I}{H^{0,0}\Psi_I}
 \leq\sum_{r\in I}p_r
              \dua{\Phi_r}{H_{\geq\kappa}^{0,0}\Phi_r}
       +\left(\kappa+\frac{3Ne^2}{2}b_\kappa^2\right)
                 \sum_{r\in I}p_r.
\end{equation}
Applying this bound to tails, together with
\cref{many-electron-soft-vacuum-energy}, proves convergence of
\cref{many-electron-soft-tag-state} in the full form norm.  Hence
$\Psi_\kappa\in \mathcal Q_\PF$ is normalized and retains electronic
antisymmetry.

Letting $I$ increase to all indices in
\cref{many-electron-soft-tag-hard-state}, by continuity of the
partial trace in trace norm, gives
\[
 \Tr_{<\kappa}|\Psi_\kappa\rangle\langle\Psi_\kappa|
 =\gamma_\kappa
 =\Tr_{<\kappa}\vGamma_\kappa.
\]
So $\Psi_\kappa$ and $\vGamma_\kappa$ have the same reduced state on
electrons and hard photons.

We next prove that $\Psi_\kappa$ and $\vGamma_\kappa$ have the same
density and combined current density.  Put
$\Psi_r^0=\Phi_r\otimes\Omega_{<\kappa}$, so that
$\vGamma_\kappa=\sum_rp_r|\Psi_r^0\rangle\langle\Psi_r^0|$, and let
$\vGamma_I=\sum_{r\in I}p_r|\Psi_r^0\rangle\langle\Psi_r^0|$.
Let $v\in C_c^\infty(\RR^3)$ and $\vA\in C_c^\infty(\RR^3;\RR^3)$.
Integrating \cref{density-def,jc-def} against $v$ and $\vA$ gives,
for every $\Psi\in\mathcal Q_\PF$,
\begin{equation}\label{many-electron-soft-tag-tested}
 \int_{\RR^3}v\rho_\Psi=N\dua{\Psi}{v(\vx_1)\Psi},
 \qquad
 \int_{\RR^3}\vA\cdot\vjc_\Psi
 =N\Re\dua{\Psi}{\vA(\vx_1)\cdot\vD_1\Psi}.
\end{equation}
Write $\vD_1=\vD_{1,\geq\kappa}-e\AF_{<\kappa}(\vx_1)$.  The operators
$v(\vx_1)$ and $\vA(\vx_1)\cdot\vD_{1,\geq\kappa}$ act only on
$\Hil_\el^N\otimes\Fock_{\geq\kappa}$.  Since the tags are
orthonormal and $\Omega_{<\kappa}$ is normalized,
\begin{align*}
 \dua{\Psi_I}{v(\vx_1)\Psi_I}
 &=\sum_{r\in I}p_r\dua{\Psi_r^0}{v(\vx_1)\Psi_r^0},\\
 \dua{\Psi_I}{\vA(\vx_1)\cdot\vD_{1,\geq\kappa}\Psi_I}
 &=\sum_{r\in I}p_r
   \dua{\Psi_r^0}{\vA(\vx_1)\cdot\vD_{1,\geq\kappa}\Psi_r^0}.
\end{align*}
The soft field contributes to neither state.  The vector
$\AF_{<\kappa}(\vx_1)\Psi_I$ has zero or two soft photons while
$\Psi_I$ has one, and $\AF_{<\kappa}(\vx_1)\Psi_r^0$ has one soft
photon while $\Psi_r^0$ has none.  Hence
\[
 \dua{\Psi_I}{\vA(\vx_1)\cdot\AF_{<\kappa}(\vx_1)\Psi_I}
 =\dua{\Psi_r^0}{\vA(\vx_1)\cdot\AF_{<\kappa}(\vx_1)\Psi_r^0}
 =0.
\]
By \cref{many-electron-soft-tag-tested} and linearity in the state,
the last three displays give
\begin{equation}\label{many-electron-soft-tag-partial-constraints}
 \int_{\RR^3}v\rho_{\Psi_I}=\int_{\RR^3}v\rho_{\vGamma_I},
 \qquad
 \int_{\RR^3}\vA\cdot\vjc_{\Psi_I}
 =\int_{\RR^3}\vA\cdot\vjc_{\vGamma_I}.
\end{equation}
Now let $I$ increase to all indices.  Then
$|\Psi_I\rangle\langle\Psi_I|\to|\Psi_\kappa\rangle\langle\Psi_\kappa|$
and $\vGamma_I\to\vGamma_\kappa$ in trace norm.  All these operators
have trace at most one, and their energies are bounded uniformly in
$I$, by \cref{many-electron-soft-tag-partial-energy} and by
$\Tr(H^{0,0}\vGamma_I)\leq\Tr(H^{0,0}\vGamma_\kappa)$.  Hence
\cref{mixed-density-passage,mixed-current-passage} of
\cref{lem-trace-compactness-and-limits} pass both sides of
\cref{many-electron-soft-tag-partial-constraints} to the limit.
Since $v$ and $\vA$ are arbitrary, $\Psi_\kappa$ and $\vGamma_\kappa$
have the same density and combined current density.  Together with
\cref{lem-exact-soft-removal}, this gives
\begin{equation}\label{many-electron-soft-tag-constraints}
 (\rho_{\Psi_\kappa},\vjc_{\Psi_\kappa})
 =(\rho_{\vGamma_\kappa},\vjc_{\vGamma_\kappa})
 =(\rho_\vGamma,\vjc_\vGamma).
\end{equation}
Passing to the limit in \cref{many-electron-soft-tag-partial-energy} and subtracting
the vacuum contribution in \cref{many-electron-soft-vacuum-energy}
gives
\begin{equation}\label{many-electron-soft-tag-cost}
 \dua{\Psi_\kappa}{H^{0,0}\Psi_\kappa}
 \leq\Tr(H^{0,0}\vGamma_\kappa)
       +\kappa+Ne^2b_\kappa^2.
\end{equation}
The last two terms tend to zero as $\kappa\downarrow0$, since
$b_\kappa\to0$ for every form factor. 

Finally, let
$\NC_{\geq\kappa}=\dd\varGamma(\iden_{\{\omega\geq\kappa\}})$.
The soft tags have exactly one photon, so
$\dua{\Psi_\kappa}{\NC\Psi_\kappa}=1+\Tr(\NC_{\geq\kappa}\gamma_\kappa)$.
Tracing \cref{exact-soft-removal-marginal} over $\Fock_{<\kappa}$ gives
$\Tr_{\el,<\kappa}\vGamma_\kappa=\Tr_{\el,<\kappa}\vGamma$, so
$\Tr(\NC_{\geq\kappa}\gamma_\kappa)=\Tr(\NC_{\geq\kappa}\vGamma)$.  Thus
\begin{equation}\label{many-electron-soft-tag-number}
 \dua{\Psi_\kappa}{\NC\Psi_\kappa}
 =1+\Tr(\NC_{\geq\kappa}\vGamma)
 \leq1+\kappa^{-1}\Tr(H_\rad\vGamma)<\infty.
\end{equation}
Together,
\cref{many-electron-soft-removal-limit,many-electron-soft-tag-constraints,many-electron-soft-tag-cost}
prove \cref{many-electron-pure-approximation} by choosing
$\kappa$ sufficiently small.

Fix $(\rho,\vjc)\in X\times Y$, and let $\vGamma\in\dm$ satisfy
$(\rho_\vGamma,\vjc_\vGamma)=(\rho,\vjc)$.  The pure states in
\cref{many-electron-pure-approximation} are admissible in
\cref{FLL-def}, so
$F_\LL^{e,\chi}(\rho,\vjc)\leq\Tr(H^{0,0}\vGamma)+\delta$ for every
$\delta>0$.  Taking the infimum over $\vGamma$ and using
\cref{L-cs-relations} proves
$F_\LL^{e,\chi}(\rho,\vjc)=\wt F_\LI^{e,\chi}(\rho,\vjc)$.
If no such $\vGamma$ exists, then no admissible pure state exists
either, since $|\Psi\rangle\langle\Psi|\in\dm$ for every normalized
$\Psi\in\mathcal Q_\PF$.  Then
$F_\LL^{e,\chi}(\rho,\vjc)=\wt F_\LI^{e,\chi}(\rho,\vjc)=+\infty$.
The equality with $F_\LI^{e,\chi}(\rho,\vjc)$ follows from
\cref{thm-unrestricted-attainment}.  If
$F_\LL^{e,\chi}(\rho,\vjc)<\infty$, equality of the two searches
identifies a pure minimizer at $(\rho,\vjc)$ with a rank-one
minimizer of the total mixed-state problem, in both directions.
\end{proof}

\phantomsection
\label{proof:cor-many-electron-pure-functional}
\begin{proof}[Proof of \cref{cor-many-electron-pure-functional}]
Properness, convexity, and norm lower semicontinuity follow from
\cref{many-electron-pure-ensemble-equality} and the corresponding
properties of $F_\LI^{e,\chi}(\rho,\vjc)$.  Every pure state defines a
rank-one density matrix, and \cref{many-electron-pure-approximation}
provides, for every mixed state, a finite-energy pure state with the
same density and combined current density.  This proves the first
equality in \cref{unrestricted-pure-mixed-PF-class}.  Since
$\Tr(H^{0,0}\vGamma)<\infty$ for every $\vGamma\in\dm$, the middle set in
\cref{unrestricted-pure-mixed-PF-class} is the domain of
$\wt{F}_\LI^{e,\chi}$, and \cref{unrestricted-Lieb-equality} identifies
$\wt{F}_\LI^{e,\chi}$ with $F_\LI^{e,\chi}$.  This proves the second
equality.  The pure-state class is therefore the domain of the convex
functional $F_\LI^{e,\chi}$, hence convex.

Fix any mixed-state minimizer $\vGamma_*$ at $(\rho,\vjc)$.  By
\cref{thm-unrestricted-attainment}, it has finite mean photon number.
For $0<\kappa\leq1$, let $\Psi_\kappa$ be the pure state
\cref{many-electron-soft-tag-state} built from $\vGamma_*$ in the
proof of \cref{thm-many-electron-soft-purification}.  That is, first
remove the soft photons from $\vGamma_*$ by
\cref{lem-exact-soft-removal}, then attach orthonormal one-photon soft
tags to the eigenvectors of the result.  Choose a sequence
$\kappa_j\downarrow0$ and put $\Psi_j=\Psi_{\kappa_j}$.  By
\cref{many-electron-soft-tag-constraints},
$(\rho_{\Psi_j},\vjc_{\Psi_j})
=(\rho_{\vGamma_*},\vjc_{\vGamma_*})=(\rho,\vjc)$ for every $j$.  By
\cref{many-electron-soft-removal-limit,many-electron-soft-tag-cost}
and $b_{\kappa_j}\to0$,
\[
 \limsup_{j\to\infty}\dua{\Psi_j}{H^{0,0}\Psi_j}
 \leq\Tr(H^{0,0}\vGamma_*)
 =F_\LL^{e,\chi}(\rho,\vjc),
\]
where the equality holds because $\vGamma_*$ is a minimizer, by
\cref{unrestricted-Lieb-equality,many-electron-pure-ensemble-equality}.
Each $\Psi_j$ is admissible in \cref{FLL-def}, so
$\dua{\Psi_j}{H^{0,0}\Psi_j}\geq F_\LL^{e,\chi}(\rho,\vjc)$, and
hence $\dua{\Psi_j}{H^{0,0}\Psi_j}\to F_\LL^{e,\chi}(\rho,\vjc)$.
Since $\NC_{\geq\kappa_j}\leq\NC$,
\cref{many-electron-soft-tag-number} and the photon-number bound
in \cref{thm-unrestricted-attainment} give
\cref{pure-minimizing-sequence-number-bound}.
\end{proof}

\phantomsection
\label{proof:prop-zero-current-density-functional}
\begin{proof}[Proof of \cref{prop-zero-current-density-functional}]
Let $\Theta$ be the time reversal \cref{time-reversal-def}, with the
properties in \cref{prop-time-reversal}, and let
$\rdens_N:=\{\rho\geq0:\ \int_{\RR^3}\rho=N,\ \sqrt\rho\in H^1(\RR^3)\}$.
We first show that, for either $F\in\{F_\LL,F_\LI\}$, every
$\rho\in\rdens_N$, and every $\vjc\in Y$,
\begin{equation}\label{zero-current-minimum}
 F^{e,\chi}(\rho,\vjc)=F^{e,\chi}(\rho,-\vjc),\qquad
 F^{e,\chi}(\rho,0)\leq F^{e,\chi}(\rho,\vjc).
\end{equation}

Evenness of both functionals in the current density is
\cref{F-time-reversal}.  For the PFL functional, convexity then gives
\[
 F_\LI^{e,\chi}(\rho,0)
 \leq\frac12\big[F_\LI^{e,\chi}(\rho,\vjc)
                  +F_\LI^{e,\chi}(\rho,-\vjc)\big]
 =F_\LI^{e,\chi}(\rho,\vjc).
\]
This is \cref{zero-current-minimum} for $F_\LI$.
By \cref{time-reversal-laws}, $\bar E(u,\vA)=\bar E(u,-\vA)$, and
concavity of $\bar E$ gives $\bar E(u,\vA)\leq\bar E(u,0)=E(u,0)$, proving 
\cref{zero-current-scalar-dual} directly from \cref{FL-def}.
For the mixed-state search, replace any $\vGamma$ by
$\tfrac12(\vGamma+\Theta\vGamma\Theta^{-1})$.  This preserves its
density and energy and sets its combined current density to zero.
\Cref{thm-unrestricted-attainment} then gives
\cref{zero-current-density-search}, including the mixed-state minimum.
The minimizing density matrix can be chosen time-reversal invariant.
The density bound makes $\rho\in\rdens_N$ necessary for finiteness.
Conversely, every $\rho\in\rdens_N$ has a finite-kinetic-energy
electronic representative \cite{Lieb1983}.  Vacuum embedding and the
same averaging then give a finite-energy density matrix at $(\rho,0)$.

For the PFLL functional, to prove minimality at zero current density,
take any normalized
$\Psi\in\mathcal Q_\PF$ with pair $(\rho,\vjc)$ and form
\[
 \vGamma_\Psi=\frac12\bigl(
 |\Psi\rangle\langle\Psi|
 +|\Theta\Psi\rangle\langle\Theta\Psi|\bigr).
\]
This density matrix has pair $(\rho,0)$ and energy
$\dua{\Psi}{H^{0,0}\Psi}$.  For every $\delta>0$, the approximation
statement \cref{many-electron-pure-approximation} supplies a normalized
pure state $\Phi_\delta$ with the exact pair $(\rho,0)$ and
\[
 \dua{\Phi_\delta}{H^{0,0}\Phi_\delta}
 \leq\dua{\Psi}{H^{0,0}\Psi}+\delta.
\]
Letting $\delta\downarrow0$ and taking the infimum over $\Psi$ proves
\cref{zero-current-minimum} for $F_\LL$, without using its convexity or
its equality with $F_\LI$.  Taking the infimum over all pure states of
density $\rho$ gives \cref{zero-current-pure-density-search}.
The density bound and the same construction applied to a vacuum-embedded
electronic representative give the finite domain $\rdens_N$.

Finally, apply \cref{many-electron-pure-approximation} to a PFL optimizer
at $(\rho,0)$.  Letting the energy error tend to zero gives
$F_\LL^{e,\chi}(\rho,0)\leq F_\LI^{e,\chi}(\rho,0)$.
The reverse inequality follows from \cref{L-cs-relations}.
Only the constructive approximation is used here, not the general
equality of the two functionals.
\end{proof}

\subsection{Properties of optimizers}

\phantomsection
\label{proof:prop-optimizer-mean-fields}
\begin{proof}[Proof of \cref{prop-optimizer-mean-fields}]
For $h\in\Hil_\ph$ with $h/\sqrt\omega\in\Hil_\ph$, the
annihilation estimate and Cauchy--Schwarz in a spectral decomposition
of $\vGamma$ give
\begin{equation}\label{optimizer-annihilation-expectation-bound}
 |\Tr(a(h)\vGamma)|
 \leq\|h/\sqrt\omega\|\,
             \big(\Tr(H_\rad\vGamma)\big)^{1/2}.
\end{equation}
Thus all linear traces below are well defined, even without finite mean
photon number.

For (i), put
$g_j=\sum_\mu\int u_j^\mu(\vx)G^\mu_{\vx}\dd\vx$ and
$\vec\AF=(\int\vu_1\cdot\AF,\ldots,\int\vu_n\cdot\AF)$, so that
$\int\vu_j\cdot\AF=\varPhi(g_j)$ and, as in
Section~\ref{sec-free-field}, $g_j\in D(\omega)$.
By the symmetry of the form factor, all scalar
products among the kernels $g_j$, $\omega g_j$, and $G^\mu_{\vx}$ are
real.  By \cref{prop-weyl-algebra}(iii), the components of $\vec\AF$
therefore strongly commute with each other and with every
$\Af^\mu(\vx)$, so
\[
 U_t=e^{\i t\phi(\vec\AF)},\qquad t\in\RR,
\]
is a photon unitary commuting with every $\Af^\mu(\vx)$.  It commutes
with each $\vD_n$ and leaves the density, the paramagnetic current
density, and the conditional field expectations unchanged, hence also
$\vjc$.  For $\xi\in\RR^n$, $e^{\i\xi\cdot\vec\AF}=W(\i\sum_j\xi_jg_j)$,
and the pointwise shift \cref{weyl-weighted-point-shift} gives
\[
 a(\vk,\lambda)e^{\i\xi\cdot\vec\AF}\eta
 =e^{\i\xi\cdot\vec\AF}\Big(a(\vk,\lambda)\eta
   +\i\sum_{j=1}^n\xi_jg_j(\vk,\lambda)\eta\Big),
 \qquad\eta\in D(\sqrt{H_\rad}).
\]
Since $e^{\i t\phi}-1\in\mathcal S(\RR^n)$, integrating this identity
against its Fourier transform, and using closedness of the
annihilation map, gives
\[
 a(\vk,\lambda)U_t\eta
 =U_t\Big(a(\vk,\lambda)\eta
   +\i t\sum_{j=1}^ng_j(\vk,\lambda)(\partial_j\phi)(\vec\AF)\eta\Big).
\]
As $g_j\in D(\omega)$, $U_t$ preserves $D(\sqrt{H_\rad})$, hence
$\mathcal Q_\PF$, and only the radiation energy changes.  Recall that
$\int\vu_j\cdot\EF_\perp=-\i\big(a^\dag(\omega g_j)-a(\omega g_j)\big)$.
Squaring and integrating against $\omega$ gives the expansion below.
In its linear term we used that the commutator
\[
 [a(\omega g_j),(\partial_j\phi)(\vec\AF)]
 =\sum_{l=1}^n\dua{\omega g_j}{g_l}(\partial_l\partial_j\phi)(\vec\AF)
\]
is self-adjoint:
\[
 \begin{aligned}
 \Tr(H^{0,0}U_t\vGamma U_t^\dag)
 &=\Tr(H^{0,0}\vGamma)
 -t\sum_{j=1}^n\Re\Tr\Big((\partial_j\phi)(\vec\AF)
   \int\vu_j\cdot\EF_\perp\,\vGamma\Big)\\
 &\quad+t^2\sum_{j,l=1}^n\dua{\sqrt\omega g_j}{\sqrt\omega g_l}
   \Tr\big((\partial_j\phi)(\vec\AF)(\partial_l\phi)(\vec\AF)
   \vGamma\big).
 \end{aligned}
\]
Optimality at $t=0$ proves \cref{optimizer-field-statistics}.

For (ii), apply (i) with $n=1$, $\vu_1=\vu$, and $\phi(q)=m\psi(q/m)$,
where $\psi\in\mathcal S(\RR)$ is real with $\psi'(0)=1$.  Then
$\phi'(\int\vu\cdot\AF)=\psi'(m^{-1}\int\vu\cdot\AF)$ is bounded
uniformly in $m$ and converges strongly to $\iden$ as $m\to\infty$.
Since $\vGamma$ has finite radiation energy, dominated convergence over
a spectral decomposition of $\vGamma$ gives
$\Re\Tr\big(\int\vu\cdot\EF_\perp\,\vGamma\big)=0$.  This trace is real,
because $\int\vu\cdot\EF_\perp$ is self-adjoint, which proves
\cref{optimizer-mean-electric-zero}.

For (iii), suppose first that $\vjc=0$.  By
\cref{zero-current-density-search}, $\vGamma$ also minimizes the
internal energy among all density matrices with density $\rho$.
For any $f\in D(\omega)$, \cref{cor-weyl-PF-shifts} shows that
$W(tf)\vGamma W(tf)^\dag$ has density $\rho$, and that its internal
energy is the quadratic polynomial \cref{weyl-PF-energy-shift} in $t$,
with linear coefficient
$2\Re\Tr(a(\omega f)\vGamma)-e\int\vc_f\cdot\vjc$.
This coefficient vanishes.  Since $\vjc=0$, using both $f$
and $\i f$ gives $\Tr(a(\omega f)\vGamma)=0$.
For any $h$ as in \cref{optimizer-annihilation-expectation-bound}, set
\[
 h_m=\iden_{\{m^{-1}\leq\omega\leq m\}}h,
 \qquad f_m=h_m/\omega\in D(\omega).
\]
Then $\|(h_m-h)/\sqrt\omega\|\to0$, so
\cref{optimizer-annihilation-expectation-bound} shows that
$\Tr(a(h)\vGamma)=0$.
Taking $h=G^\mu_{\vx}$ proves the pointwise assertion for $\AF$,
and taking $h=g_{\grad\times\vu}$ proves it for $\BF$.

Finally, if $\vjc=\rho\grad S$, the gauge covariance identities
\cref{current-gauge,energy-gauge,F-gauge}, applied to spectral
decompositions, show that
$U_{-S}\vGamma U_{-S}^\dag$ is an optimizer at $(\rho,0)$.
The electronic unitary $U_{-S}$ leaves the photon reduced state
unchanged.  All coherent photon amplitudes therefore agree with those
of this zero-current optimizer.
\end{proof}

\phantomsection
\label{proof:prop-optimizer-counterexamples}
\begin{proof}[Proof of \cref{prop-optimizer-counterexamples}]
(i) The vacuum product $\sqrt\rho\otimes\Omega$ has finite energy and pair
$(\rho,0)$.  By \cref{thm-unrestricted-attainment} an optimizer
$\vGamma_0$ at this pair exists.  Gauge covariance with
$S(\vx)=\vq\cdot\vx$ maps it to an optimizer at $(\rho,\rho\vq)$.
\Cref{prop-optimizer-mean-fields} gives the claimed fields for
every optimizer there, and direct differentiation gives the divergence.

(ii) Write $\phi=\sqrt\rho$.  For any finite-energy density matrix
$\vGamma=\sum_r p_r|\Psi_r\rangle\langle\Psi_r|$ with density $\rho$,
integration by parts gives
\[
 \Tr(H^{0,0}\vGamma)
 =\frac34+\frac12\sum_{r,\mu}p_r
           \|(\partial_\mu+x^\mu)\Psi_r\|^2
           +\Tr(H_\rad\vGamma).
\]
Indeed, $\int|\vx|^2\rho=3/2$.  The identity follows first on smooth
vectors and then by approximation in $H^1\cap D(|\vx|)$.
The state $\phi\otimes\Omega$ attains $3/4$ at $(\rho,0)$.
Equality forces each occupied vector to be $\phi\otimes\xi_r$,
and $\ker H_\rad=\Span\{\Omega\}$ then forces the density matrix to be
$|\phi\otimes\Omega\rangle\langle\phi\otimes\Omega|$.
Gauge covariance is a bijection of the optimizer sets and proves
\cref{unique-gaussian-optimizer} and its uniqueness.

Suppose that $\Psi_\vq$ were an eigenvector of some admissible
$H^{v,\vA}$ with eigenvalue $E$.  Removing its electronic phase
makes $\phi\otimes\Omega$ an eigenvector with vector potential
$\vA'=\vA+\vq\in L^3+L^\infty$.
At $e=0$ the photon vacuum decouples.  Taking the real part of the
weak electronic eigenvalue equation, tested against real compactly
supported functions, gives
\[
 -\frac12\Delta\phi+\left(v+\frac12|\vA'|^2\right)\phi=E\phi.
\]
Since $\phi>0$, this implies
\[
 v+\frac12|\vA'|^2
 =E+\frac{\Delta\phi}{2\phi}
 =E+\frac{|\vx|^2-3}{2}.
\]
The left-hand side belongs to $L^{3/2}+L^\infty$, whereas the
quadratically growing right-hand side does not.  This is a
contradiction.  Finally, time reversal maps the constraint $(\rho,\rho\vq)$ to
$(\rho,-\rho\vq)$, so uniqueness at $(\rho,\rho\vq)$ does not
imply time-reversal invariance.
\end{proof}

\phantomsection
\label{proof:prop-optimizer-symmetry}
\begin{proof}[Proof of \cref{prop-optimizer-symmetry}]
(i) By \cref{prop-time-reversal}, $\Theta$ commutes with every
$U_\vR$, and $\Theta^2=\iden$.  Thus $g\mapsto V_g$ is a strongly continuous
representation of $G$ by unitary and antiunitary operators.  By
\cref{H-rotation,rho-rotation,jc-rotation,time-reversal-laws}, each
$V_g$ preserves $\mathcal Q_\PF$ and the
energy, and maps a density matrix with pair $(\rho',\vj')$ to one with
pair $(\rho'\circ\vR^\top,s\vR\vj'\circ\vR^\top)$.  For $g\in G$ the
pair $(\rho,\vjc)$ is fixed, so $V_g$ maps optimizers to optimizers,
with inverse given by $V_{g^{-1}}$.

Let $\vGamma$ be an optimizer, and let $\dd g$ be the normalized Haar
measure on $G$.  Since $V_g$ commutes with $H^{0,0}$, the map
\[
 g\mapsto\sqrt{H^{0,0}+1}\,V_g\vGamma V_g^{-1}\sqrt{H^{0,0}+1}
 =V_g\sqrt{H^{0,0}+1}\,\vGamma\sqrt{H^{0,0}+1}\,V_g^{-1}
\]
is continuous in trace norm.  Hence
\[
 \vGamma_G=\int_G V_g\vGamma V_g^{-1}\dd g
\]
is a density matrix with $\sqrt{H^{0,0}+1}\,\vGamma_G\sqrt{H^{0,0}+1}$
trace class.  As in the proof of \cref{cor-weyl-PF-shifts}, the energy
and the pairings of the density and of the combined current density
with bounded test functions are continuous linear functionals of
$\sqrt{H^{0,0}+1}\,\vGamma\sqrt{H^{0,0}+1}$ in trace norm.  They are constant along the orbit, so they pass to the average,
and $\vGamma_G$ is an optimizer at $(\rho,\vjc)$.  Invariance of Haar
measure makes it $G$-invariant.

(ii) Every $G^\mu_{\vx}$ lies in $\Hil_S^\perp$.  By
\cref{weyl-factorization}, the field acts only on the first factor,
and $H_\rad$ is the sum of the radiation energies of the two factors.
Replacing an optimizer $\vGamma$ by the partial trace of $\vGamma$
over the second factor, tensored with $|\Omega_S\rangle\langle\Omega_S|$,
therefore preserves the density, the combined current density, and
every term of the energy except the radiation energy of the second
factor, which it removes.  Optimality forces that term to vanish.
Since $\omega>0$ almost everywhere, the radiation energy of the second
factor has kernel spanned by $\Omega_S$, so every eigenvector of
$\vGamma$ with positive eigenvalue has $\Omega_S$ as its second
factor.
\end{proof}

\begin{proof}[Proof of \cref{prop-optimizer-soft-spectrum}]
\phantomsection
\label{proof:prop-optimizer-soft-spectrum}
The competitor in \cref{unrestricted-soft-energy-gain} has exactly
the same density and combined current density as $\vGamma_*$.  Optimality
therefore gives
\[
 0\leq\Tr(H^{0,0}(\vGamma_{*,\kappa}-\vGamma_*))
 \leq-\tfrac12E_*(\kappa)+C(\kappa^2+b_\kappa^2),
\]
which proves \cref{optimizer-soft-energy-bound} after changing $C$.
No supporting external potential is needed.
Tonelli's theorem and
$\mathcal U_*(\varepsilon)=\varepsilon\mathcal N_*(\varepsilon)$
give the exact identity
\begin{equation*}
 N_*(\kappa)=\frac{E_*(\kappa)}{\kappa}
             +\int_0^\kappa\frac{E_*(t)}{t^2}\dd t.
\end{equation*}
The form factor contribution obeys a second Tonelli identity,
\[
 \frac{b_\kappa^2}{\kappa}
       +\int_0^\kappa\frac{b_t^2}{t^2}\dd t
 =\int_{\omega<\kappa}\frac{|\wh\chi(\vk)|^2}{\omega(\vk)}\dd\vk
 =a_\kappa^2.
\]
Together these prove \cref{optimizer-soft-number-bound}.  Finally,
for the hard cutoff and $\kappa\leq\Lambda$,
$b_\kappa^2=2\pi\kappa^2$ and $a_\kappa^2=4\pi\kappa$.
\end{proof}

\subsection{Zero coupling}
\label{proof:thm-soft-photon-relaxation}
\begin{proof}[Proof of \cref{thm-soft-photon-relaxation}]
Throughout the proof, $H^{0,0}$ is evaluated at $e=0$.
The equality $F_\LL^{0,\chi}=F_\LI^{0,\chi}$ follows from
\cref{thm-many-electron-soft-purification}.  It remains to prove
$F_\LL^{0,\chi}=F_\VR$ and the attainment criterion.

\noindent\textbf{Step 1. The lower bound $F_\LL^{0,\chi}\geq F_\VR$.}
Let $\Psi\in \mathcal Q_\PF$ be normalized and suppose that
$(\rho_\Psi,\vjp_\Psi)=(\rho,\vjp)$.  Its electronic and photon reduced
states are
\[
 \Gamma_\el
 :=\Tr_\ph|\Psi\rangle\langle\Psi|,\qquad
 \Gamma_\ph
 :=\Tr_\el|\Psi\rangle\langle\Psi|.
\]
The partial-trace identity for positive observables
\cref{partial-trace-positive-observables}, first for bounded spectral
truncations and then by monotone convergence, gives
\[
 \dua{\Psi}{H^{0,0}\Psi}
 =\Tr(H_\el^0\Gamma_\el)
  +\Tr(H_\rad\Gamma_\ph).
\]
In particular, $\Gamma_\el$ has finite electronic energy.  It has
the same density as $\Psi$: for $v\in C_c^\infty(\RR^3)$ and the
bounded operator $V=\sum_{n=1}^Nv(\vx_n)$ on $\Hil_\el^N$,
\[
 \int_{\RR^3}v\rho_\Psi
 =\dua{\Psi}{(V\otimes\iden_{\Fock_\ph})\Psi}
 =\Tr(V\Gamma_\el)
 =\int_{\RR^3}v\rho_{\Gamma_\el}.
\]
It also has the same paramagnetic
current density.  Indeed, insert a Schmidt decomposition
$\Psi=\sum_j\sqrt{p_j}\,\Phi_j\otimes\Xi_j$, for which
$\Gamma_\el=\sum_jp_j|\Phi_j\rangle\langle\Phi_j|$ by
\cref{Schmidt-reduced-states}, into \cref{jp-def}.  The gradient acts
only on the electronic factor, and $\dua{\Xi_j}{\Xi_k}=\delta_{jk}$
removes all cross terms, so
\[
 \vjp_\Psi=\sum_jp_j\vjp_{\Phi_j}=\vjp_{\Gamma_\el}.
\]
The convergence of these series is checked in the proof of
\cref{thm-photon-purification}, which holds for every $e\geq0$.
Consequently,
$(\rho_{\Gamma_\el},\vjp_{\Gamma_\el})=(\rho,\vjp)$, and
\[
 \dua{\Psi}{H^{0,0}\Psi}
 \geq\Tr(H_\el^0\Gamma_\el)
 \geq F_\VR(\rho,\vjp).
\]
Taking the infimum over admissible $\Psi$ gives
$F_\LL^{0,\chi}(\rho,\vjp)\geq F_\VR(\rho,\vjp)$.

\noindent\textbf{Step 2. The upper bound $F_\LL^{0,\chi}\leq F_\VR$.}
We purify an arbitrary admissible electronic density matrix with soft
photons.  Each of its eigenvectors is paired with its own one-photon
state, the photon states being orthonormal and supported at momenta below
$\epsilon$.  The resulting pure state has this density matrix as its electronic
reduced state, and hence the prescribed density and current density.  Its energy
exceeds the electronic energy only by the photon energy, which
is at most $\epsilon$ and can therefore be made arbitrarily small.
This is the one-photon purification \cref{one-photon-purification} used
in the proof of \cref{thm-photon-purification} to prove the reverse
inclusions in \cref{electronic-projection,zero-jph-slice}.  There the
purification only had to realize the prescribed pair with vanishing
field-induced current density.  Here, at $e=0$, the energy form factorizes, so we
can also compute the energy of the purified state exactly.

More precisely, let $\Gamma$ be any admissible electronic density matrix of finite energy.
Writing only its nonzero eigenvalues, its spectral decomposition is
\[
 \Gamma=\sum_{j=1}^r p_j|\Phi_j\rangle\langle\Phi_j|,\qquad
 p_j>0,\qquad \sum_{j=1}^r p_j=1,\qquad
 r\in\NN\cup\{\infty\}.
\]
The vectors $\Phi_j$ are orthonormal, belong to $Q(H_\el^0)$, and satisfy
\[
 \sum_{j=1}^r p_j\dua{\Phi_j}{H_\el^0\Phi_j}
 =\Tr(H_\el^0\Gamma)<\infty.
\]
Fix $\epsilon>0$.  The one-photon space over the region
$\{0<|\vk|<\epsilon\}$ is infinite dimensional, so it contains an
orthonormal family $\{f_j^\epsilon\}_{j=1}^r$.  Put
\[
 \Xi_j^\epsilon=a^\dag(f_j^\epsilon)\Omega.
\]
Then the $\Xi_j^\epsilon$ are orthonormal one-photon vectors and
\[
 \dua{\Xi_j^\epsilon}{H_\rad\Xi_j^\epsilon}
 =\dua{f_j^\epsilon}{|\vk|f_j^\epsilon}\leq\epsilon.
\]
When $r=\infty$, consider first the finite sums
\[
 \Psi_{\epsilon,M}
 =\sum_{j=1}^M\sqrt{p_j}\,
  \Phi_j\otimes\Xi_j^\epsilon.
\]
For $L>M$, orthogonality in both tensor factors and the form
factorization at $e=0$ give
\[
 \begin{split}
 &\|\Psi_{\epsilon,L}-\Psi_{\epsilon,M}\|^2
 +\dua{\Psi_{\epsilon,L}-\Psi_{\epsilon,M}}{
 H^{0,0}(\Psi_{\epsilon,L}-\Psi_{\epsilon,M})}\\
 &\quad=\sum_{M<j\leq L}p_j
 \left(1+\dua{\Phi_j}{H_\el^0\Phi_j}
 +\dua{\Xi_j^\epsilon}{H_\rad\Xi_j^\epsilon}\right).
 \end{split}
\]
The series on the right is summable.  Hence the partial sums converge in
the form norm to a normalized vector.  When $r<\infty$, this vector is
simply the corresponding finite sum.  In either case,
\[
 \Psi_\epsilon=\sum_{j=1}^r\sqrt{p_j}\,
 \Phi_j\otimes\Xi_j^\epsilon\in \mathcal Q_\PF.
\]
Hilbert-space convergence implies trace-norm convergence of the
corresponding rank-one density matrices, and the partial trace is
trace-norm continuous.  Taking the partial trace first for the finite sums
and then passing to the limit therefore yields
\[
 \Tr_\ph|\Psi_\epsilon\rangle\langle\Psi_\epsilon|
 =\sum_{j=1}^r p_j|\Phi_j\rangle\langle\Phi_j|=\Gamma.
\]
Thus $\Psi_\epsilon$ has precisely the prescribed density and
paramagnetic current density.  Its energy is
\[
 \begin{split}
 \dua{\Psi_\epsilon}{H^{0,0}\Psi_\epsilon}
 &=\Tr(H_\el^0\Gamma)
 +\sum_{j=1}^r p_j
  \dua{f_j^\epsilon}{|\vk|f_j^\epsilon}\\
 &\leq\Tr(H_\el^0\Gamma)+\epsilon.
 \end{split}
\]
Since $\Psi_\epsilon$ is admissible for $F_\LL^{0,\chi}(\rho,\vjp)$,
taking the infimum first over admissible $\Gamma$ and then letting
$\epsilon\downarrow0$ gives
$F_\LL^{0,\chi}(\rho,\vjp)\leq F_\VR(\rho,\vjp)$.

\noindent\textbf{Step 3. Characterization of attainment.}
Suppose that a normalized $\Psi_*$ attains the infimum on the left-hand
side of \cref{soft-photon-equality}.  Its electronic marginal is
admissible for the electronic minimization, and hence
\[
 \begin{split}
 0={}&\dua{\Psi_*}{H^{0,0}\Psi_*}
       -F_\VR(\rho,\vjp)\\
 ={}&\left[\Tr(H_\el^0\Gamma_{\el,\Psi_*})
       -F_\VR(\rho,\vjp)\right]
       +\Tr(H_\rad\Gamma_{\ph,\Psi_*}).
 \end{split}
\]
Both terms on the last line are nonnegative, so they vanish separately.
Thus $\Gamma_{\el,\Psi_*}$ is an electronic minimizer and
\[
 \Tr(H_\rad\Gamma_{\ph,\Psi_*})
 =\|\sqrt{H_\rad}\sqrt{\Gamma_{\ph,\Psi_*}}\|_{\mathfrak S_2}^2=0.
\]
It follows that the range of $\sqrt{\Gamma_{\ph,\Psi_*}}$ lies in
$\ker H_\rad$.  Sector by sector, $H_\rad$ is multiplication on the
$m$-photon sector by $\sum_{j=1}^m|\vk_j|$, which is strictly positive
almost everywhere for every $m\geq1$.  Therefore
$\ker H_\rad=\CC\Omega$ and
$\Gamma_{\ph,\Psi_*}=|\Omega\rangle\langle\Omega|$.

Since a pure state with a pure marginal is a product state,
\cref{pure-marginal-product} gives $\Psi_*=\Phi\otimes\Omega$ for a
normalized $\Phi\in Q(H_\el^0)$, and its electronic marginal
$|\Phi\rangle\langle\Phi|$ is a rank-one minimizer.  Conversely, if
$|\Phi\rangle\langle\Phi|$ is a rank-one electronic minimizer, then
$\Phi\otimes\Omega$ is admissible and attains
$F_\LL^{0,\chi}(\rho,\vjp)$.
\end{proof}

\phantomsection
\label{proof:thm-zero-coupling-convergence}
\begin{proof}[Proof of \cref{thm-zero-coupling-convergence}]
Throughout the proof, the coupling and form factor are displayed in the
notation $H_{e,\chi}^{0,0}$.

\noindent\textbf{Step 1. Comparison of the forms at couplings $e$ and $0$.}
By \cref{PF-canonical-form-domain}, all these forms have the same domain
when the form factor is fixed.  Expanding the covariant kinetic energy on this
domain gives
\[
 \begin{split}
 &\dua{\Psi}{H_{e,\chi}^{0,0}\Psi}
 -\dua{\Psi}{H_{0,\chi}^{0,0}\Psi}\\
 &\quad=-e\sum_{n=1}^N
 \Re\dua{(-\i\grad_{\vx_n})\Psi}{\AF(\vx_n)\Psi}
 +\frac{e^2}{2}\sum_{n=1}^N\|\AF(\vx_n)\Psi\|^2.
 \end{split}
\]
For normalized $\Psi$, positivity of the zero-coupling form and
\cref{A-Hrad-bound} imply
\[
 \sum_{n=1}^N\|(-\i\grad_{\vx_n})\Psi\|^2
 \leq2\dua{\Psi}{H_{0,\chi}^{0,0}\Psi},
\]
and
\[
 \sum_{n=1}^N\|\AF(\vx_n)\Psi\|^2
 \leq NC_\chi\left(
 \dua{\Psi}{H_{0,\chi}^{0,0}\Psi}+1\right).
\]
Cauchy--Schwarz therefore gives a constant $C$, depending on $N$ and the
fixed form factor but not on $e$ or $\Psi$, such that
\[
 \left|
 \dua{\Psi}{H_{e,\chi}^{0,0}\Psi}
 -\dua{\Psi}{H_{0,\chi}^{0,0}\Psi}\right|
 \leq C(e+e^2)
 \left(\dua{\Psi}{H_{0,\chi}^{0,0}\Psi}+1\right).
\]
For all sufficiently small $e$, the same estimate can be rearranged as
\[
 \dua{\Psi}{H_{0,\chi}^{0,0}\Psi}+1
 \leq C\left(
 \dua{\Psi}{H_{e,\chi}^{0,0}\Psi}+1\right).
\]
It follows in particular that a sequence with bounded
$H_{e_k,\chi}^{0,0}$ energy has bounded zero-coupling energy and that
the difference of the two form expectations tends to zero along that
sequence.

\noindent\textbf{Step 2. The lower bound (i): electronic marginals.}
It suffices to consider a subsequence on which the liminf in (i) is
finite.  Choose normalized $\Psi_k$ satisfying
\[
 (\rho_{\Psi_k},\vjc_{\Psi_k})=(\rho_k,\vj_k),\qquad
 \dua{\Psi_k}{H_{e_k,\chi}^{0,0}\Psi_k}
 \leq F_\LL^{e_k,\chi}(\rho_k,\vj_k)+\frac1k.
\]
The preceding comparison gives a uniform zero-coupling energy bound.
Set
\[
 \Gamma_k=\Tr_\ph|\Psi_k\rangle\langle\Psi_k|.
\]
The zero-coupling factorization used in
\cref{thm-soft-photon-relaxation} yields
\[
 \dua{\Psi_k}{H_{0,\chi}^{0,0}\Psi_k}
 =\Tr(H_\el^0\Gamma_k)
 +\dua{\Psi_k}{H_\rad\Psi_k}.
\]
In particular, the electronic form traces and the photon energies are
uniformly bounded.

The electronic marginal has density $\rho_k$.  Its paramagnetic current density
is not exactly $\vj_k$, but \cref{jc-def} gives
\[
 \vjp_{\Gamma_k}=\vjp_{\Psi_k}
 =\vj_k-\vjph_{\Psi_k}.
\]
By \cref{ph-current-bound},
\[
 \int_{\RR^3}\frac{|\vjph_{\Psi_k}|^2}{\rho_k}\leq Ce_k^2.
\]
Since $\int_{\RR^3}\rho_k=N$ and the weakly convergent sequence
$(\rho_k)$ is bounded in $X$, hence in $L^3$, H\"older's inequality
gives
\[
 \|\vjph_{\Psi_k}\|_1
 \leq N^{1/2}\left(
 \int_{\RR^3}\frac{|\vjph_{\Psi_k}|^2}{\rho_k}\right)^{1/2},
\qquad
 \|\vjph_{\Psi_k}\|_{3/2}
 \leq\|\rho_k\|_3^{1/2}\left(
 \int_{\RR^3}\frac{|\vjph_{\Psi_k}|^2}{\rho_k}\right)^{1/2}.
\]
Consequently,
\[
 \vjph_{\Psi_k}\longrightarrow0\quad\text{in }Y,
\qquad
 \vjp_{\Gamma_k}\wconv\vj\quad\text{weakly in }Y.
\]

\noindent\textbf{Step 3. The lower bound (i): compactness of the marginals.}
We now use compactness only on the electronic space.  The densities
are uniformly tight: with $B_R=\{\vx:|\vx|<R\}$,
\[
 \lim_{R\to\infty}\sup_k\int_{B_R^c}\rho_k=0.
\]
Indeed, since $L^\infty\subset X^*$, testing $\rho_k\wconv\rho$
against the indicator of $B_R^c$ gives
\[
 \int_{B_R^c}\rho_k\longrightarrow\int_{B_R^c}\rho
\]
for each fixed $R$.  Choose $R$ first to make the latter integral
small, and then enlarge it to handle the finitely many initial indices;
the integrals on the left decrease as $R$ grows.
As in the proof of \cref{thm-continuum-compactness}, the elementary
inequality
\[
 \iden-\prod_{n=1}^N\iden_{B_R}(\vx_n)
 =\iden_{\{\max_n|\vx_n|\geq R\}}
 \leq\sum_{n=1}^N\iden_{B_R^c}(\vx_n),
\]
tested against $\Gamma_k$, converts this into a bound on the probability
that some electron lies outside $B_R$:
\[
 \Tr\!\left[
 \left(\iden-\prod_{n=1}^N\iden_{B_R}(\vx_n)\right)\Gamma_k\right]
 \leq\int_{B_R^c}\rho_k.
\]
Equip $Q(H_\el^0)$ with the Hilbert norm
generated by $\Phi\mapsto\dua{\Phi}{(H_\el^0+1)\Phi}$.  The uniform
electronic form bound means that, for every orthonormal basis $(\psi_j)$ of
$\Hil_\el^N$,
\begin{equation}\label{electronic-form-HS-bound}
 \sum_j\|\sqrt{\Gamma_k}\psi_j\|_{Q(H_\el^0)}^2
 =1+\Tr(H_\el^0\Gamma_k)\leq C.
\end{equation}
Choose $0\leq\eta_R\leq1$ smooth, equal to one on $B_R$ and supported in
$B_{2R}$, and put
$\chi_R(\vx_1,\ldots,\vx_N)=\prod_{n=1}^N\eta_R(\vx_n)$.  Then
\[
 \|(1-\chi_R)\sqrt{\Gamma_k}\|_{\mathfrak S_2}^2
 \leq\Tr((1-\chi_R^2)\Gamma_k)
 \leq\int_{B_R^c}\rho_k.
\]
For fixed $R$, local Rellich compactness makes multiplication by $\chi_R$
a compact map from $Q(H_\el^0)$ to $\Hil_\el^N$.  Hence, for every
$\delta>0$, there is a finite-rank orthogonal projection $S$ such that
\[
 \|(1-S)\chi_R\|_{Q(H_\el^0)\to\Hil_\el^N}<\delta.
\]
Now split
\[
 (1-S)\sqrt{\Gamma_k}
 =(1-S)(1-\chi_R)\sqrt{\Gamma_k}
 +(1-S)\chi_R\sqrt{\Gamma_k}.
\]
Since $\|1-S\|\leq1$, the Hilbert--Schmidt norm of the first term is at
most $\|(1-\chi_R)\sqrt{\Gamma_k}\|_{\mathfrak S_2}$.  For the second
term,
each $\sqrt{\Gamma_k}\psi_j$ lies in $Q(H_\el^0)$ by
\cref{electronic-form-HS-bound}, so the operator-norm bound on
$(1-S)\chi_R$ gives
\[
 \|(1-S)\chi_R\sqrt{\Gamma_k}\|_{\mathfrak S_2}^2
 =\sum_j\|(1-S)\chi_R\sqrt{\Gamma_k}\psi_j\|^2
 \leq\delta^2\sum_j\|\sqrt{\Gamma_k}\psi_j\|_{Q(H_\el^0)}^2
 \leq C\delta^2.
\]
Together,
\[
 \sup_k\|(1-S)\sqrt{\Gamma_k}\|_{\mathfrak S_2}
 \leq\sup_k\|(1-\chi_R)\sqrt{\Gamma_k}\|_{\mathfrak S_2}
 +C^{1/2}\delta.
\]
First choosing $R$ large and then $\delta$ small makes the right-hand side
arbitrarily small.  Since $1-S$ is an orthogonal projection,
\[
 \Tr((1-S)\Gamma_k)=\|(1-S)\sqrt{\Gamma_k}\|_{\mathfrak S_2}^2,
\]
so for every $\varepsilon>0$ there is a single finite-rank orthogonal
projection $S$ with $\sup_k\Tr((1-S)\Gamma_k)<\varepsilon$.  This is the
hypothesis of \cref{lem-trace-compression}.  By the compression estimate
\cref{density-matrix-compression}, every $\Gamma_k$ lies within
$2\sqrt{\varepsilon}$ in trace norm of its compression $S\Gamma_kS$.
These compressions are positive and have trace at most one, and each
$A=S\Gamma_kS$ satisfies $A=SAS$, that is, it vanishes on
$(\ran S)^\perp$ and maps into $\ran S$.
They are therefore operators on the fixed finite-dimensional space
$\ran S$, extended by zero, and so lie in a compact set.  Since $\varepsilon$ is arbitrary, $(\Gamma_k)$ is
precompact in trace norm.  After passing to a subsequence,
\[
 \Gamma_k\longrightarrow\Gamma\quad\text{in }\mathfrak S_1(\Hil_\el^N),
\qquad
 \Gamma\geq0,\qquad\Tr\Gamma=1.
\]

\noindent\textbf{Step 4. The lower bound (i): passage to the limit.}
We pass both the full electronic form and the constraints to this
limit.  The Powers--St\o rmer inequality gives
$\sqrt{\Gamma_k}\to\sqrt{\Gamma}$ in $\mathfrak S_2(\Hil_\el^N)$.  By
\cref{electronic-form-HS-bound}, $\sqrt{\Gamma_k}$, regarded as a Hilbert--Schmidt
operator with values in the form domain, has a weakly convergent
subsequence.  The continuous embedding
$Q(H_\el^0)\hookrightarrow\Hil_\el^N$ and the strong $\mathfrak S_2$
convergence identify its weak limit with $\sqrt{\Gamma}$.  Weak lower semicontinuity
therefore yields
\[
 \Tr(H_\el^0\Gamma)
 \leq\liminf_{k\to\infty}\Tr(H_\el^0\Gamma_k).
\]
This applies to the full electronic form, including the Coulomb
many-body interaction.

Moreover, $-\i\grad_{\vx_1}$ is a bounded map from $Q(H_\el^0)$ to
$(\Hil_\el^N)^3$.  Hence
\[
 (-\i\grad_{\vx_1})\sqrt{\Gamma_k}
 \rightharpoonup(-\i\grad_{\vx_1})\sqrt{\Gamma}
 \quad\text{in the corresponding Hilbert--Schmidt space}.
\]
Trace-norm convergence passes the density against every bounded
multiplication operator.  For every real
$\vA\in C_c^\infty(\RR^3,\RR^3)$, strong convergence of
$\vA(\vx_1)\sqrt{\Gamma_k}$ and the preceding weak convergence give
\[
 \begin{split}
 \int_{\RR^3}\vA\cdot\vjp_{\Gamma_k}
 &=N\Re\big(\vA(\vx_1)\sqrt{\Gamma_k},
 (-\i\grad_{\vx_1})\sqrt{\Gamma_k}\big)_{\mathfrak S_2}\\
 &\longrightarrow
 N\Re\big(\vA(\vx_1)\sqrt{\Gamma},
 (-\i\grad_{\vx_1})\sqrt{\Gamma}\big)_{\mathfrak S_2}
 =\int_{\RR^3}\vA\cdot\vjp_\Gamma.
 \end{split}
\]
Since $L^\infty\subset X^*$ and
$C_c^\infty(\RR^3,\RR^3)\subset Y^*$, the weak convergences
$\rho_k\wconv\rho$ in $X$ and $\vjp_{\Gamma_k}\wconv\vj$ in $Y$ also
hold against these test functions.  Since $\rho_k=\rho_{\Gamma_k}$,
uniqueness of limits therefore gives
\[
 (\rho_\Gamma,\vjp_\Gamma)=(\rho,\vj).
\]
Using the approximate minimality, the form comparison, and the
nonnegativity of the photon energy, we conclude that
\[
 \begin{split}
 \liminf_{k\to\infty}
 F_\LL^{e_k,\chi}(\rho_k,\vj_k)
 &\geq\liminf_{k\to\infty}\left(
 \dua{\Psi_k}{H_{e_k,\chi}^{0,0}\Psi_k}-\frac1k\right)\\
 &=\liminf_{k\to\infty}
 \dua{\Psi_k}{H_{0,\chi}^{0,0}\Psi_k}\\
 &\geq\Tr(H_\el^0\Gamma)
 \geq F_\VR(\rho,\vj).
 \end{split}
\]
This proves (i).

\noindent\textbf{Step 5. The recovery bound (ii).}
Suppose $F_\VR(\rho,\vj)<\infty$ and fix $\delta>0$.
Choose an admissible electronic density matrix $\Gamma_\delta$ with
\[
 \Tr(H_\el^0\Gamma_\delta)
 \leq F_\VR(\rho,\vj)+\delta.
\]
Apply the construction in \cref{one-photon-purification} as detailed in
Step 2 of the proof of \cref{thm-soft-photon-relaxation}, choosing all
photon tags in $\{0<|\vk|<\delta\}$.  The resulting normalized vector
$\Psi_\delta$ has electronic marginal $\Gamma_\delta$ and photon energy
at most $\delta$.  It lies entirely in the one-photon sector: for an
infinite-rank $\Gamma_\delta$, this follows because every partial sum has
that property and the one-photon sector is closed.  Since $\AF(\vx)$ maps
this sector into the orthogonal sum of the vacuum and two-photon sectors,
$\dua{\Psi_\delta}{\AF(\vx)\Psi_\delta}_{\vx}=0$ almost everywhere.  Thus
\[
 \rho_{\Psi_\delta}=\rho,\qquad
 \vjp_{\Psi_\delta}=\vj,\qquad
 \vjph_{\Psi_\delta}=0,\qquad
 \dua{\Psi_\delta}{H_\rad\Psi_\delta}\leq\delta.
\]
It therefore represents the same combined-current pair $(\rho,\vj)$ at
every coupling and belongs to $Q(H_{e,\chi}^{0,0})$ for every $e\geq0$
by the common form-domain identity.  The electronic derivative preserves
photon number, whereas $\AF(\vx_n)$ changes it by one, and hence
\[
 \Re\dua{(-\i\grad_{\vx_n})\Psi_\delta}{
 \AF(\vx_n)\Psi_\delta}=0.
\]
For almost every $\vX$, the photon factor is
$\Psi_\delta(\vX)=a^\dag(\eta)\Omega$ for a one-photon vector $\eta$
supported in $\{\omega<\delta\}$, with $\|\eta\|=\|\Psi_\delta(\vX)\|$.
The canonical commutation relations give
\[
 \|\Af^\mu(\vx)a^\dag(\eta)\Omega\|^2
 =\|G^\mu_{\vx}\|^2\|\eta\|^2+2|\dua{G^\mu_{\vx}}{\eta}|^2
 \leq\big(\|G^\mu_{\vx}\|^2
   +2\|\iden_{\{\omega<\delta\}}G^\mu_{\vx}\|^2\big)\|\eta\|^2.
\]
Summing over $\mu$ by \cref{Gx-norms}, applied also to
$\iden_{\{\omega<\delta\}}\wh\chi$, and integrating over the electronic
variables gives
\[
 \|\AF(\vx_n)\Psi_\delta\|^2
 \leq\|\wh\chi\|^2+2\|\iden_{\{\omega<\delta\}}\wh\chi\|^2.
\]
Expanding the positive-coupling form now gives
\[
 \begin{split}
 F_\LL^{e,\chi}(\rho,\vj)
 &\leq
 \dua{\Psi_\delta}{H_{e,\chi}^{0,0}\Psi_\delta}\\
 &=\Tr(H_\el^0\Gamma_\delta)
 +\dua{\Psi_\delta}{H_\rad\Psi_\delta}
 +\frac{e^2}{2}\sum_{n=1}^N
 \|\AF(\vx_n)\Psi_\delta\|^2\\
 &\leq F_\VR(\rho,\vj)+2\delta
 +\frac{e^2N}{2}\big(\|\wh\chi\|^2
   +2\|\iden_{\{\omega<\delta\}}\wh\chi\|^2\big).
 \end{split}
\]
Since $\wh\chi\in L^2(\RR^3)$, letting $\delta\downarrow0$ at fixed $e$
gives explicitly
\[
 F_\LL^{e,\chi}(\rho,\vj)
 \leq F_\VR(\rho,\vj)+\frac{N}{2}\|\wh\chi\|^2e^2,
\]
which is the quantitative estimate in (ii).  The same construction proves
$(\rho,\vj)\in\rcdens_N^{e,\chi}$ for every $e\geq0$.

\noindent\textbf{Step 6. The zero-coupling limit.}
Applying (i) to the constant constraints and combining it with (ii) gives
the asserted pointwise limit on the effective domain.  If
$F_\VR(\rho,\vj)=+\infty$, (i) applied to the constant
sequence shows that the limit is $+\infty$.  Finally, since
$\vjc=\vjp$ at $e=0$, \cref{electronic-projection} gives
$\rcdens_N^{0,\chi}=\rcdens_{N,\mathrm{mix}}^{\el}$, and
\cref{soft-photon-equality} identifies the limit with
$F_\LL^{0,\chi}$.
\end{proof}

\subsection{The long-wavelength (dipole) limit}

\phantomsection
\label{proof:thm-dipole-limit}
\begin{proof}[Proof of \cref{thm-dipole-limit}]
For a finite-rank positive operator $\vGamma$ whose
range lies in $\mathcal Q_\PF$, the density $\rho_\vGamma$ and the
current densities $\vjcmodel{\varepsilon}_{\vGamma}$ and
$\vjcdip_\vGamma$ are defined as for density matrices, also when
$\Tr\vGamma<1$.  All constants
below may depend on $N,e$, and the fixed form factor, but not on
$\varepsilon\in[0,1]$.

\noindent\textbf{Step 1. Uniform bounds and fixed-state convergence.}
The field bounds \cref{Gx-norms,A-Hrad-bound} are uniform in the
spatial argument.  The proof of \cref{PF-form-norm-equivalence}
therefore gives
\begin{equation}\label{dipole-proof-uniform-form-bound}
 c\|\Psi\|_{\mathcal Q_\PF}^2
 \leq \dua{\Psi}{H_\varepsilon^{0,0}\Psi}+\|\Psi\|^2
 \leq C\|\Psi\|_{\mathcal Q_\PF}^2.
\end{equation}
In particular, the form-domain identification is uniform in
$\varepsilon$.

Define the modulus
\begin{equation}\label{dipole-kernel-modulus}
 d(r)=\sum_{\mu=1}^3\sup_{|\vy|\leq r}
 \left(\|G^\mu_{\vy}-G^\mu_{0}\|
       +\left\|\frac{G^\mu_{\vy}-G^\mu_{0}}{\sqrt\omega}\right\|\right).
\end{equation}
The continuity following \cref{Gx-norms} implies $d(r)\to0$ as
$r\downarrow0$.  If
$\Psi$ is supported where $|\vx_n|\leq2L$ for all $n$, then
\begin{equation}\label{dipole-local-field-error}
 \sum_{n=1}^N
 \|(\AF(\varepsilon\vx_n)-\AF(0))\Psi\|^2
 \leq C d(2\varepsilon L)^2
 \big(\|\sqrt{H_\rad}\Psi\|^2+\|\Psi\|^2\big).
\end{equation}
Indeed, for almost every $\vX$ in the support of $\Psi$, we have
$\Psi(\vX)\in D(\sqrt{H_\rad})$ and $|\varepsilon\vx_n|\leq2\varepsilon L$
for every $n$.  Fix such $\vX$, an index $n$, and a component $\mu$, and
put $g=G^\mu_{\varepsilon\vx_n}-G^\mu_{0}$.  Since the Segal field is
real-linear in its argument,
\[
 \big(\Af^\mu(\varepsilon\vx_n)-\Af^\mu(0)\big)\Psi(\vX)
 =\big(a^\dag(g)+a(g)\big)\Psi(\vX),
\]
and the definition \cref{dipole-kernel-modulus} gives
$\|g\|+\|g/\sqrt\omega\|\leq d(2\varepsilon L)$.  The bounds
\cref{ann-Hrad-bound,cre-Hrad-bound}, applied with $g$ in place of
$G^\mu_{\vx}$, therefore yield
\[
 \begin{split}
 \big\|\big(\Af^\mu(\varepsilon\vx_n)-\Af^\mu(0)\big)\Psi(\vX)\big\|
 &\leq2\|g/\sqrt\omega\|\,\|\sqrt{H_\rad}\Psi(\vX)\|+\|g\|\,\|\Psi(\vX)\|\\
 &\leq2d(2\varepsilon L)
 \big(\|\sqrt{H_\rad}\Psi(\vX)\|+\|\Psi(\vX)\|\big).
 \end{split}
\]
Integrating over $\vX$, and using
$\|\sqrt{H_\rad}\Psi\|^2=\int\|\sqrt{H_\rad}\Psi(\vX)\|^2\dd\vX$ because
$H_\rad$ acts only on the photon variables, gives
\cref{dipole-local-field-error} with $C=24N$.
Expanding the differences of the squared covariant-derivative norms,
and using \cref{dipole-proof-uniform-form-bound}, gives
\begin{equation}\label{dipole-local-form-error}
 |\dua{\Psi}{H_\varepsilon^{0,0}\Psi}-\dua{\Psi}{H_\dip \Psi}|
 \leq C\big(d(2\varepsilon L)+d(2\varepsilon L)^2\big)
             \big(\dua{\Psi}{H_\varepsilon^{0,0}\Psi}+\|\Psi\|^2\big).
\end{equation}
The same estimates hold after summing over any finite-rank positive operator
with this electronic support.

For a fixed $\Psi\in\mathcal Q_\PF$, the field difference instead
converges without a support assumption:
\begin{equation}\label{dipole-fixed-state-form-convergence}
 \|(\AF(\varepsilon\vx_n)-\AF(0))\Psi\|\longrightarrow0,
 \qquad \dua{\Psi}{H_\varepsilon^{0,0}\Psi}\longrightarrow \dua{\Psi}{H_\dip \Psi}.
\end{equation}
Indeed, for almost every electronic configuration $\vX$,
$\Psi(\vX)\in D(\sqrt{H_\rad})$.  The estimates
\cref{ann-Hrad-bound,cre-Hrad-bound} and the continuity of the kernels give pointwise convergence on these fibers, dominated
in squared norm by
$C(\|\sqrt{H_\rad}\Psi(\vX)\|^2+\|\Psi(\vX)\|^2)$.
Integration and then expansion of the kinetic form prove
\cref{dipole-fixed-state-form-convergence}.

The current convergence is in the full space $Y$.  For a normalized
pure state, put
\[
 b_{\varepsilon,\Psi}(\vx)
 =N\|(\AF(\varepsilon\vx)-\AF(0))\Psi\|_{\vx}^2,
\]
where the partial norm $\|\cdot\|_{\vx}$ is summed over the three
field components.  Pointwise Cauchy--Schwarz and H\"older's inequality give
\begin{align}
 |\vjcmodel{\varepsilon}_{\Psi}-\vjcdip_{\Psi}|
 &\leq e\sqrt{\rho_\Psi}\sqrt{b_{\varepsilon,\Psi}},
 \notag\\
 \|\vjcmodel{\varepsilon}_{\Psi}-\vjcdip_{\Psi}\|_Y
 &\leq e\big(\sqrt N+\|\sqrt{\rho_\Psi}\|_6\big)
                 \|\sqrt{b_{\varepsilon,\Psi}}\|.
 \label{dipole-current-difference-bound}
\end{align}
For a density matrix, define $b_{\varepsilon,\vGamma}$ by linear extension.
Cauchy--Schwarz over the state labels proves the same estimates with
$\Psi$ replaced by $\vGamma$.  The Hilbert-space-valued Sobolev
inequality bounds $\|\sqrt{\rho_\vGamma}\|_6$ by the square root of
the ordinary electronic kinetic energy.  Consequently
\begin{equation}\label{dipole-fixed-ensemble-convergence}
 \vjcmodel{\varepsilon}_{\vGamma}\longrightarrow
 \vjcdip_\vGamma\quad\text{in }Y,
 \qquad \Tr(H_\varepsilon^{0,0}\vGamma)\longrightarrow \Tr(H_\dip \vGamma)
\end{equation}
for every fixed finite-rank density matrix of finite form energy.
Moreover, by \cref{dipole-local-field-error}, the current difference
in \cref{dipole-current-difference-bound} tends to zero uniformly on
electronically supported density matrices of bounded energy whenever
$\varepsilon L\to0$.

We will also use the uniform bound
\begin{equation}\label{dipole-energy-density-current-bounds}
 \|\rho_\vGamma\|_3+
 \|\vjcmodel{\varepsilon}_{\vGamma}\|_{3/2}
 +\|\vjcmodel{\varepsilon}_{\vGamma}\|_1
 \leq C\big(1+\Tr(H_\varepsilon^{0,0}\vGamma)\big)
\end{equation}
for every density matrix $\vGamma$ with
$\Tr(H_\varepsilon^{0,0}\vGamma)<\infty$.
To see this, regard the weighted state vectors of a density matrix as a
single vector in a Hilbert-space direct sum.  Its electronic derivative
is controlled by \cref{dipole-proof-uniform-form-bound}, and its
covariant derivative by $\Tr(H_\varepsilon^{0,0}\vGamma)$.  Sobolev and H\"older then
give \cref{dipole-energy-density-current-bounds}, exactly as for pure
states in \cref{prop-current-bounds}.

\noindent\textbf{Step 2. Finite-rank approximation of the convex relaxation.}
By \cref{dipole-convex-envelope}, the epigraph of
$F_{\LI,\varepsilon}$ is the norm-closed convex hull of the epigraph
of $F_{\LL,\varepsilon}$.  Thus, whenever
$F_{\LI,\varepsilon}(\rho,\vjc)<\infty$ and $\delta>0$, there is a finite-rank
density matrix $\vGamma$, made from states in
$\mathcal Q_\PF$, such that
\begin{equation}\label{dipole-epigraph-approximation}
 \|\rho_\vGamma-\rho\|_X
 +\|\vjcmodel{\varepsilon}_{\vGamma}-\vjc\|_Y<\delta,
 \qquad
 \Tr(H_\varepsilon^{0,0}\vGamma)
 <F_{\LI,\varepsilon}(\rho,\vjc)+\delta.
\end{equation}
To obtain this, approximate an epigraph point by a finite convex
combination of epigraph points of $F_{\LL,\varepsilon}$, and choose
an almost minimizer in each of the finitely many pure-state searches.
No attainment of either search is used.  Conversely, convexity and
$F_{\LI,\varepsilon}\leq F_{\LL,\varepsilon}$ give
\begin{equation}\label{dipole-finite-ensemble-upper-bound}
 F_{\LI,\varepsilon}(\rho_\vGamma,\vjcmodel{\varepsilon}_{\vGamma})
 \leq \Tr(H_\varepsilon^{0,0}\vGamma)
\end{equation}
for every such finite-rank density matrix.

\noindent\textbf{Step 3. Lower bound along weakly convergent sequences.}
Suppose $\varepsilon_k\to0$ and $(\rho_k,\vjc_k)\rightharpoonup(\rho,\vjc)$
weakly in $X\times Y$.  If the energy liminf is infinite, there is nothing to prove.
Otherwise pass to a subsequence on which the liminf is a finite limit
and all the functional values are finite.  The case of infinitely
many $\varepsilon_k=0$ follows directly from weak lower
semicontinuity of $F_\dip$, so we may assume
$\varepsilon_k>0$.  By \cref{dipole-epigraph-approximation}, choose
finite-rank density matrices $\vGamma_k$ such that
\begin{equation}\label{dipole-liminf-ensemble-choice}
 \|\rho_{\vGamma_k}-\rho_k\|_X
 +\|\vjcmodel{\varepsilon_k}_{\vGamma_k}-\vjc_k\|_Y
 \leq k^{-1},
 \qquad
 \Tr(H_{\varepsilon_k}^{0,0}\vGamma_k)
 \leq F_{\LI,\varepsilon_k}(\rho_k,\vjc_k)+k^{-1}.
\end{equation}
Their energies are uniformly bounded.  Since $L^\infty\subset X^*$,
$\rho_{\vGamma_k}$ converges weakly in $L^1$ to $\rho$.  Positivity
and the test function $1$ show
$\rho\geq0$ and $\int_{\RR^3}\rho=N$.  Testing against the indicator of
the complement of a fixed ball gives
\[
 \int_{|\vx|>R}\rho_{\vGamma_k}\longrightarrow
 \int_{|\vx|>R}\rho
\]
for every fixed $R>0$.  Now let $\eta>0$.  Since $\rho\in L^1$, there is
$R_0$ with $\int_{|\vx|>R_0}\rho<\eta$, and the convergence just
displayed, with $R=R_0$, gives an index $K$ such that
$\int_{|\vx|>R_0}\rho_{\vGamma_k}<\eta$ for all $k\geq K$.  Each of
the finitely many densities $\rho_{\vGamma_1},\dots,\rho_{\vGamma_{K-1}}$
is integrable, so there is $R_1\geq R_0$ such that
$\int_{|\vx|>R_1}\rho_{\vGamma_k}<\eta$ also for $k<K$.  Since these
tail integrals do not increase with the radius,
$\sup_k\int_{|\vx|>R}\rho_{\vGamma_k}\leq\eta$ for every $R\geq R_1$.
As $\eta>0$ was arbitrary, this proves
\begin{equation}\label{dipole-liminf-density-tightness}
 \lim_{R\to\infty}\sup_k\int_{|\vx|>R}\rho_{\vGamma_k}=0.
\end{equation}

Choose a smooth real $h:[0,\infty)\to[0,\pi/2]$, equal to zero
on $[0,1]$ and to $\pi/2$ on $[2,\infty)$.  With
$|\vX|^2=\sum_{n=1}^N|\vx_n|^2$, define, for $\varepsilon>0$,
\[
 \theta_\varepsilon(\vX)=\cos h(\sqrt\varepsilon\,|\vX|),\qquad
 \wt\theta_\varepsilon(\vX)=\sin h(\sqrt\varepsilon\,|\vX|).
\]
Both multipliers are smooth and permutation invariant, and
$\theta_\varepsilon$ equals $1$ where $|\vX|\leq\varepsilon^{-1/2}$ and
vanishes where $|\vX|\geq2\varepsilon^{-1/2}$.  They satisfy
\[
 \theta_\varepsilon^2+\wt\theta_\varepsilon^2=1,\qquad
 \sum_{n=1}^N\big(|\grad_{\vx_n}\theta_\varepsilon|^2+
       |\grad_{\vx_n}\wt\theta_\varepsilon|^2\big)\leq C\varepsilon.
\]
Since the multipliers commute with the field and the interaction,
the covariant IMS localization formula gives, for every $\Psi\in\mathcal Q_\PF$,
\begin{equation}\label{dipole-IMS-identity}
 \begin{split}
 &\dua{\theta_\varepsilon\Psi}{H_\varepsilon^{0,0}\theta_\varepsilon\Psi}
  +\dua{\wt\theta_\varepsilon\Psi}{H_\varepsilon^{0,0}\wt\theta_\varepsilon\Psi}\\
 &\qquad=\dua{\Psi}{H_\varepsilon^{0,0}\Psi}
 +\frac12\int_{\RR^{3N}}
 \sum_{n=1}^N\big(|\grad_{\vx_n}\theta_\varepsilon|^2+
                 |\grad_{\vx_n}\wt\theta_\varepsilon|^2\big)
                     \|\Psi(\vX)\|^2\dd\vX.
 \end{split}
\end{equation}
The identity follows first on a form core and then by form-norm
approximation.

We now show that the $\vGamma_k$ may be assumed to be supported where
every $|\vx_n|\leq2\varepsilon_k^{-1/2}$.  By
\cref{dipole-liminf-ensemble-choice},
\begin{equation}\label{dipole-liminf-weak-properties}
 (\rho_{\vGamma_k},\vjcmodel{\varepsilon_k}_{\vGamma_k})
 \rightharpoonup(\rho,\vjc)\quad\text{in }X\times Y,
 \qquad
 \Tr(H_{\varepsilon_k}^{0,0}\vGamma_k)
 \leq F_{\LI,\varepsilon_k}(\rho_k,\vjc_k)+o(1),
\end{equation}
and we check that both properties survive localization by
$\theta_{\varepsilon_k}$ followed by normalization.  If
$|\vX|>\varepsilon_k^{-1/2}$, then $|\vx_n|>(N\varepsilon_k)^{-1/2}$ for
some $n$.  Hence \cref{dipole-liminf-density-tightness} gives
\begin{equation*}
 0\leq\Tr(\wt\theta_{\varepsilon_k}^2\vGamma_k)
 \leq\int_{|\vx|>(N\varepsilon_k)^{-1/2}}\rho_{\vGamma_k}(\vx)\dd\vx
 \longrightarrow0.
\end{equation*}
In particular, $\Tr(\theta_{\varepsilon_k}^2\vGamma_k)
=1-\Tr(\wt\theta_{\varepsilon_k}^2\vGamma_k)$ is positive for all large
$k$, and we discard the remaining indices.  Positivity of the
complementary localized form in \cref{dipole-IMS-identity} yields
\[
 \frac{\Tr(H_{\varepsilon_k}^{0,0}\theta_{\varepsilon_k}\vGamma_k\theta_{\varepsilon_k})}
      {\Tr(\theta_{\varepsilon_k}^2\vGamma_k)}
 \leq\frac{\Tr(H_{\varepsilon_k}^{0,0}\vGamma_k)+C\varepsilon_k}
          {\Tr(\theta_{\varepsilon_k}^2\vGamma_k)}
 =\Tr(H_{\varepsilon_k}^{0,0}\vGamma_k)+o(1),
\]
since the energies are uniformly bounded.  Combined with the energy bound in
\cref{dipole-liminf-weak-properties}, this shows that the normalized
localized density matrices satisfy that energy bound as well.

For the constraints, the density difference
$\rho_{\vGamma_k}-\rho_{\theta_{\varepsilon_k}\vGamma_k\theta_{\varepsilon_k}}$
is nonnegative and has mass $N\Tr(\wt\theta_{\varepsilon_k}^2\vGamma_k)$.
For the current density, differentiation of the real multiplier
$\theta_{\varepsilon_k}$ contributes a purely imaginary term, which
disappears from its defining real part.  Cauchy--Schwarz, including the
state labels, thus gives
\begin{equation*}
 \begin{split}
 &\|\vjcmodel{\varepsilon_k}_{\vGamma_k}
   -\vjcmodel{\varepsilon_k}_{\theta_{\varepsilon_k}\vGamma_k\theta_{\varepsilon_k}}\|_1\\
 &\qquad\leq\big(N\Tr(\wt\theta_{\varepsilon_k}^2\vGamma_k)\big)^{1/2}
 \left(\sum_{n=1}^N
       \Tr(\vD_{n,\varepsilon_k}^\dag\vD_{n,\varepsilon_k}\vGamma_k)
 \right)^{1/2}\longrightarrow0.
 \end{split}
\end{equation*}
The trace in this display denotes the squared-derivative form trace.
Division by $\Tr(\theta_{\varepsilon_k}^2\vGamma_k)$ changes the
density by another $N\Tr(\wt\theta_{\varepsilon_k}^2\vGamma_k)$ in
$L^1$ and the current density by $o(1)$ in $L^1$, by the energy bounds.
The original and localized densities are uniformly bounded in $L^3$,
and the corresponding current densities in $L^{3/2}$, by
\cref{dipole-energy-density-current-bounds}.

If $f_k\to0$ in $L^1$ and $(f_k)$ is bounded in $L^p$, $1<p<\infty$,
then $f_k\rightharpoonup0$ in $L^1\cap L^p$.  Indeed, the
$L^\infty$ part of a dual test is controlled by $L^1$ convergence.
For the $L^{p'}$ part, approximate it in that norm by bounded functions and
use the uniform $L^p$ bound.  Applying this observation to the
localization errors proves that they tend weakly to zero in $X\times Y$.
Thus the normalized localized density matrices
$\theta_{\varepsilon_k}\vGamma_k\theta_{\varepsilon_k}
/\Tr(\theta_{\varepsilon_k}^2\vGamma_k)$ also satisfy
\cref{dipole-liminf-weak-properties}, and they are supported where
every $|\vx_n|\leq2\varepsilon_k^{-1/2}$.  Replacing $\vGamma_k$ by
them, we may assume that $\vGamma_k$ has this support.

Apply \cref{dipole-local-form-error} with $L=\varepsilon_k^{-1/2}$, so
that $\varepsilon_kL=\sqrt{\varepsilon_k}\to0$.  Since the energies
are uniformly bounded, this gives
\begin{equation}\label{dipole-localized-energy-comparison}
 \Tr(H_\dip\vGamma_k)\leq \Tr(H_{\varepsilon_k}^{0,0}\vGamma_k)+o(1).
\end{equation}
Similarly, \cref{dipole-local-field-error,dipole-current-difference-bound}
give
\[
 \|\vjcmodel{\varepsilon_k}_{\vGamma_k}-\vjcdip_{\vGamma_k}\|_Y\longrightarrow0.
\]
Together with \cref{dipole-liminf-weak-properties}, this establishes
\begin{equation*}
 (\rho_{\vGamma_k},\vjcdip_{\vGamma_k})\rightharpoonup(\rho,\vjc)
 \quad\text{in }X\times Y.
\end{equation*}
The convex norm-lower-semicontinuous functional $F_\dip$ is weakly
lower semicontinuous.  Hence, using
\cref{dipole-localized-energy-comparison} and the energy bound in
\cref{dipole-liminf-weak-properties},
\[
 \begin{split}
 F_\dip(\rho,\vjc)
 &\leq\liminf_kF_\dip(\rho_{\vGamma_k},\vjcdip_{\vGamma_k})
 \leq\liminf_k\Tr(H_\dip\vGamma_k)\\
 &\leq\liminf_k\Tr(H_{\varepsilon_k}^{0,0}\vGamma_k)
 \leq\liminf_kF_{\LI,\varepsilon_k}(\rho_k,\vjc_k).
 \end{split}
\]
This proves \cref{dipole-functional-liminf}.

\noindent\textbf{Step 4. Recovery sequences converging in norm.}
Let $F_\dip(\rho,\vjc)<\infty$.  By
\cref{dipole-epigraph-approximation}, choose finite-rank density
matrices $\vGamma^m$ such that
\[
 \|\rho_{\vGamma^m}-\rho\|_X+\|\vjcdip_{\vGamma^m}-\vjc\|_Y\leq m^{-1},\qquad
 \Tr(H_\dip \vGamma^m)\leq F_\dip(\rho,\vjc)+m^{-1}.
\]
Using \cref{dipole-fixed-ensemble-convergence}, choose a strictly
decreasing sequence $\delta_m\downarrow0$ such that, for
$0<\varepsilon\leq\delta_m$,
\[
 \|\vjcmodel{\varepsilon}_{\vGamma^m}-\vjcdip_{\vGamma^m}\|_Y
       \leq m^{-1},\qquad
 |\Tr(H_\varepsilon^{0,0}\vGamma^m)-\Tr(H_\dip \vGamma^m)|\leq m^{-1}.
\]
For $\delta_{m+1}<\varepsilon\leq\delta_m$, put
$(\rho_\varepsilon,\vjc_\varepsilon)
=(\rho_{\vGamma^m},\vjcmodel{\varepsilon}_{\vGamma^m})$.
Define $(\rho_\varepsilon,\vjc_\varepsilon)$ arbitrarily for the
remaining, larger parameters.  Then
$(\rho_\varepsilon,\vjc_\varepsilon)\to(\rho,\vjc)$ in $X\times Y$, and
\cref{dipole-finite-ensemble-upper-bound} gives
\[
 \limsup_{\varepsilon\downarrow0}
 F_{\LI,\varepsilon}(\rho_\varepsilon,\vjc_\varepsilon)
 \leq\limsup_{\varepsilon\downarrow0}
                  \Tr(H_\varepsilon^{0,0}\vGamma^{m(\varepsilon)})
 \leq F_\dip(\rho,\vjc).
\]
If $F_\dip(\rho,\vjc)=+\infty$, the constant sequence is a recovery
sequence for the stated upper inequality.  This proves
\cref{dipole-functional-recovery}.  The lower bound of Step~3 applies in
particular to norm-convergent sequences, so the result also implies
norm-Gamma convergence.
\end{proof}

\phantomsection
\label{proof:prop-dipole-electronic-comparison}
\begin{proof}[Proof of \cref{prop-dipole-electronic-comparison}]
Let $\Psi\in\mathcal Q_\PF$ be normalized, with $U_{\mathrm d}$ defined
in \cref{dipole-length-unitary}.  By \cref{prop-dipole-length-gauge},
$U_{\mathrm d}^\dag\Psi$ has finite electronic kinetic energy,
$\rho_{U_{\mathrm d}^\dag\Psi}=\rho_\Psi$, and
$\vjp_{U_{\mathrm d}^\dag\Psi}=\vjcdip_\Psi$.
The Coulomb multiplication form is unchanged.  Hence the electronic
marginal
$\gamma=\Tr_\ph|U_{\mathrm d}^\dag\Psi\rangle\langle U_{\mathrm d}^\dag\Psi|$
satisfies
\begin{equation}\label{dipole-electronic-lower-proof}
 \begin{split}
 F_\VR(\rho_\Psi,\vjcdip_{\Psi})
 &\leq\Tr(H_\el^0\gamma)\\
 &=\frac12\sum_{n=1}^N\|\vD_{n,\dip}\Psi\|^2
   +\dua{\Psi}{W\Psi}
 \leq\dua{\Psi}{H_\dip\Psi},
 \end{split}
\end{equation}
where $W=\sum_{n<m}|\vx_n-\vx_m|^{-1}$.  The equality uses
\cref{dipole-length-derivative}, and the last inequality drops the
nonnegative radiation energy $\|\sqrt{H_\rad}\Psi\|^2$.

Taking the pure constrained-search infimum in
\cref{dipole-electronic-lower-proof} gives
$F_\VR\leq F_{\LL,\dip}$.  The electronic mixed-state functional is
convex by mixing, and norm lower semicontinuous by the electronic
compactness argument in the proof of
\cref{thm-zero-coupling-convergence}.  It is therefore bounded above
by the norm-lower-semicontinuous convex envelope of $F_{\LL,\dip}$.
By \cref{dipole-convex-envelope}, this proves
$F_\VR\leq F_\dip$.

For the upper bound, take any finite-energy electronic density matrix
$\gamma$ representing $(\rho,\vjcdip)$ and tensor it with the photon
vacuum:
\[
 \vGamma=\gamma\otimes|\Omega\rangle\langle\Omega|.
\]
The vacuum field expectation vanishes, so
$(\rho_\vGamma,\vjcdip_\vGamma)=(\rho,\vjcdip)$.  The cross terms in the
covariant kinetic form also vanish.  Since
$\Af^\mu(0)\Omega=a^\dag(G^\mu_{0})\Omega$,
\cref{Gx-norms} gives exactly
\begin{equation}\label{dipole-vacuum-upper-proof}
 \Tr(H_\dip\vGamma)=\Tr(H_\el^0\gamma)
       +\frac{Ne^2}{2}\sum_{\mu=1}^3\|G^\mu_{0}\|^2
 =\Tr(H_\el^0\gamma)+\frac{Ne^2}{2}\|\wh\chi\|^2.
\end{equation}
In particular, $\vGamma$ has finite energy, so it is a competitor in
the mixed-state search \cref{dipole-ensemble-search} at
$(\rho,\vjcdip)$.  Hence \cref{thm-dipole-ensemble-realization} gives
\[
 F_\dip(\rho,\vjcdip)=\wt F_\dip(\rho,\vjcdip)\leq\Tr(H_\dip\vGamma).
\]
Taking the infimum over $\gamma$ in
\cref{dipole-vacuum-upper-proof} proves the upper inequality in
\cref{dipole-electronic-comparison}.  The two inequalities imply
\cref{dipole-effective-domain}, and the hard-cutoff value follows from
\cref{harduv-norms}.
\end{proof}

\begin{proof}[Proof of \cref{thm-dipole-ensemble-realization}]
\phantomsection
\label{proof:thm-dipole-ensemble-realization}
Fix a finite particle number $N\in\NN$.  First consider the mixed-state
search \cref{dipole-ensemble-search}.  Factor the photon Fock space at
$\omega=\kappa$, and write
\[
 \AF(0)=\AF_{\geq\kappa}(0)+\AF_{<\kappa}(0),\qquad
 H_\rad=H_{\rad,\geq\kappa}+H_{\rad,<\kappa}.
\]
As in the paragraph preceding \cref{dipole-length-unitary}, \cref{prop-weyl-algebra}(iii)
shows that the three components of each constant field strongly
commute, and fields acting on the low- and high-frequency factors also
commute.  With $\vR=\sum_{n=1}^N\vx_n$, put
\begin{equation}\label{dipole-exact-soft-channel}
 V_\kappa=\exp(-\i e\vR\cdot\AF_{<\kappa}(0)),\qquad
 \vGamma_\kappa=
 \big(\Tr_{<\kappa}V_\kappa\vGamma V_\kappa^\dag\big)
 \otimes|\Omega_{<\kappa}\rangle\langle\Omega_{<\kappa}|.
\end{equation}
This is unitary conjugation followed by a partial trace and vacuum
preparation, and hence defines a positive trace-preserving channel.
The multiplier $V_\kappa$ is permutation invariant, so this channel
preserves the antisymmetric electronic space.

The weak-derivative argument used for
\cref{dipole-unitary-gradient-proof} gives
\[
 (-\i\grad_{\vx_n}-e\AF_{\geq\kappa}(0))V_\kappa\Psi
   =V_\kappa\vD_{n,\dip}\Psi,\qquad 1\leq n\leq N.
\]
The high-mode radiation energy commutes with $V_\kappa$.  Pointwise
unitarity preserves the full electronic configuration density and hence
the Coulomb energy.  Taking the low-mode partial trace preserves the
density and combined current density, while vacuum preparation adds the low-mode
vacuum variance to each of the $N$ kinetic terms.  Writing
$b_\kappa=\|\iden_{\{\omega<\kappa\}}\wh\chi\|$ and
$E_{<\kappa}(\vGamma)=\Tr(H_{\rad,<\kappa}\vGamma)$, we obtain the
exact identities
\begin{equation}\label{dipole-exact-soft-cost}
 \begin{split}
 \rho_{\vGamma_\kappa}&=\rho_\vGamma,
 \qquad\vjcdip_{\vGamma_\kappa}=\vjcdip_{\vGamma},\\
 \Tr(H_\dip\vGamma_\kappa)
 &=\Tr(H_\dip\vGamma)-E_{<\kappa}(\vGamma)
                      +\frac{Ne^2}{2}b_\kappa^2.
 \end{split}
\end{equation}
In particular the output has finite form energy.  No low-mode
radiation form bound on $V_\kappa\Psi$ is needed, since that factor
is traced out.  This observation avoids imposing any electronic
position moment.

Let $m=\widetilde F_\dip(\rho,\vjcdip)<\infty$, and let $\vGamma$
represent $(\rho,\vjcdip)$ with $\Tr(H_\dip\vGamma)\leq m+\delta$.
Since the channel preserves the density and combined current density,
\cref{dipole-exact-soft-cost} implies, for every $t>0$,
\begin{equation*}
 E_{<t}(\vGamma)\leq\delta+\frac{Ne^2}{2}b_t^2.
\end{equation*}
The high-mode photon marginal is unchanged by the channel.  Thus,
for $0<\kappa\leq1$, integration of the one-photon energy measure
against $\omega^{-1}\iden_{\{\omega\geq\kappa\}}$ gives
\begin{align}
 \Tr(\NC\vGamma_\kappa)
 &=\Tr\!\left(\dd\varGamma(\iden_{\{\omega\geq\kappa\}})\vGamma\right)
 \notag\\
 &\leq\Tr(H_\rad\vGamma)
       +\int_\kappa^1\frac{E_{<t}(\vGamma)}{t^2}\dd t
 \notag\\
 &\leq\Tr(H_\dip\vGamma)+\frac\delta\kappa
       +\frac{Ne^2}{2}
        \left\|\frac{\iden_{\{\omega<1\}}\wh\chi}{\sqrt\omega}\right\|^2.
 \label{dipole-soft-number-bound}
\end{align}
For the last step use Tonelli's theorem on
$\int_0^1b_t^2t^{-2}\dd t$.

Choose $\kappa_l\downarrow0$ and almost minimizers with
$\delta_l\leq\kappa_l$.  Their images under
\cref{dipole-exact-soft-channel} have the prescribed density and
combined current density,
uniformly bounded mean photon number, and energies at most
$m+\delta_l+(Ne^2/2)b_{\kappa_l}^2\to m$.
The photon-loss compactness argument of
\cref{thm-continuum-compactness} applies also to the constant
coupling kernels $G^\mu_{0}$.  To check this directly, its form
bounds and electronic localization are unchanged, and the kernel
compactness step uses the singleton $\{G^\mu_{0}\}$ in place of
$\{G^\mu_{\vx}:|\vx|\leq2L\}$, so the same photon-localization error
tends to zero.  The uniform mean-number bound supplies
its local-sector tightness condition.  The fixed electronic density
supplies electronic tightness.  Consequently it produces a normalized
finite-energy density matrix with density $\rho$, combined current
density $\vjcdip$, and energy at most $m$.  This proves attainment.  The retained limit also has finite
mean photon number by lower semicontinuity through the localized
trace-norm limits.

The same argument proves norm lower semicontinuity.  Given pairs
$(\rho_l,\vjcdip_l)$ converging in $X\times Y$ with bounded mixed-state
infima, choose an almost minimizer at each of them with error $\delta_l\leq\kappa_l\to0$ and apply the
channel at $\kappa_l$.  The output constraints are unchanged, the
energies increase by at most $(Ne^2/2)b_{\kappa_l}^2\to0$, and
\cref{dipole-soft-number-bound} is uniform.  Norm convergence of the
densities gives electronic tightness, so the same compactness
argument produces a density matrix at the limiting pair with energy at
most the original liminf.  Thus $\widetilde F_\dip$ is norm
lower semicontinuous.  It is convex by mixing, and
\[
 F_\dip\leq\widetilde F_\dip\leq F_{\LL,\dip}.
\]
The closed-convex-envelope identity
\cref{dipole-convex-envelope} now gives
$\widetilde F_\dip=F_\dip$.

It remains to compare pure-state and mixed-state infima.  The dipole
field leaves enough photon modes uncoupled to purify a density matrix at
arbitrarily small energy.  Indeed, in radial coordinates write
\[
 \Hil_\ph=L^2((0,\infty),r^2\dd r;
                   L^2(S^2\times\{1,2\})).
\]
Let $p_{\rm b}(r)$ be the measurable orthogonal projection onto
$\Span\{G^\mu_{0}(r,\cdot):\mu=1,2,3\}$, and set
$\Hil_{\rm b}=\operatorname{Ran}p_{\rm b}$ and
$\Hil_{\rm d}=\Hil_{\rm b}^{\perp}$.
Both spaces reduce multiplication by $\omega=r$, and every coupling
kernel belongs to $\Hil_{\rm b}$.  Consequently, in the Fock
factorization over $\Hil_{\rm b}\oplus\Hil_{\rm d}$, the field acts
only on the first factor.  Each angular fiber of $\Hil_{\rm d}$
is infinite dimensional, because $p_{\rm b}(r)$ has rank at most
three.  Thus, for every $a>0$, the dark subspace with
$0<\omega<a$ contains an orthonormal sequence $(\xi_l)$.

Given a finite-energy density matrix, trace out its dark factor.  This
preserves its density and combined current density and can only
decrease its energy.  Write the
resulting electronic--bright density matrix on
$\Hil_\el^N\otimes\Fock_s(\Hil_{\rm b})$ spectrally as
$\sum_l p_l|\Phi_l\rangle\langle\Phi_l|$ and purify it by
\[
 \Psi_a=\sum_l\sqrt{p_l}\,\Phi_l\otimes
                  a^\dag(\xi_l)\Omega_{\rm d}.
\]
Each $\Phi_l$ is antisymmetric in the electronic variables, and so is
$\Psi_a$.  Orthogonality of the photon tags preserves the density, the combined
current density, and the electronic interaction energy exactly.  Their
energies are at most $a$, so $\Psi_a$ has form energy at most the
original energy plus $a$.  Letting $a\downarrow0$ gives
$F_{\LL,\dip}\leq\widetilde F_\dip$, and the reverse inequality follows
by including rank-one density matrices.  Therefore
$F_{\LL,\dip}=\widetilde F_\dip=F_\dip$.
The construction proves equality of the infima, without asserting
that a pure minimizer exists.
\end{proof}

\phantomsection
\label{proof:cor-dipole-length-gauge}
\begin{proof}[Proof of \cref{cor-dipole-length-gauge}]
\noindent\textbf{Step 1. The radiation-energy form.}
Let $U_{\mathrm d}$ be the unitary \cref{dipole-length-unitary}.  By
\cref{prop-dipole-length-gauge}, $\Phi\mapsto\Psi=U_{\mathrm d}^\dag\Phi$
maps $\mathcal Q_\PF$ bijectively onto the vectors $\Psi$ with
$U_{\mathrm d}\Psi\in\mathcal Q_\PF$, with
$\rho_\Psi=\rho_\Phi$, $\vjp_\Psi=\vjcdip_\Phi$, and, by
\cref{dipole-length-derivative} and invariance of the Coulomb form,
$\dua{\Phi}{H_\dip\Phi}
=\dua{\Psi}{H_\el^0\Psi}+\|\sqrt{H_\rad}U_{\mathrm d}\Psi\|^2$.
The forms of $H_\dip$, $H^{0,0}$ and $H_\el^0+H_\rad$ are all
equivalent to the norm of $\mathcal Q_\PF$
(\cref{PF-form-norm-equivalence}), so the admissible set of
\cref{dipole-ensemble-search} consists of the $\vGamma\in\dm$ with the
prescribed density and combined current density.  Applied to the eigenvectors of a density matrix, the
above shows that $\vGamma\mapsto U_{\mathrm d}^\dag\vGamma U_{\mathrm d}$
maps $\dm$ bijectively onto the density matrices $\vGamma$ with
$U_{\mathrm d}\vGamma U_{\mathrm d}^\dag\in\dm$.  It preserves the
density, takes $\vjcdip$ to $\vjp$, and
\begin{equation}\label{dipole-length-trace-identity}
 \Tr(H_\dip U_{\mathrm d}\vGamma U_{\mathrm d}^\dag)
 =\Tr(H_\el^0\vGamma)+\Tr(H_\rad U_{\mathrm d}\vGamma U_{\mathrm d}^\dag).
\end{equation}
By \cref{thm-dipole-ensemble-realization}, $\wt F_\dip=F_\dip$, and
every finite infimum in \cref{dipole-ensemble-search} is attained.
Hence, for all $(\rho,\vjcdip)\in X\times Y$,
\begin{equation}\label{dipole-length-functional}
 F_\dip(\rho,\vjcdip)
 =\inf_{\substack{U_{\mathrm d}\vGamma U_{\mathrm d}^\dag\in\dm\\
 \rho_\vGamma=\rho,\ \vjp_\vGamma=\vjcdip}}
 \Big[\Tr(H_\el^0\vGamma)
 +\Tr(H_\rad U_{\mathrm d}\vGamma U_{\mathrm d}^\dag)\Big],
\end{equation}
and every finite infimum here is attained.  This step uses neither
$\sqrt\omega\,\wh\chi\in L^2$ nor the moment assumption.

\noindent\textbf{Step 2. Expansion of the radiation energy.}
Let $\vGamma=\sum_j\lambda_j|\Psi_j\rangle\langle\Psi_j|$, with
$\lambda_j>0$ and orthonormal $\Psi_j$, have density $\rho$.  Since
$|\vR|^2\leq N\sum_n|\vx_n|^2$,
\[
 \sum_j\lambda_j\dua{\Psi_j}{|\vR|^2\Psi_j}
 \leq N\int_{\RR^3}|\vx|^2\rho<\infty.
\]
At each electronic configuration take $h=e\sum_\mu R^\mu G^\mu_{0}$;
then $\|\sqrt\omega h\|^2=e^2\vR^\top\vM\vR\leq e^2\|\vM\||\vR|^2$.  The
bound \cref{weyl-radiation-bound}, applied with $\pm\i h$ and integrated,
gives, for every $\Psi$ with $\dua{\Psi}{|\vR|^2\Psi}<\infty$,
\[
 \|\sqrt{H_\rad}U_{\mathrm d}^{\pm1}\Psi\|^2
 \leq2\|\sqrt{H_\rad}\Psi\|^2
 +2e^2\|\vM\|\dua{\Psi}{|\vR|^2\Psi}.
\]
Applying this to $\Psi_j$ and to $U_{\mathrm d}\Psi_j$ shows that
$U_{\mathrm d}\Psi_j\in D(\sqrt{H_\rad})$ if and only if
$\Psi_j\in D(\sqrt{H_\rad})$.  By \cref{prop-dipole-length-gauge},
applied also with $-e$ in place of $e$, $U_{\mathrm d}\Psi_j$ and
$\Psi_j$ have finite electronic kinetic energy together, so
$U_{\mathrm d}\Psi_j\in\mathcal Q_\PF$ if and only if
$\Psi_j\in\mathcal Q_\PF$.  Summing the displayed bound over $j$ and
using \cref{dipole-length-trace-identity} and the form-norm equivalence
of Step~1 then shows that $U_{\mathrm d}\vGamma U_{\mathrm d}^\dag\in\dm$
if and only if $\vGamma\in\dm$.  Thus the admissible sets in
\cref{dipole-length-functional} and
\cref{dipole-length-hamiltonian-functional} coincide.  For each $j$,
the Weyl shift \cref{dipole-length-Weyl-shift} with $\eta=\Psi_j(\vX)$,
integrated over $\vX$, gives
\[
 \|\sqrt{H_\rad}U_{\mathrm d}\Psi_j\|^2
 =\dua{\Psi_j}{\big[H_\rad-e\vR\cdot\EF_\perp(0)
   +e^2\vR^\top\vM\vR\big]\Psi_j},
\]
where $e\sum_\mu R^\mu\varPi(\omega G^\mu_{0})=-e\vR\cdot\EF_\perp(0)$ by
\cref{dipole-electric-field-origin}.  The bound on $a_\omega$ following
\cref{weyl-weighted-field} gives
$|\varPi(\omega h)[\eta,\eta]|
\leq2\|\sqrt\omega h\|\,\|\sqrt{H_\rad}\eta\|\,\|\eta\|$, so by the
Cauchy--Schwarz inequality the weighted sums over $j$ of the three terms
converge absolutely.  Summing over $j$ turns
\cref{dipole-length-functional} into
\cref{dipole-length-hamiltonian-functional}, and the attainment
statement carries over.  Restricting to rank-one $\vGamma$ and using
$F_{\LL,\dip}=F_\dip$ from \cref{dipole-search-equality} gives the
pure-state version.
\end{proof}

\appendix

\section{Transformation properties}\label{app-transformation-properties}

We work in the solenoidal one-photon representation and, unless stated
otherwise, specialize to the hard cutoff \cref{harduv}.
Second quantization is understood whenever a
one-photon unitary is applied to a Fock-space state.  The displayed
$e,\Lambda$ labels on $\AF_\Lambda$, $H_{e,\Lambda}^{0,0}$,
$(\vjc_\Psi)^{e,\Lambda}$, and $F^{e,\Lambda}$ record parameter dependence
that was suppressed earlier.  Here $\Lambda$ abbreviates the form-factor label
$\chi_\Lambda$, so that, for instance, $F^{e,\Lambda}=F^{e,\chi_\Lambda}$.
An unsuperscripted $F$ carries the parameters
shown in the surrounding formula.  All state identities below apply to
normalized vectors in the indicated form domains, and Hamiltonian
identities are identities of self-adjoint operators established through
their closed quadratic forms, unless explicitly stated only as form
identities.

\subsection{Isometries}

The field momentum is the second quantization
\[
 \vP_\ph=\dd\varGamma(\vk)
 =\sum_{\lambda=1}^2\int_{\RR^3}\vk\,
 a^\dag(\vk,\lambda)a(\vk,\lambda)\dd\vk.
\]
For $\vb\in\RR^3$, define the photon translation
\[
 (t_\vb\vf)(\vk)=e^{-\i\vb\cdot\vk}\vf(\vk),\qquad
 T_\ph(\vb)=\varGamma(t_\vb)=e^{-\i\vb\cdot\vP_\ph},
\]
and the electron translation
\[
 (T_\el(\vb)\Phi)(\vx_1,\ldots,\vx_N)
 =\Phi(\vx_1-\vb,\ldots,\vx_N-\vb).
\]
The composite system is translated by
\[
T(\vb)=T_\el(\vb)\otimes T_\ph(\vb).
\]
\begin{proposition}
\label[proposition]{prop-translation-covariance}
 $T(\vb)$ is a unitary preserving $\mathcal Q_\PF$, and
\begin{equation}\label{A-translation}
 T_\ph(\vb)^\dag\AF_\Lambda(\vx)T_\ph(\vb)
 =\AF_\Lambda(\vx-\vb).
\end{equation}
The internal Hamiltonian, density, and combined current density satisfy
\begin{equation}\label{translation-laws}
 \begin{gathered}
 T(\vb)^\dag H^{0,0}T(\vb)=H^{0,0},\\
 \rho_{T(\vb)\Psi}(\vx)=\rho_\Psi(\vx-\vb),\qquad
 \vjc_{T(\vb)\Psi}(\vx)=\vjc_\Psi(\vx-\vb).
 \end{gathered}
\end{equation}
Consequently, for every $(\rho,\vjc)\in X\times Y$ and
$F\in\{F_\LL,F_\LI\}$,
\begin{equation}\label{F-translation}
 F(\rho(\,\cdot-\vb),\vjc(\,\cdot-\vb))=F(\rho,\vjc).
\end{equation}
\end{proposition}

\begin{proof}
The solenoidal coupling functions in \cref{Gx-solenoidal} obey
$t_\vb^\dag\vG^\mu_{\vx;\Lambda}
=\vG^\mu_{\vx-\vb;\Lambda}$.  Covariance of creation and annihilation
operators under second quantization gives \cref{A-translation}.
Moreover, $t_\vb$ commutes with $\omega$, so $T_\ph(\vb)$ commutes with
$H_\rad$.  Electronic translation preserves weak derivatives and all
interparticle distances.  Thus $T(\vb)$ preserves $\mathcal Q_\PF$,
and the electronic and photon translations cancel in the field term:
\[
 T(\vb)^\dag\AF_\Lambda(\vx_n)T(\vb)=\AF_\Lambda(\vx_n),
 \qquad T(\vb)^\dag\vD_nT(\vb)=\vD_n.
\]
These identities first hold on test vectors and extend by the field bound
\cref{A-Hrad-bound} and form-norm approximation.  The kinetic, Coulomb,
and radiation forms are therefore invariant.  Changing the electronic
integration variables in \cref{density-def,jc-def} gives the remaining
identities in \cref{translation-laws}.

The inverse $T(-\vb)$ makes this an energy-preserving bijection between
the corresponding constraint sets, proving \cref{F-translation} for
$F_\LL$, including empty constraint sets.  Translation is a linear
homeomorphism of $X\times Y$, so taking the norm-lower-semicontinuous
convex envelope in \cref{L-vs-LL} proves the identity for $F_\LI$.
\end{proof}

For $\vR\in{\rm O}(3)$, define on $\Hil_\sol$
\begin{equation}\label{rotation-one-photon}
 (\Upsilon_\vR\vf)(\vk)=\vR\vf(\vR^\top\vk),
\end{equation}
and on $\HC_N$ define
\[
 (U_\vR\Psi)(\vx_1,\ldots,\vx_N)
 =\Upsilon_\vR\Psi(\vR^\top\vx_1,\ldots,\vR^\top\vx_N).
\]

\begin{proposition}
\label[proposition]{prop-orthogonal-covariance}

These maps are unitary, and $U_\vR$ preserves $\mathcal Q_\PF$.
The creation, annihilation, and field operators transform according to
\begin{align}
 \Upsilon_\vR a^\#(\vf)\Upsilon_\vR^\dag
 &=a^\#(\Upsilon_\vR\vf),
 \label{rotation-creation}\\
 \Upsilon_\vR^\dag\AF_\Lambda(\vx)\Upsilon_\vR
 &=\vR\AF_\Lambda(\vR^\top\vx).
 \label{A-rotation}
\end{align}
The corresponding state identities are
\begin{align}
 U_\vR^\dag H^{0,0}U_\vR&=H^{0,0},
 \label{H-rotation}\\
 \rho_{U_\vR\Psi}(\vx)&=\rho_\Psi(\vR^\top\vx),
 \label{rho-rotation}\\
 \vjc_{U_\vR\Psi}(\vx)&=\vR\vjc_\Psi(\vR^\top\vx).
 \label{jc-rotation}
\end{align}
Consequently, for every $(\rho,\vjc)\in X\times Y$ and
$F\in\{F_\LL,F_\LI\}$,
\begin{equation}\label{F-rotation}
 F(\rho\circ\vR^\top,\vR\vjc\circ\vR^\top)=F(\rho,\vjc).
\end{equation}
\end{proposition}

\begin{proof}
Orthogonal changes of variables preserve Lebesgue measure.  For solenoidal
$\vf$, they also preserve transversality, since
\[
 \vk\cdot\vR\vf(\vR^\top\vk)
 =(\vR^\top\vk)\cdot\vf(\vR^\top\vk)=0.
\]
Thus $\Upsilon_\vR$ is unitary on $\Hil_\sol$ with inverse
$\Upsilon_{\vR^\top}$, and \cref{rotation-creation} is the
second-quantization identity.  By \cref{polarization-projector},
\[
 P_\sol(\vR\vk)=\vR P_\sol(\vk)\vR^\top.
\]
Using radiality of the hard cutoff in \cref{Gx-solenoidal}, we obtain
\[
 \Upsilon_\vR^\dag\vG^\mu_{\vx;\Lambda}
 =\sum_{\nu=1}^3R^{\mu\nu}
                  \vG^\nu_{\vR^\top\vx;\Lambda},
\]
which proves \cref{A-rotation}.  Radiality of $\omega$ also gives
$\Upsilon_\vR^\dag H_\rad\Upsilon_\vR=H_\rad$.

The electronic chain rule and \cref{A-rotation} yield
$U_\vR^\dag\vD_nU_\vR=\vR\vD_n$.
Orthogonality preserves the squared covariant-derivative norms and the
Coulomb interaction.  These facts prove form-domain preservation and
\cref{H-rotation}, first on test vectors and then by form approximation.
Changing variables in the partial inner products gives
\cref{rho-rotation,jc-rotation}.  Finally, $U_{\vR^\top}$ is the inverse
map on constraint sets, so \cref{F-rotation} holds for $F_\LL$.
The induced map on $X\times Y$ is a linear homeomorphism, so
\cref{L-vs-LL} proves the result for $F_\LI$.
\end{proof}

\subsection{Dilations}

\begin{proposition}[Dilation identities]
\label[proposition]{prop-dilation-covariance}
For $\lambda>0$, define on $\Hil_\sol$
\begin{equation*}
 (\Upsilon_\lambda\vf)(\vk)=\lambda^{3/2}\vf(\lambda\vk).
\end{equation*}
This is unitary, and
\begin{align}
 \Upsilon_\lambda a^\#(\vf)\Upsilon_\lambda^\dag
 &=a^\#(\Upsilon_\lambda\vf),
 \label{a-scaling}\\
 \Upsilon_\lambda H_\rad\Upsilon_\lambda^\dag
 &=\lambda H_\rad,
 \label{Hrad-scaling}\\
 \Upsilon_\lambda\AF_\Lambda(\vx)\Upsilon_\lambda^\dag
 &=\lambda\AF_{\Lambda/\lambda}(\lambda\vx).
 \label{A-scaling}
\end{align}
Let
\[
 (S_\lambda\Phi)(\vX)=\lambda^{3N/2}\Phi(\lambda\vX),
 \qquad V_\lambda=S_\lambda\otimes\Upsilon_\lambda^\dag,
 \qquad \Psi_\lambda=V_\lambda\Psi.
\]
The map $V_\lambda$ is unitary on $\HC_N$
and maps $Q(H_{e,\Lambda/\lambda}^{0,0})$ bijectively onto
$Q(H_{e,\Lambda}^{0,0})$.  The density and current density satisfy
\begin{align}
 \rho_{\Psi_\lambda}(\vx)
 &=\lambda^3\rho_\Psi(\lambda\vx),
 \label{rho-scaling}\\
 (\vjc_{\Psi_\lambda})^{e,\Lambda}(\vx)
 &=\lambda^4(\vjc_\Psi)^{e,\Lambda/\lambda}(\lambda\vx).
 \label{jc-scaling}
\end{align}
The internal Hamiltonian satisfies the operator identity
\begin{equation}\label{energy-scaling}
 \begin{split}
 V_\lambda^\dag H_{e,\Lambda}^{0,0}V_\lambda
 &=\frac{\lambda^2}{2}\sum_{n=1}^N
 \big(-\i\grad_{\vx_n}-e\AF_{\Lambda/\lambda}(\vx_n)\big)^2\\
 &\quad+\lambda\sum_{1\leq n<m\leq N}
 \frac{1}{|\vx_n-\vx_m|}+\lambda H_\rad.
 \end{split}
\end{equation}
The right-hand side denotes the self-adjoint operator associated with
the displayed sum of nonnegative forms on $\mathcal Q_\PF$.
\end{proposition}

\begin{proof}
A change of momentum variables proves unitarity of
$\Upsilon_\lambda$, and transversality is preserved because
$\vk\cdot\vf(\lambda\vk)=\lambda^{-1}(\lambda\vk)\cdot
\vf(\lambda\vk)$.  Second quantization gives \cref{a-scaling}, while
$\Upsilon_\lambda\omega\Upsilon_\lambda^\dag=\lambda\omega$
gives \cref{Hrad-scaling}.  The hard cutoff obeys
\[
 P_\sol(\lambda\vk)=P_\sol(\vk),\qquad
 \wh\chi_\Lambda(\lambda\vk)
 =\lambda^{-1/2}\wh\chi_{\Lambda/\lambda}(\vk).
\]
Consequently, \cref{Gx-solenoidal} gives
$\Upsilon_\lambda\vG^\mu_{\vx;\Lambda}
=\lambda\vG^\mu_{\lambda\vx;\Lambda/\lambda}$,
which proves \cref{A-scaling}.

The electronic chain rule and \cref{A-scaling} imply
\[
 \vD_n^{e,\Lambda}V_\lambda\Psi
 =\lambda V_\lambda\vD_n^{e,\Lambda/\lambda}\Psi.
\]
The electronic derivatives scale by $\lambda$, the radiation form by
$\lambda$, and $V_\lambda^{-1}=V_{1/\lambda}$.  Thus the reference
domain \cref{PF-reference-form-domain} is preserved in both directions.
A change of electronic variables gives \cref{rho-scaling,jc-scaling},
with the extra factor in the current density coming from the covariant derivative.
The Coulomb interaction scales by $\lambda$, so the squared-derivative
identity gives equality of the quadratic forms on the two sides of
\cref{energy-scaling}.  The calculation extends from test vectors to
$\mathcal Q_\PF$ by \cref{PF-form-norm-equivalence}.  Since
$\lambda>0$, the form on the right is closed on this domain, with a
norm equivalent to the reference form norm.  Polarization and uniqueness
in the representation theorem for closed forms therefore give the
asserted equality of self-adjoint operators.
\end{proof}

The covariant kinetic term scales quadratically.  The Coulomb and photon
terms scale linearly.

\subsection{Time reversal}
\label{app-time-reversal}

Time reversal needs only \cref{cutoff-assumption}, not the hard
cutoff.  In the solenoidal representation, recall the conjugation
$(J\vf)(\vk)=\ol{\vf(-\vk)}$.  This is an antiunitary involution
commuting with $\omega$ and with the rotations
\cref{rotation-one-photon}, and \cref{cutoff-assumption} gives
$J\vG^\mu_{\vx}=\vG^\mu_{\vx}$.  Let $\mathcal C_J$ be the induced
antiunitary involution of $\Fock_\ph$, characterized by
$\mathcal C_Ja^\#(\vf)\mathcal C_J=a^\#(J\vf)$, and define the time
reversal
\begin{equation}\label{time-reversal-def}
 (\Theta\Psi)(\vX)=(-1)^\NC\mathcal C_J\big(\Psi(\vX)\big),
\end{equation}
where $(-1)^\NC=\varGamma(-\iden)=e^{\i\pi\NC}$ multiplies the $m$-photon
component by $(-1)^m$.  Equivalently, $\Theta=C\otimes\varGamma(-J)$,
where $C$ is complex conjugation of electronic wave functions.

\begin{proposition}[Time reversal]
\label[proposition]{prop-time-reversal}
$\Theta$ is an antiunitary involution of $\HC_N$ that preserves
$\mathcal Q_\PF$ and commutes with every $U_\vR$ of
\cref{prop-orthogonal-covariance}.  The field and the covariant
derivatives are odd and the radiation energy is even:
\begin{equation}\label{A-time-reversal}
 \Theta\AF(\vx)\Theta^{-1}=-\AF(\vx),\qquad
 \Theta\vD_n\Theta^{-1}=-\vD_n,\qquad
 \Theta H_\rad\Theta^{-1}=H_\rad.
\end{equation}
Consequently, $\EF_\perp$ is even and $\BF$ is odd, and for external
potentials in \cref{potential-spaces},
\begin{equation}\label{H-time-reversal}
 \Theta H^{v,\vA}\Theta^{-1}=H^{v,-\vA}.
\end{equation}
The corresponding state identities are
\begin{equation}\label{time-reversal-laws}
 \rho_{\Theta\Psi}=\rho_\Psi,\qquad
 \vjc_{\Theta\Psi}=-\vjc_\Psi,\qquad
 \dua{\Theta\Psi}{H^{0,0}\Theta\Psi}=\dua{\Psi}{H^{0,0}\Psi}.
\end{equation}
Consequently, for every $(\rho,\vjc)\in X\times Y$ and
$F\in\{F_\LL,F_\LI\}$,
\begin{equation}\label{F-time-reversal}
 F(\rho,-\vjc)=F(\rho,\vjc).
\end{equation}
\end{proposition}

\begin{proof}
Second quantization of $J\vG^\mu_{\vx}=\vG^\mu_{\vx}$ and $J\omega=\omega J$
gives $\mathcal C_J\AF(\vx)\mathcal C_J=\AF(\vx)$ and
$\mathcal C_JH_\rad\mathcal C_J=H_\rad$, while
$(-1)^\NC a^\#(\vf)(-1)^\NC=-a^\#(\vf)$.  Applied pointwise in $\vX$,
the antilinear map $\mathcal C_J$ also conjugates the electronic
dependence, so $-\i\grad_{\vx_n}$ changes sign.  This proves
\cref{A-time-reversal} on test vectors.  The field bound
\cref{A-Hrad-bound} and form-norm approximation extend it, and show
that $\Theta$ preserves $\mathcal Q_\PF$.  Since $\mathcal C_J^2=\iden$
and $(-1)^\NC$ commutes with $\mathcal C_J$, $\Theta^2=\iden$.  The
pointwise action preserves fermionic antisymmetry.  Because $J$
commutes with the one-photon rotations, $\mathcal C_J$ and $(-1)^\NC$
commute with their second quantizations, so $\Theta U_\vR=U_\vR\Theta$.

As $\Theta$ is antilinear, $\EF_\perp=-\i[H_\rad,\AF]$ is even, and
$\BF=\grad\times\AF$ is odd.  The real multiplication operators $v$,
$\vA(\vx_n)$, and the Coulomb interaction commute with $\Theta$, which
gives \cref{H-time-reversal} as an identity of closed forms.  Pointwise
antiunitarity gives $\rho_{\Theta\Psi}=\rho_\Psi$ and, by
\cref{A-time-reversal},
\[
 \vjc_{\Theta\Psi}(\vx)
 =N\Re\dua{\Theta\Psi}{\vD_1\Theta\Psi}_\vx
 =-N\Re\ol{\dua{\Psi}{\vD_1\Psi}_\vx}
 =-\vjc_\Psi(\vx).
\]
The energy identity is \cref{H-time-reversal} with $v=0$ and $\vA=0$.
Finally, $\Theta$ is an energy-preserving bijection between the pure
constraint sets at $(\rho,\vjc)$ and $(\rho,-\vjc)$, which proves
\cref{F-time-reversal} for $F_\LL$.  The map
$(\rho,\vjc)\mapsto(\rho,-\vjc)$ is a linear homeomorphism of
$X\times Y$, so \cref{L-vs-LL} gives the identity for $F_\LI$.
\end{proof}

\subsection{Gauge covariance}

\begin{proposition}[Gauge covariance]
\label[proposition]{prop-gauge-covariance}
Let $g\in W^{1,1}_{\rm loc}(\RR^3,\RR)$ satisfy
\[
 \grad g\in L^3(\RR^3,\RR^3)+L^\infty(\RR^3,\RR^3),
\]
and define
\begin{equation}\label{gauge-unitary}
 (U_g\Psi)(\vX)=e^{\i\sum_{n=1}^Ng(\vx_n)}\Psi(\vX).
\end{equation}
Then $U_g$ is unitary and preserves $\mathcal Q_\PF$.  For external
potentials in \cref{potential-spaces},
\begin{align}
 H^{v,\vA-\grad g}&=U_gH^{v,\vA}U_g^\dag,
 \label{H-gauge}\\
 \rho_{U_g\Psi}&=\rho_\Psi,
 \qquad
 \vjc_{U_g\Psi}=\vjc_\Psi+\rho_\Psi\grad g.
 \label{current-gauge}
\end{align}
In particular, the physical current density $\vjc_\Psi+\rho_\Psi\vA$ is
invariant under the simultaneous transformation
$(\Psi,\vA)\mapsto(U_g\Psi,\vA-\grad g)$.
The internal energy satisfies
\begin{equation}\label{energy-gauge}
 \dua{U_g\Psi}{H^{0,0}U_g\Psi}
 =\dua{\Psi}{H^{0,0}\Psi}+\int_{\RR^3}\grad g\cdot\vjc_\Psi
 +\frac12\int_{\RR^3}\rho_\Psi|\grad g|^2.
\end{equation}
For every $(\rho,\vjc)\in X\times Y$ and $F\in\{F_\LL,F_\LI\}$,
\begin{equation}\label{F-gauge}
 F(\rho,\vjc+\rho\grad g)
 =F(\rho,\vjc)+\int_{\RR^3}\grad g\cdot\vjc
 +\frac12\int_{\RR^3}\rho|\grad g|^2.
\end{equation}
\end{proposition}

\begin{proof}
The Sobolev chain and product rules give
\[
 \vD_nU_g\Psi=U_g\big(\vD_n+(\grad g)(\vx_n)\big)\Psi.
\]
By the vector-valued Sobolev inequality, multiplication by $(\grad g)(\vx_n)$
is bounded from $\mathcal Q_\PF$ to $\HC_N$, so this identity is valid
for every $\Psi\in\mathcal Q_\PF$.
The phase commutes with $H_\rad$, and its inverse is $U_{-g}$, so it
preserves the form domain in both directions.  Adding
$\vA(\vx_n)-(\grad g)(\vx_n)$ to the transformed derivative proves
\cref{H-gauge} as a closed-form identity.  Pointwise unitarity and the
same derivative formula give \cref{current-gauge} and the asserted
physical-current invariance.

Expanding the squared covariant-derivative norms, the cross term is
\[
 \sum_{n=1}^N\Re\dua{\vD_n\Psi}{(\grad g)(\vx_n)\Psi}
 =\int_{\RR^3}\grad g\cdot\vjc_\Psi,
\]
and the quadratic term is $\tfrac12\int_{\RR^3}\rho_\Psi|\grad g|^2$.
The Coulomb and radiation terms are unchanged, proving \cref{energy-gauge}.
The bijection $U_g$ between the two constraint sets therefore proves
\cref{F-gauge} for $F_\LL$.

For $F_\LI$ the argument is the same, applied to density matrices.
By \cref{current-gauge,energy-gauge}, applied to spectral
decompositions, $\vGamma\mapsto U_g\vGamma U_g^\dag$ is a bijection of
$\dm$ onto itself that maps the constraint set of
\cref{FLcs-def} at $(\rho,\vjc)$ onto the one at
$(\rho,\vjc+\rho\grad g)$ and shifts the energy by the last two terms
in \cref{F-gauge}.  This proves \cref{F-gauge} for $\wt F_\LI$, and
hence for $F_\LI$ by \cref{unrestricted-Lieb-equality}.
\end{proof}

\subsection{Weyl displacement operators}
\label{app-weyl-displacements}

We collect the displacement identities used throughout the paper.
The algebraic statements
hold on the bosonic Fock space over any complex one-particle space
$\Hil$.
Define the exponential vector and the Weyl operator by
\begin{equation}\label{weyl-definitions}
 \boxed{\mathcal E(f):=\bigoplus_{n=0}^\infty\frac{f^{\otimes n}}{\sqrt{n!}},
 \qquad
 W(f):=\exp\bigl(-\i\varPhi(\i f)\bigr)
       =\exp\bigl(a^\dag(f)-a(f)\bigr).}
\end{equation}
The exponential is defined through the self-adjoint Segal field
$\varPhi(\i f)$.  This convention gives
$W(\i f)=\exp(\i\varPhi(f))$.

\begin{proposition}[Weyl relations and field shifts]
\label[proposition]{prop-weyl-algebra}
For $f,g,h\in\Hil$, the following properties hold true.
\begin{enumerate}[(i)]
\item The Weyl operators act on exponential vectors and compose
according to
\begin{align}
 W(f)\mathcal E(g)
 &=e^{-\|f\|^2/2-\dua{f}{g}}\mathcal E(f+g),
 \label{weyl-exponential-action}\\
 W(f)^\dag=W(-f),\qquad
 W(f)W(g)
 &=e^{-\i\Im\dua{f}{g}}W(f+g).
 \label{weyl-relations}
\end{align}
\item The map $f\mapsto W(f)$ is strongly continuous.  On the linear
span of exponential vectors, $W(f)$ has the normally ordered expression
\[
 W(f)=e^{-\|f\|^2/2}e^{a^\dag(f)}e^{-a(f)}.
\]
Here the two normally ordered exponentials act by their convergent
power series on exponential vectors.
\item The closed creation and annihilation operators and the Segal
fields satisfy
\begin{align}
 W(f)^\dag a(h)W(f)
 &=a(h)+\dua{h}{f}\iden,
 \label{weyl-annihilation-shift}\\
 W(f)^\dag a^\dag(h)W(f)
 &=a^\dag(h)+\dua{f}{h}\iden,
 \label{weyl-creation-shift}\\
 W(f)^\dag\varPhi(h)W(f)
 &=\varPhi(h)+2\Re\dua{h}{f}\iden,
 \label{weyl-field-shift}\\
 W(f)^\dag\varPi(h)W(f)
 &=\varPi(h)+2\Im\dua{h}{f}\iden.
 \label{weyl-conjugate-field-shift}
\end{align}
Each identity includes invariance of the domain of the operator
being conjugated.  If $\Im\dua{h}{g}=0$, then $\varPhi(h)$ and
$\varPhi(g)$ strongly commute.
\item For any unitary $T$ on $\Hil$,
\begin{equation}\label{weyl-second-quantization}
 \varGamma(T)W(f)\varGamma(T)^\dag=W(Tf).
\end{equation}
\item Under the canonical Fock-space factorization over
$\Hil=\Hil_1\oplus\Hil_2$,
\begin{equation}\label{weyl-factorization}
 W(f_1\oplus f_2)=W(f_1)\otimes W(f_2).
\end{equation}
\end{enumerate}
\end{proposition}

\begin{proof}
(i) The defining series gives
$\dua{\mathcal E(g)}{\mathcal E(h)}=e^{\dua{g}{h}}$, and exponential
vectors are eigenvectors of the annihilation operators:
\begin{equation}\label{exponential-annihilation-eigenvector}
 a(f)\mathcal E(g)=\dua{f}{g}\mathcal E(g).
\end{equation}
Exponential vectors have dense linear span.  Indeed, if
$\Psi=(\psi^{(n)})_{n\geq0}$ is orthogonal to every $\mathcal E(tg)$,
$t\in\RR$, then the power series
$\sum_nt^n\dua{g^{\otimes n}}{\psi^{(n)}}/\sqrt{n!}$ vanishes identically,
so $\dua{g^{\otimes n}}{\psi^{(n)}}=0$ for all $g$ and $n$.  The
polarization identity
\[
 \mathrm{Sym}(g_1\otimes\cdots\otimes g_n)
 =\frac{1}{2^nn!}\sum_{\epsilon\in\{\pm1\}^n}\epsilon_1\cdots\epsilon_n
 (\epsilon_1g_1+\cdots+\epsilon_ng_n)^{\otimes n},
\]
where $\mathrm{Sym}$ averages over permutations of the factors, then
gives $\dua{\mathrm{Sym}(g_1\otimes\cdots\otimes g_n)}{\psi^{(n)}}=0$.
These symmetrizations span a dense subspace of the $n$-particle sector,
so $\psi^{(n)}=0$.
The right-hand side of \cref{weyl-exponential-action} preserves
inner products of exponential vectors.  Its inverse is the same
formula with $-f$, so it defines a unitary.  Composing the formulas
gives \cref{weyl-relations}, and norm continuity on exponential
vectors, followed by density, gives strong continuity in $f$.
In particular, $t\mapsto W(tf)$ is a strongly continuous one-parameter
unitary group.  Differentiating at $t=0$ gives
$a^\dag(f)-a(f)=-\i\varPhi(\i f)$ as its generator.
The exponential span is a core for this Segal field, so this unitary
is precisely \cref{weyl-definitions}.

(ii) Strong continuity was shown in (i).  The normally ordered
formula follows from \cref{exponential-annihilation-eigenvector} and
$e^{a^\dag(f)}\mathcal E(g)=\mathcal E(f+g)$.

(iii) The eigenvector identity \cref{exponential-annihilation-eigenvector} proves
\cref{weyl-annihilation-shift} on the exponential span.
This span is an operator core for $a(h)$ and $a^\dag(h)$:
finite symmetric tensors are obtained by differentiating
exponential vectors in their graph norms, and finite-particle
truncation gives a core for the closed operators.
Since $W(f)$ and $W(-f)$ map the exponential span onto itself,
closure gives \cref{weyl-annihilation-shift} with equality of
domains.  Taking adjoints gives \cref{weyl-creation-shift}.
Applying \cref{weyl-relations} to
$W(\i t h)=e^{\i t\varPhi(h)}$ and using uniqueness of the
self-adjoint generator proves \cref{weyl-field-shift}.
Replacing $h$ by $\i h$ gives
\cref{weyl-conjugate-field-shift}.
When $\Im\dua{h}{g}=0$, the Weyl relations make the two unitary
groups $e^{\i t\varPhi(h)}$ and $e^{\i s\varPhi(g)}$ commute for all
$s,t\in\RR$, which is strong commutation of their generators.

(iv) The identity $\varGamma(T)\mathcal E(g)=\mathcal E(Tg)$ and
\cref{weyl-exponential-action} prove \cref{weyl-second-quantization}
on exponential vectors, and density extends it.

(v) The canonical factorization sends
$\mathcal E(g_1\oplus g_2)$ to
$\mathcal E(g_1)\otimes\mathcal E(g_2)$, and applying
\cref{weyl-exponential-action} proves \cref{weyl-factorization}.
\end{proof}

The radiation form is expressed through the annihilation operators.
The sectorwise annihilation formulas define a closed map
\[
 (\mathcal T\eta)(\vk,\lambda)
 :=\sqrt{\omega(\vk)}\,a(\vk,\lambda)\eta,\qquad
 \|\mathcal T\eta\|=\|\sqrt{H_\rad}\eta\|,
\]
from $D(\sqrt{H_\rad})$ to $L^2(\RR^3\times\{1,2\};\Fock_\ph)$.  It lets
us extend the annihilation operator to kernels that are not square
integrable.  This is needed below for $g=\omega f$ with only
$f\in D(\sqrt\omega)$, in which case $g$ need not lie in $\Hil_\ph$ and
$a(g)$ is not defined.  For a measurable kernel $g$ with
$g/\sqrt\omega\in\Hil_\ph$ and $\eta,\zeta\in D(\sqrt{H_\rad})$, define
\begin{align}
 a_\omega(g)\eta
 &:=\sum_{\lambda=1}^2\int_{\RR^3}
   \overline{\frac{g(\vk,\lambda)}{\sqrt{\omega(\vk)}}}\,
   (\mathcal T\eta)(\vk,\lambda)\,\dd\vk,
 \label{weyl-weighted-annihilation}\\
 \varPhi_\omega(g)[\eta,\zeta]
 &:=\dua{a_\omega(g)\eta}{\zeta}
      +\dua{\eta}{a_\omega(g)\zeta}.
 \label{weyl-weighted-field}
\end{align}
The integrand pairs two square-integrable factors, so the
Cauchy--Schwarz inequality gives
\[
 \|a_\omega(g)\eta\|
 \leq\|g/\sqrt\omega\|\,\|\sqrt{H_\rad}\eta\|.
\]
Thus $a_\omega(g)$ is a bounded operator from the radiation form domain
to $\Fock_\ph$, and $\varPhi_\omega(g)$ is a continuous symmetric form on
that domain.  If also $g\in\Hil_\ph$, then $a_\omega(g)=a(g)$ on
$D(\sqrt{H_\rad})$, and $\varPhi_\omega(g)$ is the form of $\varPhi(g)$
there.

\begin{proposition}[Radiation energy and photon number under displacement]
\label[proposition]{prop-weyl-radiation}
On $\Fock_\ph$, let $f\in\Hil_\ph$.  The following hold.
\begin{enumerate}[(i)]
\item If $f\in D(\sqrt\omega)$, then $W(f)$ maps $D(\sqrt{H_\rad})$
bijectively onto itself, and
\begin{equation}\label{weyl-radiation-bound}
 \|\sqrt{H_\rad}W(f)\eta\|
 \leq\|\sqrt{H_\rad}\eta\|
       +\|\sqrt\omega f\|\,\|\eta\|.
\end{equation}
\item If $f\in D(\sqrt\omega)$, the radiation form shifts exactly: for
all $\eta,\zeta\in D(\sqrt{H_\rad})$,
\begin{equation}\label{weyl-radiation-form-shift}
 \dua{W(f)\eta}{H_\rad W(f)\zeta}
 =\dua{\eta}{H_\rad\zeta}
   +\varPhi_\omega(\omega f)[\eta,\zeta]
   +\|\sqrt\omega f\|^2\dua{\eta}{\zeta},
\end{equation}
where $\dua{\eta}{H_\rad\zeta}
=\dua{\sqrt{H_\rad}\eta}{\sqrt{H_\rad}\zeta}$ is the radiation form.
\item If $f\in D(\omega)$, then $W(f)$ preserves
$D(H_\rad)$, and the operator identity on this domain is
\begin{equation}\label{weyl-radiation-operator-shift}
 W(f)^\dag H_\rad W(f)
 =H_\rad+\varPhi(\omega f)+\|\sqrt\omega f\|^2\iden.
\end{equation}
\item The photon number satisfies
\begin{equation}\label{weyl-number-bound}
 W(f)D(\sqrt\NC)=D(\sqrt\NC),\qquad
 \|\sqrt\NC W(f)\eta\|
 \leq\|\sqrt\NC\eta\|+\|f\|\,\|\eta\|.
\end{equation}
\item The normalized coherent vector $W(f)\Omega$ satisfies
\begin{equation}\label{weyl-coherent-energy}
 \dua{W(f)\Omega}{\NC W(f)\Omega}=\|f\|^2,\qquad
 \dua{W(f)\Omega}{H_\rad W(f)\Omega}=\|\sqrt\omega f\|^2.
\end{equation}
The second equality allows the value $+\infty$.  Thus
$f\in D(\sqrt\omega)$ is also necessary for preservation of the
radiation form domain.
\end{enumerate}
\end{proposition}

\begin{proof}
We use the closed map $\mathcal T$ defined before the proposition.
For $h\in D(\sqrt\omega)$,
\cref{weyl-exponential-action} gives, first on $\eta=\mathcal E(h)$,
\begin{equation}\label{weyl-weighted-point-shift}
 \mathcal T W(f)\eta
 =(\iden\otimes W(f))
      \bigl(\mathcal T\eta+\sqrt\omega f\otimes\eta\bigr).
\end{equation}
The span of these exponential vectors is a radiation form core.
Indeed, finite symmetric tensors with factors in
$D(\sqrt\omega)$ form such a core by particle truncation and
one-particle spectral truncation.  They are in turn limits, in the
form norm, of finite differences of exponential vectors, obtained
by polarizing and differentiating their defining series.

The triangle inequality in \cref{weyl-weighted-point-shift} proves
\cref{weyl-radiation-bound} on this core.
For a form-convergent sequence, the right-hand side of
\cref{weyl-weighted-point-shift} converges and $W(f)\eta_n$ converges
in Fock space.  Closedness of $\mathcal T$ extends the identity and
the bound to the full form domain.
Applying the same argument to $-f$ proves invariance in both
directions.

Taking the inner product of the two sides of
\cref{weyl-weighted-point-shift}, for $\eta$ and $\zeta$, and using
\cref{weyl-weighted-annihilation} with $g=\omega f$, so that
$g/\sqrt\omega=\sqrt\omega f$, gives \cref{weyl-radiation-form-shift}.
For $f\in D(\omega)$ and $\zeta\in D(H_\rad)$, the right-hand side of
\cref{weyl-radiation-form-shift} is represented by the vector
$(H_\rad+\varPhi(\omega f)+\|\sqrt\omega f\|^2)\zeta$.
The representation theorem for closed forms therefore gives
$W(f)\zeta\in D(H_\rad)$ and
\cref{weyl-radiation-operator-shift}.  Repeating this with $-f$
proves equality of the operator domains.

The same proof with weight $1$ in place of $\omega$ proves
\cref{weyl-number-bound}.  Applying the identities to $\Omega$
proves \cref{weyl-coherent-energy} when the right-hand side is
finite.  In general the coherent annihilation formula
$a(\vk,\lambda)W(f)\Omega=f(\vk,\lambda)W(f)\Omega$ gives the energy equality
as a nonnegative integral, including the value $+\infty$.
\end{proof}

\begin{corollary}[Displacement of density, current, and energy]
\label[corollary]{cor-weyl-PF-shifts}
Assume \cref{cutoff-assumption}, let $f\in D(\sqrt\omega)$, and set
\begin{equation}\label{weyl-classical-field}
 c_f^\mu(\vx)=2\Re\dua{G^\mu_{\vx}}{f}.
\end{equation}
The field $\vc_f$ is real, bounded, continuous, and divergence-free.
The photon unitary $W(tf)$ preserves $\mathcal Q_\PF$ for every
$t\in\RR$.  For $\vGamma\in\dm$, let
\[
\boxed{\vGamma_t=W(tf)\vGamma W(tf)^\dag,}
\]
with the electronic identity
suppressed.  Then the following hold.
\begin{enumerate}[(i)]
\item $\vGamma_t\in\dm$, its electronic reduced state is unchanged, and
\begin{equation}\label{weyl-density-current-shift}
 \rho_{\vGamma_t}=\rho_\vGamma,\qquad
 \vjp_{\vGamma_t}=\vjp_\vGamma,\qquad
 \vjc_{\vGamma_t}=\vjc_\vGamma-et\rho_\vGamma\vc_f.
\end{equation}
\item For admissible external potentials, the exact energy change is
\begin{equation}\label{weyl-PF-energy-shift}
 \begin{split}
 \Tr(H^{v,\vA}\vGamma_t)
 &=\Tr(H^{v,\vA}\vGamma)\\
 &\quad+t\left[\Tr\bigl(\varPhi_\omega(\omega f)\vGamma\bigr)
       -e\int\vc_f\cdot(\vjc_\vGamma+\rho_\vGamma\vA)\right]\\
 &\quad+t^2\left[\|\sqrt\omega f\|^2
       +\frac{e^2}{2}\int\rho_\vGamma|\vc_f|^2\right].
 \end{split}
\end{equation}
Here the trace of the weighted field is the sum of its forms over
any spectral decomposition of $\vGamma$.  It is finite, with
\begin{equation}\label{weyl-field-trace-bound}
 \left|\Tr\bigl(\varPhi_\omega(\omega f)\vGamma\bigr)\right|
 \leq2\|\sqrt\omega f\|
             \bigl(\Tr(H_\rad\vGamma)\bigr)^{1/2}.
\end{equation}
For $f\in D(\omega)$, this trace equals
$2\Re\Tr(a(\omega f)\vGamma)$.
\end{enumerate}
\end{corollary}

\begin{proof}
The continuity and uniform bounds of $G^\mu_{\vx}$ give
$\|\vc_f\|_\infty\leq2\|\wh\chi\|\|f\|$ and continuity.
The weak Coulomb-gauge identity gives $\div\vc_f=0$.
Since $W(tf)$ acts independently of the electronic positions, it
commutes with electronic weak derivatives.
\Cref{prop-weyl-radiation} then gives preservation of
$\mathcal Q_\PF$.
Partial trace over photons is unchanged by a photon unitary, so
the density and paramagnetic current density are unchanged.
By \cref{weyl-field-shift},
\[
 W(tf)^\dag\AF(\vx)W(tf)=\AF(\vx)+t\vc_f(\vx).
\]
The definitions of $\vjph$ and $\vjc$ now give
\cref{weyl-density-current-shift}.
Expanding the kinetic form replaces $\vD_n+\vA$ by
$\vD_n+\vA-et\vc_f$.  Its linear term is
$-et\int\vc_f\cdot(\vjc_\vGamma+\rho_\vGamma\vA)$ and its quadratic
term is $\tfrac12e^2t^2\int\rho_\vGamma|\vc_f|^2$.
The scalar and Coulomb forms are unchanged.
Adding \cref{weyl-radiation-form-shift} proves
\cref{weyl-PF-energy-shift}, first for pure states.
The weighted annihilation bound and Cauchy--Schwarz in a spectral
decomposition prove \cref{weyl-field-trace-bound} and justify
summing the identities for density matrices.
For independence of the spectral decomposition, put
$K=(1+H_\rad)^{1/2}$ and let $B_f$ be the bounded operator with
\[
 \dua{\eta}{B_f\zeta}
 =\varPhi_\omega(\omega f)[K^{-1}\eta,K^{-1}\zeta].
\]
The form trace equals $\Tr(B_fK\vGamma K)$, where
$K\vGamma K$ is the positive trace-class operator defined by the
finite radiation energy.  Hence it does not depend on the chosen
decomposition.
All remaining terms are
integrable since $\vc_f$ is bounded and
$\vjc_\vGamma,\rho_\vGamma\vA\in L^1$.
\end{proof}

In particular, for real $\vu\in C_c^\infty(\RR^3;\RR^3)$ and
$g_\vu=\sum_\mu\int u^\mu(\vx)G^\mu_{\vx}\dd\vx$, symmetry of the form factor gives
$\dua{G^\mu_{\vx}}{g_\vu}\in\RR$.  Consequently,
\[
 \vc_{\i g_\vu}=0,\qquad
 \varPhi(\i\omega g_\vu)=-\int\vu\cdot\EF_\perp.
\]
These are the displacements preserving both constraints in the proof
of \cref{prop-optimizer-mean-fields}.  General displacements instead
give the current change in \cref{weyl-density-current-shift}.

\begin{proposition}[Coherent vectors and position-dependent displacements]
\label[proposition]{prop-weyl-coherent-derivatives}
Let $\mathcal O\subset\RR^m$ be open.  The following hold.
\begin{enumerate}[(i)]
\item For $f,g,h\in\Hil$, the normalized coherent vectors satisfy
\begin{gather}
 W(f)\Omega=e^{-\|f\|^2/2}\mathcal E(f),\qquad
 a(h)W(f)\Omega=\dua{h}{f}W(f)\Omega,
 \label{weyl-coherent-vector}\\
 \dua{W(f)\Omega}{W(g)\Omega}
 =\exp\Bigl(-\frac12\|f-g\|^2+\i\Im\dua{f}{g}\Bigr).
 \label{weyl-coherent-overlap}
\end{gather}
\item As a map of real Hilbert spaces, $f\mapsto W(f)\Omega$ is
smooth.  Its derivative at $f$ in the direction $h$ is
\begin{equation}\label{weyl-coherent-derivative}
 \frac{\dd}{\dd s}W(f+sh)\Omega\Big|_{s=0}
 =\bigl(a^\dag(h)-\Re\dua{f}{h}\bigr)W(f)\Omega,
\end{equation}
and this vector satisfies
\begin{equation}\label{weyl-coherent-metric}
 \begin{split}
 \dua{W(f)\Omega}{\bigl(a^\dag(h)-\Re\dua{f}{h}\bigr)W(f)\Omega}
 &=\i\Im\dua{f}{h},\\
 \big\|\bigl(a^\dag(h)-\Re\dua{f}{h}\bigr)W(f)\Omega\big\|^2
 &=\|h\|^2+\bigl(\Im\dua{f}{h}\bigr)^2.
 \end{split}
\end{equation}
\item Let $f\in H^1_{\rm loc}(\mathcal O;\Hil)$ and
$S\in H^1_{\rm loc}(\mathcal O;\RR)$.  Then
$e^{\i S}W(f)\Omega\in W^{1,1}_{\rm loc}$ and, almost everywhere,
\begin{equation}\label{weyl-coherent-phase-derivatives}
 \begin{split}
 \Im\dua{e^{\i S}W(f)\Omega}{\partial_j\bigl(e^{\i S}W(f)\Omega\bigr)}
 &=\partial_jS+\Im\dua{f}{\partial_jf},\\
 \big\|\partial_j\bigl(e^{\i S}W(f)\Omega\bigr)\big\|^2
 &=\|\partial_jf\|^2+\bigl|\partial_jS+\Im\dua{f}{\partial_jf}\bigr|^2.
 \end{split}
\end{equation}
In particular, $e^{\i S}W(f)\Omega\in H^1_{\rm loc}$ if
$\grad S+\Im\dua{f}{\grad f}\in L^2_{\rm loc}$.
\item If $J$ is a conjugation on $\Hil$, the map $f\mapsto W(f)\Omega$
is globally Lipschitz on $\Hil_J:=\{f:Jf=f\}$:
\begin{equation}\label{weyl-real-coherent-Lipschitz}
 \|W(f)\Omega-W(g)\Omega\|^2
 =2\bigl(1-e^{-\|f-g\|^2/2}\bigr)\leq\|f-g\|^2
 \qquad(f,g\in\Hil_J).
\end{equation}
Thus for $f\in H^1_{\rm loc}(\mathcal O;\Hil_J)$,
$W(f)\Omega\in H^1_{\rm loc}$ and
\[
 \dua{W(f)\Omega}{\partial_j\bigl(W(f)\Omega\bigr)}=0,\qquad
 \|\partial_j\bigl(W(f)\Omega\bigr)\|^2=\|\partial_j f\|^2.
\]
\end{enumerate}
\end{proposition}

\begin{proof}
\Cref{weyl-exponential-action} applied to the vacuum gives
\cref{weyl-coherent-vector}, and the exponential-vector inner
product gives \cref{weyl-coherent-overlap}.
The norm-convergent exponential series and its derivatives converge
uniformly on bounded sets of $f$, so differentiating gives
\cref{weyl-coherent-derivative}.
The eigenvector identity \cref{exponential-annihilation-eigenvector} and the CCR yield
\[
 \dua{W(f)\Omega}{a^\dag(h)W(f)\Omega}=\dua{f}{h},\qquad
 \|a^\dag(h)W(f)\Omega\|^2=\|h\|^2+|\dua{h}{f}|^2.
\]
Substitution proves \cref{weyl-coherent-metric}.
Differentiating the scalar phase then proves
\cref{weyl-coherent-phase-derivatives} for smooth $f,S$.

For the stated Sobolev regularity, first compose the map
$g\mapsto W(g)\Omega$ with smooth radial truncations of $f$.  On
bounded sets its derivative is bounded, and
\cref{weyl-coherent-metric} gives the global estimate
\[
 \big\|\bigl(a^\dag(h)-\Re\dua{f}{h}\bigr)W(f)\Omega\big\|
 \leq(1+\|f\|)\|h\|.
\]
Since $(1+\|f\|)\|\grad f\|$ is locally integrable, the truncated
chain rules pass to the limit in $W^{1,1}_{\rm loc}$.
The product rule with $e^{\i S}$ proves the displayed identities
almost everywhere.  Their right-hand sides give square-integrable
derivatives when $\grad S+\Im\dua{f}{\grad f}\in L^2_{\rm loc}$.
Finally, scalar products of $J$-real vectors are real, so
\cref{weyl-coherent-overlap} gives
\cref{weyl-real-coherent-Lipschitz}.  In this case
$\Im\dua{f}{\partial_jf}=0$, and the last assertions follow from the
same derivative identities with $S=0$.
\end{proof}

These formulas imply \cref{photonic-real-coherent-derivatives} and
apply to the tags in \cref{lem-coherent-tag}.  The distinction between
$W^{1,1}_{\rm loc}$ and $H^1_{\rm loc}$ matters for complex
displacements: $f\in H^1_{\rm loc}$ alone does not control the
square of $\Im\dua{f}{\grad f}$.

\subsection{Length-gauge transformation}
\label{app-dipole-length-gauge}

In this subsection the form factor need only satisfy
\cref{cutoff-assumption}.  The additional ultraviolet moment needed to
expand the transformed radiation form will be stated separately.  Put
$\vR=\sum_{n=1}^N\vx_n$, and let $H_\dip$ be the dipole Hamiltonian
in \cref{dipole-internal-hamiltonian}.

Note first that by \cref{Gx-def,polarization-projector},
\[
 \dua{G^\mu_{0}}{G^\nu_{0}}
 =\frac12\int_{\RR^3}|\wh\chi(\vk)|^2
   \left(\delta^{\mu\nu}-\frac{k^\mu k^\nu}{|\vk|^2}\right)
   \,\dd\vk
 \in\RR.
\]
Since $\Af^\mu(0)=\varPhi(G^\mu_{0})$, \cref{prop-weyl-algebra}(iii)
shows that the components of $\AF(0)$ strongly commute.  At each
electronic configuration $\vX$, the operator $e\vR\cdot\AF(0)$ is the
Segal field $\varPhi\bigl(e\sum_{\mu=1}^3R^\mu G^\mu_{0}\bigr)$.  We
define the multiplication operator
\begin{equation}\label{dipole-length-unitary}
 U_{\mathrm d}:=\exp(\i e\vR\cdot\AF(0)),\qquad\text{that is,}\qquad
 U_{\mathrm d}(\vX)=W\bigl(\i e\textstyle\sum_{\mu=1}^3R^\mu G^\mu_{0}\bigr).
\end{equation}

\begin{proposition}[Length-gauge transformation]
\label[proposition]{prop-dipole-length-gauge}
Let $\mathcal Q_{\mathrm d}:=U_{\mathrm d}^\dag\mathcal Q_\PF$ be the
transported form domain.  The following hold.
\begin{enumerate}[(i)]
\item $U_{\mathrm d}$ is unitary on $\HC_N$.
\item Every $\Psi\in\mathcal Q_{\mathrm d}$ has finite ordinary
electronic kinetic energy, and on $\mathcal Q_{\mathrm d}$
\begin{equation}\label{dipole-length-derivative}
 U_{\mathrm d}^\dag\vD_{n,\dip}U_{\mathrm d}=-\i\grad_{\vx_n},
 \qquad 1\leq n\leq N.
\end{equation}
\item For $\Psi\in\mathcal Q_{\mathrm d}$, the density and current
density satisfy
\begin{equation}\label{dipole-length-current}
 \rho_\Psi=\rho_{U_{\mathrm d}\Psi},\qquad
 \vjp_\Psi=\vjcdip_{U_{\mathrm d}\Psi}.
\end{equation}
\item If $\sqrt\omega\,\wh\chi\in L^2(\RR^3)$, then
$\mathscr C_\PF\subset\mathcal Q_{\mathrm d}$, and on this algebraic
test space the following identity holds in the quadratic-form sense:
\begin{equation}\label{dipole-length-form}
 U_{\mathrm d}^\dag H_\dip U_{\mathrm d}
 =H_\el^0+H_\rad
   +e\sum_{\mu=1}^3R^\mu\varPi(\omega G^\mu_{0})
   +e^2\vR^\top\vM\vR,
\end{equation}
where $\vM$ is the matrix \cref{dipole-polarization-matrix}.
For $f/\sqrt\omega\in\Hil_\ph$, $\varPi(f)$ denotes the form
$\varPhi_\omega(\i f)$ of \cref{weyl-weighted-field} on
$D(\sqrt{H_\rad})$.  Since $\varPi(f)=\varPhi(\i f)$, this extends the
conjugate Segal field \cref{conjugate-segal-field-def}.  An
interpretation as a Segal operator requires $f\in\Hil_\ph$.
\end{enumerate}
\end{proposition}

\begin{proof}
By \cref{dipole-length-unitary}, $U_{\mathrm d}$ acts at each
electronic configuration by a Weyl operator, which is unitary and, by
\cref{prop-weyl-algebra}(ii), strongly continuous in $\vX$.  Its
dependence on $\vX$ is permutation invariant, and hence it preserves
fermionic antisymmetry.

For $\Psi\in\mathscr C_\PF$, differentiation of this multiplier gives
\begin{equation}\label{dipole-unitary-gradient-proof}
 -\i\grad_{\vx_n}U_{\mathrm d}^\dag\Psi
 =U_{\mathrm d}^\dag\vD_{n,\dip}\Psi,
 \qquad 1\leq n\leq N.
\end{equation}
Here the photon factors in the core belong to the domains of all three
field components, and their strong commutation justifies differentiation
of the corresponding unitary groups.  For general
$\Psi\in\mathcal Q_\PF$, approximate by $\mathscr C_\PF$ in the common
form norm.  The field bound \cref{A-Hrad-bound} makes
$\vD_{n,\dip}:\mathcal Q_\PF\to\HC_N^3$ continuous.  Unitarity and
closedness of the electronic weak derivative then extend
\cref{dipole-unitary-gradient-proof} to the limit.  This proves both the
finite kinetic energy of every vector in $\mathcal Q_{\mathrm d}$ and
\cref{dipole-length-derivative}, but not that
$\mathcal Q_{\mathrm d}\subset\mathcal Q_\PF$.

Pointwise unitarity preserves the electronic configuration density, and
\cref{dipole-unitary-gradient-proof} preserves the local pairings that
define the current density.  Integrating over the remaining electronic
variables gives \cref{dipole-length-current}.

It remains to prove (iv), so assume $\sqrt\omega\,\wh\chi\in L^2(\RR^3)$.
Then $G^\mu_{0}\in D(\sqrt\omega)$ by \cref{Gx-def}, and the linear term
in \cref{dipole-length-form} is defined, since
$\omega G^\mu_{0}/\sqrt\omega=\sqrt\omega\,G^\mu_{0}\in\Hil_\ph$.
For $h\in D(\sqrt\omega)$, \cref{prop-weyl-radiation}(i) with
$f=\i h$ shows that $W(\i h)$ maps $D(\sqrt{H_\rad})$ bijectively onto
itself, and \cref{prop-weyl-radiation}(ii) gives, for
$\eta\in D(\sqrt{H_\rad})$,
\begin{equation}\label{dipole-length-Weyl-shift}
 \|\sqrt{H_\rad}W(\i h)\eta\|^2
 =\|\sqrt{H_\rad}\eta\|^2
   +\varPi(\omega h)[\eta,\eta]
   +\|\sqrt\omega h\|^2\|\eta\|^2.
\end{equation}

At each electronic configuration take
$h=e\sum_\mu R^\mu G^\mu_{0}$, so that
$U_{\mathrm d}(\vX)=W(\i h)$,
$\varPi(\omega h)=e\sum_\mu R^\mu\varPi(\omega G^\mu_{0})$, and
$\|\sqrt\omega h\|^2=e^2\vR^\top\vM\vR$.
For $\Phi\in\mathscr C_\PF$ the electronic support is compact, so
\cref{weyl-radiation-bound} gives finite radiation energy of
$U_{\mathrm d}\Phi$.  Differentiation of the unitary multiplier and
its commutation with $\AF(0)$ give finite electronic kinetic energy as
well.  Thus $U_{\mathrm d}\Phi\in\mathcal Q_\PF$, that is,
$\Phi\in\mathcal Q_{\mathrm d}$.  By \cref{dipole-length-derivative}
and invariance of the Coulomb multiplication form, the kinetic and
Coulomb parts of
$\dua{\Phi}{U_{\mathrm d}^\dag H_\dip U_{\mathrm d}\Phi}$ equal
$\dua{\Phi}{H_\el^0\Phi}$.  Integrating \cref{dipole-length-Weyl-shift}
over $\vX$ gives the radiation part and proves \cref{dipole-length-form}.
\end{proof}

The integral form of the matrix \cref{dipole-polarization-matrix}
follows from \cref{Gx-def,polarization-projector}.
For the hard cutoff \cref{harduv}, the radial weight in this integral
is the indicator of the ball of radius $\Lambda$.  Since
$\int_{\mathbb S^2}(\delta^{\mu\nu}-\theta^\mu\theta^\nu)
\,\dd\theta=(8\pi/3)\delta^{\mu\nu}$ and
$\int_0^\Lambda r^2\,\dd r=\Lambda^3/3$, it follows that
\begin{equation*}
 \vM=\frac{4\pi\Lambda^3}{9}\iden.
\end{equation*}

The last term in \cref{dipole-length-form} is the dipole
self-polarization energy and is part of the unitary transformation.
Its role in stability is discussed in \cite{rokaj2018lightmatter}.
The expanded formula is asserted on the stated test space.  On the full
transported domain $\mathcal Q_{\mathrm d}$, the radiation energy is
$\|\sqrt{H_\rad}U_{\mathrm d}\Psi\|^2$, which is nonnegative and
requires neither the separate terms in the expansion to be finite nor
an electronic position moment.

Even though $\AF(0)$ is spatially uniform, its conditional expectation in
an entangled state can depend on $\vx$.  Thus the field-induced current density
need not equal $\rho(\vx)$ times the unconditional field expectation.
Moreover, this limit keeps the photon continuum: the span of the three
$G^\mu_{0}$ is generally not invariant under $\omega$.  The coupled
one-photon space is the closed span of $f(\omega)G^\mu_{0}$, with $f$
bounded Borel.  A finite-mode model requires a further approximation.
Likewise, fixing photon displacement expectations as additional density
variables would define a different constrained-search functional.

\section{Symmetric tensor lifts}\label{app-symmetric-tensor-lift}

In this appendix we construct the permutation-invariant multipliers
used in Appendix~\ref{app-current-shift} and in the proof of
\cref{thm-photonic-representability}.
We use the following notation.  For a
separable complex Hilbert space $\mathcal V$, the permutation
group $\mathscr S_N$ acts on configurations
$\vX=(\vx_1,\ldots,\vx_N)\in(\RR^3)^N$ and on $\mathcal V^{\otimes N}$
by
\[
 (\pi\vX)^i=\vx^{\pi^{-1}(i)},\qquad
 R_\pi(z_1\otimes\cdots\otimes z_N)
      =z_{\pi^{-1}(1)}\otimes\cdots\otimes z_{\pi^{-1}(N)},
\]
and $\Pi_N=\frac1{N!}\sum_{\pi\in \mathscr S_N}R_\pi$ is the projection onto the
fully symmetric tensors.  Let $\mathcal D_N\subset(\RR^3)^N$ be the open
set of configurations with pairwise distinct points.  For
$t:\RR^3\to\mathcal V$ put 
\[
T(\vX)=\bigotimes_{i=1}^Nt(\vx_i),
\]
 so that
$T(\pi\vX)=R_\pi T(\vX)$.  For a map $F$ on $\mathcal D_N$,
$\dd F=\sum_{i,\mu}\partial_{x_i^\mu}F\,\dd x_i^\mu$ denotes its
differential in $\vX$.

The product $T$ has the connection and derivatives we need, but it is
not permutation invariant.  The next lemma provides isometries
$U_N(\vX)$ that absorb the permutations, so that $U_NT$ is invariant.
The flatness condition $U_N^\dag\dd U_N=0$ ensures that composition
with $U_N$ leaves the connection of $T$ unchanged.  The construction
glues local maps with a partition of unity whose gradients blow up
when all electrons come together.  There $T$ is almost fully
symmetric, which is why the fully symmetric part is kept fixed.  The
resulting derivative bound is uniform when $t$ is Lipschitz, even when
electron coordinates coincide.  For general $t$ it is controlled by the
inverse squared pair distances, which are integrable against
finite-energy states by Hardy's inequality.

\begin{lemma}[Flat symmetric isometries]
\label[lemma]{lem-flat-symmetric-isometries}
For every $N\in\NN$ and every separable complex Hilbert space
$\mathcal V$, there are a separable complex Hilbert space
$\mathcal K_N$, finite-dimensional if $\mathcal V$ is, a smooth map
$U_N$ from $\mathcal D_N$ to the isometries
$\mathcal V^{\otimes N}\to\mathcal K_N$, and a constant $b_N$ depending
only on $N$ with the following properties.  For $\vX\in\mathcal D_N$ and
$\pi\in \mathscr S_N$,
\begin{equation}\label{symmetric-tensor-lift-flat}
 U_N(\pi\vX)R_\pi=U_N(\vX),\qquad
 U_N^\dag\dd U_N=0,\qquad U_N(\vX)\Pi_N\text{ is independent of }\vX,
\end{equation}
and for every map $t:\RR^3\to\mathcal V$ with $\|t(\vx)\|=1$,
\begin{equation}\label{symmetric-tensor-lift-extra-metric}
 \sum_{i=1}^N\sum_{\mu=1}^3
   \|(\partial_{x_i^\mu}U_N(\vX))T(\vX)\|^2
 \leq b_N\min\Big\{\sum_{i<j}\frac1{|\vx_i-\vx_j|^2},\
   \|\grad t\|_\infty^2\Big\},
\end{equation}
where $\|\grad t\|_\infty=\infty$ unless $t$ is Lipschitz.
\end{lemma}
\phantomsection
\label{proof:lem-flat-symmetric-isometries}
\begin{proof}
We argue by induction on $N$.  For $N=1$ take $\mathcal K_1=\mathcal V$,
$U_1=\iden$, and $b_1=0$.  Let $N\geq2$ and suppose that
$\mathcal K_{N'}$, $U_{N'}$, and $b_{N'}$ have been constructed for all
$N'<N$.

\noindent\textbf{Step 1. An invariant partition of unity.}
Put
\[
 \bar\vx=\frac1N\sum_i\vx_i,\qquad
 r(\vX)=\Big(\sum_i|\vx_i-\bar\vx|^2\Big)^{1/2},\qquad
 s(\vX)=\frac{(\vx_1-\bar\vx,\ldots,\vx_N-\bar\vx)}{r(\vX)}.
\]
Since $N\geq2$, $r>0$ on $\mathcal D_N$, and $s$ is smooth there with
values in the compact centered unit sphere
\[
 \mathbb S_N=\Big\{\vY\in(\RR^3)^N:\ \sum_i\vy_i=0,\
 \sum_i|\vy_i|^2=1\Big\}.
\]
Moreover $r(\pi\vX)=r(\vX)$ and $s(\pi\vX)=\pi s(\vX)$.  The stabilizer
of a point $\vY\in\mathbb S_N$ is the product of the permutation groups
of the blocks of indices $i$ with equal $\vy_i$.  Every block has fewer
than $N$ elements, because $\mathbb S_N$ excludes total coincidence.
Choose a ball $O$ in $\mathbb S_N$ centered at $\vY$, so small that
$\pi O\cap O=\emptyset$ for every $\pi$ outside the stabilizer.  Since
permutations act isometrically, $O$ is invariant under the stabilizer,
and any two translates $\pi O$, $\pi'O$ are either equal or disjoint.
By compactness, finitely many such balls $O_a$, with blocks $I_{a,j}$,
cover $\mathbb S_N$.  Take smooth bump functions supported in the
$O_a$ whose positivity sets still cover $\mathbb S_N$, average each
over the stabilizer of $O_a$, extend it to the translates $\pi O_a$ by
invariance, and normalize by the square root of the sum of their
squares.  This gives smooth permutation-invariant functions $\chi_a$
on $\mathbb S_N$, supported compactly inside $\bigcup_\pi\pi O_a$, with
$\sum_a\chi_a^2=1$.  We write $\chi_a$ also for $\chi_a\circ s$.  Then
$\sum_a\chi_a\dd\chi_a=0$, and since the derivatives of $s$ are
$O(r^{-1})$,
\[
 \sum_{a,i,\mu}|\partial_{x_i^\mu}\chi_a(\vX)|^2
 \leq A_Nr(\vX)^{-2},
\]
where $A_N<\infty$ depends only on the chosen atlas, hence only on $N$.

\noindent\textbf{Step 2. Local maps.}
Fix a chart $a$.  For $\vX\in\mathcal D_N$ with $s(\vX)\in O_a$ put
\[
 V_a(\vX)=\bigotimes_jU_{|I_{a,j}|}(\vX_{I_{a,j}}),\qquad
 \vX_I=(\vx_i)_{i\in I},
\]
where $\mathcal V^{\otimes N}$ is identified with
$\bigotimes_j\mathcal V^{\otimes I_{a,j}}$ by a fixed reordering of the
tensor factors.  By the inductive hypothesis, $V_a$ is a smooth
isometry into the separable space
$\mathcal K^{(a)}=\bigotimes_j\mathcal K_{|I_{a,j}|}$, it satisfies
$V_a(\tau\vX)R_\tau=V_a(\vX)$ for $\tau$ in the stabilizer of $O_a$, and
$V_a^\dag\dd V_a=0$ by the product rule.  Moreover, $V_a\Pi_N$ is
constant: a fully symmetric tensor is symmetric within each block, so
$\Pi_N=(\bigotimes_j\Pi_{|I_{a,j}|})\Pi_N$, and each
$U_{|I_{a,j}|}\Pi_{|I_{a,j}|}$ is constant.  Extend $V_a$ to the
translates by
\[
 \widetilde V_a(\vX)=V_a(\pi^{-1}\vX)R_{\pi^{-1}},\qquad
 s(\vX)\in\pi O_a.
\]
If $\pi O_a=\pi'O_a$, then $\pi^{-1}\pi'$ lies in the stabilizer of
$O_a$, and the invariance of $V_a$ under the stabilizer shows that
$\widetilde V_a$ does not depend on the choice of $\pi$.  By
construction, $\widetilde V_a(\pi\vX)R_\pi=\widetilde V_a(\vX)$.  The
map $\widetilde V_a$ inherits $\widetilde V_a^\dag\dd\widetilde V_a=0$,
and $\widetilde V_a\Pi_N$ is constant because $R_{\pi^{-1}}\Pi_N=\Pi_N$.

\noindent\textbf{Step 3. Definition of $U_N$.}
Let $\mathcal K_N=\operatorname{Ran}\Pi_N\oplus\bigoplus_a\mathcal K^{(a)}$,
an orthogonal direct sum, and define
\[
 U_N(\vX)u=\Pi_Nu\ \oplus\
   \bigoplus_a\chi_a(\vX)\widetilde V_a(\vX)(\iden-\Pi_N)u,
\]
where each term is extended by zero outside the support of $\chi_a$.
Since $\sum_a\chi_a^2=1$ and each $\widetilde V_a$ is an isometry,
$\|U_N(\vX)u\|^2=\|\Pi_Nu\|^2+\|(\iden-\Pi_N)u\|^2=\|u\|^2$.  The first
identity in \cref{symmetric-tensor-lift-flat} follows from the
invariance of $\chi_a$ and $\widetilde V_a$ and from
$R_\pi \Pi_N=\Pi_NR_\pi=\Pi_N$.  The third holds because
$U_N(\vX)\Pi_Nu=\Pi_Nu\oplus0$.  For the second,
\[
 U_N^\dag\dd U_N
 =(\iden-\Pi_N)\Big(\sum_a\chi_a\dd\chi_a\,
      \widetilde V_a^\dag\widetilde V_a
   +\sum_a\chi_a^2\,\widetilde V_a^\dag\dd\widetilde V_a\Big)
   (\iden-\Pi_N)=0.
\]

\noindent\textbf{Step 4. The derivative bound.}
Let $t$ be as in \cref{symmetric-tensor-lift-extra-metric}, and let
$m(\vX)$ denote the minimum on its right-hand side, without the factor
$b_N$.  Since $T$ is a unit vector, $\|(\iden-\Pi_N)T\|\leq1$.  If $t$
is Lipschitz, then, since $t(\bar\vx)^{\otimes N}\in\operatorname{Ran}\Pi_N$
and all $t(\vx_i)$ are unit vectors, telescoping the tensor product
gives
\[
 \|(\iden-\Pi_N)T\|
 \leq\|T-t(\bar\vx)^{\otimes N}\|
 \leq\sum_i\|t(\vx_i)-t(\bar\vx)\|
 \leq\sqrt N\,r(\vX)\|\grad t\|_\infty.
\]
Moreover, $r^2=N^{-1}\sum_{i<j}|\vx_i-\vx_j|^2$, so
$r^{-2}\leq N|\vx_k-\vx_l|^{-2}$ for every pair $k<l$.  Using
$\|(\iden-\Pi_N)T\|\leq1$ with this for the first entry of the
minimum, and the telescoping bound for the second,
\begin{equation}\label{symmetric-tensor-lift-remainder}
 r(\vX)^{-2}\|(\iden-\Pi_N)T(\vX)\|^2\leq N\,m(\vX).
\end{equation}
Since $\widetilde V_a\Pi_N$ is constant,
$(\partial\widetilde V_a)(\iden-\Pi_N)T=(\partial\widetilde V_a)T$ for
$\partial=\partial_{x_i^\mu}$, and therefore
\[
 (\partial U_N)T
 =0\oplus\bigoplus_a\Big((\partial\chi_a)\widetilde V_a(\iden-\Pi_N)T
   +\chi_a(\partial\widetilde V_a)T\Big).
\]
The cross terms in the squared norm vanish by
$\widetilde V_a^\dag\partial\widetilde V_a=0$, so
\[
 \sum_{i,\mu}\|(\partial_{x_i^\mu}U_N)T\|^2
 =\sum_{a,i,\mu}|\partial_{x_i^\mu}\chi_a|^2\|(\iden-\Pi_N)T\|^2
  +\sum_a\chi_a^2\sum_{i,\mu}\|(\partial_{x_i^\mu}\widetilde V_a)T\|^2.
\]
By Step 1 and \cref{symmetric-tensor-lift-remainder}, the first sum is
at most $NA_Nm(\vX)$.  For the second, let $s(\vX)\in\pi O_a$ and
$\vY=\pi^{-1}\vX$.  Since $R_{\pi^{-1}}T(\vX)=T(\vY)$ and the sum
over $i,\mu$ is invariant under relabeling the particles,
\[
 \sum_{i,\mu}\|(\partial_{x_i^\mu}\widetilde V_a)(\vX)T(\vX)\|^2
 =\sum_{i,\mu}\|(\partial_{y_i^\mu}V_a)(\vY)T(\vY)\|^2.
\]
The vector $V_a(\vY)T(\vY)$ is the tensor product over $j$ of the
unit vectors
\[
 U_{|I_{a,j}|}(\vY_{I_{a,j}})\,T(\vY_{I_{a,j}}),
\]
and a derivative in a coordinate from block $I_{a,j}$ acts only on the
$j$th factor.  The
inductive bound therefore estimates the right-hand side by
$\sum_jb_{|I_{a,j}|}m(\vX)$, since the pairs within a block are among
all pairs and the sum over all pairs is permutation invariant.  Hence
\cref{symmetric-tensor-lift-extra-metric} holds with
\[
 b_N=NA_N+\max_a\sum_jb_{|I_{a,j}|},
\]
which depends only on $N$.  This completes the induction.
\end{proof}

\begin{lemma}[A symmetric tensor lift with controlled derivatives]
\label[lemma]{lem-symmetric-tensor-lift}
Let $N\in\NN$, let $\mathcal V$ be a separable complex Hilbert space, and
let $\mathcal K_N$ and $b_N$ be as in
\cref{lem-flat-symmetric-isometries}.
If $t\in H^1_{\mathrm{loc}}(\RR^3;\mathcal V)$ and $\|t(\vx)\|=1$, there
is a permutation-invariant map
$K\in H^1_{\mathrm{loc}}((\RR^3)^N;\mathcal K_N)$ such that, almost
everywhere,
\begin{align}
 \|K(\vX)\|^2&=1,\qquad
 \langle K(\vX),\partial_{x_i^\mu}K(\vX)\rangle
   =\langle t(\vx_i),\partial_\mu t(\vx_i)\rangle,
 \label{symmetric-tensor-lift-connection}\\
 \sum_{i=1}^N\sum_{\mu=1}^3
       \|\partial_{x_i^\mu}K(\vX)\|^2
   &\leq\sum_{i=1}^N\|\grad t(\vx_i)\|^2
   +b_N\min\Big\{\sum_{i<j}\frac1{|\vx_i-\vx_j|^2},\
     \|\grad t\|_\infty^2\Big\}.
 \label{symmetric-tensor-lift-metric}
\end{align}
If $t\in W^{1,\infty}(\RR^3;\mathcal V)$, then
$K\in W^{1,\infty}((\RR^3)^N;\mathcal K_N)$, and the right-hand side of
\cref{symmetric-tensor-lift-metric} is at most $C_N\|\grad t\|_\infty^2$
with $C_N=N+b_N$.  Here $\vX=(\vx_1,\ldots,\vx_N)$ and the inner
product is linear in its second argument.
\end{lemma}
\phantomsection
\label{proof:lem-symmetric-tensor-lift}
\begin{proof}
Let $U_N$ be as in \cref{lem-flat-symmetric-isometries}, take the
Lipschitz representative of $t$ if $t$ is Lipschitz, and set
$K(\vX)=U_N(\vX)T(\vX)$ on $\mathcal D_N$.  Because $t$ is bounded and
in $H^1_{\mathrm{loc}}$, so is $T$.  Since
$T(\pi\vX)=R_\pi T(\vX)$, the first identity in
\cref{symmetric-tensor-lift-flat} makes $K$ permutation invariant, and
$\|K\|=\|T\|=1$.  For $\partial=\partial_{x_i^\mu}$,
\[
 \partial K=(\partial U_N)T+U_N\partial T
\]
almost everywhere.  Flatness gives
$\langle K,(\partial U_N)T\rangle
=\langle T,U_N^\dag(\partial U_N)T\rangle=0$.  Since $U_N$ is an
isometry and the other factors of $T$ are unit vectors,
\[
 \langle K,\partial K\rangle
 =\langle T,\partial T\rangle
 =\langle t(\vx_i),\partial_\mu t(\vx_i)\rangle,
\]
which is \cref{symmetric-tensor-lift-connection}.  Flatness also removes
the cross term in $\|\partial K\|^2$, so
\[
 \sum_{i,\mu}\|\partial_{x_i^\mu}K\|^2
 =\sum_i\|\grad t(\vx_i)\|^2
   +\sum_{i,\mu}\|(\partial_{x_i^\mu}U_N)T\|^2,
\]
and \cref{symmetric-tensor-lift-extra-metric} gives
\cref{symmetric-tensor-lift-metric} on $\mathcal D_N$.  The collision
diagonals have codimension three, so $|\vx_i-\vx_j|^{-2}$ is locally
integrable and the right-hand side of
\cref{symmetric-tensor-lift-metric} lies in
$L^1_{\mathrm{loc}}((\RR^3)^N)$.  Almost every line parallel to a
coordinate axis misses the diagonals, so the absolute-continuity
criterion on coordinate lines extends $K$ to
$H^1_{\mathrm{loc}}((\RR^3)^N;\mathcal K_N)$, and to
$W^{1,\infty}((\RR^3)^N;\mathcal K_N)$ when $t$ is Lipschitz, since the
bound is then uniform.  The extension remains permutation invariant and
has norm one.  In the Lipschitz case, the right-hand side of
\cref{symmetric-tensor-lift-metric} is at most
$(N+b_N)\|\grad t\|_\infty^2$.
\end{proof}

\section{Paramagnetic current density shift}\label{app-current-shift}
The next lemma shifts the paramagnetic (and hence the combined) current density
by combining six
contributions to the mixed state, corresponding to momentum kicks in
the positive and negative coordinate directions.
Their position-dependent weights make the average kick equal to the
prescribed velocity shift $\vv$, while preserving the electronic
density and photon reduced state.

\begin{lemma}[Exact current shift for one electron]
\label[lemma]{lem-one-electron-current-repair}
Let $N=1$, $\vGamma\in\dm$, and
$\vv\in W^{1,\infty}(\RR^3,\RR^3)$ with $\vv\neq0$.  There is a finite
mixture $\vGamma_{\vv}\in\dm$ with
\begin{equation}\label{one-electron-repaired-constraints}
 \rho_{\vGamma_{\vv}}=\rho_\vGamma,\qquad
 \vjc_{\vGamma_{\vv}}=\vjc_\vGamma+\rho_\vGamma\vv,
 \qquad
 \Tr_\el\vGamma_{\vv}=\Tr_\el\vGamma,
\end{equation}
and
\begin{equation*}
 \Tr(H^{0,0}\vGamma_{\vv})
 \leq\Tr(H^{0,0}\vGamma)+\int_{\RR^3}\vjc_\vGamma\cdot\vv
       +36\|\vv\|_\infty^2+\left(\int_{\RR^3}\rho_\vGamma
                         |\grad\vv|^2\right)^{1/2}.
\end{equation*}
\end{lemma}
\phantomsection
\label{proof:lem-one-electron-current-repair}
\begin{proof}
For $q>3\|\vv\|_\infty$, let $K_{\mu,\sigma}$ be electronic multiplication by
\[
 K_{\mu,\sigma}(\vx)
 =\frac1{\sqrt6}
       \sqrt{1+\frac{3\sigma v^\mu(\vx)}q}\,
                   e^{\i\sigma qx^\mu},
 \qquad \mu=1,2,3,\quad\sigma\in\{-1,1\},
\]
and set 
\[
\vGamma_{\vv,q}
=\sum_{\mu,\sigma}K_{\mu,\sigma}\vGamma K_{\mu,\sigma}^\dag, \qquad \vGamma_{\vv,q}(\vx,\vx')
=\sum_{\mu,\sigma} K_{\mu,\sigma}(\vx) \, \vGamma(\vx,\vx') \, \ol{K_{\mu,\sigma}(\vx')}
\]
These multipliers are bounded with bounded weak derivatives and hence
act boundedly on $\mathcal Q_\PF$.  In particular,
$\vGamma_{\vv,q}$ has finite form energy.  Directly from their definition,
\begin{align*}
 \sum_{\mu,\sigma}|K_{\mu,\sigma}(\vx)|^2
 &=\frac16\sum_{\mu,\sigma}
       \left(1+\frac{3\sigma v^\mu(\vx)}{q}\right)=1,\\
 \sum_{\mu,\sigma}\ol{K_{\mu,\sigma}(\vx)} \grad K_{\mu,\sigma}(\vx)
 &=\frac12\grad\sum_{\mu,\sigma}|K_{\mu,\sigma}(\vx)|^2
   +\i q\sum_{\mu,\sigma}\sigma|K_{\mu,\sigma}(\vx)|^2\ve_\mu
 =\i\vv(\vx),
\end{align*}
where $\ve_\mu$ is the $\mu$th coordinate unit vector.  The above
scalar relation means that
$\sum_{\mu,\sigma}K_{\mu,\sigma}^\dag K_{\mu,\sigma}=\iden$ as an operator.
Each summand defining $\vGamma_{\vv,q}$ is positive and trace class,
and cyclicity of the trace gives
\[
 \Tr\vGamma_{\vv,q}
 =\Tr\!\left(\vGamma\sum_{\mu,\sigma}
             K_{\mu,\sigma}^\dag K_{\mu,\sigma}\right)
 =\Tr\vGamma=1.
\]
Consequently, $\vGamma_{\vv,q}\in\dm$.

For the density, write a spectral decomposition
$\vGamma=\sum_rp_r|\Psi_r\rangle\langle\Psi_r|$ with $p_r>0$.
We use the density and current formulas also for unnormalized vectors,
with their usual quadratic scaling.  Since $N=1$ and each multiplier
is scalar in the photon variables,
$\rho_{K_{\mu,\sigma}\Psi_r}
=|K_{\mu,\sigma}|^2\rho_{\Psi_r}$.  Hence, almost everywhere,
\[
 \rho_{\vGamma_{\vv,q}}
 =\sum_rp_r\sum_{\mu,\sigma}
        |K_{\mu,\sigma}|^2\rho_{\Psi_r}
 =\sum_rp_r\rho_{\Psi_r}=\rho_\vGamma.
\]
To check the photon reduced state, let $B\in\mathcal B(\Fock_\ph)$.
Each $K_{\mu,\sigma}$ acts only on the electronic factor and therefore
commutes with $\iden_\el\otimes B$.  By the definition
\cref{partial-trace-duality} of the partial trace and cyclicity of the
trace,
\begin{align*}
 \Tr\!\left((\Tr_\el\vGamma_{\vv,q})B\right)
 &=\sum_{\mu,\sigma}\Tr\!\left(
   \vGamma K_{\mu,\sigma}^\dag(\iden_\el\otimes B)K_{\mu,\sigma}\right)\\
 &=\Tr\!\left(\vGamma(\iden_\el\otimes B)
          \sum_{\mu,\sigma}K_{\mu,\sigma}^\dag K_{\mu,\sigma}\right)\\
 &=\Tr\!\left((\Tr_\el\vGamma)B\right).
\end{align*}
Since this holds for every bounded $B$, it proves
$\Tr_\el\vGamma_{\vv,q}=\Tr_\el\vGamma$.

For the current density, fix one multiplier $K=K_{\mu,\sigma}$.  The field
operator commutes with electronic scalar multiplication, so the weak
product rule gives
\[
 \vD_1(K\Psi)=K\vD_1\Psi-\i(\grad K)\Psi.
\]
Using $\vjc_\Psi=\Re\dua{\Psi}{\vD_1\Psi}_{\vx}$ for $N=1$, we obtain
\[
 \vjc_{K\Psi}
 =\Re\dua{K\Psi}{K\vD_1\Psi-\i(\grad K)\Psi}_{\vx}
 =|K|^2\vjc_\Psi+\Im(\ol K\grad K)\rho_\Psi.
\]
Summing over the six multipliers therefore yields
\[
 \sum_{\mu,\sigma}\vjc_{K_{\mu,\sigma}\Psi}
 =\vjc_\Psi
  +\rho_\Psi\Im\!\left(
       \sum_{\mu,\sigma}\ol K_{\mu,\sigma}\grad K_{\mu,\sigma}\right)
 =\vjc_\Psi+\rho_\Psi\vv.
\]
Since the multipliers commute with $\AF(\vx)$, the field-induced
current density is unchanged and the shift is entirely paramagnetic.
Apply this to each $\Psi_r$ and sum with weights $p_r$ to conclude
\[
 \vjc_{\vGamma_{\vv,q}}
 =\sum_rp_r\big(\vjc_{\Psi_r}+\rho_{\Psi_r}\vv\big)
 =\vjc_\vGamma+\rho_\vGamma\vv.
\]
The density series converges in $L^1$ by positivity and normalization,
and the current series by Cauchy--Schwarz and the finite
mean form energy.  Thus the identities hold also for an infinite-rank
$\vGamma$.

Expanding the squared norm in the same product rule gives
\begin{align}
 \Tr(H^{0,0}\vGamma_{\vv,q})-\Tr(H^{0,0}\vGamma)
 &=\int_{\RR^3}\vjc_\vGamma\cdot\vv+\frac{q^2}{2}
 \notag\\
 &\quad+\frac{3}{8q^2}\sum_{\mu=1}^3
      \int_{\RR^3}\rho_\vGamma
        \frac{|\grad v^\mu|^2}{1-9(v^\mu)^2/q^2}.
 \label{one-electron-repair-exact-energy}
\end{align}
The radiation energy is unchanged.  These identities extend by
form approximation, or termwise in a spectral decomposition of
$\vGamma$.
Choose
\[
 q^2=\max\left\{36\|\vv\|_\infty^2,
 \left(\int_{\RR^3}\rho_\vGamma|\grad\vv|^2\right)^{1/2}\right\},
\]
so that $q\geq6\|\vv\|_\infty>3\|\vv\|_\infty$ since $\vv\neq0$.
The denominators in \cref{one-electron-repair-exact-energy} are at
least $3/4$, and the last two terms are at most
\[
 \frac{q^2}{2}
 +\frac{1}{2q^2}\int_{\RR^3}\rho_\vGamma|\grad\vv|^2
 \leq q^2
 \leq36\|\vv\|_\infty^2
 +\left(\int_{\RR^3}\rho_\vGamma|\grad\vv|^2\right)^{1/2}.
\]
Set $\vGamma_{\vv}=\vGamma_{\vv,q}$ with this choice of $q$.  Photon
localization acts only
on the photon factor, so equality of photon marginals also preserves
the localized photon-number distributions.
\end{proof}

To lift this construction to the antisymmetric $N$-electron space, we
replace tensor products of the one-electron multipliers by symmetric
multipliers with the same connection.  These are constructed in
Appendix~\ref{app-symmetric-tensor-lift}, and only
\cref{lem-symmetric-tensor-lift} is used below.  The lift applies to
any unit-vector field with connection $\i\vv$, not only to the six
multipliers above.  Part (ii) of the next lemma uses this to remove the
boundedness assumptions on $\vv$.

\begin{lemma}[Exact current shift for arbitrary particle number]
\label[lemma]{lem-fermionic-current-repair}
Let $N\in\NN$ and $\vGamma\in\dm$.
\begin{enumerate}[(i)]
\item If $\vv\in W^{1,\infty}(\RR^3,\RR^3)$ and $\vv\neq0$, there is a
finite electronic multiplication mixture $\vGamma_{\vv}\in\dm$
satisfying \cref{one-electron-repaired-constraints} and
\begin{equation}\label{fermionic-repair-cost}
 \Tr(H^{0,0}\vGamma_{\vv})
 \leq\Tr(H^{0,0}\vGamma)+\int_{\RR^3}\vjc_\vGamma\cdot\vv
       +C_N\big(\|\vv\|_\infty^2+\|\grad\vv\|_\infty\big).
\end{equation}
\item Let $\mathcal V$ be a separable complex Hilbert space, let
$t\in H^1_{\mathrm{loc}}(\RR^3;\mathcal V)$ satisfy $\|t\|=1$ and
$\int_{\RR^3}\rho_\vGamma\|\grad t\|^2<\infty$, and put
$\vv=\Im\dua{t}{\grad t}$.  Then there is a countable electronic
multiplication mixture $\vGamma_{\vv}\in\dm$ satisfying
\cref{one-electron-repaired-constraints} and
\begin{equation}\label{fermionic-repair-cost-general}
 \Tr(H^{0,0}\vGamma_{\vv})
 \leq(1+C_N)\Tr(H^{0,0}\vGamma)+\int_{\RR^3}\vjc_\vGamma\cdot\vv
   +\frac12\int_{\RR^3}\rho_\vGamma\|\grad t\|^2.
\end{equation}
\end{enumerate}
Here $C_N$ depends only on $N$.  In both cases, $\vGamma_{\vv}$ and
$\vGamma$ have the same Coulomb energy.
\end{lemma}
\phantomsection
\label{proof:lem-fermionic-current-repair}
\begin{proof}
\emph{(i)} Let $K_{\mu,\sigma}$ be the six multipliers from the
proof of \cref{lem-one-electron-current-repair}, with
$q^2=\max\{36\|\vv\|_\infty^2,\|\grad\vv\|_\infty\}$, and collect
them in the unit vector
$t(\vx)=(K_{\mu,\sigma}(\vx))_{\mu,\sigma}\in\CC^6$.
By the identities established there, $\dua{t}{\grad t}=\i\vv$ and
\[
 |\grad t|^2
 =q^2+\frac{3}{4q^2}\sum_{\mu=1}^3
       \frac{|\grad v^\mu|^2}{1-9(v^\mu)^2/q^2}
 \leq q^2+\frac{\|\grad\vv\|_\infty^2}{q^2}
 \leq36\|\vv\|_\infty^2+2\|\grad\vv\|_\infty.
\]
Apply \cref{lem-symmetric-tensor-lift} with $\mathcal V=\CC^6$ to $t$,
fix an orthonormal basis $(e_\alpha)_{\alpha=1}^{\dim\mathcal K_N}$ of
$\mathcal K_N$, and let $K_\alpha=\langle e_\alpha,K\rangle$.  These
are finitely many permutation-invariant scalar multipliers with
\begin{gather*}
 \sum_\alpha|K_\alpha|^2=1,\qquad
 \sum_\alpha\ol K_\alpha\grad_{\vx_i}K_\alpha=\i\vv(\vx_i),\\
 \sum_{i,\alpha}|\grad_{\vx_i}K_\alpha|^2
 \leq C_N\big(\|\vv\|_\infty^2+\|\grad\vv\|_\infty\big).
\end{gather*}
Define $\vGamma_{\vv}
=\sum_\alpha K_\alpha\vGamma K_\alpha^\dag$.
Symmetry and bounded weak derivatives preserve the fermionic form
domain.  Since the $K_\alpha$ act by multiplication in $\vX$ and
$\sum_\alpha|K_\alpha|^2=1$, the first identity preserves the density,
the Coulomb energy, and the photon marginal.  The second identity gives
the current shift, and the third
bounds the kinetic remainder after expanding each covariant derivative.
The radiation energy is unchanged.

\emph{(ii)} Apply \cref{lem-symmetric-tensor-lift} to $t$, fix an
orthonormal basis $(e_\alpha)$ of $\mathcal K_N$, and let
$K_\alpha=\langle e_\alpha,K\rangle$.  These are countably many
permutation-invariant scalar multipliers with
\begin{gather*}
 \sum_\alpha|K_\alpha|^2=1,\qquad
 \sum_\alpha\ol K_\alpha\grad_{\vx_i}K_\alpha=\i\vv(\vx_i),\\
 \sum_{i,\alpha}|\grad_{\vx_i}K_\alpha|^2
 \leq\sum_i\|\grad t(\vx_i)\|^2
   +b_N\sum_{i<j}\frac1{|\vx_i-\vx_j|^2}.
\end{gather*}
Since $\sum_\alpha\Tr(K_\alpha\vGamma K_\alpha^\dag)=1$, the series
$\vGamma_{\vv}=\sum_\alpha K_\alpha\vGamma K_\alpha^\dag$ converges in
trace norm.  The multipliers need not have bounded derivatives.
Still, for every eigenvector $\Psi$ of
$\vGamma$ with positive eigenvalue, the last bound keeps
$\sum_\alpha\|\grad_{\vx_n}(K_\alpha\Psi)\|^2$ finite.  Its first part
contributes $\int\rho_\Psi\|\grad t\|^2$, which is finite because
$\int\rho_\vGamma\|\grad t\|^2<\infty$, and its second part is
controlled by the Hardy bound below.  Hence each $K_\alpha\Psi$ lies in
the form
domain, and the density, the Coulomb energy, the photon marginal, and
the current shift follow as in (i).  Expanding each covariant
derivative gives exactly
\[
 \Tr(H^{0,0}\vGamma_{\vv})
 =\Tr(H^{0,0}\vGamma)+\int_{\RR^3}\vjc_\vGamma\cdot\vv
  +\frac12\Tr\Big(\vGamma\sum_{i,\alpha}|\grad_{\vx_i}K_\alpha|^2\Big),
\]
where the middle integral converges absolutely by
\cref{joint-density-current-bound} and the Cauchy--Schwarz inequality,
since $|\vv|\leq\|\grad t\|$.  The first part of the multiplier bound
contributes $\frac12\int\rho_\vGamma\|\grad t\|^2$.  For the second,
we use the diamagnetic inequality
$|\grad_{\vx_n}\|\Psi(\vX)\||\leq\|(\vD_n\Psi)(\vX)\|$ and Hardy's
inequality in the relative coordinate $\vx_i-\vx_j$.  They give
\[
 \int\frac{\|\Psi(\vX)\|^2}{|\vx_i-\vx_j|^2}\,\dd\vX
 \leq2\big(\|\vD_i\Psi\|^2+\|\vD_j\Psi\|^2\big),
\]
and hence
\begin{equation*}
 \Tr\Big(\vGamma\sum_{i<j}\frac1{|\vx_i-\vx_j|^2}\Big)
 \leq2(N-1)\sum_{n=1}^N\Tr(\vD_n^2\vGamma)
 \leq4(N-1)\Tr(H^{0,0}\vGamma).
\end{equation*}
This proves \cref{fermionic-repair-cost-general} with
$C_N^{\mathrm{(ii)}}=2(N-1)b_N$.  Taking the maximum of this constant
and the constant obtained in part (i) gives a common $C_N$ for both
estimates.
\end{proof}

Part (ii) needs a unit-vector field with prescribed connection.  For
connections of the form $\grad S+\vw$ with $\vw$ in a weighted Sobolev
space, coherent states provide one.

\begin{lemma}[Coherent-state tags]
\label[lemma]{lem-coherent-tag}
Let $\mathcal V=\Fock_s(\CC^3)$, with vacuum $\Omega_{\mathcal V}$ and
Weyl operators $W(z)$, $z\in\CC^3$, as in \cref{weyl-definitions}.  Let
$S\in H^1_{\mathrm{loc}}(\RR^3;\RR)$,
$\vw\in H^1_{\mathrm{loc}}(\RR^3;\RR^3)$ and $\lambda>0$, and put
\[
 h(\vx)=\frac1{\sqrt2}\sum_{k=1}^3
   \Big(\lambda w_k(\vx)+\frac{\i x_k}{\lambda}\Big)\ve_k,\qquad
 t(\vx)=e^{\i(S(\vx)+\frac12\vx\cdot\vw(\vx))}\,
   W(h(\vx))\Omega_{\mathcal V},
\]
where $(\ve_k)$ is the standard basis of $\CC^3$.  Then
$t\in H^1_{\mathrm{loc}}(\RR^3;\mathcal V)$, $\|t\|=1$, and almost
everywhere
\begin{equation}\label{coherent-tag-identities}
 \Im\dua{t}{\grad t}=\grad S+\vw,\qquad
 \|\grad t\|^2=|\grad S+\vw|^2
   +\frac12\big(\lambda^2|\grad\vw|^2+3\lambda^{-2}\big).
\end{equation}
\end{lemma}
\phantomsection
\label{proof:lem-coherent-tag}
\begin{proof}
The phase $S+\frac12\vx\cdot\vw$ lies in $H^1_{\mathrm{loc}}$, and
$h\in H^1_{\mathrm{loc}}(\RR^3;\CC^3)$.  Since
$\sum_kx_k\partial_jw_k=\partial_j(\vx\cdot\vw)-w_j$,
\begin{gather*}
 \Im\dua{h}{\partial_jh}
 =\frac12\sum_k\big(w_k\delta_{jk}-x_k\partial_jw_k\big)
 =w_j-\frac12\partial_j(\vx\cdot\vw),\\
 \sum_j|\partial_jh|^2
 =\frac12\big(\lambda^2|\grad\vw|^2+3\lambda^{-2}\big).
\end{gather*}
Hence
$\partial_j(S+\frac12\vx\cdot\vw)+\Im\dua{h}{\partial_jh}=\partial_jS+w_j$, and
\cref{prop-weyl-coherent-derivatives}(iii) gives
\cref{coherent-tag-identities}.  Unitarity of $W(h)$ gives $\|t\|=1$.  The right-hand side of the second identity in \cref{coherent-tag-identities} is
locally integrable, so $t\in H^1_{\mathrm{loc}}$ by the same
proposition.
\end{proof}

With this tag, the last term in \cref{fermionic-repair-cost-general}
is $\frac12\int\rho_\vGamma\|\grad t\|^2$, which involves only the
weighted norms $\int\rho_\vGamma(|\grad S+\vw|^2+|\grad\vw|^2)$ and $N$.
For instance, the rigid rotation $\vw=\vomega\times\vx$ is allowed for
every $\vGamma$ with $\int\rho_\vGamma|\vx|^2<\infty$.

\section{Gaussian representation of the field}
\label{app-soft-gaussian}

In this appendix we represent the field in a subspace of the photon
modes by multiplication operators and condition states on its values.

In the solenoidal representation, recall the conjugation
$(J\vf)(\vk)=\ol{\vf(-\vk)}$.  The symmetry
$\wh\chi(\vk)=\ol{\wh\chi(-\vk)}$ in \cref{cutoff-assumption} gives
$J\vG^\mu_{\vx}=\vG^\mu_{\vx}$.  Fix an infinite-dimensional closed
subspace $\mathfrak h\subset\Hil_\ph$ with $J\mathfrak h\subset\mathfrak h$,
and let $P_{\mathfrak h}$ be the orthogonal projection onto it.  Since
$J$ is an antiunitary involution, also
$J\mathfrak h^\perp\subset\mathfrak h^\perp$, so $P_{\mathfrak h}$
commutes with $J$ and every $P_{\mathfrak h}G^\mu_{\vx}$ is $J$-real.
The exponential law for Fock spaces gives
$\Fock_\ph\simeq\Fock_s(\mathfrak h^\perp)\otimes\Fock_s(\mathfrak h)$.
Put $\mathcal K=\Hil_\el^N\otimes\Fock_s(\mathfrak h^\perp)$, so that
$\HC_N\simeq\mathcal K\otimes\Fock_s(\mathfrak h)$, and let
$\AF_{\mathfrak h}(\vx)$ have components
$\Af^\mu_{\mathfrak h}(\vx)=\varPhi(P_{\mathfrak h}G^\mu_{\vx})$.
We use two cases.  For the whole field, $\mathfrak h=\Hil_\ph$ and
$\mathcal K=\Hil_\el^N$.  For the soft field, fix $\kappa>0$ and use
the notation introduced before \cref{lem-exact-soft-removal}: then
$\mathfrak h=\Hil_{<\kappa}$, $P_{\mathfrak h}=\iden_{\{\omega<\kappa\}}$,
and $\mathcal K=\Hil_\el^N\otimes\Fock_{\geq\kappa}$.  This subspace is
$J$-invariant because $\omega(-\vk)=\omega(\vk)$.

Scalar products of $J$-real vectors are real, so all components of
$\AF_{\mathfrak h}$ commute strongly by \cref{prop-weyl-algebra}(iii).
Let $(\xi_n)_{n\geq1}$ be a $J$-real orthonormal basis
of $\mathfrak h$, equip $\mathcal Q=\RR^{\NN}$ with its product
$\sigma$-algebra and the product measure
\[
 \mu=\bigotimes_{n\geq1}\gamma,\qquad
 \dd\gamma(t)=(2\pi)^{-1/2}e^{-t^2/2}\dd t,
\]
and write $u=(u_n)_{n\geq1}\in\mathcal Q$.  Mapping the occupation basis
associated with $(\xi_n)$ to products of normalized Hermite polynomials in
the variables $u_n$ gives a unitary
$\mathcal G:\Fock_s(\mathfrak h)\to L^2(\mathcal Q,\mu)$, the
Wiener--It\^o--Segal isomorphism
\cite{segal1956tensor,simon1974pphi2,janson1997gaussian}.  For its use
with the Pauli--Fierz model, see \cite{lorinczi2011feynmankac}.  Under
$\mathcal G$, the field $\varPhi(g)$ with $g\in\mathfrak h$ and $Jg=g$
becomes multiplication by
the real Gaussian variable $\sum_n\dua{\xi_n}{g}u_n$ of variance
$\|g\|^2$.  In particular, $\varPhi(\xi_n)$ becomes multiplication by $u_n$.
Hence $\AF_{\mathfrak h}(\vx)$ becomes multiplication by the real random
field $\va_u(\vx)$ with components
\[
 a_u^\mu(\vx)=\sum_{n\geq1}
 \dua{\xi_n}{P_{\mathfrak h}G^\mu_{\vx}}\,u_n,
\]
the series converging in $L^2(\mathcal Q,\mu)$.

Now let $\vGamma$ be a positive trace-class operator on $\HC_N$.  We
write $\mathcal G$ also for $\iden\otimes\mathcal G$, which maps $\HC_N$
unitarily onto $L^2(\mathcal Q,\mu;\mathcal K)$.  If $(\Psi_j)_{j\geq1}$ is any
orthonormal basis of $\HC_N$, put
\[
 \Theta(u)=\sum_{j\geq1}
 |(\mathcal G\sqrt{\vGamma}\Psi_j)(u)\rangle
 \langle(\mathcal G\sqrt{\vGamma}\Psi_j)(u)|.
\]

\begin{lemma}[Disintegration over the field]
\label[lemma]{lem-soft-disintegration}
Let $\vGamma$ and $\Theta$ be as above.
\begin{enumerate}[(i)]
\item For $\mu$-almost every $u$, the series defining $\Theta(u)$
converges in trace norm to a positive trace-class operator on
$\mathcal K$.
\item The family $(\Theta(u))_{u\in\mathcal Q}$ disintegrates $\vGamma$
over the values of $\AF_{\mathfrak h}$: for every bounded operator $B$ on
$\mathcal K$ and every bounded measurable function $f$ on $\mathcal Q$,
\begin{equation}\label{soft-disintegration}
 \Tr\big(\vGamma\,(B\otimes\mathcal G^{-1}M_f\mathcal G)\big)
 =\int_{\mathcal Q}f(u)\Tr\big(B\Theta(u)\big)\dd\mu(u),
\end{equation}
where $M_f$ is multiplication by $f$ on $L^2(\mathcal Q,\mu)$.  In
particular, $\int_{\mathcal Q}\Theta(u)\dd\mu(u)$ is the partial trace
of $\vGamma$ over $\Fock_s(\mathfrak h)$.
\item Every family of trace-class operators on $\mathcal K$ satisfying
\cref{soft-disintegration} for all such $B$ and $f$ agrees with
$\Theta(u)$ for $\mu$-almost every $u$.  In particular, $\Theta$ does
not depend on the choice of the basis $(\Psi_j)_{j\geq1}$, up to a
$\mu$-null set.
\end{enumerate}
\end{lemma}
\begin{proof}
(i) Each term of the series is positive and of rank one, and its trace is
the squared norm of its vector.  By Tonelli's theorem and unitarity of
$\mathcal G$,
\[
 \int_{\mathcal Q}\sum_j\|(\mathcal G\sqrt{\vGamma}\Psi_j)(u)\|^2\dd\mu(u)
 =\sum_j\|\sqrt{\vGamma}\Psi_j\|^2=\Tr\vGamma<\infty.
\]
Hence the traces are summable for $\mu$-almost every $u$.

(ii) Expanding the trace in the basis $(\Psi_j)$, the left-hand side of \cref{soft-disintegration} equals
\begin{align*}
 &\sum_j\dua{\sqrt{\vGamma}\Psi_j}
     {(B\otimes\mathcal G^{-1}M_f\mathcal G)\sqrt{\vGamma}\Psi_j}\\
 &\qquad=\sum_j\int_{\mathcal Q}f(u)
   \dua{(\mathcal G\sqrt{\vGamma}\Psi_j)(u)}
       {B(\mathcal G\sqrt{\vGamma}\Psi_j)(u)}\dd\mu(u).
\end{align*}
The integrands are bounded by
$\|B\|\|f\|_\infty\|(\mathcal G\sqrt{\vGamma}\Psi_j)(u)\|^2$, so the bound
in (i) allows
the sum and the integral to be exchanged, which gives
\cref{soft-disintegration}.  For $f=1$, the definition
\cref{partial-trace-duality} of the partial trace yields the last
identity, the integral converging in trace norm by the same bound.

(iii) Let $(\Theta'(u))_{u\in\mathcal Q}$ be another family of
trace-class operators satisfying \cref{soft-disintegration}.  Its
left-hand side does not involve the family, so for every bounded $B$,
\[
 \int_{\mathcal Q}f(u)\Tr\big(B(\Theta(u)-\Theta'(u))\big)\dd\mu(u)=0
\]
for all bounded measurable $f$.  Hence
$\Tr(B(\Theta(u)-\Theta'(u)))=0$ outside a $\mu$-null set depending on
$B$.  Since $\mathcal K$ is separable, there is a countable family of
finite-rank operators that is norm dense in the compact operators on
$\mathcal K$.  Outside the union of the corresponding null sets,
$\Tr(B(\Theta(u)-\Theta'(u)))=0$ for every compact $B$, and since the
trace class is the dual of the compact operators,
$\Theta(u)=\Theta'(u)$.
\end{proof}

\begingroup
\renewcommand{\bibliofont}{\normalsize}

\endgroup

\begin{thebibliography}{59}

\bibitem{bach1998confined}
Bach, V., Fröhlich, J., and Sigal, I.~M. ``Quantum Electrodynamics of Confined Nonrelativistic Particles''. \emph{Advances in Mathematics} 137.2 (1998), 299--395. \href{https://doi.org/10.1006/aima.1998.1734}{doi:\nolinkurl{10.1006/aima.1998.1734}}.

\bibitem{bach1999spectral}
Bach, V., Fröhlich, J., and Sigal, I.~M. ``Spectral Analysis for Systems of Atoms and Molecules Coupled to the Quantized Radiation Field''. \emph{Communications in Mathematical Physics} 207.2 (1999), 249--290. \href{https://doi.org/10.1007/s002200050726}{doi:\nolinkurl{10.1007/s002200050726}}.

\bibitem{bakkestuen2025dicke}
Bakkestuen, V.~H., Csirik, M.~A., Laestadius, A., and Penz, M. ``Density-Functional Theory for the {Dicke} {H}amiltonian''. \emph{Journal of Statistical Physics} 192.5 (2025), 61. \href{https://doi.org/10.1007/s10955-025-03442-9}{doi:\nolinkurl{10.1007/s10955-025-03442-9}}.

\bibitem{bakkestuen2025rabi}
Bakkestuen, V.~H., Falm{å}r, V., Lotfigolian, M., Penz, M., Ruggenthaler, M., and Laestadius, A. ``Quantum-Electrodynamical Density-Functional Theory Exemplified by the Quantum {Rabi} Model''. \emph{The Journal of Physical Chemistry A} 129.9 (2025), 2337--2360. \href{https://doi.org/10.1021/acs.jpca.4c07690}{doi:\nolinkurl{10.1021/acs.jpca.4c07690}}.

\bibitem{Basov2020}
Basov, D.~N., Asenjo-Garcia, A., Schuck, P.~J., Zhu, X., and Rubio, A. ``Polariton panorama''. \emph{Nanophotonics} 10.1 (2020), 549--577. \href{https://doi.org/10.1515/nanoph-2020-0449}{doi:\nolinkurl{10.1515/nanoph-2020-0449}}.

\bibitem{Bloch2022}
Bloch, J., Cavalleri, A., Galitski, V., Hafezi, M., and Rubio, A. ``Strongly correlated electron--photon systems''. \emph{Nature} 606.7912 (2022), 41--48. \href{https://doi.org/10.1038/s41586-022-04726-w}{doi:\nolinkurl{10.1038/s41586-022-04726-w}}.

\bibitem{bratteli-robinson-2}
Bratteli, O. and Robinson, D.~W. \emph{Operator Algebras and Quantum Statistical Mechanics 2: Equilibrium States. {M}odels in Quantum Statistical Mechanics}. 2nd ed. Berlin: Springer, 1997.

\bibitem{capelle2002nonuniqueness}
Capelle, K. and Vignale, G. ``Nonuniqueness and derivative discontinuities in density-functional theories for current-carrying and superconducting systems''. \emph{Physical Review B} 65.11 (2002), 113106. \href{https://doi.org/10.1103/physrevb.65.113106}{doi:\nolinkurl{10.1103/physrevb.65.113106}}.

\bibitem{coleman1963structure}
Coleman, A.~J. ``Structure of Fermion Density Matrices''. \emph{Reviews of Modern Physics} 35.3 (1963), 668. \href{https://doi.org/10.1103/RevModPhys.35.668}{doi:\nolinkurl{10.1103/RevModPhys.35.668}}.

\bibitem{csirik2024thermodynamic}
Csirik, M.~A., Laestadius, A., and Tellgren, E.~I. ``Thermodynamic limit for the magnetic uniform electron gas and representability of density-current pairs''. \emph{Journal of Mathematical Physics} 65.11 (2024), 113301. \href{https://doi.org/10.1063/5.0198786}{doi:\nolinkurl{10.1063/5.0198786}}.

\bibitem{dinezza2012hitchhiker}
Di Nezza, E., Palatucci, G., and Valdinoci, E. ``Hitchhiker's Guide to the Fractional {S}obolev Spaces''. \emph{Bulletin des Sciences Mathématiques} 136.5 (2012), 521--573. \href{https://doi.org/10.1016/j.bulsci.2011.12.004}{doi:\nolinkurl{10.1016/j.bulsci.2011.12.004}}.

\bibitem{ebbesen2016hybrid}
Ebbesen, T.~W. ``Hybrid Light--Matter States in a Molecular and Material Science Perspective''. \emph{Accounts of Chemical Research} 49.11 (2016), 2403--2412. \href{https://doi.org/10.1021/acs.accounts.6b00295}{doi:\nolinkurl{10.1021/acs.accounts.6b00295}}.

\bibitem{flick2015kohn}
Flick, J., Ruggenthaler, M., Appel, H., and Rubio, A. ``{Kohn}--{Sham} approach to quantum electrodynamical density-functional theory: Exact time-dependent effective potentials in real space''. \emph{Proceedings of the National Academy of Sciences} 112.50 (2015), 15285--15290. \href{https://doi.org/10.1073/pnas.1518224112}{doi:\nolinkurl{10.1073/pnas.1518224112}}.

\bibitem{frank2008hardy}
Frank, R.~L., Lieb, E.~H., and Seiringer, R. ``{Hardy--Lieb--Thirring} Inequalities for Fractional {Schrödinger} Operators''. \emph{Journal of the American Mathematical Society} 21 (2008), 925--950. \href{https://doi.org/10.1090/S0894-0347-07-00582-6}{doi:\nolinkurl{10.1090/S0894-0347-07-00582-6}}.

\bibitem{garciavidal2021manipulating}
Garcia-Vidal, F.~J., Ciuti, C., and Ebbesen, T.~W. ``Manipulating matter by strong coupling to vacuum fields''. \emph{Science} 373.6551 (2021), eabd0336. \href{https://doi.org/10.1126/science.abd0336}{doi:\nolinkurl{10.1126/science.abd0336}}.

\bibitem{georgescu2004spectral}
Georgescu, V., Gérard, C., and M{ø}ller, J.~S. ``Spectral Theory of Massless {Pauli--Fierz} Models''. \emph{Communications in Mathematical Physics} 249.1 (2004), 29--78. \href{https://doi.org/10.1007/s00220-004-1111-x}{doi:\nolinkurl{10.1007/s00220-004-1111-x}}.

\bibitem{gerard2000ground}
Gérard, C. ``On the Existence of Ground States for Massless {Pauli--Fierz} Hamiltonians''. \emph{Annales Henri Poincaré} 1.3 (2000), 443--459. \href{https://doi.org/10.1007/s000230050002}{doi:\nolinkurl{10.1007/s000230050002}}.

\bibitem{griesemer2004exponential}
Griesemer, M. ``Exponential Decay and Ionization Thresholds in Non-Relativistic Quantum Electrodynamics''. \emph{Journal of Functional Analysis} 210.2 (2004), 321--340. \href{https://doi.org/10.1016/j.jfa.2003.06.001}{doi:\nolinkurl{10.1016/j.jfa.2003.06.001}}.

\bibitem{griesemer2001ground}
Griesemer, M., Lieb, E.~H., and Loss, M. ``Ground States in Non-Relativistic Quantum Electrodynamics''. \emph{Inventiones Mathematicae} 145.3 (2001), 557--595. \href{https://doi.org/10.1007/s002220100159}{doi:\nolinkurl{10.1007/s002220100159}}.

\bibitem{hasler2008absence}
Hasler, D. and Herbst, I. ``Absence of Ground States for a Class of Translation Invariant Models of Non-Relativistic {QED}''. \emph{Communications in Mathematical Physics} 279.3 (2008), 769--787. \href{https://doi.org/10.1007/s00220-008-0444-2}{doi:\nolinkurl{10.1007/s00220-008-0444-2}}.

\bibitem{hasler2008selfadjoint}
Hasler, D. and Herbst, I. ``On the Self-Adjointness and Domain of {Pauli--Fierz} Type Hamiltonians''. \emph{Reviews in Mathematical Physics} 20.7 (2008), 787--800. \href{https://doi.org/10.1142/S0129055X08003389}{doi:\nolinkurl{10.1142/S0129055X08003389}}.

\bibitem{hiroshima2002selfadjoint}
Hiroshima, F. ``Self-Adjointness of the {Pauli--Fierz} Hamiltonian for Arbitrary Values of Coupling Constants''. \emph{Annales Henri Poincaré} 3.1 (2002), 171--201. \href{https://doi.org/10.1007/s00023-002-8615-8}{doi:\nolinkurl{10.1007/s00023-002-8615-8}}.

\bibitem{janson1997gaussian}
Janson, S. \emph{Gaussian {H}ilbert Spaces}. Vol.~129. Cambridge Tracts in Mathematics. Cambridge: Cambridge University Press, 1997.

\bibitem{kvaal2021lower}
Kvaal, S., Laestadius, A., Tellgren, E.~I., and Helgaker, T. ``Lower Semicontinuity of the Universal Functional in Paramagnetic Current--Density Functional Theory''. \emph{The Journal of Physical Chemistry Letters} 12.5 (2021), 1421--1425. \href{https://doi.org/10.1021/acs.jpclett.0c03422}{doi:\nolinkurl{10.1021/acs.jpclett.0c03422}}.

\bibitem{laestadius2014density}
Laestadius, A. ``Density functionals in the presence of magnetic field''. \emph{International Journal of Quantum Chemistry} 114.21 (2014), 1445--1456. \href{https://doi.org/10.1002/qua.24707}{doi:\nolinkurl{10.1002/qua.24707}}.

\bibitem{laestadius2014hohenberg}
Laestadius, A. and Benedicks, M. ``{Hohenberg}--{Kohn} theorems in the presence of magnetic field''. \emph{International Journal of Quantum Chemistry} 114.12 (2014), 782--795. \href{https://doi.org/10.1002/qua.24668}{doi:\nolinkurl{10.1002/qua.24668}}.

\bibitem{MY-CDFTpaper2019}
Laestadius, A., Tellgren, E.~I., Penz, M., Ruggenthaler, M., Kvaal, S., and Helgaker, T. ``{Kohn--Sham theory with paramagnetic currents: compatibility and functional differentiability}''. \emph{Journal of Chemical Theory and Computation} 15.7 (2019), 4003--4020. \href{https://doi.org/10.1021/acs.jctc.9b00141}{doi:\nolinkurl{10.1021/acs.jctc.9b00141}}.

\bibitem{levy1979universal}
Levy, M. ``Universal variational functionals of electron densities, first-order density matrices, and natural spin-orbitals and solution of the $v$-representability problem''. \emph{Proceedings of the National Academy of Sciences} 76.12 (1979), 6062--6065. \href{https://doi.org/10.1073/pnas.76.12.6062}{doi:\nolinkurl{10.1073/pnas.76.12.6062}}.

\bibitem{levy1982}
Levy, M. ``Electron densities in search of {H}amiltonians''. \emph{Physical Review A} 26.3 (1982), 1200--1208. \href{https://doi.org/10.1103/PhysRevA.26.1200}{doi:\nolinkurl{10.1103/PhysRevA.26.1200}}.

\bibitem{lewin2011geometric}
Lewin, M. ``Geometric Methods for Nonlinear Many-Body Quantum Systems''. \emph{Journal of Functional Analysis} 260.12 (2011), 3535--3595. \href{https://doi.org/10.1016/j.jfa.2010.11.017}{doi:\nolinkurl{10.1016/j.jfa.2010.11.017}}.

\bibitem{lewin2024spectral}
Lewin, M. \emph{Spectral Theory and Quantum Mechanics}. Universitext. Cham: Springer, 2024. \href{https://doi.org/10.1007/978-3-031-66878-4}{doi:\nolinkurl{10.1007/978-3-031-66878-4}}.

\bibitem{lewin2023universal}
Lewin, M., Lieb, E.~H., and Seiringer, R. ``Universal Functionals in Density Functional Theory''. In: \emph{Density Functional Theory: Modeling, Mathematical Analysis, Computational Methods, and Applications}. Ed. by E.~Cancès and G.~Friesecke. Cham: Springer, 2023, 115--182. \href{https://doi.org/10.1007/978-3-031-22340-2_3}{doi:\nolinkurl{10.1007/978-3-031-22340-2_3}}.

\bibitem{Lieb1983}
Lieb, E.~H. ``Density Functionals for {C}oulomb Systems''. \emph{International Journal of Quantum Chemistry} 24.3 (1983), 243--277. \href{https://doi.org/10.1002/qua.560240302}{doi:\nolinkurl{10.1002/qua.560240302}}.

\bibitem{lieb2003existence}
Lieb, E.~H. and Loss, M. ``Existence of Atoms and Molecules in Non-Relativistic Quantum Electrodynamics''. \emph{Advances in Theoretical and Mathematical Physics} 7.4 (2003), 667--710. \href{https://doi.org/10.4310/ATMP.2003.v7.n4.a3}{doi:\nolinkurl{10.4310/ATMP.2003.v7.n4.a3}}.

\bibitem{lieb2013current}
Lieb, E.~H. and Schrader, R. ``Current densities in density-functional theory''. \emph{Physical Review A} 88.3 (2013), 032516. \href{https://doi.org/10.1103/physreva.88.032516}{doi:\nolinkurl{10.1103/physreva.88.032516}}.

\bibitem{lieb2010stability}
Lieb, E.~H. and Seiringer, R. \emph{The Stability of Matter in Quantum Mechanics}. Cambridge: Cambridge University Press, 2010.

\bibitem{lorinczi2011feynmankac}
Lőrinczi, J., Hiroshima, F., and Betz, V. \emph{Feynman--{K}ac-Type Theorems and {G}ibbs Measures on Path Space: With Applications to Rigorous Quantum Field Theory}. De Gruyter Studies in Mathematics. Berlin: De Gruyter, 2011. \href{https://doi.org/10.1515/9783110203738}{doi:\nolinkurl{10.1515/9783110203738}}.

\bibitem{matte2017paulifierz}
Matte, O. ``{Pauli--Fierz} Type Operators with Singular Electromagnetic Potentials on General Domains''. \emph{Mathematical Physics, Analysis and Geometry} 20.2 (2017), 18. \href{https://doi.org/10.1007/s11040-017-9249-x}{doi:\nolinkurl{10.1007/s11040-017-9249-x}}.

\bibitem{Novokreschenov2023}
Novokreschenov, D., Kudlis, A., Iorsh, I., and Tokatly, I.~V. ``Quantum electrodynamical density functional theory for generalized Dicke model''. \emph{Physical Review B} 108.23 (2023), 235424. \href{https://doi.org/10.1103/physrevb.108.235424}{doi:\nolinkurl{10.1103/physrevb.108.235424}}.

\bibitem{pauli1938theorie}
Pauli, W. and Fierz, M. ``Zur Theorie der Emission langwelliger {Lichtquanten}''. \emph{Il Nuovo Cimento} 15.3 (1938), 167--188. \href{https://doi.org/10.1007/bf02958939}{doi:\nolinkurl{10.1007/bf02958939}}.

\bibitem{penz2023structureI}
Penz, M., Tellgren, E.~I., Csirik, M.~A., Ruggenthaler, M., and Laestadius, A. ``The Structure of Density-Potential Mapping. Part {I}: Standard Density-Functional Theory''. \emph{ACS Physical Chemistry Au} 3.4 (2023), 334--347. \href{https://doi.org/10.1021/acsphyschemau.2c00069}{doi:\nolinkurl{10.1021/acsphyschemau.2c00069}}.

\bibitem{penz2023structureII}
Penz, M., Tellgren, E.~I., Csirik, M.~A., Ruggenthaler, M., and Laestadius, A. ``The Structure of the Density-Potential Mapping. Part {II}: Including Magnetic Fields''. \emph{ACS Physical Chemistry Au} 3.6 (2023), 492--511. \href{https://doi.org/10.1021/acsphyschemau.3c00006}{doi:\nolinkurl{10.1021/acsphyschemau.3c00006}}.

\bibitem{reed-simon-2}
Reed, M. and Simon, B. \emph{Methods of Modern Mathematical Physics. II: Fourier Analysis, Self-Adjointness}. Vol.~2. Elsevier, 1975.

\bibitem{rokaj2018lightmatter}
Rokaj, V., Welakuh, D.~M., Ruggenthaler, M., and Rubio, A. ``Light--matter interaction in the long-wavelength limit: no ground-state without dipole self-energy''. \emph{Journal of Physics B: Atomic, Molecular and Optical Physics} 51.3 (2018), 034005. \href{https://doi.org/10.1088/1361-6455/aa9c99}{doi:\nolinkurl{10.1088/1361-6455/aa9c99}}.

\bibitem{Ruggenthaler2017}
Ruggenthaler, M. \emph{Ground-State Quantum-Electrodynamical Density-Functional Theory}. 2017. \href{https://doi.org/10.48550/arXiv.1509.01417}{doi:\nolinkurl{10.48550/arXiv.1509.01417}}. arXiv: \href{https://arxiv.org/abs/1509.01417}{\nolinkurl{1509.01417} [quant-ph]}.

\bibitem{ruggenthaler2014quantum}
Ruggenthaler, M., Flick, J., Pellegrini, C., Appel, H., Tokatly, I.~V., and Rubio, A. ``Quantum-electrodynamical density-functional theory: Bridging quantum optics and electronic-structure theory''. \emph{Physical Review A} 90.1 (2014), 012508. \href{https://doi.org/10.1103/physreva.90.012508}{doi:\nolinkurl{10.1103/physreva.90.012508}}.

\bibitem{ruggenthaler2023understanding}
Ruggenthaler, M., Sidler, D., and Rubio, A. ``Understanding Polaritonic Chemistry from ab initio Quantum Electrodynamics''. \emph{Chemical Reviews} 123.19 (2023), 11191--11229. \href{https://doi.org/10.1021/acs.chemrev.2c00788}{doi:\nolinkurl{10.1021/acs.chemrev.2c00788}}.

\bibitem{ruggenthaler2018quantum}
Ruggenthaler, M., Tancogne-Dejean, N., Flick, J., Appel, H., and Rubio, A. ``From a quantum-electrodynamical light--matter description to novel spectroscopies''. \emph{Nature Reviews Chemistry} 2.3 (2018), 0118. \href{https://doi.org/10.1038/s41570-018-0118}{doi:\nolinkurl{10.1038/s41570-018-0118}}.

\bibitem{schafer2021making}
Schäfer, C., Buchholz, F., Penz, M., Ruggenthaler, M., and Rubio, A. ``Making ab initio {QED} functional(s): Nonperturbative and photon-free effective frameworks for strong light--matter coupling''. \emph{Proceedings of the National Academy of Sciences} 118.41 (2021), e2110464118. \href{https://doi.org/10.1073/pnas.2110464118}{doi:\nolinkurl{10.1073/pnas.2110464118}}.

\bibitem{schafer2020relevance}
Schäfer, C., Ruggenthaler, M., Rokaj, V., and Rubio, A. ``Relevance of the Quadratic Diamagnetic and Self-Polarization Terms in Cavity Quantum Electrodynamics''. \emph{ACS Photonics} 7.4 (2020), 975--990. \href{https://doi.org/10.1021/acsphotonics.9b01649}{doi:\nolinkurl{10.1021/acsphotonics.9b01649}}.

\bibitem{segal1956tensor}
Segal, I.~E. ``Tensor algebras over {H}ilbert spaces. {I}''. \emph{Transactions of the American Mathematical Society} 81.1 (1956), 106--134. \href{https://doi.org/10.1090/S0002-9947-1956-0076317-8}{doi:\nolinkurl{10.1090/S0002-9947-1956-0076317-8}}.

\bibitem{simon1974pphi2}
Simon, B. \emph{The {$P(\phi)_2$} {E}uclidean (Quantum) Field Theory}. Princeton Series in Physics. Princeton, NJ: Princeton University Press, 1974.

\bibitem{solovej2014many}
Solovej, J.~P. \emph{Many Body Quantum Mechanics}. Lecture notes, version of March 5, 2014, with corrections and additions by P.~T.~Nam. 2014.

\bibitem{spohn2004dynamics}
Spohn, H. \emph{Dynamics of Charged Particles and their Radiation Field}. Cambridge: Cambridge University Press, 2004. \href{https://doi.org/10.1017/cbo9780511535178}{doi:\nolinkurl{10.1017/cbo9780511535178}}.

\bibitem{tellgren2014fermion}
Tellgren, E.~I., Kvaal, S., and Helgaker, T. ``Fermion {$N$}-representability for prescribed density and paramagnetic current density''. \emph{Physical Review A} 89.1 (2014), 012515. \href{https://doi.org/10.1103/physreva.89.012515}{doi:\nolinkurl{10.1103/physreva.89.012515}}.

\bibitem{tellgren2012choice}
Tellgren, E.~I., Kvaal, S., Sagvolden, E., Ekström, U., Teale, A.~M., and Helgaker, T. ``Choice of basic variables in current-density-functional theory''. \emph{Physical Review A} 86.6 (2012), 062506. \href{https://doi.org/10.1103/physreva.86.062506}{doi:\nolinkurl{10.1103/physreva.86.062506}}.

\bibitem{tokatly2013time}
Tokatly, I.~V. ``Time-Dependent Density Functional Theory for Many-Electron Systems Interacting with Cavity Photons''. \emph{Physical Review Letters} 110.23 (2013), 233001. \href{https://doi.org/10.1103/physrevlett.110.233001}{doi:\nolinkurl{10.1103/physrevlett.110.233001}}.

\bibitem{vignale1987density}
Vignale, G. and Rasolt, M. ``Density-functional theory in strong magnetic fields''. \emph{Physical Review Letters} 59.20 (1987), 2360--2363. \href{https://doi.org/10.1103/physrevlett.59.2360}{doi:\nolinkurl{10.1103/physrevlett.59.2360}}.

\bibitem{vignale1988current}
Vignale, G. and Rasolt, M. ``Current- and spin-density-functional theory for inhomogeneous electronic systems in strong magnetic fields''. \emph{Physical Review B} 37.18 (1988), 10685--10696. \href{https://doi.org/10.1103/physrevb.37.10685}{doi:\nolinkurl{10.1103/physrevb.37.10685}}.

\end{thebibliography}
\end{document}